\documentclass[twoside,floatfix,12pt]{mbook} %\documentclass[floatfix,14pt]{book} %\usepackage{14pt}
\usepackage{graphicx}
\usepackage{amssymb}
\usepackage{amsmath}
\usepackage[T1]{fontenc}
\usepackage[cp 1250]{inputenc}
\usepackage{textcomp}
\usepackage{amsfonts}
\usepackage{epsfig}
\usepackage{here}
\usepackage{cite}
\usepackage{esint}
\usepackage{fancyhdr}
\usepackage{color}
\usepackage{multirow}
\title{{\bf Merging and splitting of clusters with the full simulation of 
the barrel electromagnetic 
calorimeter of the KLOE detector with FLUKA.  
}}
\author{Jarosław Zdebik}
\date{17.03.2008 r.}
\begin{document}
%\maketitle
%%%%%%%%%%%%%%%%%%%%%%%%%%%%%%%%%%%%%%%%%%%%
%        TEKST WLASCIWY DOKUMENTU          %
%%%%%%%%%%%%%%%%%%%%%%%%%%%%%%%%%%%%%%%%%%%%
\thispagestyle{empty}
\newpage
\thispagestyle{empty}
\begin{center}
% Faculty of Physics, Astronomy and Applied Computer Science
{\large INSTITUTE OF PHYSICS  }\\
{\large FACULTY OF PHYSICS, ASTRONOMY  }\\
{\large AND APPLIED COMPUTER SCIENCE}\\
{\large JAGIELLONIAN UNIVERSITY}\\
\end{center}
\begin{center}\end{center}
%\begin{center}\end{center}
%\begin{center}\end{center}
\begin{center}\end{center}
%
%%%%%%%%%%%%%%%%55
%{\Large{\bf Examination of the merging and splitting effects of the }}
%\end{center}
%\begin{center}
%{\Large{\bf $ \phi \to \eta \gamma \to 3 \gamma $  }}
%\end{center}
%\begin{center}
%{\Large{\bf and}}
%\end{center}
%\begin{center}
%{\Large{\bf $\phi \to \eta \gamma \to \pi^{0}\pi^{0}\pi^{0} \to 7 \gamma $}}
%\end{center}
%\begin{center}
%{\Large{\bf decays on the KLOE detector }}
%\end{center}
\begin{center}
{\Large{\bf Merging and splitting of clusters in }}
\end{center}
\begin{center}
{\Large{\bf the electromagnetic calorimeter }}
\end{center}
\begin{center}
{\Large{\bf of the KLOE detector }}
\end{center}
%\begin{center}
%{\Large{\bf In the case of the study of the final states with large}}
%\end{center}
%\begin{center}
%{\Large{\bf multiplicity of gamma quanta }}
%\end{center}
%
%
%
\vspace{1.5cm}
\begin{center}\end{center}
\begin{center}
{\large Jaros{\l}aw Zdebik}
\end{center}
\begin{center}\end{center}
\begin{center}\end{center}
\begin{center}
{\normalsize Master Thesis  }\\
{\normalsize supervised by } 
\end{center}
\begin{center}
{\large Prof. Dr habil. Pawe{\l} Moskal}
\end{center}
\begin{center}
{\normalsize prepared in Nuclear Physics Department\\}
{\normalsize of the Jagiellonian University}
\end{center}
\begin{center}\end{center}
\begin{center}
{\normalsize improved version\\}
\end{center}
\begin{center}\end{center}
\begin{center}
\begin{figure}[h]
\hspace{6.7cm}
\vspace{-3.cm}
\parbox{0.10\textwidth}{\centerline{\epsfig{file=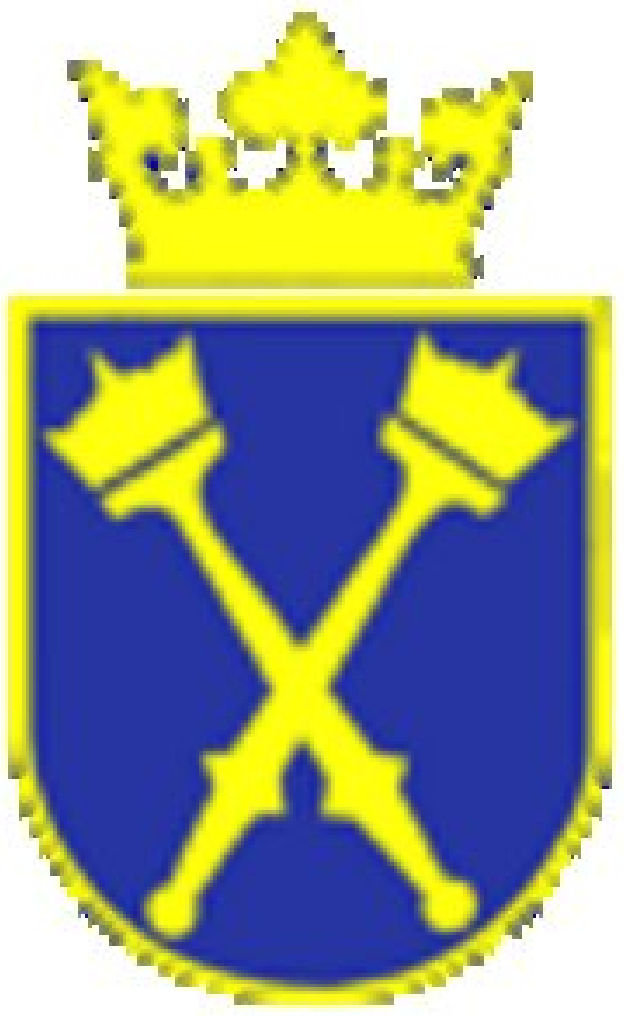,width=0.15\textwidth}}}
% \parbox{0.10\textwidth}{\centerline{\includegraphics{uj_herb_00.eps,width=0.15\textwidth}}}
\end{figure}
\end{center}
\begin{center}
{\vspace{1.7cm}\normalsize KRAKÓW 2008}
\end{center}
%%%%%%%%%%%%%%%%%%%%%%%%%%%%%%%%%%%%%%%%%%%
% Podziekowania
%\newpage
%\thispagestyle{empty}
%\begin{center}\end{center}
%\begin{center}\end{center}
%\begin{flushright}
%{\Large {\bf Podziękowania}}
%\end{flushright}
%\begin{center}\end{center}

% Tu podziekowania
\newpage
~\\

\thispagestyle{empty}
\newpage
\thispagestyle{empty}
\begin{center}
% Faculty of Physics, Astronomy and Applied Computer Science
{\large INSTYTUT FIZYKI}\\
{\large WYDZIA{\L} FIZYKI, ASTRONOMII  }\\
{\large I INFORMATYKI STOSOWANEJ}\\
{\large UNIWERSYTET JAGIELLO\'NSKI}\\
\end{center}
\begin{center}\end{center}
%\begin{center}\end{center}
%\begin{center}\end{center}
\begin{center}\end{center}
\begin{center}
{\Large{\bf Efekty {\l}\c{a}czenia i separacji klastr\'ow }}
\end{center}
\begin{center}
{\Large{\bf w kalorymatrze elektromagnetycznym }}
\end{center}
\begin{center}
{\Large{\bf detektora KLOE }}
\end{center}
%\begin{center}
%{\Large{\bf In the case of the study of the final states with large}}
%\end{center}
%\begin{center}
%{\Large{\bf multiplicity of gamma quanta }}
%\end{center}
%
%
%
\vspace{1.5cm}
\begin{center}\end{center}
\begin{center}
{\large Jarosław Zdebik}
\end{center}
\begin{center}\end{center}
\begin{center}\end{center}
\begin{center}
{\normalsize Praca magisterska  }\\
{\normalsize pod kierunkiem } 
\end{center}
\begin{center}
{\large Prof. dr hab. Paw{\l}a Moskala}
\end{center}
\begin{center}
{\normalsize przygotowana w Zak{\l}adzie Fizyki J\c{a}drowej\\}
{\normalsize Uniwersytetu Jagiello\'nskiego}
\end{center}
\begin{center}\end{center}
\begin{center}
{\normalsize wersja poprawiona\\}
\end{center}
\begin{center}\end{center}
\begin{center}
\begin{figure}[h]
\hspace{6.7cm}
\vspace{-3.cm}
\parbox{0.10\textwidth}{\centerline{\epsfig{file=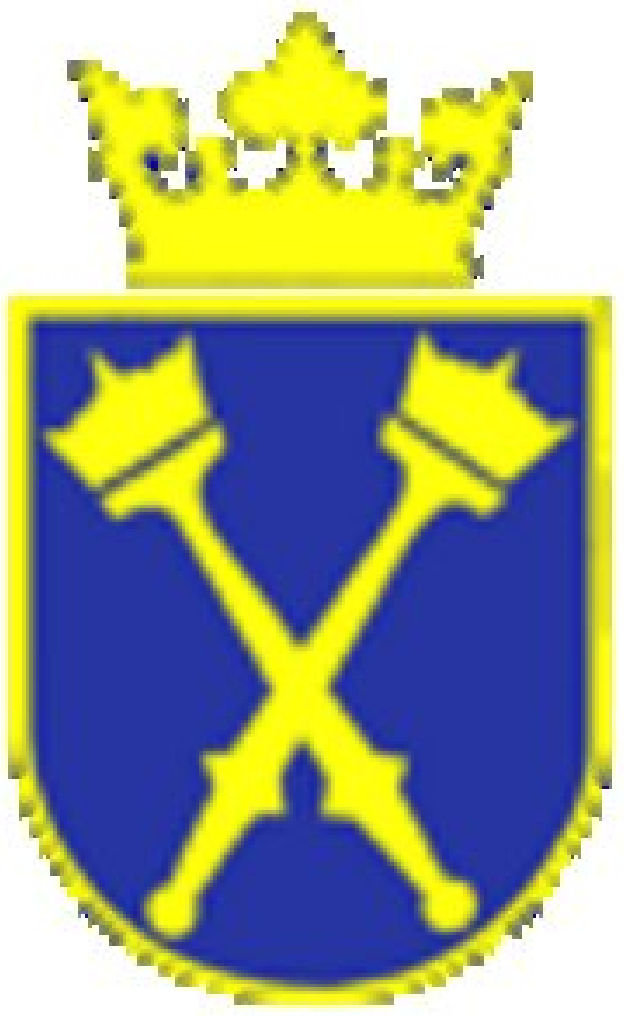,width=0.15\textwidth}}}
% \parbox{0.10\textwidth}{\centerline{\includegraphics{uj_herb_00.eps,width=0.15\textwidth}}}
\end{figure}
\end{center}
\begin{center}
{\vspace{1.7cm}\normalsize KRAKÓW 2008}
\end{center}

\newpage
\thispagestyle{empty}
~\\
~\\
~\\

\newpage
\thispagestyle{empty}
~\\
~\\
~\\
~\\
~\\
~\\
~\\
~\\
~\\
~\\
~\\
~\\
~\\
~\\
~\\
~\\
~\\
~\\
~\\
~\\
~\\
~\\
~\\
~\\
~\\
~\\
~\\
~\\
~\\
~\\
%\parbox[l]{0.5\textwidth}{
%The next question was - what makes planets go around the sun? At the time of Kepler some people answered 
%this problem by saying that there were angels behind them beating their wings and pushing the planets around 
%an orbit. As you will see, the answer is not very far from the truth. The only difference is that the angels 
%sit in a different direction and their wings push inward. 
%}
%
%
\hspace{-1.5cm}
\mbox{
\parbox[l]{0.5\textwidth}{
"The next question was - what makes planets go around the sun? At the time of Kepler some people answered 
this problem by saying that there were angels behind them beating their wings and pushing the planets around 
an orbit. As you will see, the answer is not very far from the truth. The only difference is that the angels 
sit in a different direction and their wings push inward." 
}
\hspace{1.0cm}
\parbox[r]{0.5\textwidth}{
"Nast\c{e}pnym pytaniem by{\l}o - co sprawia, {\.z}e planety kr\c{a}{\.z}\c{a} wok\'o{\l} s{\l}o\'nca? 
W czasach Keplera niekt\'orzy ludzie 
odpowiadali, {\.z}e za planetami s\c{a} anio{\l}owie, kt\'orzy popychaj\c{a} je machaj\c{a}c skrzyd{\l}ami. 
Jak widzisz, 
odpowied\'z ta nie jest daleka od prawdy. Jedyna r\'o{\.z}nica jest taka, {\.z}e anio{\l}owie zasiadaj\c{a} 
w innej pozycji 
i ich skrzyd{\l}a pchaj\c{a} do \'srodka." 
}
}
~\\
~\\
~\\
\mbox{
\parbox[l]{0.5\textwidth}{
~~~
}
\parbox[r]{0.5\textwidth}{
~~~~~~~~~~~~~~~~~~~~~~ Richard Feynman (1918 - 1988),
}
}
% ~\\
\mbox{
\parbox[l]{0.5\textwidth}{
~~~
}
\parbox[r]{0.5\textwidth}{
~~~~~~~~~~~~~~~~~~~~~~~ {\it Character of Physical Law}
}
}

\newpage
~ \\
\thispagestyle{empty}

\newpage
~ \\
\thispagestyle{empty}
\begin{center}
\LARGE{
\textbf{Abstract}
}
\end{center}

\hspace{\parindent}
~\\
~\\
~\\
~\\
\indent The work was carried out in the framework of the KLOE collaboration studying the decays of the $\phi$ meson produced in 
the DA$\Phi$NE accelerator in the collisions of electron and positron. \\
\indent The main aim of this thesis was investigation of the influence of the merging and splitting of clusters in decays with the high 
multiplicity of $\gamma$ quanta, which are at most biased by these effects. 
For this aim we implemented the full geometry and realistic material composition of the barrel electromagnetic calorimeter in 
FLUKA package. The prepared Monte Carlo based simulation program permits to achieve a fast generation of the 
detector response separately for each interested reaction. 
 The program was used to study the reconstruction efficiency with the KLOE clustering algorithm as a 
function of the photocathode quantum efficiency. \\
\indent It was also used to investigate  
  merging and splitting probabilities as a function of the quantum efficiency.  
The conducted studies indicated that the increase of quantum efficiency does not improve significantly the identification of clusters.  
 The influence of these effects was estimated for $\eta$ meson decays into 3$\pi^{0}$ and K$_{short}$ meson into 2$\pi^{0}$.

\newpage
~ \\
\thispagestyle{empty}

\newpage
~ \\
\thispagestyle{empty}
\begin{center}
\LARGE{
\textbf{Streszczenie}
}
\end{center}
\hspace{\parindent}
~\\
~\\
~\\
~\\
\indent Praca zosta{\l}a wykonana w ramach kolaboracji KLOE, kt\'ora zajmuje si\c{e} badaniem rozpad\'ow mezonu $\phi$, 
produkowanego w kolizjach elektronu i pozytonu na akceleratorze DA$\Phi$NE. \\
\indent G{\l}\'ownym celem pracy by{\l}o oszacowanie wp{\l}ywu efekt\'ow "merging" i "splitting" na mo{\.z}liwo\'sci rekonstrukcji 
     rozpadów mezonu $\phi$ w kana{\l}y z fotonami. W{\l}a\'snie rekonstrukcja topologii reakcji z du{\.z}\c{a} zawarto\'sci\c{a} 
     foton\'ow w kanale wyj\'sciowym jest najbardziej obci\c{a}{\.z}ona przez te efekty. 
     W tym celu zaimplementowali\'smy ca{\l}\c{a} geometri\c{e} kalorymetru elektromagnetycznego detektora KLOE 
     wraz z realistycznym opisem materia{\l}\'ow do programu symulacyjnego opartego na metodzie 
 Monte Carlo. Program symulacyjny pozwala na studiowanie wybranych kana{\l}\'ow 
     rozpadu oddzielnie. R\'ownie{\.z} wysymulowali\'smy odpowied\'z fotopowielaczy na ca{\l}ym kalorymetrze detektora KLOE. \\
  \indent Zbadane zosta{\l}y tak{\.z}e efekty "merging" i "splitting" w zale{\.z}no\'sci od wydajno\'sci kwantowej w~fotopowielaczach. 
     Powy{\.z}sze studia pokaza{\l}y, {\.z}e zwi\c{e}kszenie wydajno\'sci kwantowej fotopowielaczy nie wp{\l}ywa znacz\c{a}co na 
     redukcj\c{e} tych efekt\'ow. 
     Wp{\l}yw efekt\'ow "merging" i "splitting" zosta{\l} oszacowany dla reakcji rozpadu mezonu $\eta$ na trzy neutralne 
     piony oraz kaonu K$_{short}$ na dwa neutralne piony.

\newpage
~ \\
\thispagestyle{empty}

%%%%%%%%%%%%%%%%%%%%%%%%%%%%%%%%%%%%%%%%%%
     \normalsize
     \def\contentsname{\Large \mbox{} \hspace{6cm} Contents}
     \tableofcontents
     \pagestyle{myheadings}
     \markboth{Contents}{Contents}
     \newpage
     \clearpage
     \normalsize
%%%%%%%%%%%%%%%%%%%%%%%%%%%%%%%%%%%%%%%%%%%
\pagestyle{fancy}
\fancyhead{}
\fancyfoot{}
\renewcommand{\headrulewidth}{0.9pt}
\fancyhead[RO]{\textbf{\sffamily{{{\thepage}}~}}}
\fancyhead[LO]{\bf\footnotesize{{\nouppercase{\rightmark}}}}
\fancyhead[RE]{\textbf{\sffamily{{{\thepage}}~}}}
\fancyhead[LE]{\bf\footnotesize{{\nouppercase{\rightmark}}}}
\advance\headheight by 5.3mm
\advance\headsep by -3mm
%
%
%%%%%%%%%%%%%%%%%%%%%%%%%%%%%%%%%%%%%%%%%%%%
%%%%%%%%%%%%%%% ROZDZIAŁ 1 %%%%%%%%%%%%%%%%%
%\newpage
%\pagestyle{myheadings}
%
%%%%%%%%%%%%%%%%%%%%%%%%% MAIN PHYSICS MOTIVATION %%%%%%%%%%%%%%%%%%%%%%%%%%%%%%%%%%
% \chapter{Introduction - tests of discrete symmetries: CP violating effects and ChPT for analyzing reactions on the KLOE detector}
% \hspace{\parindent}
%

\chapter{Introduction}  % opis 1 strona max
\hspace{\parindent}
Nowadays, elementary particles and interactions among   
 them are described in the framework of the Quantum Mechanics and Standard Model. Tests of these theories and of 
fundamental symmetries like C, P, T and their combinations are therefore crucially important for understanding the 
phenomena in the world of particles \cite{A_Flavor_of_KLOE}.
Experiments aiming at the determination of the particle properties and search for processes beyond the applicability of the
 domain of the Standard Model  
are conducted in many particle physics laboratories. \\
\indent Such studies are carried out also in the Laboratori Nazionali Di Frascati (LNF) by means of the electron-positron
 collider DA$\Phi$NE and the KLOE detector setup. The KLOE group has taken data at the colliding
 electron-positron center-of-mass energy corresponding to the $\phi$ meson mass. The main objectives of the KLOE experimental program 
are: investigations of the decays of kaons produced in pairs from the decays of the $\phi$ meson, the radiative 
decays of the $\phi$ meson, the rare branching ratios of the $\eta$ meson and the light mesons particle properties. 
%  
% Although, the main goal of the group were investigations of the $K_{L} \to \pi^{+}\pi^{-}$ and $K_{L} \to \pi^{0}\pi^{0}$ 
% decays. 
% The facility permits also to study the full spectrum of scalar, pseudoscalar and vector mesons orginating 
% from the radiative decays of the meson $\phi$. 
% 
%\indent One of such laboratory is Laboratori Nazionali Di Frascati (LNF) near the Rome in Italy. In this laboratory is situated 
%the electron-positron machine (Da$\Phi$ne collider) that works at energy $\approx \sqrt{s}$=1019 MeV, corresponding to the $\phi$ 
%resonance mass. The main detector on this collider is detector which was designed to investigate a violation of CP 
%symmetry in kaons decays $K_{L} \to \pi^{+}\pi^{-}$ and $K_{L} \to 2\pi^{0}$ separated from the much more abundant 
%$K_{L} \to \pi \mu \nu$ and $K_{L} \to 3\pi^{0}$ \cite{Letter_of_intent}.
%
One of the interesting example for the pseudoscalar mesons decay is the G-symmetry violating $\eta \to \pi^{0}\gamma\gamma$ process 
whose branching ratio cannot be described neither in the framework of the Chiral Perturbation Theory nor by the Vector Meson 
Dominance Model. \\
% The PDG quoted value for Branching Ratio (BR) is equal $ (7.2 \pm 1.4) \cdot 10^{-4} $, while the models 
% give values about 1/2 lower \cite{Biagio_thesis}. KLOE, with its hight pure $\eta$ sample, is able to give very precise 
% measurement of this branching ratio \cite{Biagio_thesis}. \\
% 
\indent At present KLOE detector is being upgrated in 
order to improve the possibility of the reconstruction of the decays of the $K_{S}$ mesons. In particular in order to investigate 
a CP-violating $K_{S} \to 3\pi^{0}$ decay \cite{perkins}. \\
% 
%\indent And one of the interesting pseudoskalar mesons decay is the following channel: 
%$\eta \to \pi^{0} \gamma \gamma$ which on the one hand non conserve G symmetry but also a Branching Ratio 
%value for this channel which was measured in experiment doesn't accept the teoretical calculations and 
%ChPT (Chiral Perturbation Theory) and VMD (Vector Meson Dominance) predictions.  
%
\indent The $\eta \to \pi^{0}\gamma\gamma$ and $K_{S} \to 3\pi^{0}$ are examples of highly interested channels which are however very 
challenging to study experimentally due to the large number of the $\gamma$ quanta in the final state. 
This fact is caused by difficulties of the rejection of background channels due to 
merging and splitting of clusters \cite{Letter_of_intent}. For example, for the first mentioned reaction a background channel is $\eta \to 3\pi^{0}$ with 
two merged clusters leading to the same topology as in the case of the $\eta \to \pi^{0}\gamma\gamma \to 4\gamma$. 
And in the case of the $K_{S} \to 3\pi^{0}$ the background constitues a $K_{S} \to 2\pi^{0}$ reaction with two splitted clusters. \\
%Branching Ratio for this channel is equal: 32.1\% \cite{pdg_niebieska}. 
%In this case we can reconstruct four particles in the final state, thus the reconstructed topology is the same 
%as for channel $\eta \to \pi^{0}\gamma\gamma$. \\    
% 
% It is caused by an influence from merging and splitting effects to reconstruction possibilites. \\
%  
%Investigations of these reaction are very usefull to make tests of Standard Model Theory however there are problems with
%correct reconstruction energy and position for particles in the final state of reaction for these channels during the analysis process.
%It is caused by an influence from merging and splitting effects to reconstruction possibilites for example in decays
%$\eta \to \pi^{0}\gamma\gamma$ and $K_{S} \to 3\pi^{0}$.
%
\indent In this diploma thesis we investigate the influence of the merging and splitting of clusters on decays 
with the high multiplicity of $\gamma$ quanta, which are at most biased by these effects.  
 For this aim we implemented the full geometry and realistic material composition of the barrel electromagnetic calorimeter into FLUKA package. 
The prepared Monte Carlo based simulation program permits to achieve a fast generation of the detector response separately 
for each interested reaction. \\    
%
% We estimated influence of those effects on the whole barrel calorimeter geometry which was 
% implemented into FLUKA package. \\
%
%
\indent In chapter 2 we will describe the DA$\Phi$NE collider and the KLOE detector. \\ 
\indent Chapter 3 comprises description of the vertex 
generator, used to simulate the kinematics of physical decays, and the description of the physics models used by the FLUKA Monte-Carlo program for the
 generation of nuclear, hadronic and electromagnetic reactions. This chapter includes also description of implementation of materials composition 
and geometry of the barrel calorimeter in the FLUKA Monte Carlo. \\ 
% Also simulations of the photomultipliers response and 
% clustering algorithm program are described. \\ 
%
\indent Chapter 4 presents a calibration of the DIGICLU program used for the reconstruction of the photomultiplier response 
and for the cluster recognition. This chapter includes: i) preparation of the data sample for  
$e^{+}e^{-} \to \phi \to \eta \gamma \to 3\gamma $ reaction, ii) estimation of the attenuation length for scintillating fibers, 
iii) calibration of ionization deposits and implementation of the threshold formula to the source code. \\  
\indent In chapter 5 we describe effects of merging of clusters for $\eta \to 3\pi^{0}$ as a background for the $\eta \to \pi^{0}\gamma\gamma$ 
channel and splitting of clusters for $K_{S} \to 2\pi^{0}$ as a background for  $K_{S} \to 3\pi^{0}$ reaction. Further on reconstruction 
efficiency with the KLOE clustering algorithm as a function of the photocathode quantum efficiency is presented. 
Finally, merging and splitting probabilities as a function of the quantum efficiency are studied. \\  
%
% \indent Chapter 6 comprises a description of the reconstruction efficiency as a function of azimuthal anngle and 
% we present an energy deposition efficiency of the cluster reconstruction in the barrel calorimeter. \\
%
\indent Chapter 6 summarises the whole thesis and brings the conclusions and remarks. \\
\indent This thesis is supplemented with appendics where section A presents a kinematic fit procedure and section B generally describes 
Monte Carlo Methods. In appendix C energy distribution for $\gamma$ is presented. Section D presents a time distribution 
for single and multi-gamma hits.   
The sections E and F contain an estimation of probability for multi-gamma 
hits at a single calorimeter module and example of the event reconstruction for the process 
$e^{+}e^{-} \to \phi \to \eta \gamma \to 3\pi^{0}\gamma \to 7\gamma $, 
respectively. The G section comprises a description of the energy deposition in the barrel calorimeter as a function of 
azimuthal angle and the last section presents definition of the used coordinate system.      
%
%  
%In the second section of this chapter we will describe a simulations of the response of detector with FLUKA and response of the 
%photomultipliers with clustering algorithm. One of the main part of this chapter consists of a description of the implementation of the whole 
%Kloe Barrel Calorimeter into FLUKA. \\  
%   
%\indent In Chapter 4 we presented a calibration of the clustering algorithm and implementation of the threshold formula to the source code. \\
%
%\indent Chapter 5 introduces results of simulations aiming at investigate influence of the merging and splitting effects.
%Also test of the reconstruction possibilities with a present clustering algorithm will be presented.
%In the last part of this chapter we will describe influence of quantum efficiency to scale on the reconstruction of merging and splitting effects. \\
% Chapter 8 consists description of the reconstruction possibilities of new upgrated clustering algorithm. 
%
%\indent Chapter 6 describes energy response and reconstruction of particles on the barrel calorimeter, and  
% 
%
%    
%               
%%%%%%%%%%%%%%%%%%%%%%% EXPERIMENT DESCRIPTION %%%%%%%%%%%%%%%%%%%%%%%%%%%%%%%%%%%%%%
%\vspace{-0.7cm}
\chapter{KLOE2 at DA$\Phi$NE experimental facility}
\hspace{\parindent}
% \hspace{\parindent} 
% opis eksperymentu
% concomiant - towarzyszacy, wspolistniejacy 
% meant - oznaczalo w intencjach
% appreciate - doceniac, byc wdziecznym, zyskiwac na wartosci
% frontier - granica
% occur - wystepowac
% In this chapter I would like to present the KLOE experiment. The main informations about detector's 
% structure and accelerator machine on which KLOE detector works. 
%\indent
%
The KLOE (\textbf{Klo}ng \textbf{E}xperiment) detector is installed at the interaction point of the electron and positron beams
  of the DA$\Phi$NE (\textbf{D}ouble \textbf{A}nnular
\textbf{$\phi$}-factory for \textbf{N}ice \textbf{E}xperiments) collider operating in the Laboratori Nazionali di Frascati (LNF). \\
%This machine is used in particular to study $\phi$, $\eta$ and kaons decays. 
%But the main part of investigated reactions are decays of kaons in the decays of the meson $\phi$:
%$\phi \to K^{+}K^{-}$ (BR = 49.1\%) and $\phi \to  K_{L}K_{S}$ (BR = 33.8\%). \\   
%This is the largest and oldest laboratory of the Italian 
%Institute of Nuclear Physics (INFN). \\ 
%The decision to construct an $e^{+}e^{+}$ collider meant to opearate around 1020 MeV (the mass of the $\phi$ meson) was taken in 1988 year.
\indent
It has been designed with the primary goal to measure the CP violation parameter R($\frac{\epsilon^{'}}{\epsilon}$) 
\cite{passeri_ciambore_kloe} with a sensitivity 
of one part in ten thousand by using the double ratio method \cite{palutan_kloe_trigger}. \\
%
%with accuracy of O($10^{-4}$) \cite{passeri_ciambore_kloe}. $\frac{\epsilon^{'}}{\epsilon}$ with a sensitivity
% of one part in ten thousand by using the double ratio method \cite{palutan_kloe_trigger}. The measurement of this  \\  
%
\indent
This detector was fully constructed by 
the end of the year 1998 \cite{A_Flavor_of_KLOE}, and since then it was taking data for seven years. 
The experimental program was completed with integrated luminosity of 2.5~fb$^{-1}$ obtained predominantly with the center-of-mass energy 
equal to the mass of the $\phi$ meson ($\sqrt{s}$~$\sim$~$M_{\phi}$~=~1019.456$\pm$0.020 MeV \cite{Biagio_thesis}). \\
\indent 
% The $\phi$ meson is generated in $e^{+}e^{-}$ collisions and its decay products are then analyzed 
% by the KLOE detector.
% DA$\Phi$NE works at the energy of $\sqrt{s} \approx$ 1020 MeV corresponding to the $\phi$ resonance 
% peak ($M_{\phi}$ = 1019.456$\pm$0.020 MeV) \cite{Biagio_thesis}. 
%The energy of the production cross section
% of the $e{+}e{-}$ collision reached the maximum to 
% this year the KLOE detector has taking data for the next 7 years with integrated
%luminosity $10^{32} cm^{-2}sec^{-1}$.
%KLOE detector was taking an experimental data especially for $\phi$ meson 
%decays mostly to kaons both neutral and charged, in pairs. The decays $\phi$ meson to channels with $\eta$ meson aren't so frequent 
%(lower branching ratio value is shown in Table:~\ref{branching_ratio}).
%
%
\vspace{-0.7cm}
\begin{table}[H]
\begin{center}
\begin{tabular}{|l|c|}  \hline \hline
\emph{Decay} & \multicolumn{1}{|c|}
                {\emph{BR(\%)}} \\ \hline
$\phi \to K^{+} K^{-} $ & 49.1  \\
$\phi \to K_{S} K_{L} $ & 33.8  \\
$\phi \to \rho \pi $ / $ \pi^{+} \pi^{0} \pi^{-}  $ & 15.6  \\
$\phi \to \eta \gamma $ & 1.26  \\  \hline \hline 
\end{tabular}               
\end{center}
\caption{Main decays of the $\phi$ meson \cite{pdg_niebieska}.}
\label{branching_ratio}
\end{table}
\vspace{0.5cm}
\vspace{-0.4cm}
The cross section for the production of the $\phi$ vector meson is large and amounts to \\ 
$\sigma(e^{+}e^{-}\to\phi)$~=~3.1~$\mu$b. \\  
\indent The $\phi$ meson decays predominantly into pairs of neutral or charged kaons 
(see Table~\ref{branching_ratio}).

% span - szerokosc, obejmowac
%
%The laboratory is mainly devoted to nuclear, high energy and 
%astroparticle physics research. It is an important accelerator development and technology center.       
%
% both .... and - zarowno .... jak i
%
%
%

\section{DA$\Phi$NE collider}        % Produces section heading.  Lower-level
\begin{figure}[H]
\hspace{2.8cm}
\parbox[c]{1.0\textwidth}{
\parbox[c]{0.55\textwidth}{\includegraphics[width=0.66\textwidth]{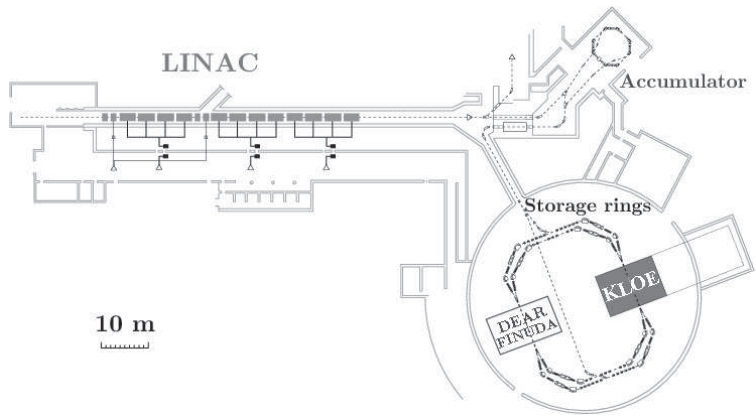}}
}
\caption{Scheme of the DA$\Phi$NE collider. The figure is adapted from \cite{A_Flavor_of_KLOE}.}
\label{dafne_accelerator}
\end{figure}

%                                        % sections are begun with similar 
%                                        % \subsection and \subsubsection commands.
\hspace{\parindent}
DA$\Phi$NE   
 at Frascati consists of two intersecting crossing
 accelerator rings, 
one for positrons and one for electrons. The main DA$\Phi$NE parameters are presented in Table~\ref{dafne_parameters},
 and scheme of the facility is shown in Fig.~\ref{dafne_accelerator}.
\begin{table}[H]
\begin{center}
\begin{tabular}{|l|c|}  \hline \hline
\emph{Parameter} & \multicolumn{1}{|c|}
                {\emph {Value}} \\ \hline    
Energy [GeV] & 0.51 \\
Trajectory length [m] & 97.69 \\
RF frequency [MHz] & 368.26 \\
Bunch length [cm] & 1-3 \\
Number of colliding bunches & 111 \\
Emittance, $\epsilon_{x}$ [mm$\cdot$mrad] & 0.34 \\  \hline \hline 
\end{tabular}               
\end{center}
\caption{The main characteristics of the DA$\Phi$NE collider \cite{dafne_arxiv}.}
\label{dafne_parameters}
\end{table} 
Positrons and electrons are accelerated in the LINAC (\textbf{Lin}ear \textbf{Ac}celerator) which   
delivers electron or positron beams
in the energy range from 25 to 725 MeV with intensities varying
from $10^{10}$ particle per pulse down to a single-electron \cite{dafne_arxiv}.
This 60 meters long accelerator is the heart of the DA$\Phi$NE injection system. 
It is an S-band machine (2.865 GHz) which 
delivers 10 ns pulses at a repetition rate of 50~Hz.
Electrons, after acceleration to
final energy in the LINAC, are accumulated and cooled in the accumulator and
transferred to a single bunch into ring. Positrons require first accelerating of 
electrons to about 250 MeV to target in the LINAC, where
positrons are created. Afterwards the positrons follow the same accelerator elements as 
electrons \cite{A_Flavor_of_KLOE}.
Positrons and electrons after acceleration and accumulation process run around in two storage rings 
and hit in the collision points.
This facility is called a Frascati $\Phi$-Factory complex because it was able to produce about 3.3 bilions of  
$\phi$ mesons during the years from 2000 to 2005. 
%Physics research are based on radiative decays of $\phi$ meson and are studied 
%in three experiments. 
The KLOE experiment was located in one of the two collision points at 
DA$\Phi$NE collider, whereas the second collision point was alternatively occupied by two other experiments: DEAR \cite{dear_homepage} and FINUDA 
\cite{finuda_homepage}. \\
\indent At present a new $e^{+}e^{-}$ interaction region is being constructed \cite{Alesini_krab} 
 in order to increase the collider luminosity up to an order of magnitude, from 10$^{32}$cm$^{-1}$s$^{-2}$ 
to $\approx 10^{33}$cm$^{-1}$s$^{-2}$ \cite{Letter_of_intent}.
%
%
%  
%
%
% repetition  - powtarzanie, powtorka
% rate        - stosunek liczbowy, wskaznik, stopa 
% band        - towarzystwo, wstazka
% run arround - obiegaja
% approach    - podchodzic, zblizac sie
% connected with - oparta na
% devoted - oddany, przywiazany
% adjust - uporzadkowac, dostosowac
% lean - opierac sie, sklaniac sie
% layout - rozmieszczenie, rozplanowanie
% counter - licznik, stoisko
%
% And one of the most important physics objectives are studies $\phi$ meson radiative decays. 
%
At the interaction point (IP) the beam pipe has the shape of a   
sphere which is made of a beryllium-aluminium alloy with 10 cm diameter and 
50 $\mu$m thickness. The beryllium, having a low atomic number, has been used to minimize the interaction of particles produced 
at the interaction point with the beam pipe material \cite{Biagio_thesis}.
%
%
%
%\hspace{.5in} 
%\vspace{.25in}
%
%
%
\section{KLOE detector}
\begin{figure}[H]
\hspace{3.9cm}
\parbox[c]{0.45\textwidth}{\includegraphics[width=0.49\textwidth]{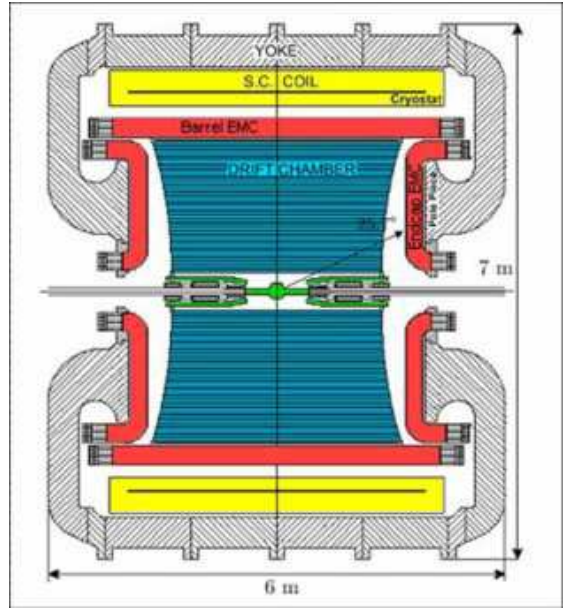}}
\caption{The KLOE detector. For the description see text. The figure is adapted from \cite{Letter_of_intent}.}
\label{the_kloe_detector}
\end{figure}

\hspace{\parindent}
The KLOE detector, which is shown schematically in Fig.~\ref{the_kloe_detector}, was designed for the study of CP violation in 
the neutral-kaon system. 
It consists of two main elements: i) an electromagnetic calorimeter (EmC) \cite{kloe_electromagnetic_nim2} 
for the detection of $\gamma$ quanta,
 charged pions and $K_{L}$ mesons and ii) a  
large drift chamber (DC) \cite{Adinolfi_Drift_Chamber} 
for the measurement of charged particles trajectories \cite{Data_handling}.   
The drift chamber and the calorimeter are inserted in the field of the superconducting coil which produces a magnetic 
field parallel to the beam axis \cite{kloe_electromagnetic_nim2}. The field intensity is equal to 0.52 T \cite{Biagio_thesis}.
%(Inner tracker \cite{Letter_of_intent}) as well as in order to tagg $\gamma\gamma$ fusion reactions 
%(with $\gamma\gamma$ tagger \cite{Letter_of_intent}).
%
%
\newpage
The radius of the active part of the KLOE detector is two meters. This size enables to register about 40\% 
decays of neutral long-lived kaons \cite{A_Flavor_of_KLOE}.         
In the near future KLOE will be upgraded by inner tracker and  $\gamma\gamma$ tagger  
in order to improve its tracking capabilities \cite{Letter_of_intent}. 
\subsection{Drift Chamber}
\hspace{\parindent}
The KLOE Drift Chamber (DC) \cite{Adinolfi_Drift_Chamber} consists of 12582 drift cells (2x2, 3x3 $cm^{2}$) arranged in 58 cylindrical 
layers surrounding the beam pipe. The diameter and length of the DC is equal to 4~m and 3.3~m, respectively \cite{Agnelo_kloe_colaboration}. 
It is filled with 90\% helium and 10\% isobutane gas mixture, giving a radiation length (gas + wires)
equal to 900 m. Charged particles traveling through the drift chamber are ionizing gas medium and then electrons
created along the particle trajectory drift to the wires with positive voltage. A multiplication mechanism
 causes detectable signal at the wire's end \cite{A_Flavor_of_KLOE}. 
The DC measures the $K_{L,S}$ charged vertex with $\sim$~1~mm accuracy. It provides fractional momentum resolution of $\frac{\sigma_{p}}{p}$~$\sim$~0.5\%
 for low momentum tracks. It is transparent to $\gamma$ down to 20 MeV and limits to acceptable levels the $K^{0}$ 
regeneration and $K^{\pm}$ multiple scattering.
% 
% 
%Observation a decay is related with rejestration the decay products or particles. 
%
\begin{figure}[H]
\hspace{3.8cm}
\parbox[c]{1.0\textwidth}{
\parbox[c]{0.45\textwidth}{\includegraphics[width=0.49\textwidth]{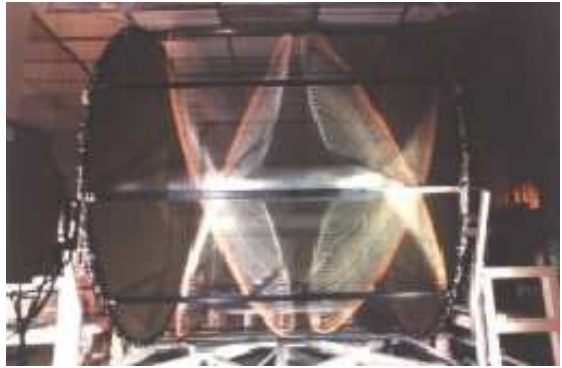}}
}
\caption{Photo of the KLOE drift chamber. The figure is adapted from \cite{strona_kloe_obrazki}.}
\label{kloe_drift_chamber}
\end{figure}

The hit position resolution is 150 $\mu$m in the central part of the cell, increasing close to the
 wire 
% (effect caused by fluctuations of the primary He ionization)
and towards the cell boundary. The spatial resolution in the "$\phi$ coordinate" (azimuthal angle) is well below 200 $\mu$m. 
%(due to longitudinal 
%diffusion and $\sigma_{t}$). 
The hit identification efficiency is larger than 99\%, whereas the efficiency for associating an existing hit to track amounts to
 about 97\% \cite{Agnelo_kloe_colaboration}. \\
%
% The shape of the electric field from these sense wires builds 12500 small  cells 
% \cite{A_Flavor_of_KLOE}. 
%             
%
% \hspace{.5in} 
% \vspace{.25in}
\vspace{-0.4cm}
\subsection{Electromagnetic Calorimeter}
\begin{figure}[H]
\hspace{2.5cm}
\parbox[c]{1.0\textwidth}{\includegraphics[width=0.69\textwidth]{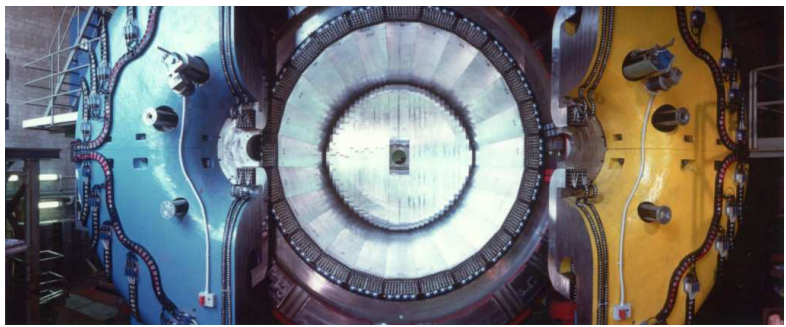}}
\caption{Photo of the KLOE calorimeter. One sees 24 modules of the barrel and the inner plane of the endcap. 
The figure is adapted from \cite{strona_kloe_obrazki}.}
\label{barrel_calorimeter}
\end{figure}

\hspace{\parindent}
The KLOE calorimeter is made of the lead layers with about 1,2 mm thickness (200 layers per 1 module) which are 
filled with scintillating fibers of 1 mm diameter. The whole electromagnetic calorimeter (EmC) consists of three main parts: 
barrel and two endcaps. Barrel (Fig.~\ref{barrel_calorimeter}) is composed of 24 
modules with trapezoid shape of 23 cm thickness aligned with the beams and surrounding 
the drift chamber detector. Endcaps are situated over the magnet 
pole pieces (see Fig.~\ref{the_kloe_detector}) and hermetically close the calorimeter with 
98\% of 4$\pi$ \cite{A_Flavor_of_KLOE}. Each of the two endcaps
calorimeters consists of 32 vertical modules with length ranging from 0.7 to 3.9 meters. The endcap modules are bent and 
their cross-section 
with plane parallel to the beam axis is rectangular with thickness of 23 cm \cite{kloe_electromagnetic_nim1}. 
%The time resolution for 
%this calorimeter is 54 picoseconds  and energy resolution is equal to 5.7\% for a one GeV photon.
%
The volume of the calorimeter consists of 50\% fiber, 40\% lead and 10\% of glue. 
The measured performances for this detector are: full efficiency for $\gamma$ quanta from 20 to 500 MeV \cite{Agnelo_kloe_colaboration},
$\sigma(x)\sim$~1~cm, $\sigma(E)\sim\frac{5.7\%}{\sqrt{E(GeV)}}$,  $\sigma(t)\sim\frac{54~\text{ps}}{\sqrt{E(GeV)}}$.
%   
%
%
%
%%%%%%%%%%%%%%%%%%%%%%%%%%%%%%%%%%%%%%%%%%%%%%%%%%%%%%%%%%%%%%%%%%%%%%%%
%%%%%%%%%%%%%%%%%%%%%%%% JJPLUTO %%%%%%%%%%%%%%%%%%%%%%%%%%%%%%%%%%%%%%%
%%%%%%%%%%%%%%%%%%%%%%%%%%%%%%%%%%%%%%%%%%%%%%%%%%%%%%%%%%%%%%%%%%%%%%%%
%
\subsection{Upgrade of the KLOE detector}
\hspace{\parindent}
At present the KLOE detector is being upgrated in view of the new experimental program which will extend the studies 
to the more precise measurement of the $K_{S}$ mesons and the production of meson in the $\gamma\gamma$ fusion 
\cite{Letter_of_intent}. For this 
aim a vertex detector and the $\gamma\gamma$-tagger \cite{domenici_kloe_upgrade} are being built \cite{roll_in_proposal}.
%
%

%In KLOE 2 detector will be installed two new elements, the first will be install a vertex detector 
%and the second a small angle tagger for $\gamma\gamma$ physics.
%
\subsubsection{Vertex Detector}
\begin{figure}[H]
\hspace{4.3cm}
\parbox[c]{0.55\textwidth}{\includegraphics[width=0.49\textwidth]{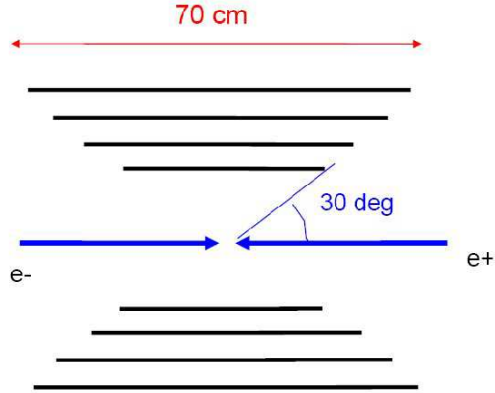}}
\caption{Schematic view of the vertex detector. The figure is adapted from \cite{Letter_of_intent}.}
\label{vertex_detector}
\end{figure}

\hspace{\parindent}
In the KLOE detector the first hit was measured by drift chamber at a radius of 28~cm from the interaction point 
(IP) \cite{Letter_of_intent}. 
%The implementation of this detector into KLOE will bring several benefits.
Therefore in order to improve the resolution of the determination 
of the $K_{S}$ and $K_{L}$ decay near the interaction point a new vertex detector is constructed \cite{domenici_kloe_upgrade}. 
This detector consists of five concentric layers of cylindrical triple-GEM (C-GEM), 
completely realized with very thin polyimide foils \cite{kloe_inner_tracker,kloe_inner_tracker2,kloe_inner_tracker3,domenici_full,domenici_hadron}. 
%   
% Insertion of a Inner Tracker. At present, first tracking layer is at 30 cm (i.e. 50 ?S) from the I.P
%
The scheme of the vertex detector is presented in Fig.~\ref{vertex_detector}.
It will be made of light materials in order to minimize $\gamma$ absorbtion.
The resolution on r-$\phi$ surface is expected to be around 200 $\mu$m and on z coordinate about  
500 $\mu$m \cite{Venanozi_teoretyk}. 
% It will be made of light materials in order to minimize a $\gamma$ absorbtion.

\subsubsection{$\gamma\gamma$ tagger}
\hspace{\parindent}
The main task of this detector is the detection of $e^{+}$ and $e^{-}$ from $\gamma\gamma$ reactions emitted at small angle 
\cite{Venanozi_teoretyk,gg_fizyka_kloe} with the widest possible energy ranges. 
%
%\begin{figure}[H]
%\hspace{5.0cm}
%\parbox[c]{0.35\textwidth}{\includegraphics[width=0.29\textwidth]{kinegg.eps}}
%\caption{Two-photon particle production in a $e^{+}e^{-}$ collider. The figure is adapted from \cite{gg_fizyka_kloe}.}
%\label{gg_fizyka}
%\end{figure}
%
$\gamma\gamma$ tagger will provide information on the angle and the energy of the scattered electrons and positrons 
\cite{gamma_gamma} and hence it will permit to study the production of mesons in $\gamma\gamma$ fusion via the reaction: 
\begin{equation}
e^{+}e^{-} \to e^{+}e^{-}\gamma^{*}\gamma^{*} \to e^{+}e^{-} + X ,   
\end{equation}
where X is some arbitrary final state allowed by conservations laws. 
The tagger consists of microstrip silicon detector and plastic scintillator hodoscope. \\
%
%\indent This detector enables to study the reactions related with $\gamma\gamma$ physics.
%The term '$\gamma\gamma$ physics' stands for the study of the following reaction
%(see Fig.~\ref{gg_fizyka}) 
%
%\begin{equation}
%e^{+}e^{-} \to e^{+}e^{-}\gamma^{*}\gamma^{*} \to e^{+}e^{-} + X   
%\end{equation}
%
%where X is some arbitrary final state allowed by conservations laws. These processes 
%show a logarithmic dependence from the energy E
%of the colliding beams that reflects in a not negligible cross section. It turns out
%that for E greater than a few GeV the $\gamma\gamma$  processes dominate with respect to the
%corresponding annihilation processes \cite{gg_fizyka_kloe}.
%

\vspace{-0.3cm}
\subsubsection{New QCAL}
\begin{figure}[H]
\hspace{3.5cm}
\parbox[c]{1.0\textwidth}{\includegraphics[width=0.59\textwidth]{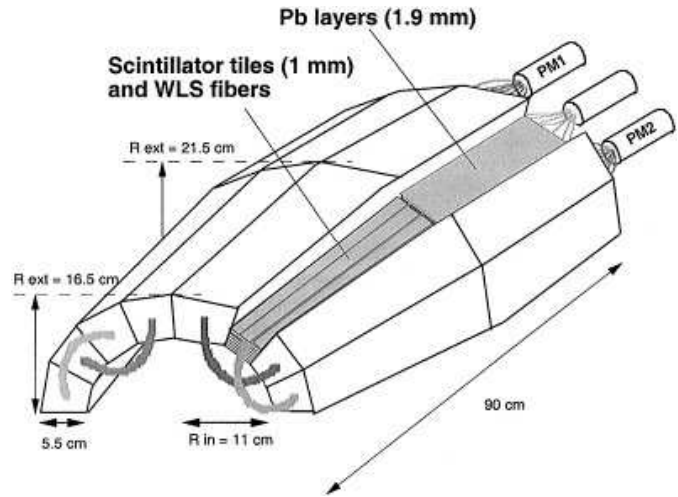}}
\caption{Schematic view of the QCAL detector. The figure is adapted from \cite{qcal_calorimeter}.}
\label{qcal_detector}
\end{figure}

\hspace{\parindent}
The upgrade of QCAL detector (Fig.~\ref{qcal_detector}) \cite{qcal_calorimeter} was needed, because  
 the interaction region was modified and in the present scheme the angle between 
colliding beams has been increased from 8 to 18 degrees, which practically excludes the possibility to use the existing 
QCAL calorimeter \cite{roll_in_proposal}. 
The schematic view of upper part of the previous QCAL detector is presented in Fig.~\ref{qcal_detector}. 
\newpage
The upgraded detector, with improved position resolution, will allow to 
extend the search for $K_{L} \to 2\pi^{0}$ events also in case when three photons are reconstructed in 
the EmC and one photon in QCALT, thus strongly reducing the correction for acceptance \cite{roll_in_proposal} 
and the contamination from $K_{L} \to 3\pi^{0}$ when some photons hit the detector.  
% The schematic view of upper part of the previous QCAL detector is presented in the figure \ref{qcal_detector}. 
%
% The details of the description of the previous quadrupole tile calorimeter of KLOE (QCAL) are presented in
% \cite{qcal_calorimeter}.       
%\subsubsection{New Calorimeter Readout}
%\hspace{\parindent}
%
%In KLOE 2 experiment will be used a photomultiliers with increased high-quantum-efficiency (HQE). 
%The advantage of HQE rests in a possible increase in time-resolution (thus longitudinal coordinate 
%resolution) and in higher detection efficiency for low-energy photons \cite{roll_in_proposal}. The photomultiplayers are 
%presented in the \ref{photomultipliers}. 
%
%\begin{figure}[H]
%\hspace{4.0cm}
%\parbox[c]{1.0\textwidth}{\includegraphics[width=0.49\textwidth]{photomultipliers.eps}}
%\caption{The view of the photomultiplyers, first three on left with HQE option, the rest are standard KLOE detectors.  
%Courtasy of Antonio Di Domenico \cite{antonio_di_domenico}.}
%\label{photomultipliers}
%\end{figure}
%

%%%%%%%%%%%%%%%%%%%%%%%%%%%%%%%%%
% SIMULATION TOOLS              %
%%%%%%%%%%%%%%%%%%%%%%%%%%%%%%%%%
\chapter{Simulation tools}
\hspace{\parindent}
In this chapter we will describe tools used for simulations of signals in the EmC. 
% were using in our simulations. 
To compute the physical response of the 
barrel calorimeter we applied the following programs: i) VERTEX GENERATOR which simulates kinematics of the physical reactions 
% with hight multiplicity of photons in the final state, 
ii) % implemented geometry of the whole Barrel Calorimeter into 
FLUKA package which reproduces particle interaction, propagation and a realistic 
light output in the scintillating fibers, and iii) DIGICLU reconstruction program which simulates the response of  
photomultipliers and reconstructs energy and time for particles hitting the calorimeter. 
FLUKA package is mostly used for low energy physic, like: hadronic interactions, medical science, neutrino beam simulation or in 
dosimetry science. Using FLUKA we can achieve accurate description of 
the physics processes in the KLOE detector. 
% hence we prepared a full simulation of the barrel calorimeter with FLUKA Monte Carlo. 
%
% because, 
% it can result in more accurate description of 
% the physics processes on this detector than KLOE Monte Carlo which based on Geant 3 
%package.   
% 
% The KLOE Monte Carlo is based on Geant 3 
%package   
%

\section{Vertex generator}
\hspace{\parindent}
In order to enable a fast simulations of the meson decays in the KLOE detector we have prepared a "vertex generator"  
for simulations of the final state momenta of the decay products\footnote{In our group this generator is called jjpluto.}.
%
%One of the most important things which are investigated on KLOE detector are mesons decays. We wanted to simulate this meson decays and
%decided to write a vertex generator. 
%
This program is compatible with FLUKA Monte Carlo and simulates decays of $\phi$ meson 
to channels with $\eta$, $\eta^{\prime}$ and kaons. It is based on the CERN library procedure GENBOD which generates  
four-momentum vectors of particles in
the final state which are homogenously distributed in the phase space \cite{James_genbod}. 
This tool permits to calculate four-momenta of final state particles in the rest frame of the decaying object. 
% Central Mass system of the main particle which decaying \cite{James_genbod}.
%The simulation is based on a calculations of the value of the $\sqrt{s}$ which is the Mandelstam parameter which is a total energy
% in Central Mass System and which is a Lorentz transformation invariant too \cite{perkins}.
Additionally we implemented an angular distribution ($\frac{1+cos^{2}\theta}{2}$) for $\gamma$ quanta from the radiative 
$\phi$ meson decay ($\phi$ is a vector meson with spin = 1).
%
%We want to investigate $\phi$ meson decays to channels with $\eta$ and $\eta^\prime$ and then decay this mesons to $ \pi^0$ mesons and $\gamma$ quanta
%in the final state. 
At present, for the purpose of this work the generator can simulate six below listed reactions:
%
%Using this generator we were able to simulate a detector response for a following reactions:
\begin{enumerate}
\item    $e^{+}e^{-} \to \phi \to \eta \gamma \to 3\gamma$ , 
\item    $e^{+}e^{-} \to \phi \to \eta \gamma \to 3\pi^0 \gamma \to 7\gamma$ ,
\item    $e^{+}e^{-} \to \phi \to \eta^\prime \gamma \to 3\gamma$ ,
\item    $e^{+}e^{-} \to \phi \to \eta^\prime \gamma \to 3\pi^0 \gamma \to 7\gamma$ ,
\item    $e^{+}e^{-} \to \phi \to K_{S}K_{L} $ ,
\item    $e^{+}e^{-} \to \phi \to \pi^{0} \gamma \to 3\gamma$ . 
\end{enumerate}
It is however constructed in a way which enables an easy implementation of any further process.
%
% --------------------------------------------------------
%
%         MODELE FIZYCZNE UZYWANE WE FLUCE 
%
% --------------------------------------------------------
%
%
\section{FLUKA - a multi-particle Monte Carlo transport code}
\hspace{\parindent}
This section is based on the informations given in references \cite{FLUKA_manual,FLUKA_homepage}.
FLUKA is a multipurpose transport Monte Carlo code, able to treat hadron-hadron, hadron-nucleus, neutrino, electromagnetic, 
and $\mu$ interactions up to 10000 TeV \cite{the_physics_models_of_fluka}. 
Also the transport of the charged particles is well recreated (handled in magnetic field too). 
This Monte Carlo is based, as far as possible, on orginal and well tested microscopic models. Due to this "microscopic" approach 
to hadronic interaction modelling, each step is self-consistent and has well confirmed physical bases \cite{the_physics_models_of_fluka}. 
% The very short description of physics models using by FLUKA will be presented in this section. 
FLUKA can be used in realistic simulations involving following  
physical effects: \\
\indent \textbf{Hadron inelastic nuclear interactions} - are based on resonance production and decay below a few GeV, and above on the Dual Parton 
model. FLUKA can also simulate photo-nuclear interactions (described by Vector Meson Dominance, Delta Resonance, Quasi-Deuteron and 
Giant Dipole Resonance models) \cite{FLUKA_manual}. \\ 
\indent  \textbf{Elastic Scattering} - this process is accomplished with tabulated nucleon-nucleus cross sections. Tabulated are also 
phase shift data for pion-proton and for kaon-proton scattering  \cite{FLUKA_manual}. To this Monte Carlo also parametrised 
nucleon-nucleon cross-sections were implemented. \\ 
% as well as detailed kinematics of elastic scattering on hydrogen nuclei. \\     
%
\indent  \textbf{Nucleus-Nucleus interactions} - are treated through interfaces to external event generators. From 0.1 to 5 GeV 
per nucleon a modified RQMD is used, whereas for higher energies DPMJET generator is applied \cite{DPMJET}. \\  
\indent  \textbf{Transport of charged hadrons and muons} - a code can handle electron backscattering and energy deposition in 
thin layers even in the few keV energy range. The energy loss is simulated with Bethe-Bloch theory 
% and optional delta-ray production 
 with account for spin effects and ionisation fluctuations \cite{FLUKA_manual}. \\   
\indent  \textbf{Low-energy neutrons} - for neutrons with energy lower than 20 MeV, FLUKA uses neutron cross-section library, containing 
more than 140 different materials. The transport of these particles is realised by simulation of the standard multigroup 
transport with photon and fission neutron generation \cite{FLUKA_manual}. 
Detailed kinematics of elastic scattering on hydrogen nuclei is implemented too. \\ 
\indent  \textbf{Electrons} - FLUKA uses an original transport algorithm for charged particles, including a complete multiple Coulomb scattering 
treatment giving the correct lateral displacement even near boundary. The Landau-Pomeranchuk-Migdal suppression effect 
and the Ter-Mikaelyan polarisation effect in the soft part of the bremsstrahlung spectrum are also implemented. Electrons are propagated 
taking into account a positron annihilation in flight and M{\"o}ller scattering effects. \\ 
\indent  \textbf{Photons} - to reproduce photons physics the following effects were implemented to this Monte Carlo: pair production 
with actual angular distribution of electrons and positrons, Rayleight scattering, photon polarisation effect, photo-hadron production, 
Compton effect with account for atomic bonds through use of inelastic Hartree-Fock form factors. \\     
\indent  \textbf{Optical photons} - generation and transport is based on Cherenkov, Scintillation and Transition Radiation. \\    
%
% interactions are implemented but independently from tracking. 
% Default they are tracked without interactions \cite{FLUKA_manual}.    
%
\indent  \textbf{Neutrinos and muons} - 
FLUKA package includes also cosmic ray physics \cite{FLUKA_neutrinos} and can simulate muons and neutrinos interactions. 
This tool is used for basic research and applied studies in space and atmospheric 
flight dosimetry and radiation damage \cite{FLUKA_neutrinos}. Interactions are implemented independly of tracking procedures \cite{FLUKA_manual}.  
By default they are tracked without interactions \cite{FLUKA_manual}.
%
%
%
%

%%%%%%%%%%%%%%%%%%%%%%%%%%%%%%%%%%%%%%%%%%%%%%%%%%%%%%%%%%%%%%%%%%%%%%%%%%
%%%%%%%%%%%%%%%%%% Struktura powstawania sygnałów w detektorze %%%%%%%%%%%
%%%%%%%%%%%%%%%%%%%%%%%%%%%%%%%%%%%%%%%%%%%%%%%%%%%%%%%%%%%%%%%%%%%%%%%%%%
%
%
\section{Implementation of the calorimeter material composition in FLUKA}
\begin{figure}[H]
\hspace{2.7cm}
\parbox[c]{0.60\textwidth}{\includegraphics[width=0.60\textwidth]{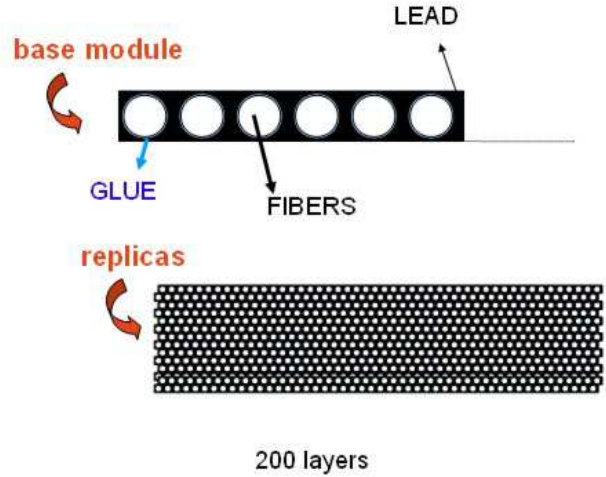}}
\caption{Details of the implementation of the calorimeter layers \cite{annaferrari2}.}
\label{base_cell}
\end{figure}

\hspace{\parindent}
%
% Szczególowa budowa detektora.
%
%
The Monte Carlo code FLUKA is used to determine the position, time and energy of the ionization deposits 
in the fibers caused by particles hitting the calorimeter.
We used the FLUKA "lattice" tool to design the fiber structure of the calorimeter
module \cite{FLUKA_homepage}. In the base module the calorimeter is simulated in detail (see Fig.~\ref{base_cell}), 
both under the geometrical 
point of view and with respect to the used materials. 
\vspace{-0.4cm}
%
%
%
% At first I would like to describe materials simulated in FLUKA.
The calorimeter is built out of lead, plexi and glue \cite{FLUKA_simulation_KLOE_EMC} which are composed as follows: 
%
% All the compounds have been carefully simulated: \\
\begin{itemize}
\item lead compound is made of Pb (95\%) and Bi (5\%),
\item the scintillating fibers are made of Polystyrene ($C_{2}H_{3}$),
\item the glue is a mixture of the epoxy resin ($C_{4}H_{4}O$, $\rho$ = 1.14 g/cm$^{3}$) 
and hardener \\ ($\rho$ = 0.95 g/cm$^{3}$) constituting 72\% and 28\% of the mixture, respectively.  
\end{itemize}
The hardener has been simulated as a mixture of several materials  which composition is given in Table~\ref{materials_fluka}. 
%     
% \vspace{0.4cm}
\vspace{-0.7cm}
\hspace{3.5cm}
\begin{table}[H]
\begin{center}
\begin{tabular}{|l|l|l|} \hline\hline
\emph{Compound} & \emph{formula} & \emph{fraction} \\ \hline
Polyoxypropylediamine   &  $C_{7}H_{20}O_{3}$   &  90\%  \\
Triethanolamine         &  $C_{6}H_{15}O_{3}$   &  7\%   \\
Aminoethylpiperazine    &  $C_{6}H_{15}N_{3}$   &  1.5\% \\
Diethylenediamine       &  $C_{4}H_{10}N_{2}$   &  1.5\% \\ \hline\hline 
\end{tabular}
\caption{Composition of the hardener mixture \cite{FLUKA_simulation_KLOE_EMC}.}
\label{materials_fluka}
\end{center}
\end{table}               
\vspace{-0.4cm}
% \vspace{0.4cm}
%
%     -  for the fibers, an average density  between
%          cladding and core has been used : $\rho$  = 1.044 g/cm$^{3}$ \\
%
%     -  glue: 72\% epoxy resin $C_{2}H_{4}O$, $\rho$=1.14 g/cm$^{3}$,
%              + 28\% hardener, $\rho$=0.95 g/cm$^{3}$ \\
%
%\begin{figure}[h]
%\hspace{3.5cm}
%\parbox[c]{0.55\textwidth}{\includegraphics[width=0.59\textwidth]{tab_materialy.eps}}
%\caption{Calorimeter materials.}
%\label{calorimeter_materials}
%\end{figure}
%
% Active material was built as: \\ 
% 1.0 mm diameter scintillating fiber emitting region:
%  lPeak \~ 460 nm. \\
% Core: polystyrene, $\rho$=1.050 g/cm$^{3}$, n=1.6 \\
%
% High sampling structure: \\
% 200 layers  of 0.5 mm grooved lead foils (95 \% Pb and 5 \% Bi).
% Glue: Bicron BC-600ML, 72 \% epoxy resin, 28 \% hardener.
% Lead:Fiber:Glue volume ratio = 42:48:10
%
%
 The module consists of two types of scintillating fibers: "Kuraray" and "Pol.hi.tech".
The first are implemented until a depth of 12 cm and the remaining part of the module is built with
"Pol.hi.tech" fibers \cite{FLUKA_simulation_KLOE_EMC}.
A proper attenuation lenght parameter for fibers material in order to calculate a light intensity
at each side of the module was taken into account. The attenuation function of fibers was described with 
the following formula 
\cite{FLUKA_simulation_KLOE_EMC}: 
\begin{equation}
B = A \cdot e^{-\frac{y}{\lambda_{1}}} + (1 - A) \cdot e^{-\frac{y}{\lambda_{2}}} ,     
\end{equation}
where y is distance between the place of deposited energy and the photocathode. 
The attenuation factor B is the ratio between detected and generated light signal which changes as a 
function of the distance (on y axis) from the generation point to the photocatode. The value of the 
parameters for "Kuraray" and "Pol.hi.tech" are shown in Table~\ref{fibers_parameters}. \\
\indent However, it is worth mentioning that, as we will show in the next chapters the data can be also well described using
this formula but with only one exponential function.
%
%    
% \vspace{0.4cm}
\hspace{3.5cm}
\vspace{-0.4cm}
\begin{table}[H]
\begin{center}
\begin{tabular}{|c|c|c|c|} \hline\hline
\emph{}           & \emph{A} & \emph{$\lambda_{1}$ cm} & \emph{$\lambda_{2}$ cm} \\ \hline
Kuraray           &  0.35    &  50                     &   430    \\
Pol.hi.tech       &  0.35    &  50                     &   330    \\ \hline\hline
\end{tabular}
\caption{Fibers parameters.}
\label{fibers_parameters}
\end{center}
\end{table}
\vspace{-0.6cm}
% \vspace{0.4cm}
% The readout simulation. 
% Fluka gives energy deposits in the fiber.
% The light is propagated by hand at the end of the fiber taking into account the attenuation. 
%
%The simulation of the Birks effect is used in simulation too.
%
The energy deposits are computed by FLUKA \cite{neutron_response_sciencedirect} taking into account the Birks effect \cite{zrodlo_birka} 
(see equation~\ref{birk_equation}), that is
the saturation of the light output (L) of a scintillating material when the energy release is high \cite{neutron_response_sciencedirect}.  
For high densities of energy deposition, due to the quenching interactions between the excited molecules along the path of
incident particles, the light output is not changing linearly with the energy deposition but 
instead it can be described as \cite{geant_manual,geant4_manual,birk_leo,birk_internet}:

\begin{equation}
\frac{dL}{dx} = \frac {k \cdot \frac{dE} {dx} }{\left[ 1 + c_{1} \cdot \frac{dE}{dx} + c_{2} \cdot \left(\frac{dE}{dx}\right)^{2}\right]} ~, 
\label{birk_equation}
\end{equation}
%\vspace{0.4cm}
%
This law describes the light output of (organic) scintillators \cite{birk_internet,birk_leo}.
The "k" constant depends on the particle type, and is in the order of 0.01 g~cm$^{2}$~MeV$^{-1}$, and    
the parameters $c_{1}$ and $c_{2}$ are equal to \cite{annaferrari2,non_linear_Birk,anna_ferrari_private_communication}: \\ 
\vspace{-0.4cm}
\begin{eqnarray}
\nonumber
  c_{1} &=& 0.013 ~ \text{cm}^{2}~\text{MeV}^{-1} ~, \\ 
\nonumber
  c_{2} &=& 9.6 \times 10^{-6} ~ \text{cm}^{3}~\text{MeV}^{-2} ~. 
\end{eqnarray}
%
%
%
% Simulation of the energy read-out: \\
%
% The light is propagated by hand at the end of the fiber using the parametrization:
%
% a) Kuraray
%
% \begin{equation}
% E_{a,b}^{(fib)} = E_{(dep)} \cdot [0.35 \cdot e^{-\frac{x(a,b)}{50}} + (1 - 0.35) \cdot e^{-\frac{x(a,b)}{430}}]     
% \end{equation}
%
%b) Politech
%
% \begin{eqnarray}
%\nonumber
%E_{a,b}^{(fib)} = E_{(dep)} \cdot [0.35 \cdot e^{-\frac{x(a,b)}{50}} + (1 - 0.35) \cdot e^{-\frac{x(a,b)}{330}} ]\\
%\nonumber
%t_{a,b}^{(fib)} = t_{(dep)} + \frac{X_{(a,b)}}{17.09}
%\end{eqnarray}
%
% in exponent function is attenuation length parameter \cite{annaferrari}.
%
% Next we would like to show a schematic view of the particle which goes through a scintilating material 
% (see Fig.~\ref{energy_deposition_schematic}).
%
% \begin{figure}[h]
% \hspace{4.5cm}
% \parbox[c]{0.275\textwidth}{\includegraphics[width=0.295\textwidth]{particle_track.eps}}
% \caption{The energy deposition in fiber.}
% \label{energy_deposition_schematic}
% \end{figure}
%
%%%%%%%%%%%%%%%%%%%%%%%%%%%%%%%%%%%%%%%%%%%%%%%%%%%%%%%%%%%%%%%%%%%%%%%%%%%%%%%%%%%%%%%%
%%%%%%%%%%%%%%%%%%%%%%%% 24 modules implementation  %%%%%%%%%%%%%%%%%%%%%%%%%%%%%%%%%%%%
%%%%%%%%%%%%%%%%%%%%%%%%%%%%%%%%%%%%%%%%%%%%%%%%%%%%%%%%%%%%%%%%%%%%%%%%%%%%%%%%%%%%%%%%
\section{Implementation of the whole geometry of the barrel calorimeter in FLUKA}
\begin{figure}[H]
\parbox{1.0\textwidth}{
\hspace{0.0cm}
\includegraphics[angle=-90,width=0.49\textwidth]{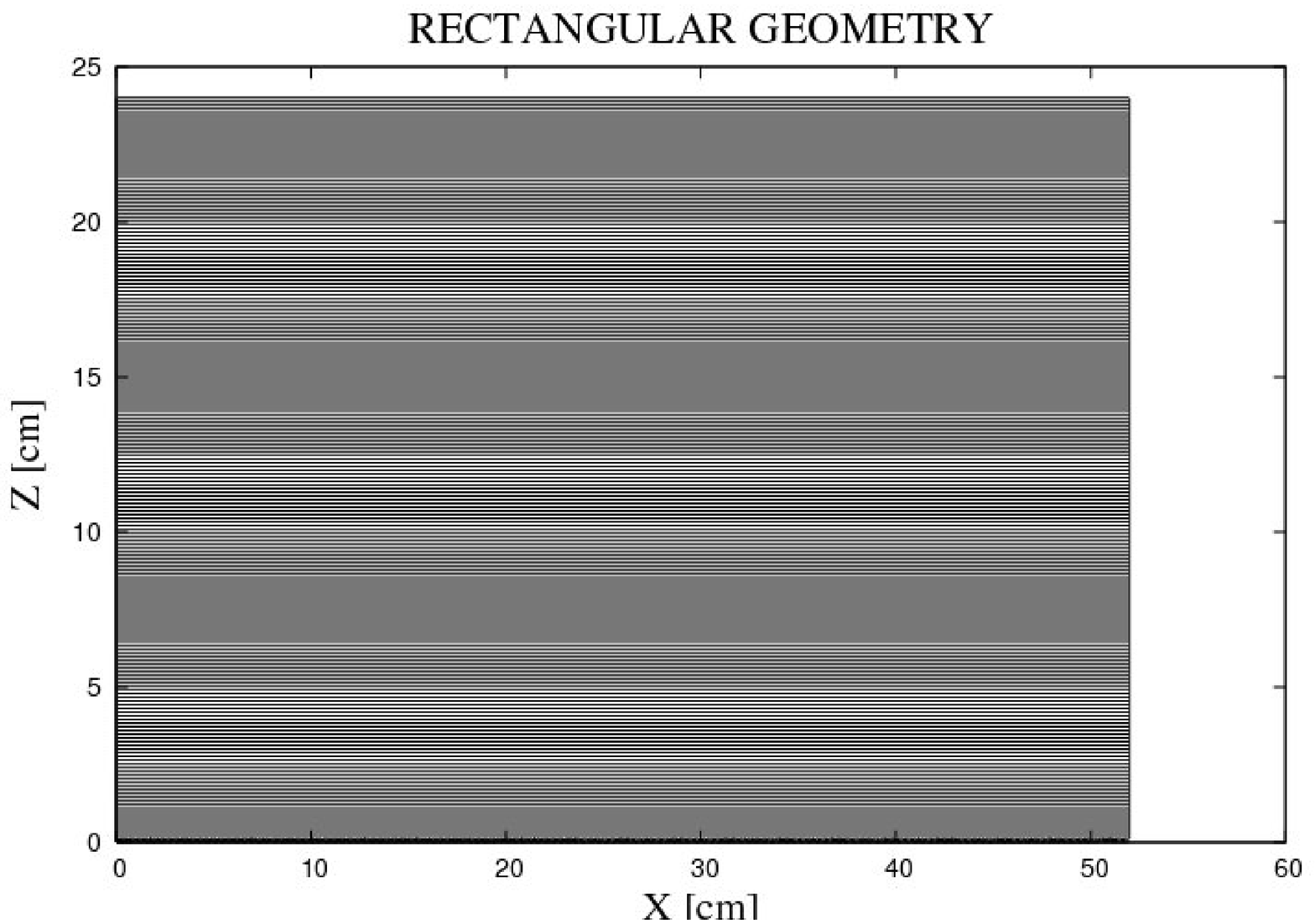}
\includegraphics[angle=-90,width=0.49\textwidth]{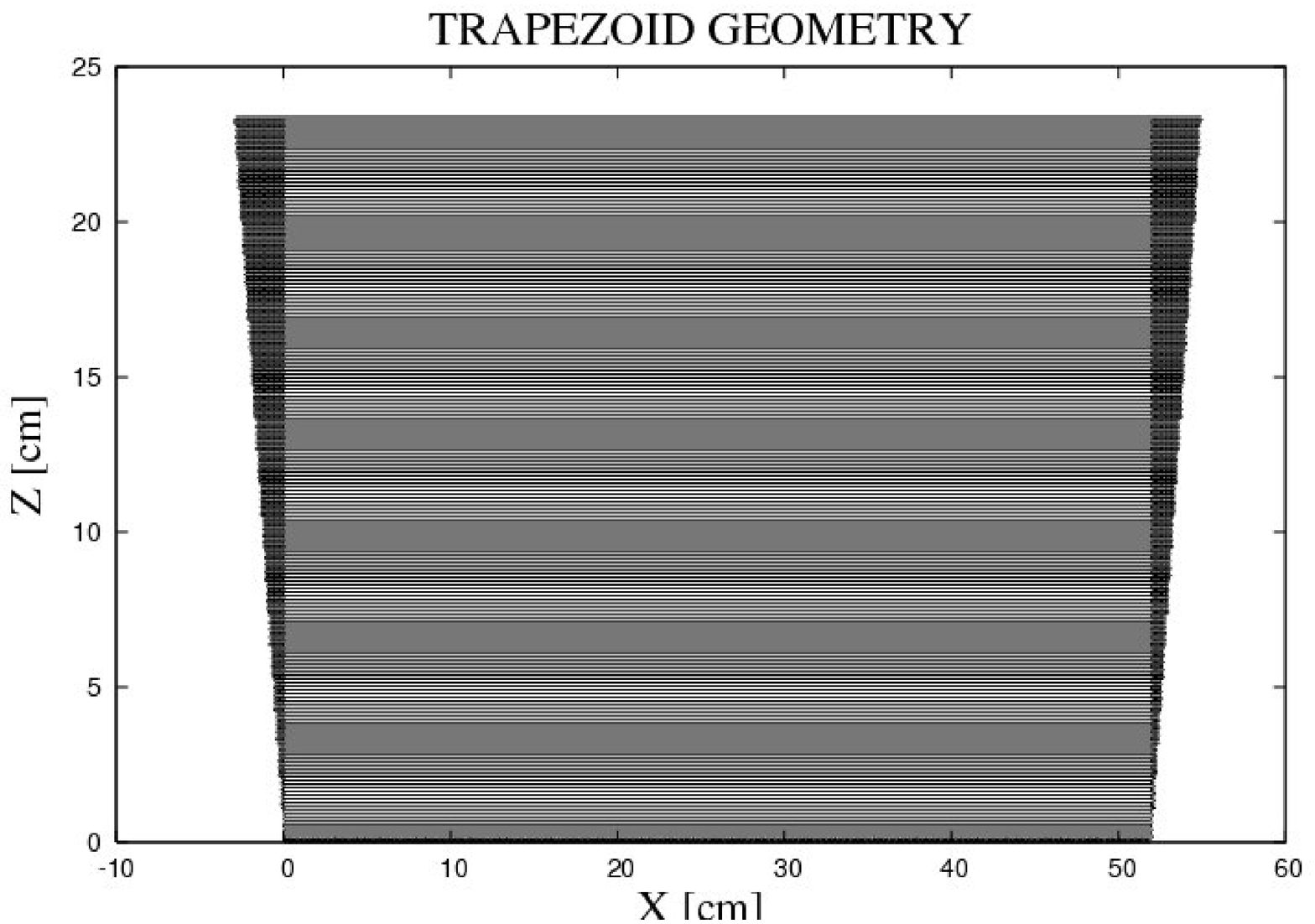}
}
\caption{Cross section of the EmC module with rectangular and trapezoid geometry. 
Visualization by the FLAIR program \cite{FLAIR_homepage} using as an input 
a geometry setup files of FLUKA.}
\label{1module_geometry_rectangular_trapezoid}
\end{figure}

\begin{figure}[H]
\hspace{3.0cm}
\parbox[c]{0.55\textwidth}{\includegraphics[width=0.59\textwidth]{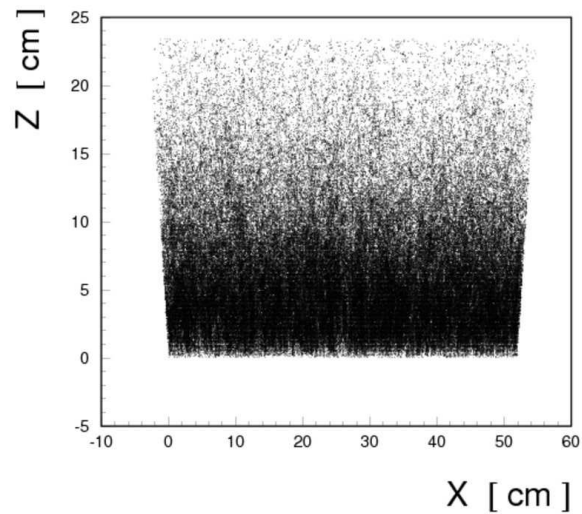}}
\caption{Distribution of energy deposits in scintillating fibers of a single unit of the electromagnetic calorimeter.}
\label{trapez_energy_deposits}
\end{figure}

\begin{figure}[H]
\hspace{4.1cm}
\parbox[c]{1.0\textwidth}{\includegraphics[width=0.49\textwidth]{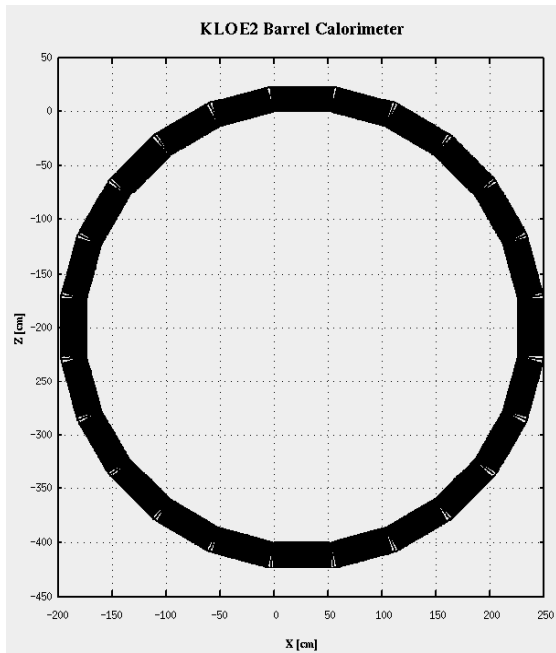}}
\caption{Visualization of the 24 trapezoid modules of the barrel calorimeter with FLAIR.}
\label{24_modules_FLAIR}
\end{figure}

\begin{figure}[H]
\hspace{0.0cm}
\parbox[c]{1.0\textwidth}{
\parbox[c]{0.5\textwidth}{\includegraphics[width=0.50\textwidth]{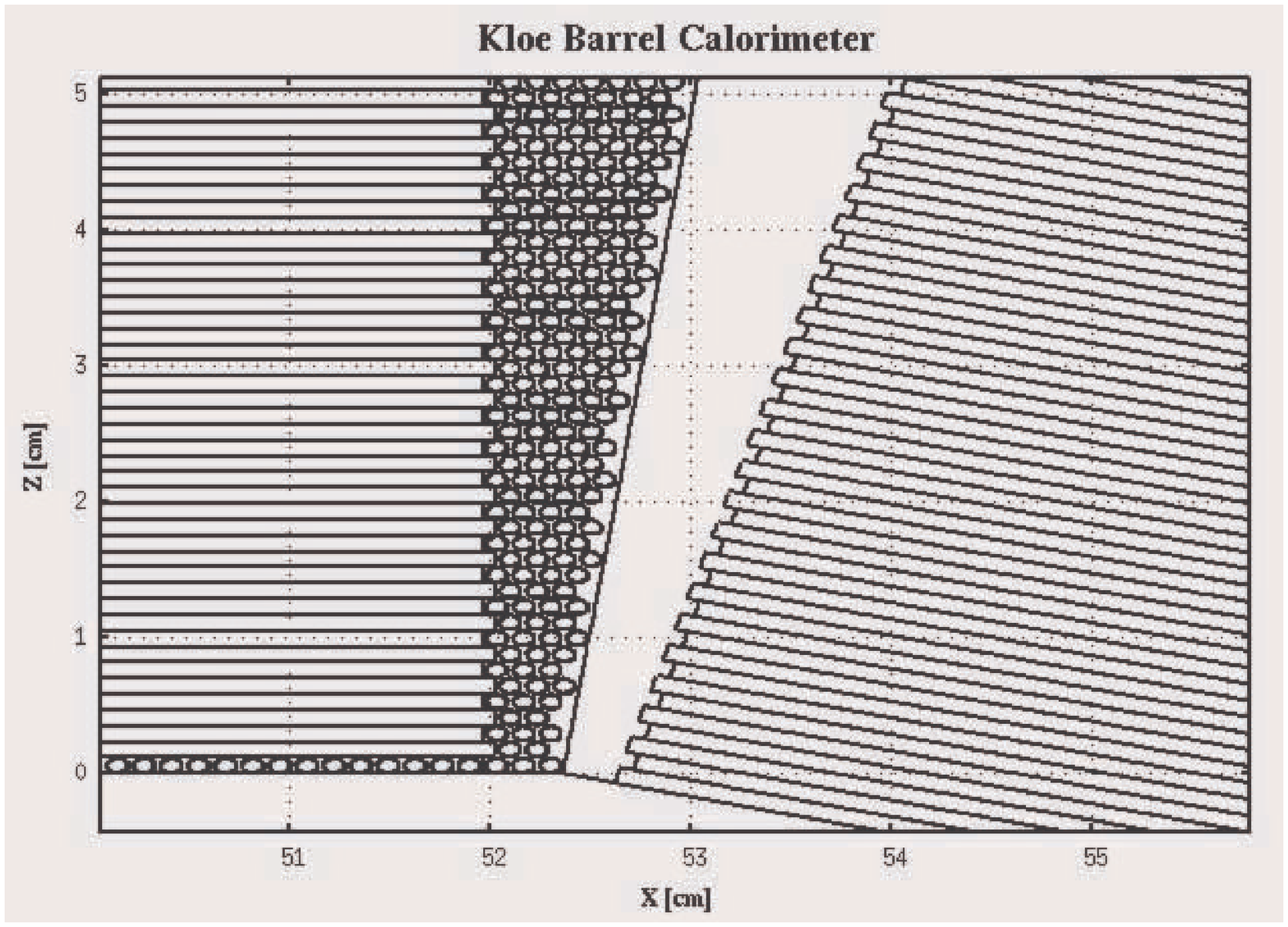}}
\hspace{+0.1 cm}
\parbox[c]{0.5\textwidth}{\includegraphics[width=0.55\textwidth]{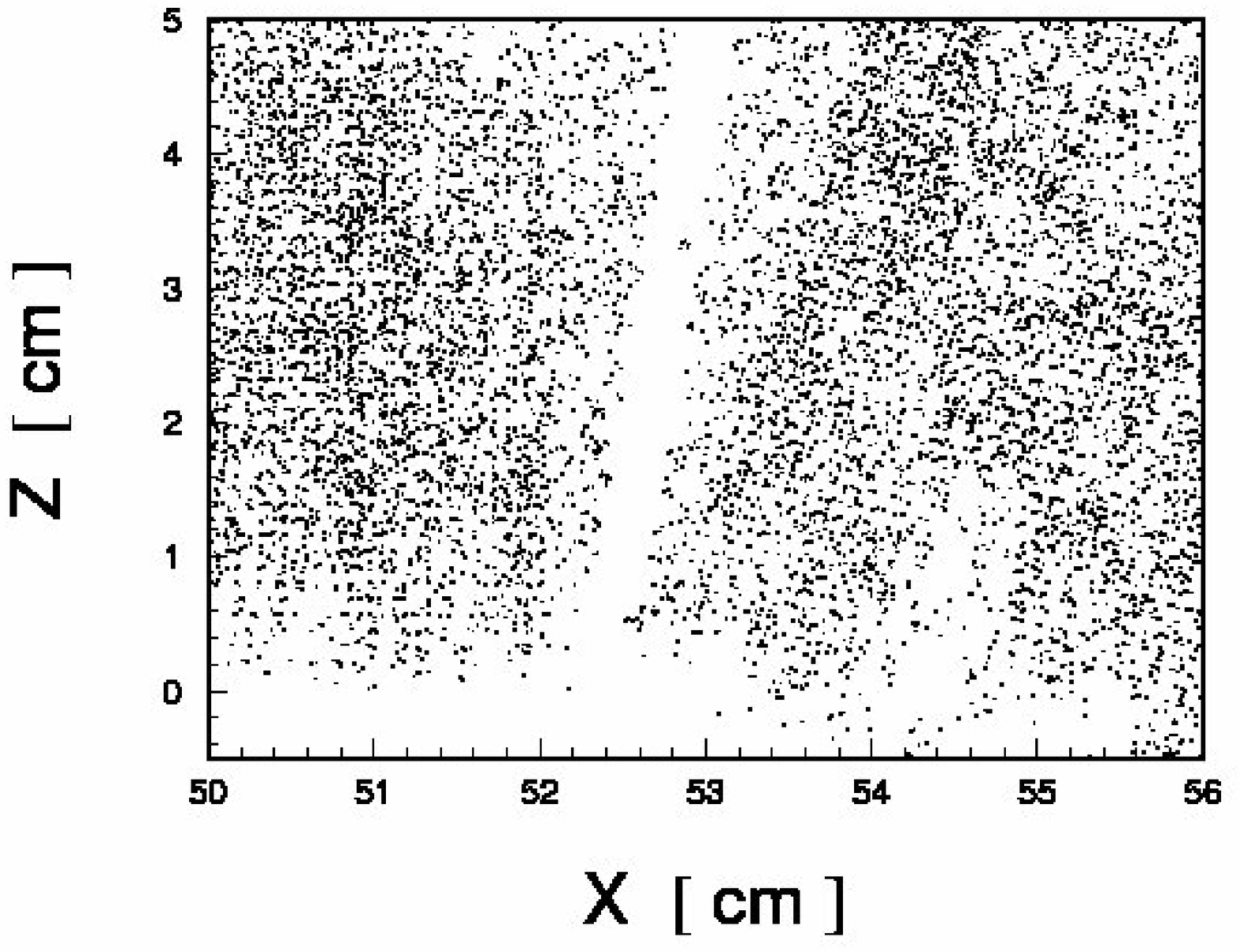}}
}
\caption{Details of the implementation on the edge of two modules.}
\label{24_details_of_implementation}
\end{figure}

\hspace{\parindent}
FLUKA simulation package is mostly used in low energy physics, for example in: 
medical science, dose evaluation, neutriono beam 
simulation and for simulations of the low energy hadronic interactions. 
Therefore, the program is especially suited for simulations of the response of the 
KLOE electromagnetic calorimeter, because at KLOE experiment we deal with particles in the rather 
low energy range up to 1 GeV. \\ 
\indent The main idea for implementation of the geometry of the KLOE detector with FLUKA 
was proposed by Giuseppe Battistoni \cite{Giuseppehomepage}.   
%We decided to implement the whole barrel calorimeter geometry with FLUKA Monte Carlo because
%with this tool we are able to achieve the more accurate description of the detector physics 
%on the KLOE detector with its particle spectra up to 1 GeV. \\
%
It was first realized for one calorimeter module with rectangular shape\footnote{Implementation to FLUKA package was done
 by: G.~Battistoni, B.~Di~Micco, A.~Ferrari, A.~Passeri and V.~Patera.}, with accurate 
description of materials. 
The base cell with dimensions of 52~cm~$\times$~1.2~mm~$\times$~430~cm consists of the lead block filled with 385 scintillating fibers and 
glue cylinders. 
%
%
%385 fibers and glue cylinders, and one lead block 
This cell was replicated 199 times (using FLUKA lattice tool) to build 200 layers forming the calorimeter module 
(see left panel in Fig.~\ref{1module_geometry_rectangular_trapezoid}). \\
\indent Our first step in the way to implement the whole barrel of electromagnetic calorimeter into FLUKA was the extension of the 
rectangular module to the realistic trapezoid shape.   
%
%
% The first step to implement the whole of the barrel calorimeter geometry was the implementation of the
% one calorimeter module with trapezoid shape. 
%
This was achieved by declaration of a new base cell consisting of a scintillating fiber and glue cylinder both inserted into 
a small block of lead. The length of the cell is equal to the length of the module (4.3~m) and its cross section 
corresponds to the square of 1.2~mm, where the fiber diameter is equal to 1~mm and width of the glue cylinder equals to 0.1~mm. 
% with one
%fiber and cylinder and small lead block which has got width and hight equal a half distance beetwen fibers
%on the z axis and on the x axis (Fig.~\ref{coordinate_system}). 
%
Then this new base cell was replicated about 4000 
times to build two triangular sections on the left and right side of the module.   
A complete visualisation of the new trapezoid 
geometry\footnote{Visualization is based on the FLUKA sources and prepared 
using a FLAIR (FLUKA advanced visualization geometry tool) \cite{FLAIR_homepage}.}
 is shown in Fig.~\ref{1module_geometry_rectangular_trapezoid}.  
% A working geometry with energy deposits response of th one trapezoid calorimeter is shown on  \ref{trapez_energy_deposits}.
%
%
%\hspace{-1.5cm}
%\mbox{
%\parbox{.38\textwidth}{
%\begin{figure}[H]
%\parbox{0.38\textwidth}{\includegraphics[width=0.43\textwidth]{rectangular.eps}}
%\caption{1 EMC module with rectangular geometry.}
%\label{1module_geometry_rectangular}
%\end{figure}
%}
%}
%\hspace{0.5cm}
%\parbox{.38\textwidth}{
%\begin{figure}[H]
%\parbox{0.38\textwidth}{\includegraphics[width=0.43\textwidth]{1trapes.eps}}
%\caption{1 EMC module with trapezoid geometry.}
%\label{1module_geometry_trapezoid}
%\end{figure}
%}
%}
%
\vspace{-0.3cm}
\vspace{-0.3cm}
%
%\hspace{0.5cm}
%\parbox{0.53\textwidth}{
%\begin{figure}[H]
%\includegraphics[angle=-90,width=0.59\textwidth]{trapez.eps}
%\caption{One EMC module with trapezoid geometry.}
%\label{1module_geometry_trapezoid}
%\end{figure}
%}
%}
%
%
%\begin{figure}[H]
%\hspace{0.5cm}
%\parbox[c]{1.0\textwidth}{
%\parbox[c]{0.43\textwidth}{\includegraphics[width=0.49\textwidth]{rectangular.eps}}
%\parbox[c]{0.43\textwidth}{\includegraphics[width=0.49\textwidth]{1trapes.eps}}
%}
%\caption{24 modules - details of implementation}
%\label{1module_geometry}
%\end{figure}
%
In order to raise our confidence to the functioning of the programs we have simulated energy deposits in a single calorimeter module 
for the reaction:
%
% The energy deposits in scintilating fibers for module with trapezoid geometry and for the whole barrel calorimeter 
% was simulated with vertex generator with 
$e^{+}e^{-} \to \phi \to \eta\gamma \to 3 \pi^{0}\gamma \to 7 \gamma$.
The results of these simulations for 20000 events are shown in Fig.~\ref{trapez_energy_deposits}. 
% One event is represented by one $\phi$ meson decay to the 7$\gamma$ quanta in the final state of reaction.
% A trapezoid shape of the module can be easily recognized.    
%
\vspace{-0.4cm}
A trapezoid structure which was built with a realistic light output from the 
scintillating fibers can be easily recognized. One can also see that as expected the most energy 
deposits are observed in the bottom part of the module.  
As a next step, it would be natural to replicate 23 times the defined geometry for one module and to build a full 
barrel of the electromagnetic calorimeter.  
%However, it was not possible to build the whole barrel calorimeter with a one trapezoid calorimeter module which would be replicated
%(using lattices tool) 23 times to reconstruct 24 modules of the barrel calorimeter. 
This task was unfortunatelly impossible to realize with present
version of FLUKA\footnote{FLUKA 2006 for GNU linux operating system.} \cite{FLUKA_homepage} because this version doesn't permit 
to replicate the   
region which had been replicated before. It is due to the fact that the lattice replication on the second 
and higher levels is not implemented yet. 
% 
% It is caused by fact that FLUKA 
%hasn't got implemented the lattice tool replication on the second level. 
This is the reason why geometry was built in another, unfortunatelly much more complicated way. \\   
%
% The main idea of this implementation was that we have built a one trapezoid region as a base cell and the rest 23 
% trapezoid containers are a replicated areas. The traezoid base cell consists of one layer filled with fibers and glue cylinders,  
%  and with lead material (the same as we used for implementation of the trapezoid module) but additionally we have built the whole 
% structure of fibers and glue cylinders for two triangular sections on the right and left side the previous rectangular module 
% \cite{zdebik_geometry_gm}. We filled these two triangular sections with lead material. The last part of the trapezoid base cell is 
% region which was built with 199 replicated layers in the previous rectangular section.  
%
\indent The barrel was first defined as 24 empty volumes (containers). In one of these volumes we implemented the base cell which described 
one layer of lead, fibers and glue. This cell was replicated 199 times in this volume, to build a rectangular section part of the main 
trapezoid module (on the top in Fig.~\ref{24_modules_FLAIR})\footnote{The same idea 
was used to build a single module with rectangular shape.}, and these base cell was replicated 200 times in each of the remaining 23 modules. 
Next in the first module we filled the full structure of lead, fibers and glue cylinder in the two triangle areas. These two regions were 
then replicated at the corresponding positions in the remaining modules \cite{zdebik_geometry_gm}. \\
%
% Each of the 23 replicated trapezoid containers 
% consists of 200 replicated layers and two triangular sections. 
%
% We used a FLUKA lattice tool to replicate the trapezoid base cells into 
% 23 empty containers. \\ 
%
%
\indent Fig.~\ref{24_modules_FLAIR} shows a visualisation of the whole barrel calorimeter geometry made with FLAIR~
\cite{FLAIR_homepage}. 
The left panel of the next figure presents  
details of the edge area, between base module on the top and the first 
replicated area on the right side of it. In the right panel of Fig.~\ref{24_details_of_implementation} the energy deposits in scintillating fibers for this area are 
presented for $e^{+}e^{-} \to \phi \to \eta\gamma \to 3 \pi^{0}\gamma \to 7 \gamma$ reaction with statistics of 1000 events.   
% (right part of this figure).
%    
Using this geometry we are able to study in details energy deposits also at the edges of the modules of the calorimeter,  
where there are small areas without fibers \cite{antonio_di_domenico, federico_nguyen}. 
% effects near boundary between of the each two modules of the KLOE calorimeter. \\
%
%
% The area in which particles didn't deposit energy exists between two modules. This effect is related with fact
% that fibers structure isn't exactly replicated to the edges of the modules (see left panel in the figure~\ref{edge_2tr}). 
% The same effect appears in the real world 
% for the real calorimeter modules \cite{antonio_di_domenico, federico_nguyen}. 
% In the left panel of figure~\ref{24_details_of_implementation} details of base cells and the first replicated lattice area are presented. \\ 
%The base cell consists of two sections with triangular shape and filled with fibers structure and lead material and one 
%(on the bottom) rectangular material layer
%with fibers. Other parts of the geometry were built as replicated areas by using FLUKA lattice tool. 
%Each replicated area is built with 200 replicated layers and 2 triangular
%shapes which are lattices areas too for each of the 23 trapezoid containers without the base container. \\
%
% \indent In the figure~\ref{24_energy_deposits} is shown a energy deposits in scintillating fibers
%  on the whole barrel calorimeter detector, in two (left panel) and three dimensions (right panel).
%
\begin{figure}[H]
\hspace{-0.2cm}
\parbox[c]{1.00\textwidth}{
\parbox[c]{0.55\textwidth}{\includegraphics[width=0.55\textwidth]{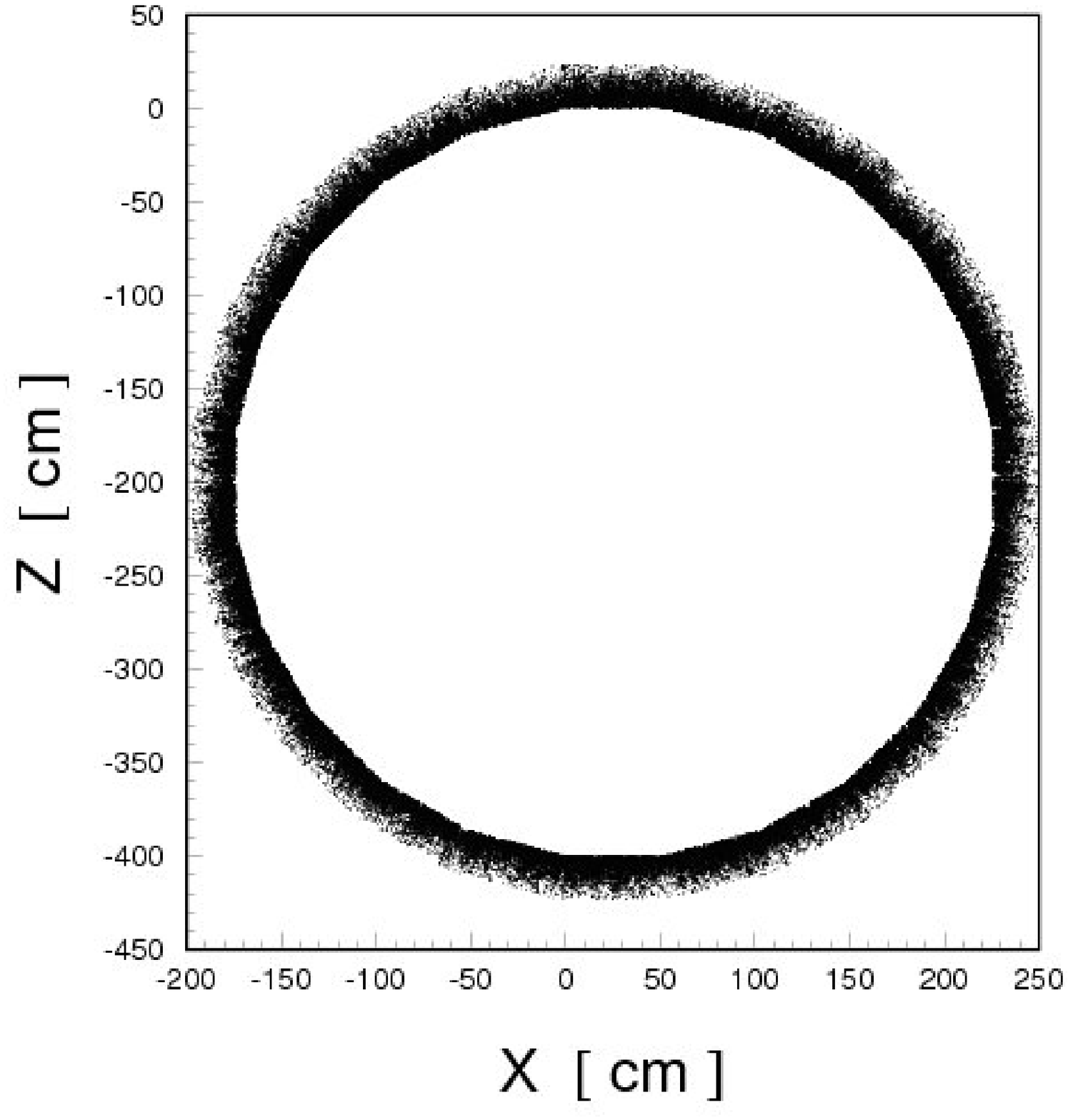}}
\hspace{-0.9 cm}
\parbox[c]{0.55\textwidth}{\includegraphics[width=0.55\textwidth]{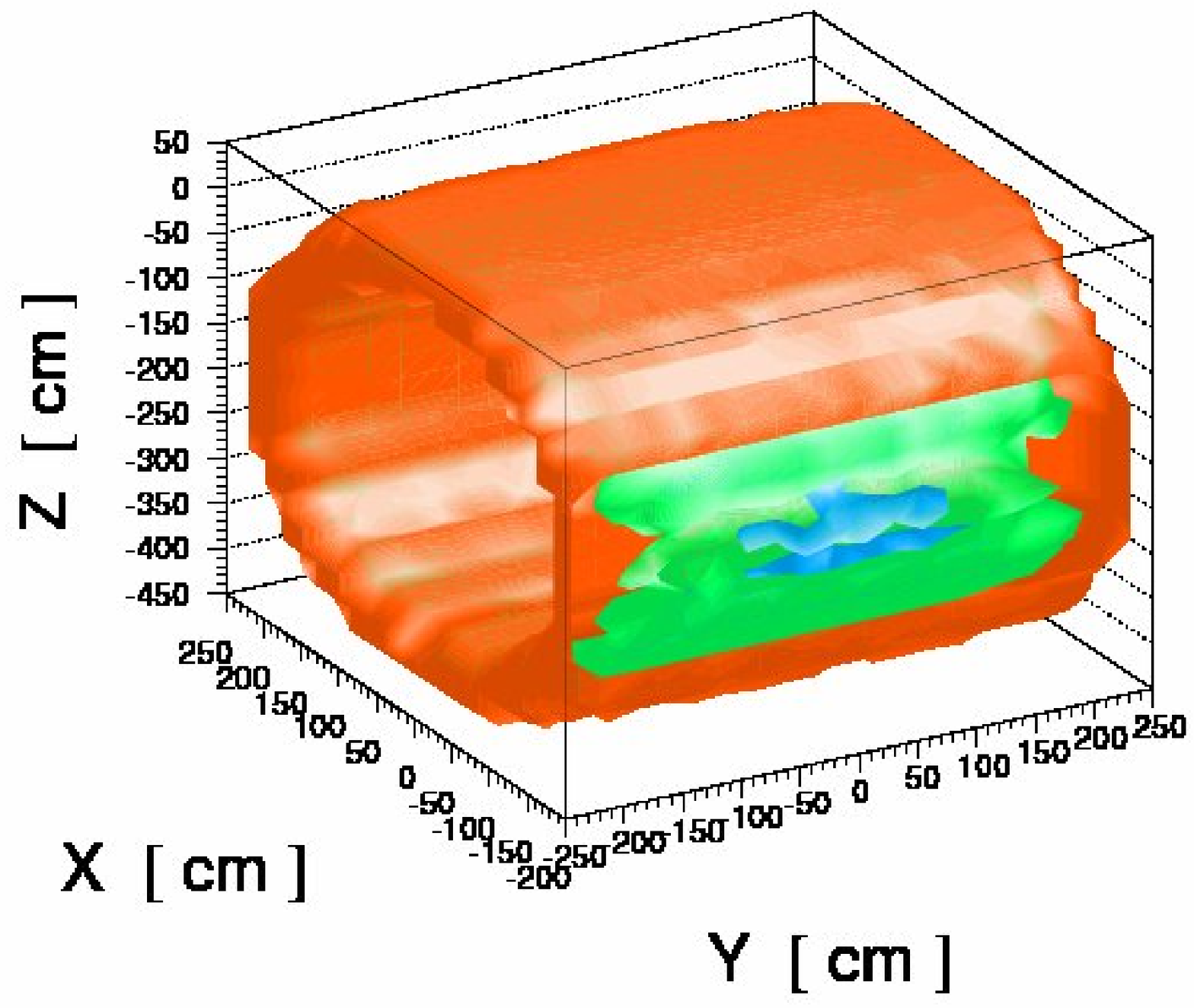}}
}
\caption{Energy depositions in fibers in the KLOE barrel calorimeter.}
\label{24_energy_deposits}
\end{figure}
\indent Finally in Fig.~\ref{24_energy_deposits}, as an example of the proper implementation of the geometry we show energy deposits 
in scintillating fibers in the whole barrel calorimeter.  

%%%%%%%%%%%%%%%%%%%%%%%%%%%%%%%%%%%%%%%%%%%%%%%%%%%%%%%%%%%%%%%%%%%%%%%%%%%%%%%%%%%%%%%%%%%%%%%%%%%%%%%%%%%%
%%%%%%%%%%%%%%%%%%%%%%%%%%%%%%%%%%%%%%%%   DIGICLU  DESCRIPTION %%%%%%%%%%%%%%%%%%%%%%%%%%%%%%%%%%%%%%%%%%%%
%%%%%%%%%%%%%%%%%%%%%%%%%%%%%%%%%%%%%%%%%%%%%%%%%%%%%%%%%%%%%%%%%%%%%%%%%%%%%%%%%%%%%%%%%%%%%%%%%%%%%%%%%%%%
%
%
\section{Simulations of the photomultipliers response}
\begin{figure}[H]
\hspace{4.2cm}
\parbox[c]{1.0\textwidth}{
%\parbox[c]{0.3\textwidth}{\includegraphics[width=0.22\textwidth]{trapes_schemat.eps}}
%\parbox[c]{0.4\textwidth}{\includegraphics[width=0.49\textwidth]{trapes_fotop_schemat.eps}}
% \includegraphics[angle=45,width=52mm]{myfig.eps}
%
\includegraphics[angle=-90,width=0.43\textwidth]{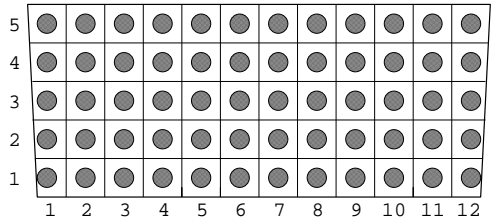}
}
\caption{Schematic view of the readout cells structure on the one side of the barrel module. Filled circles represent photomultipliers.}
\label{trapes_fotop_schemat}
\end{figure}

\hspace{\parindent}
After simulation of energy deposits in scintillating fibers with FLUKA program  
we calculated amplitude of signals in photomultipliers using DIGICLU program. 
This program also enables to reconstruct the time, position and energy for  
all readout segments (cells) of each calorimeter module \cite{kloe_electromagnetic_nim2}. 
  Each module is divided into 60 cells 
which are situated in 5 layers and 12 columns (see Fig.~\ref{trapes_fotop_schemat}). 
Each cell is read out on both sides by  
photomultiplers. This segmentation provides the determination of the position 
of energy deposits in the calorimeter.  
% Each endcap is built of 32 modules, in this thesis we will concentrate on the barrel only. \\    
Altogether, the barrel calorimeter consists of 1440 cells, which are read out by photomultiplayers on two sides 
(referred to as side A and side B in the following). 
% The structure of cells on the one side of the one module 
% is presented in the figure~\ref{trapes_fotop_schemat}.
%
%
%
% 
%
% Each black point represents one photomultiplier. \\ 
\indent For each cell 
two time signals $T^{A}$, $T^{B}$ (digitized by the Time to Digital Converter (TDC)) and two amplitude signals $S^{A}$, $S^{B}$   
(measured by Analog to Digital Converter (ADC)) are recorded. 
% from the corresponding PM's signals.
% The longitudinal position (parallel to beam axis) of the energy deposit is obtained from the difference $t^{A} - t^{B}$.
%
% charges $Q^{A,B}_{ADC}$ 
% and times $t^{A,B}_{TDC}$,
The arrival time $t$ and position $s$ of the impact point along the fiber direction (the zero being taken at the fiber center)
 is calculated with the aid of
 times measured at two ends as:
\begin{eqnarray}
  t (ns) & = & \frac{1}{2} (t^{A} + t^{B} - t^{A}_{0} - t^{B}_{0}) - \frac{L}{2v} ~, \\
  s (cm) & = & \frac{v}{2} (t^{A} - t^{B} - t^{A}_{0} + t^{B}_{0}) ~,
\end{eqnarray}
with $ t^{A,B} = c^{a,b} \cdot T^{A,B} $, where $ c^{A,B} $ are the TDC calibration constants, 
$ t^{A,B}_{0} $ denotes overall time offsets,    
 L stands for length of the cell (cm) and  
v is the light velocity in fibers (cm/ns). 
%  
%
%where: \\
%$ c^{A,B} $ are the TDC calibration constsnts, \\
%$ t^{A,B}_{0} $ are overall time offsets \\
%L is length of the each cell (cm) \\
%v is the light velocit in fibers (cm/ns). \\
%$t_{0}$ is being taken at the fiber center \\ 
%
The energy signal is calculated according to the formula:
\begin{eqnarray}
E_{i}^{A,B}(MeV) = \frac{S_{i}^{A,B} - S_{0,i}^{A,B} }{S_{M,i}} \cdot K ,  
\end{eqnarray}
where 
$ S_{0,i} $ are the zero-offsets of the amplitude scale,    
$ S_{M,i} $ corresponds to the response for the minimum ionizing particle crossing the calorimeter center and 
K factor gives the energy scale in MeV, and it is obtained from signals of particles with known energy.  

The total energy deposited in a cell is calculated as the mean of values determined at both ends for each cell \cite{kloe_electromagnetic_nim2}.
\noindent
The energy read-out has been simulated by including both the generation of photoelectrons with the Poisson 
distribution and the threshold on the constant fraction discriminator.
% threshold was using either.
%
%\begin{itemize}
%\item the generation of photoelectrons with the Poisson distribution 
%\item the constant fraction distribution 
%\item the discriminator threshold \cite{annaferrari}. 
%\end{itemize}
%
%
%

\section{Description of the KLOE clustering algorithm}
\hspace{\parindent}
As we presented in the previous section the DIGICLU program simulates photomultipliers response 
and reconstructs energy deposit, position and time for particles passing through each cell. 
These values are used to recognize groups of cells (clusters) belonging to particles 
entering the calorimeter. For this aim a clustering algorithm is used \cite{kloe_electromagnetic_nim2}. 
Ideally, to each particle it should assign exactly one cluster but in practice it is not always the 
case. After the recognision of clusters the program reconstructs the spatial coordinates and time of 
each shower with high accuracy,
needed to reconstruct the decay vertex of the $K_{L}$ \cite{KIM_MEMO}. 
In particular the algorithm is based on the following steps. First for each cell the position and energy 
of the shower is reconstructed. Next preclusters are built by connecting the neighboring cells 
in time and space in order to recreate a full shower \cite{kloe_electromagnetic_nim2}. 
Cells are taken into account in searching of preclusters only if times and energy signals are available on
both sides, otherwise these cells in the most cases are added to the already recognized clusters \cite{Biagio_thesis}. 
 Subsequently, preclusters are splitted if the spread of the time of the assigned cells is larger than 2.5 ns. 
On the other hand cells are merged in one cluster if a distance between them and the center of the precluster is less than 20~cm.  
After this check the groups of cells are defined as clusters which position and time are computed as energy-weighted averages 
of the contributing cells.

\chapter{Adjustment of the detector properties in simulations}
\hspace{\parindent}
%
% adjustment - dopasowywanie, korekta
%
In order to simulate a realistic response of photomultipliers we have adjusted parameters 
used in the programme by comparing the output of simulations to results from the experimental 
sample of events identified as $\phi \to \eta\gamma \to 3\gamma$ process. 
\vspace{-0.2cm}
%
%
%
%in the clustering
%program.  
%
%The first step to achieve this purpose was was selection of the KLOE experimental data sample.
%
%%%%%%%%%%%%%%%%%% MC VALIDATION STUDIES - zostaja podpiete do rozdzialu DIGICLU %%
% \chapter{MC validation studies}
%
% DATA SAMPLE
%

\section{Preparation of the data sample ($e^{+}e^{-} \to \phi \to \eta \gamma \to 3\gamma$)}
\hspace{\parindent}
At a first stage, an experimental data sample for the $e^{+}e^{-} \to \phi \to 3\gamma$ reaction 
has been extracted \cite{biagio} applying 
the kinematic fit procedure to the 3 $\gamma$ events with the following conditions: 
%\hspace{\parindent}
%\hspace{1.5cm} 1. $\sum E_{\gamma} = E_{\phi} $ \\
%\hspace{1.5cm} 2. $\sum p_{\gamma} = p_{\phi} $ \\
%\hspace{1.5cm} 3. $ t_{0} - \frac{r}{c} = 0 $  \\ 
%\hspace{1.5cm} 4. E1 < E2 < E3 \\
%
%
\begin{eqnarray}
\nonumber
\sum_{i=1}^{3} E_{\gamma_{i}} & = & E_{\phi} ~,\\ 
\nonumber
\sum_{i=1}^{3} p_{\gamma_{i}} & = & p_{\phi} ~,\\
\nonumber
t_{i} - \frac{r_{i}}{c} & = & 0 ~,\\
% \nonumber
% E1~ <~ E2 & < & E3
\end{eqnarray}
 where, 
 c is the light velocity,       
 $ t_{i} $ denotes the time of each reconstructed cluster and  
 r$_{i}$ stands for the distance from the vertex collision point to the cluster centroid position. 
$E_{\phi}$ and $p_{\phi}$ denote the total energy and momentum of the $\phi$ meson, respectively. 
The first and second condition results from energy and momentum conservation rules. 
The third requirement ensures that we take events which orginate only from the collision point. 
% 
%The next step in the analysis was to make cut on Dalitz plot from Kloe experimental data sample
%for the channel $ e^{+}e^{-} \to \phi \to \eta \gamma \to 3 \gamma $
%
%
\vspace{-0.4cm}
\begin{figure}[H]
\hspace{0.2cm}
\parbox[c]{1.0\textwidth}{
\parbox[c]{0.50\textwidth}{\includegraphics[width=0.44\textwidth]{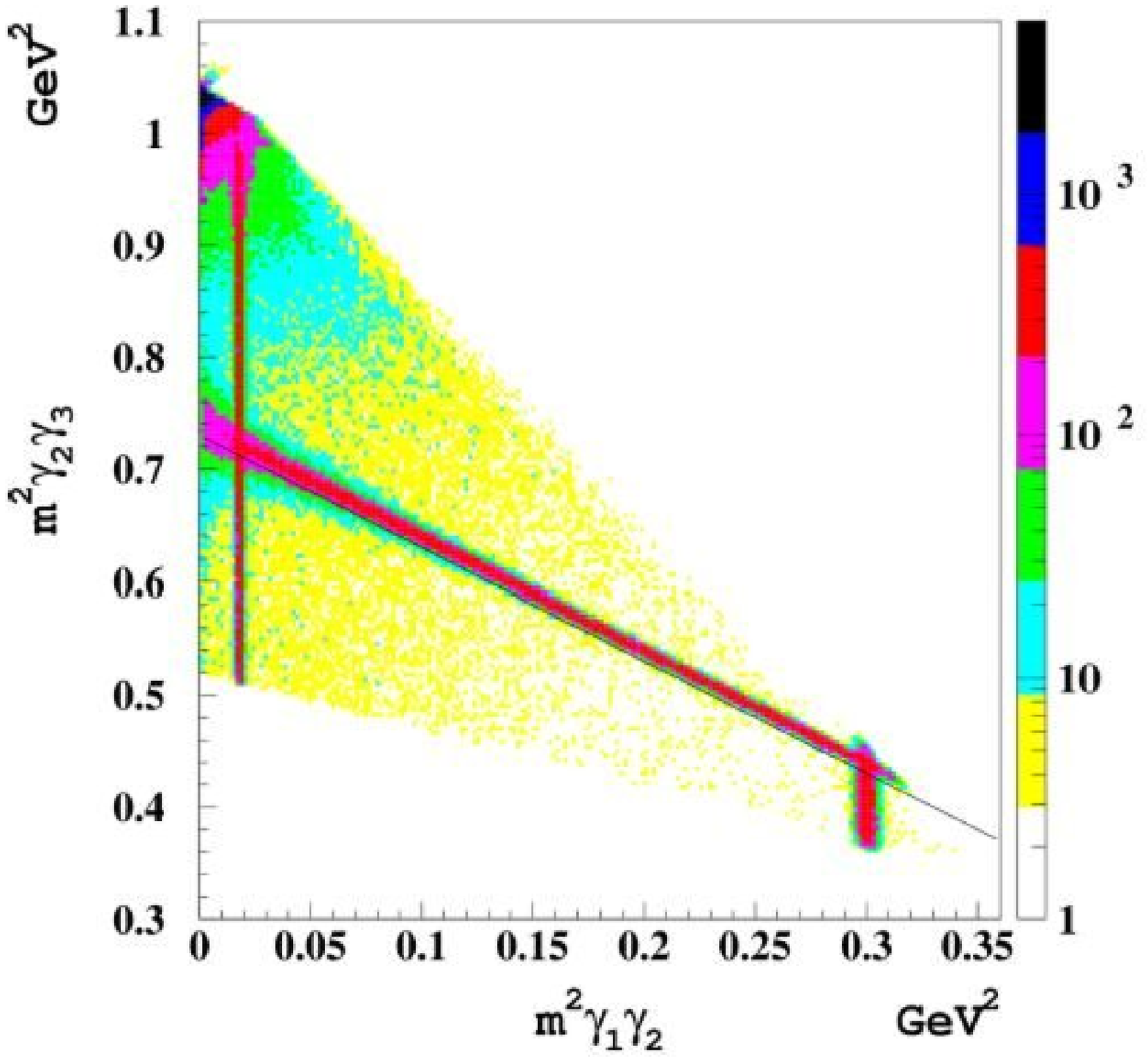}}
\parbox[c]{0.50\textwidth}{\includegraphics[width=0.49\textwidth]{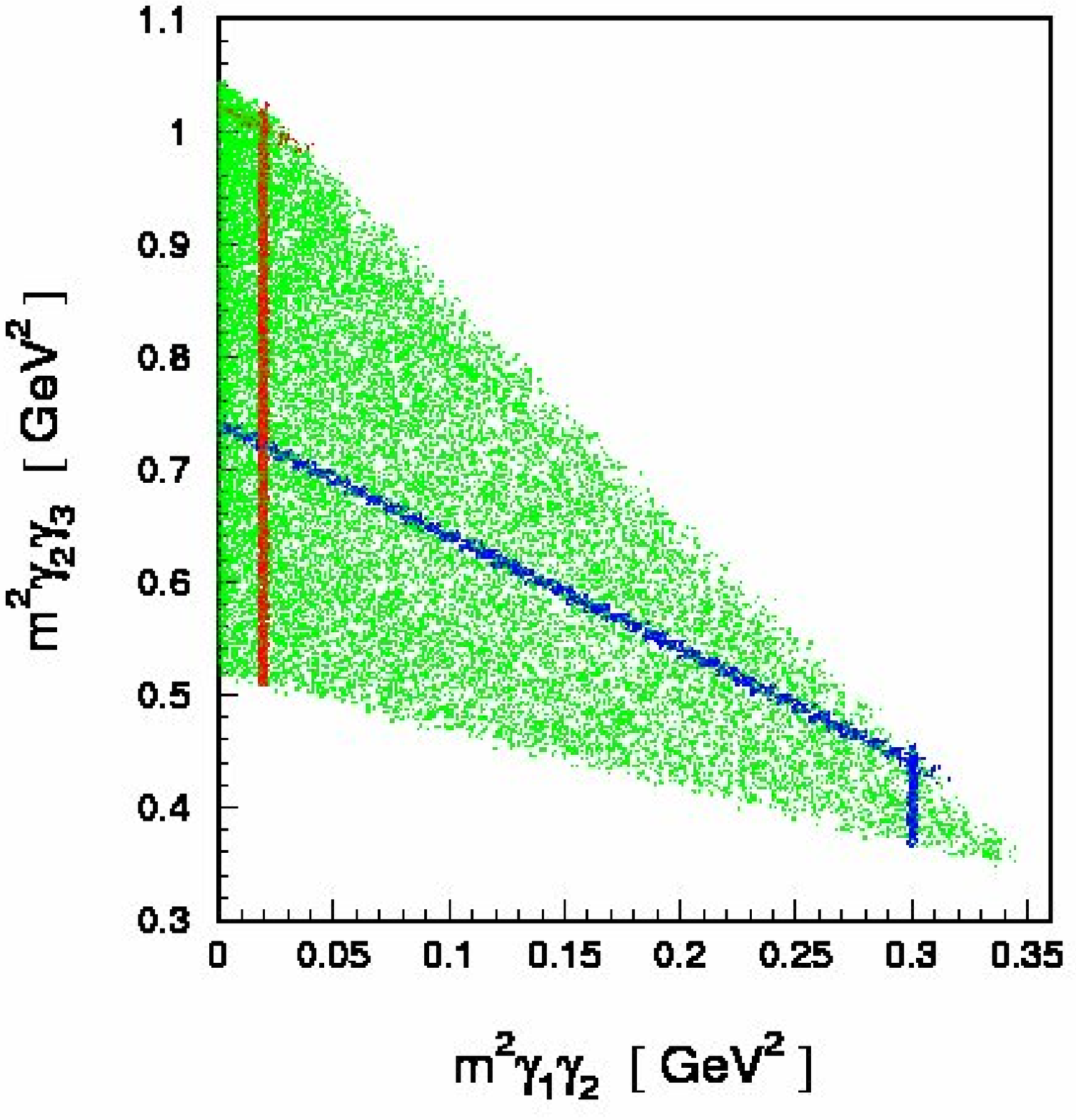}}
}
\caption{The distribution of the square of the invariant mass for $\gamma$ pairs from $ e^{+}e^{-} \to \phi \to 3 \gamma $ reaction         
from the experiment (left panel \cite{biagio}) and from the 
simulation (right panel).}
\label{biagio_data_sample}
\end{figure}
\indent As a next step, the distributions of the square of the invariant mass for pairs of the $\gamma$ quanta for 
$e^{+}e^{-} \to \phi \to 3\gamma$ process have been constructed. The results are shown in the 
left panel of Fig.~\ref{biagio_data_sample}. The numbers were assigned to the $\gamma$ quanta such  
that they are ordered according to the increasing energy: $E_{1} < E_{2} < E_{3}$. 
Thus $\gamma_{3}$ is the photon with the highest energy and $\gamma_{1}$ 
with the lowest. The right panel presents results of simulations for the following reactions: 
%
%Using this generator we were able to simulate a detector response for a following reactions:
\begin{enumerate}
\item $e^{+}e^{-} \to \phi \to \eta \gamma \to 3\gamma$ ~,
\item $e^{+}e^{-} \to \phi \to \pi^{0} \gamma \to 3\gamma$ ~,
\item $e^{+}e^{-} \to \phi \to 3\gamma$ ~.
\end{enumerate}
The blue bevel line and the vertical band at the $\eta$ mass squared denote signals from the first reaction. 
The channel where $\phi$ meson decays to $\pi^{0}\gamma$ is described by the red line with value of m$^{2}\gamma_{1}\gamma_{2}$ equal 
to 0.018 GeV$^{2}$. 
And finally the green "triangular" region orginates from reactions where $\phi$ meson decays directly to 
3$\gamma$ quanta. \\ 
\indent For the further analysis we took all events from the vertical $\eta$ band which are positioned below the red bevel line 
(see left panel  
in Fig.~\ref{biagio_data_sample}) \cite{biagio}. 
%
% On the right side are presented results for the same reaction but 
% achieved with simulations with the vertex generator.
% We took only this part from the all events \cite{biagio} in order to take into account in analysis process only 
% two $\gamma$ quanta from $\eta$ meson decay \cite{biagio}. One can see in for this reaction 
% (the blue, bevel line) gamma quanta could orginate from both $\phi$ and $\eta$ meson decays, due to this fact 
% we aren't able to distinguish from which decay each $\gamma$ orginates from. 
% That's a reason why we made cut on this events and took only the events below this line. 
%
%
%
%
%----------------------------------------
% ATTENUATION LENGTH
%
\section{Determination of the simulation parameters}
\hspace{\parindent}
%
%In this section we will describe parameters which were calibrated in our
%simulations.
% In order to perform a realistic simulation of the photomultipliers response and reconstuction of the 
% particles tracks in the module, 
In this section we describe determination of the attenuation length for the scintillating fibers, determination of the
 threshold  
function, and calibration of the ionization deposits in the electromagnetic calorimeter. 

\subsection{Estimation of the attenuation length of the scintillating fibers}
\hspace{\parindent}
\vspace{-0.3cm}
%
% estimation of the attenuation length parameter value, subsequently this new value was implemented 
% to the DIGICLU program source code \cite{erykbiagio}. 

\begin{figure}[H]
\parbox[c]{1.0\textwidth}{
\parbox[c]{0.33\textwidth}{\includegraphics[width=0.39\textwidth]{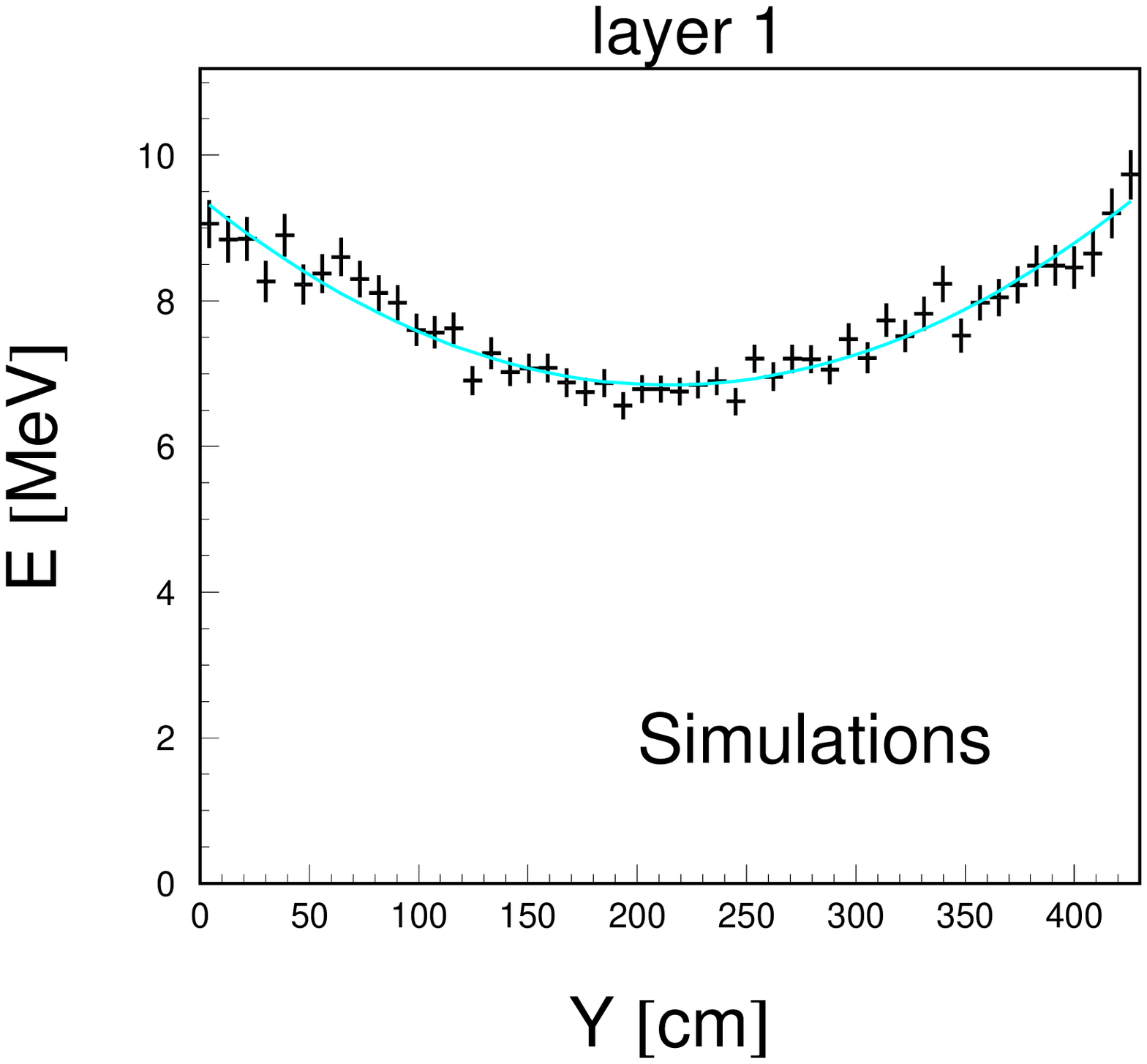}}
\parbox[c]{0.33\textwidth}{\includegraphics[width=0.39\textwidth]{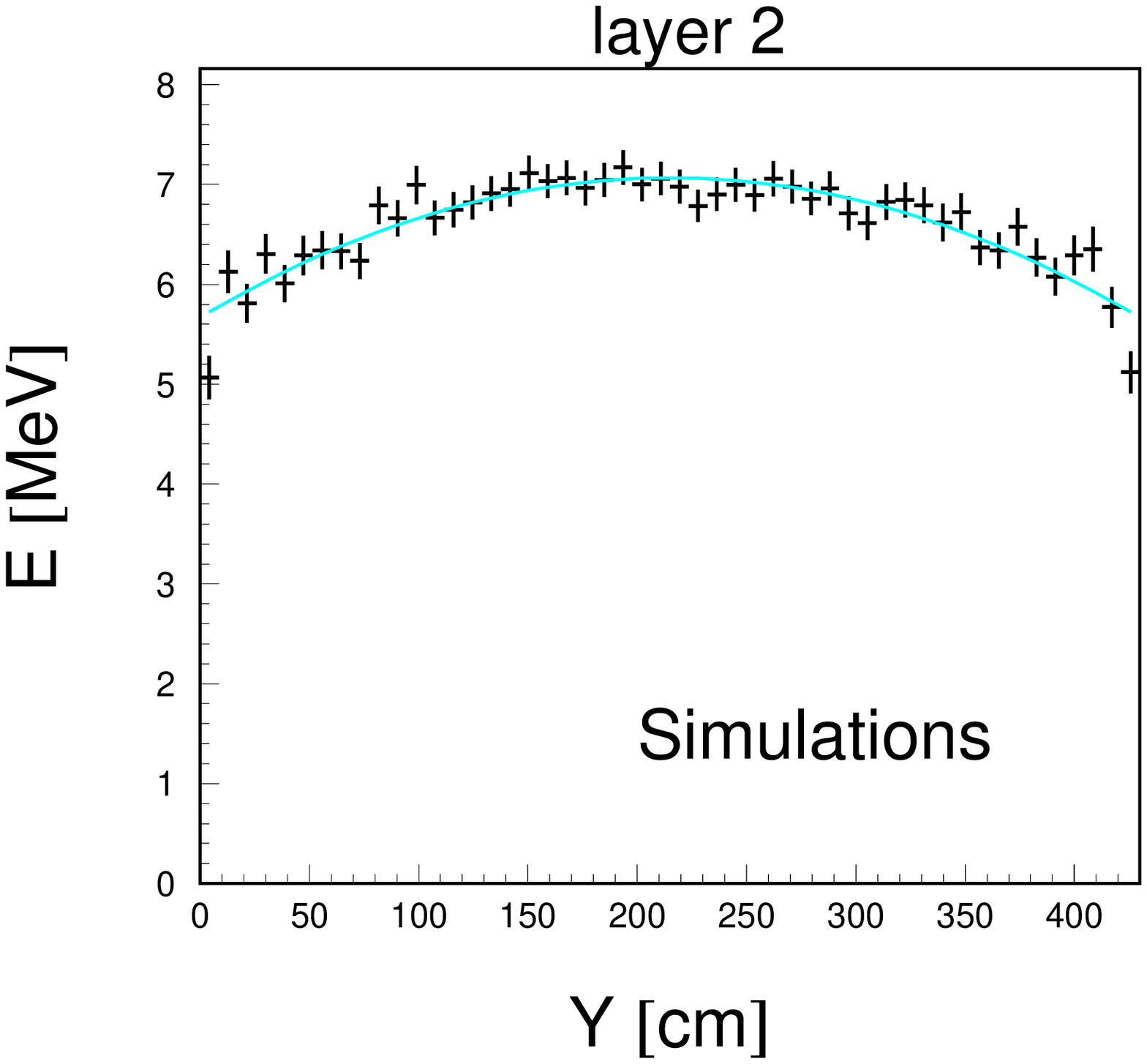}}
\parbox[c]{0.33\textwidth}{\includegraphics[width=0.39\textwidth]{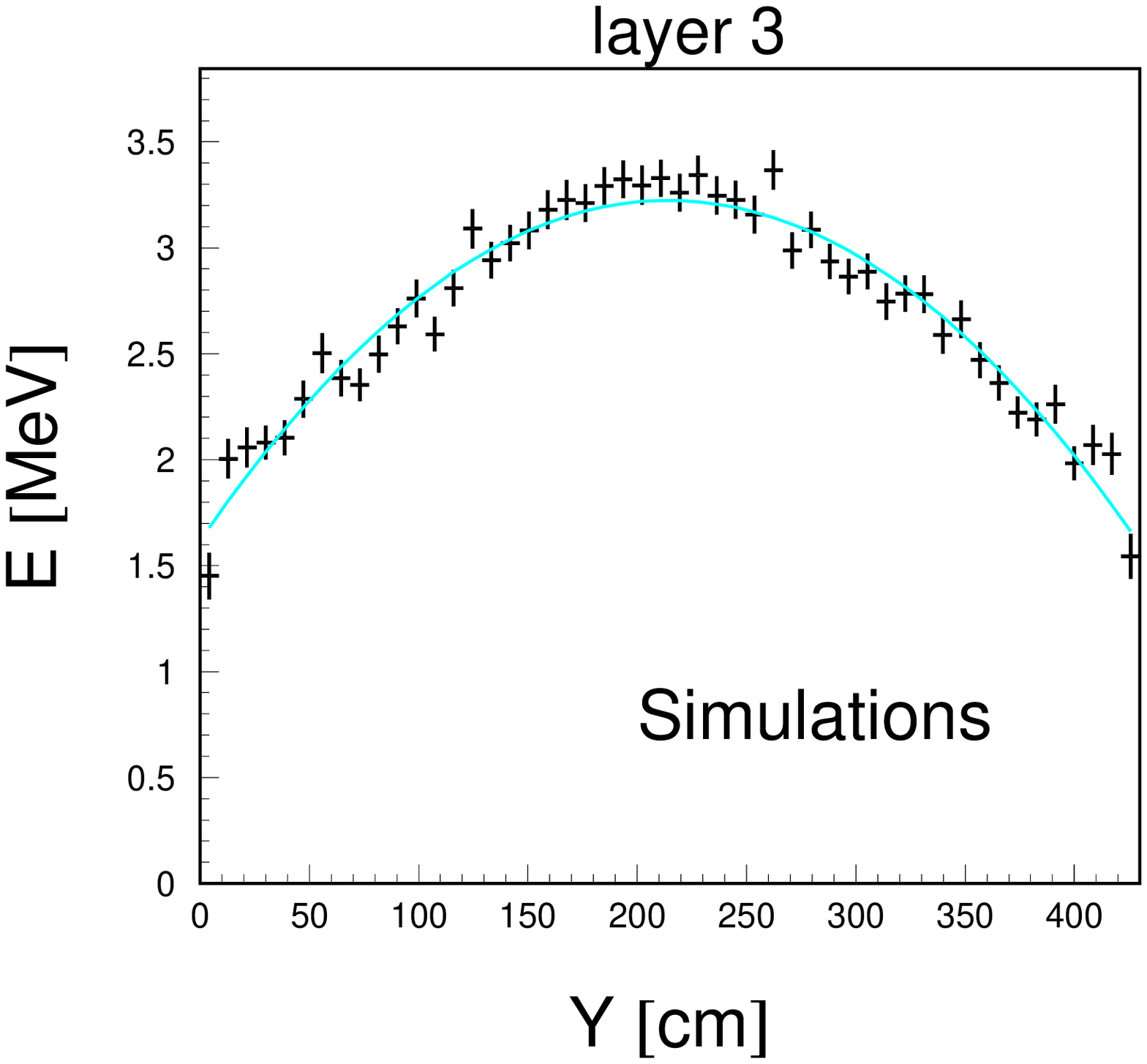}}
}
\parbox[c]{1.0\textwidth}{
\vspace{-0.5cm}
\parbox[c]{0.33\textwidth}{\includegraphics[width=0.39\textwidth]{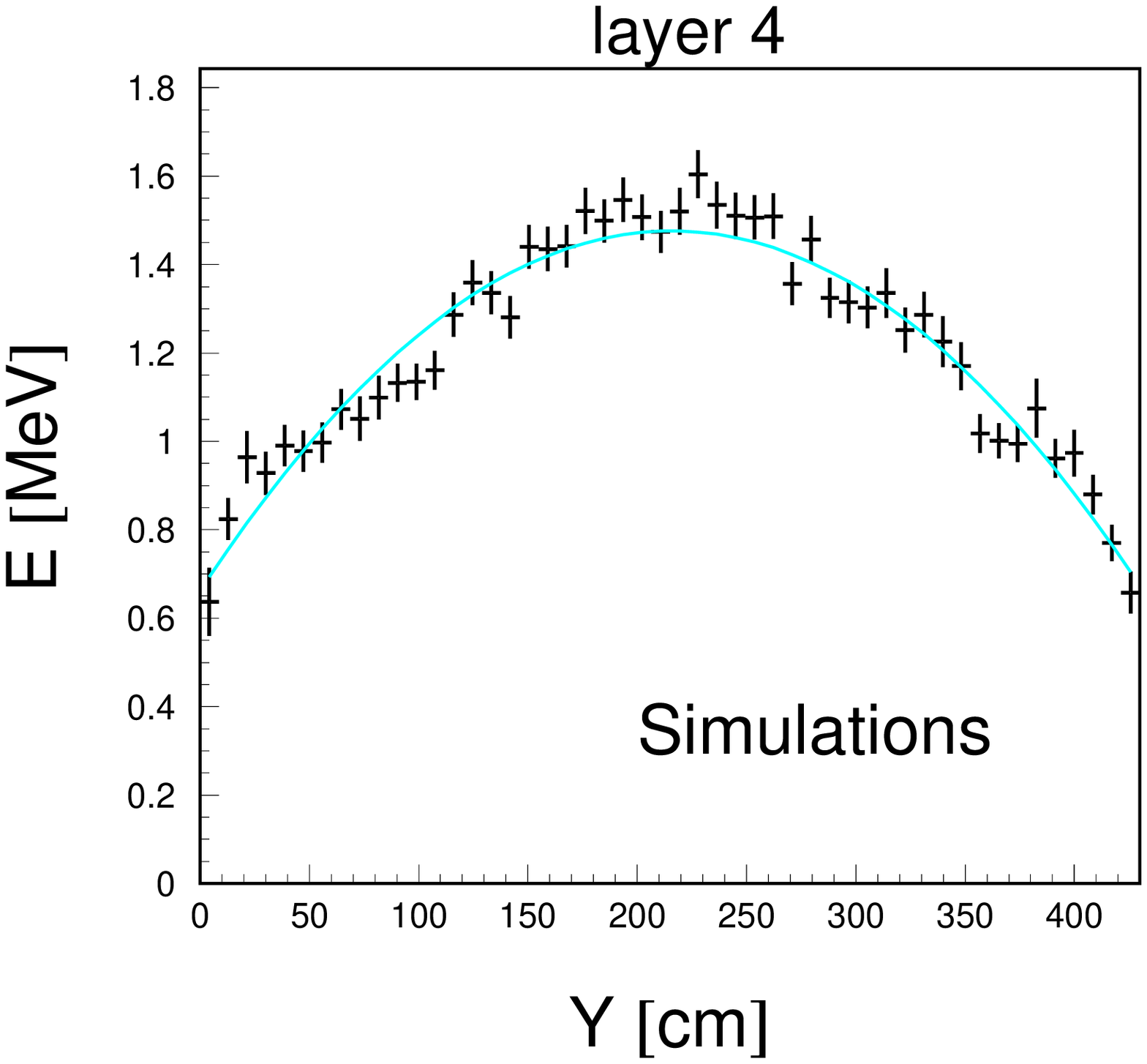}}
\parbox[c]{0.33\textwidth}{\includegraphics[width=0.39\textwidth]{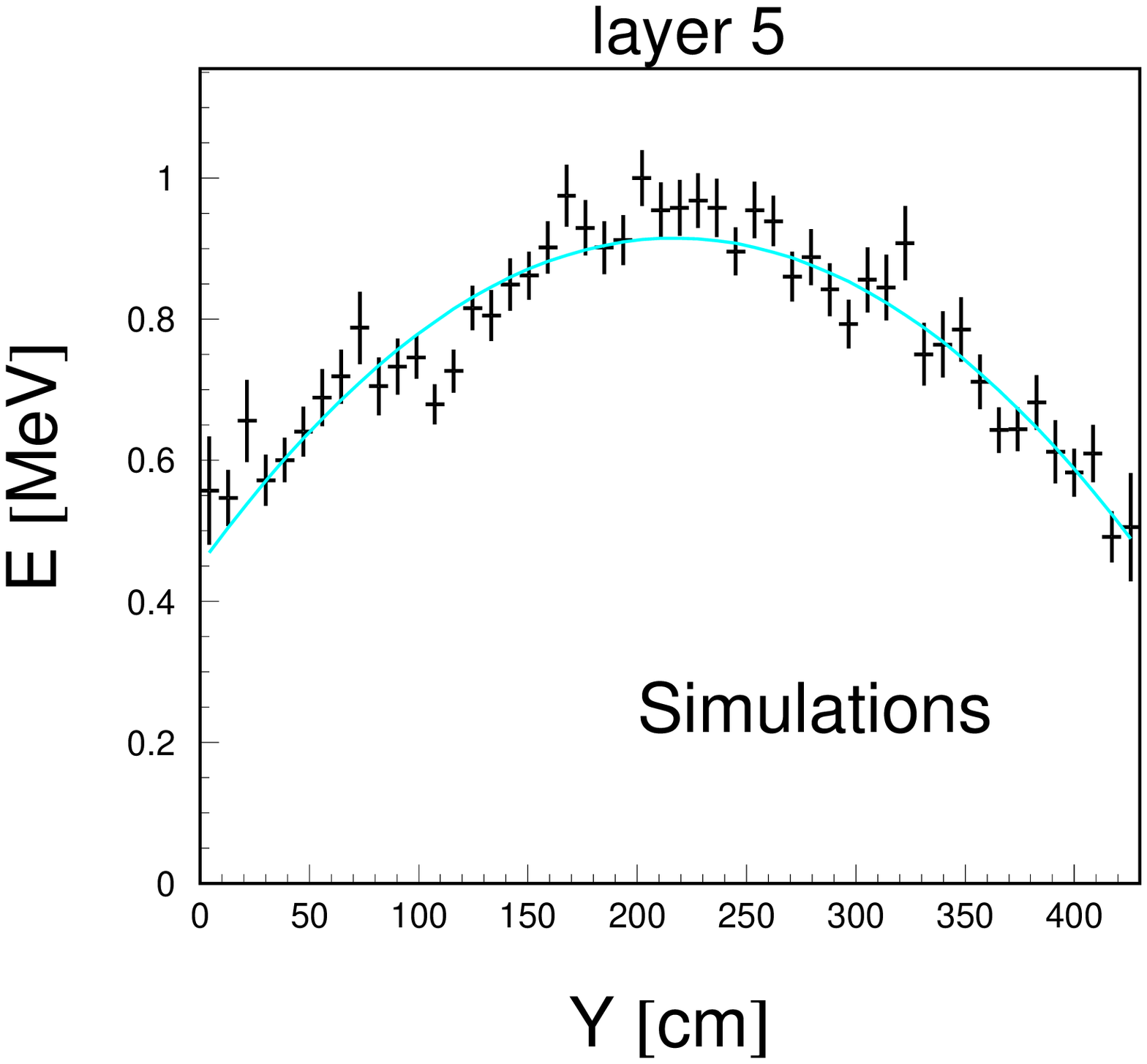}}
}
\caption{Simulation of the total energy deposits in cells for each of the 5 calorimeter layers 
for the $\gamma$ quanta from the $\phi \to \eta\gamma$ decay. Courtasy of E. Czerwi\'nski and B. Di Micco \cite{erykbiagio}.}
\label{energy_5layers_studies}
\end{figure}

\begin{figure}[H]
\parbox[c]{1.0\textwidth}{
\hspace{1.1cm}
\parbox[c]{0.375\textwidth}{
\vspace{-0.6cm}
\includegraphics[width=0.455\textwidth]{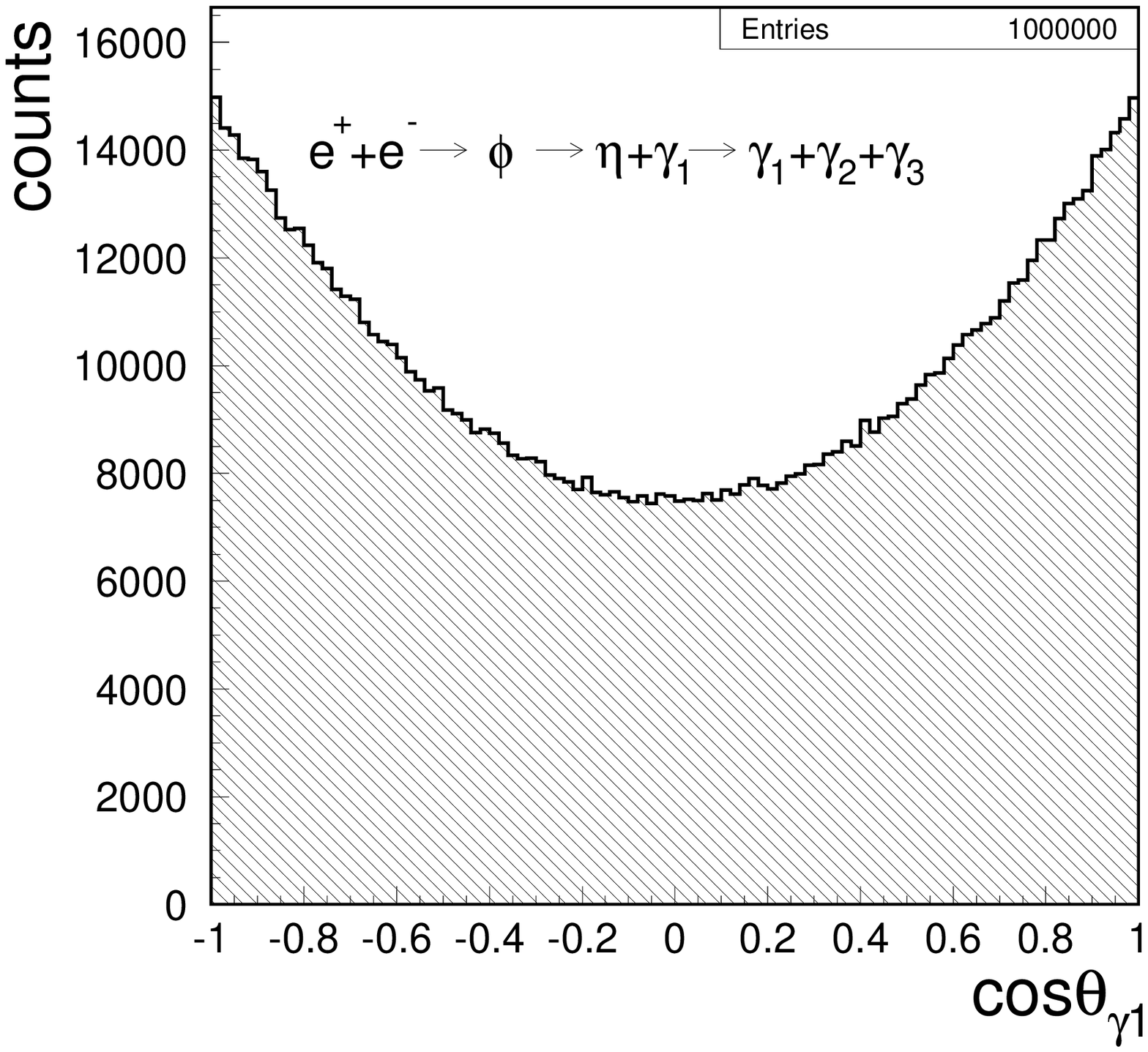}}
\hspace{0.4cm}
\parbox[c]{0.375\textwidth}{
\vspace{-0.6cm}
\includegraphics[width=0.455\textwidth]{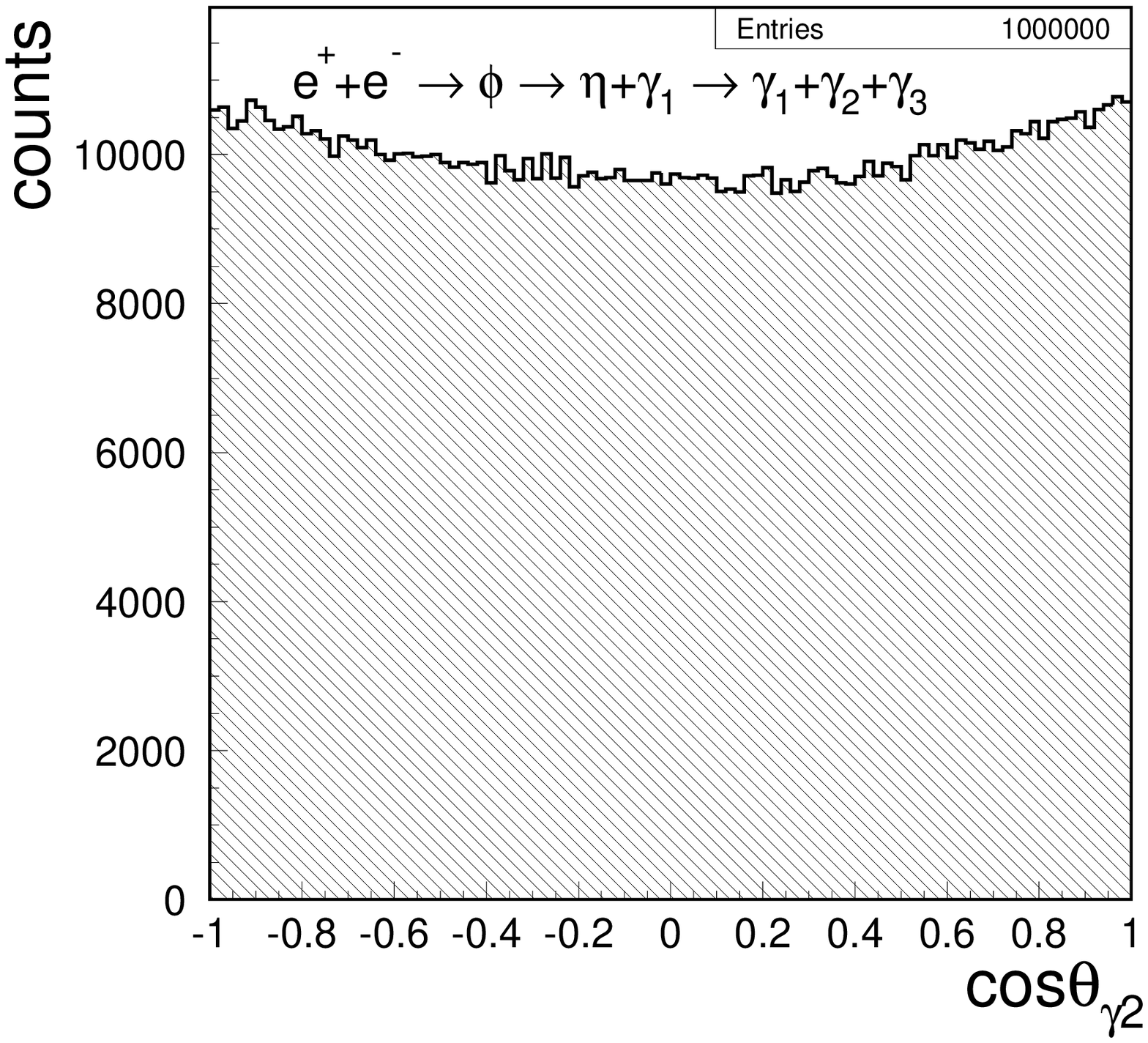}}
}
\caption{Simulated angular distributions for photons from reaction: $e^{+}e^{-} \to \phi \to \eta \gamma \to 3\gamma$.} 
\label{1reaction_parameters}
\end{figure}

\begin{figure}[H]
\hspace{3.8cm}
\parbox{0.45\textwidth}{
\vspace{-1.1cm}
\centerline{\epsfig{file=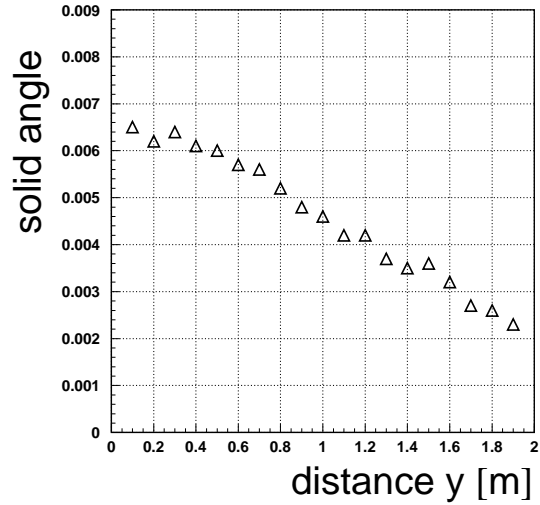,width=0.55\textwidth}}}
\caption{
The solid angle distribution as a function of the distance from the center of the module.
}
\label{solid_angle}
\end{figure}

\begin{figure}[H]
\hspace{3.5cm}
\parbox[c]{1.0\textwidth}{
\parbox[c]{0.5\textwidth}{\includegraphics[width=0.49\textwidth]{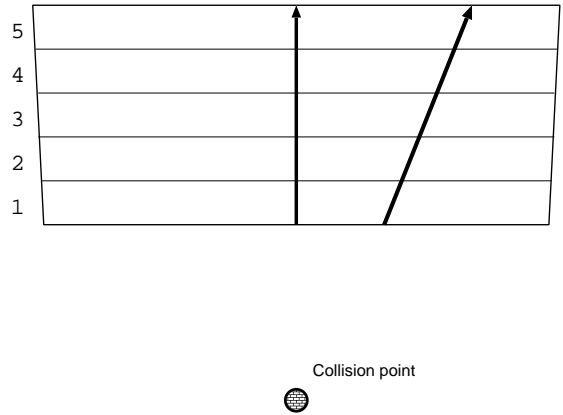}}
}
\caption{Schematic view of the particle tracks for two different entering examples.} 
\label{drogi_w_module_od_kata}
\end{figure}

The first step for estimation of the attenuation length 
 % was described in the previous section and was based on 
was the selection of events corresponding to the $\phi \to \eta \gamma \to \gamma \gamma \gamma$ reaction 
and identification of monoenergetic $\gamma$ from the $\phi \to \eta\gamma$ radiative decay.   
%
% from KLOE experimental data sample for reaction 
% $\phi \to \eta \gamma \to \gamma \gamma \gamma$. Afterwards we have selected a monoenergetic $\gamma$ using the 
% kinematic fit procedure to determine energy and entering position on the Z coordinate which is the length of the calorimeter 
% module (see Figure~\ref{coordinate_system}, reconstruction
% geometry). 
%
%
In the next step we simulated the total energy deposited in cells of a given plane as a function of Z coordinate 
\cite{eryk}. \\ 
\indent We used the following selection cuts: \\
1. Monochromatic photons (361 < $E_{\gamma}$ < 365 MeV) from $\phi$ meson decay~, \\
2. | Z$_{\gamma}^{\text{~reconstructed}}$ - $Z_{\gamma}^{\text{~generated}}$ | < 20~cm ~.  \\
\indent The first condition guarantees that we take only events with $\gamma$ quanta from $\phi$ meson decay directly.  
The second condition guarantees that we take only events where the distance between position where $\gamma$ quanta 
hit the module in simulation  
and reconstructed position of the cluster must be less than 20~cm. \\
% \indent In order to determine an attenuation factor value we needed of the uniform distribution of the deposited energy in cells 
% as a function of the distance between sides A and B of the calorimeter module. 
% In the figure~\ref{energy_5layers_studies} are shown energy deposits for 5 layers in the one module of the 
% calorimeter detector as a function of distance from one side
% of the calorimeter module. 
The results for 5 layers are shown in Fig.~\ref{energy_5layers_studies}.
One can see that distribution of the average 
energy depositions in cells are not uniform, and therefore for the studies of the attenuation length  
it is mandatory to divide the experimental data by the Monte Carlo distributions.
\newpage
\vspace{-0.2cm}
In case of the first layer, which is the nearest one with respect to the collision point,  
the shape of the energy distribution is mostly related (i) with the angular distributions of the 
$\gamma$ quanta from the $\phi \to \eta\gamma_{1} \to \gamma_{1}\gamma_{2}\gamma_{3}$ decay, (ii) with the changes of the solid angle 
distribution along the module, and (iii) with the change of the input angle into calorimeter surface. 
 The angular distributions for these particles are presented in  
Fig.~\ref{1reaction_parameters}.
The angular distribution for $\gamma$ quanta from the $\phi \to \eta\gamma$ decay is described with the  
 formula: $\frac{1 + cos^{2}\theta}{2}$. Interestingly, Fig.~\ref{1reaction_parameters} indicates that also $\gamma$ quanta 
from subsequent $\eta$ meson decay have still small contribution in the form of cos$^{2}\theta$.  
%
%
% One can see also that the $\gamma$ quantas from $\eta$ meson 
% decay also have angular distributions. \\
%
The decrease along the module of the solid angle as seen from the interaction point is shown in Fig.~\ref{solid_angle}.  
%
% \indent In the next layers the most relevant effect which forms the shape of the energy deposits is the distribution of 
% the solid angle (Figure~\ref{solid_angle}). 
%
The variation of the solid angle decrease the events population towards the edges.  
The third effect which play a role in this context is the angle at which particle hits 
the detector surface and hence the angle of the shower propagation (see Fig.~\ref{drogi_w_module_od_kata}).

% The solid angle distribution determines that the most of $\gamma$ quanta 
% hit the middle part of the each module of the barrel calorimeter. 
%The results of the distribution of hitting places 
%on the surface on the module are shown in Fig.~\ref{distribution_on_surface}. 
%
% The distribution of the places where particles hit the calorimeter on the bottom surface of this module is 
% shown in the figure~\ref{distribution_on_surface}. 
%
%\begin{figure}[H]
%\hspace{1.0cm}
%\parbox[c]{1.00\textwidth}{
%\parbox[c]{0.45\textwidth}{\includegraphics[width=0.49\textwidth]{xsrpl_2reaction.eps}}
%\parbox[c]{0.45\textwidth}{\includegraphics[width=0.49\textwidth]{ysrpl_2reaction.eps}}
%}
%\caption{Distribution of the hitting places on the surface on the calorimeter module.}
%\label{distribution_on_surface}
%\end{figure}
%
% On the left panel one can see a distibution on the x axis (width of module = 52 cm) on the right on y axis 
% (length of the module = 4.3 m). The distribution in the first case is uniform but on the bigger distance
% the most part of particles hit in the middle of module. This effect is caused by solid angle distribution 
% and a distribution of angle $\frac{1+cos^{2}\theta}{2}$ for $\gamma$ quanta which orginating from $\phi$ meson decay directly. 
%
%
Due to this fact in the first layer the energy deposited by showers in the middle of the module is lower than on the sides. 
Therefore, in subsequent layers relatively larger fraction of the energy of the shower will be deposited in the middle of the module 
and less on its sides. Hence, the shapes of distributions observed in Fig.~\ref{energy_5layers_studies} is determined by three discussed effects: 
angular distribution of $\gamma$ quanta, changes of the solid angle, and variation of the fractional length of the showers in the 
layers as a function of the impact angle. \\ 
\newpage
\begin{figure}[H]
\vspace{-1.4cm}
\hspace{0.7cm}
\includegraphics[width=0.46\textwidth]{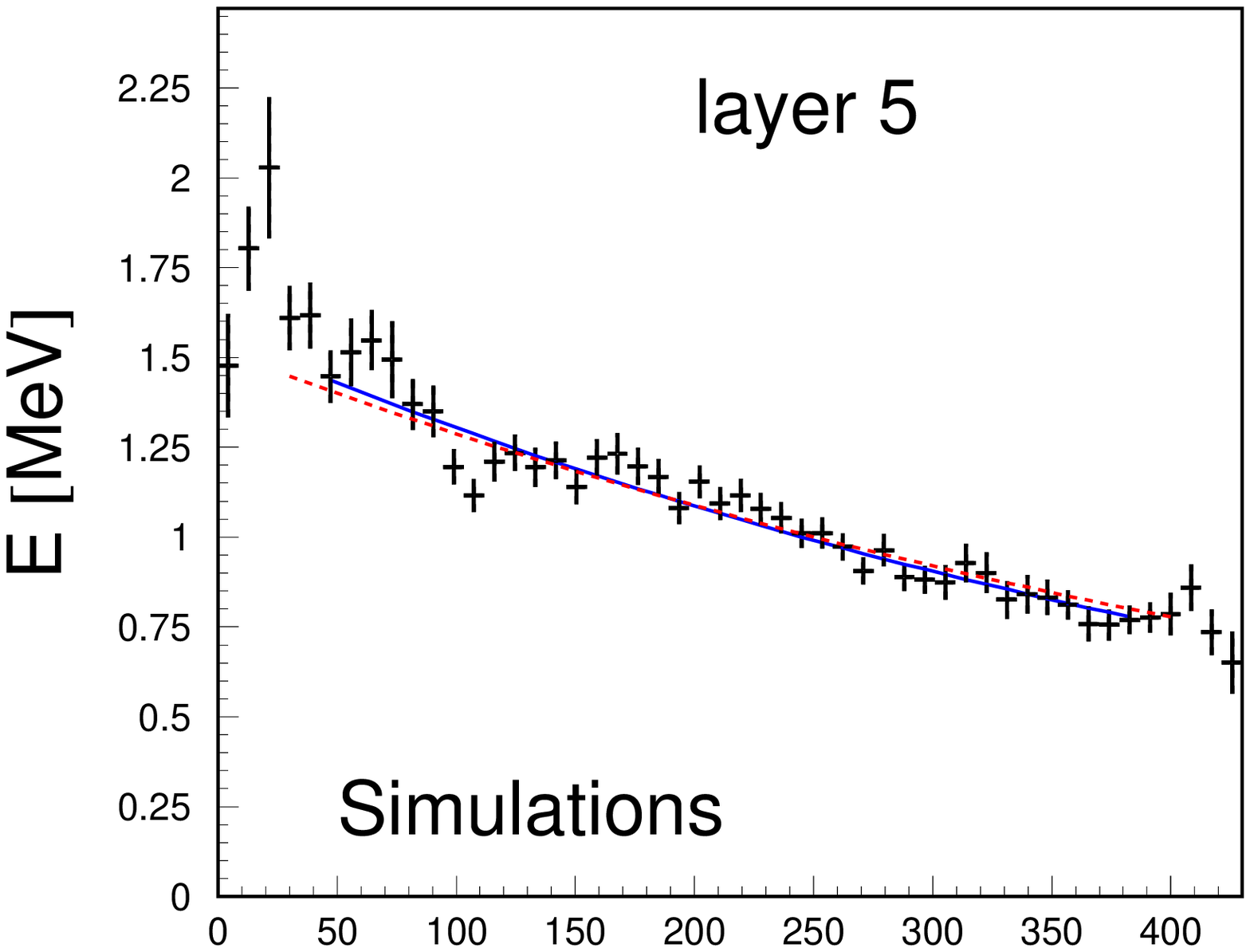}
\includegraphics[width=0.46\textwidth]{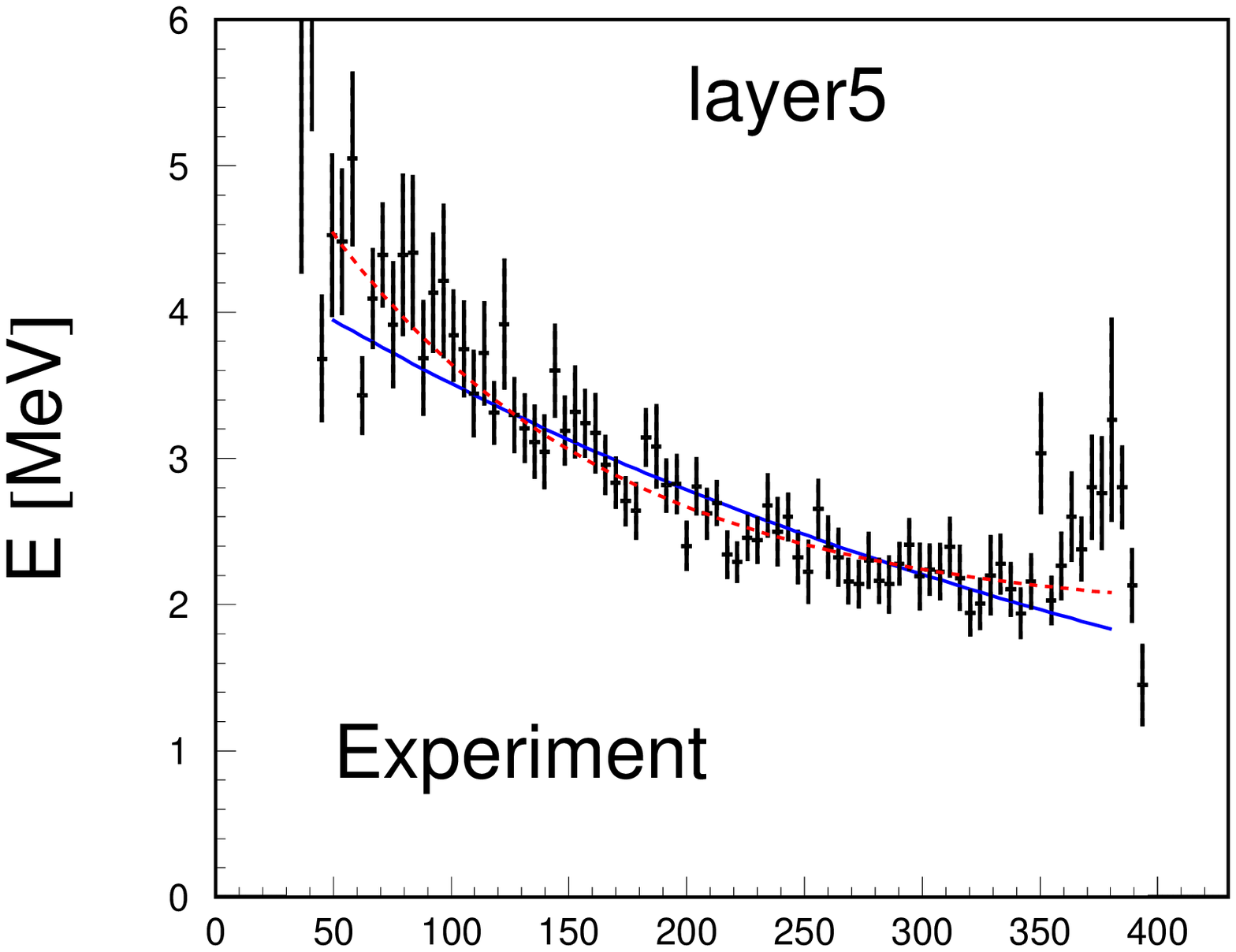} 

\vspace{-2.45cm}
\hspace{0.7cm}
\includegraphics[width=0.46\textwidth]{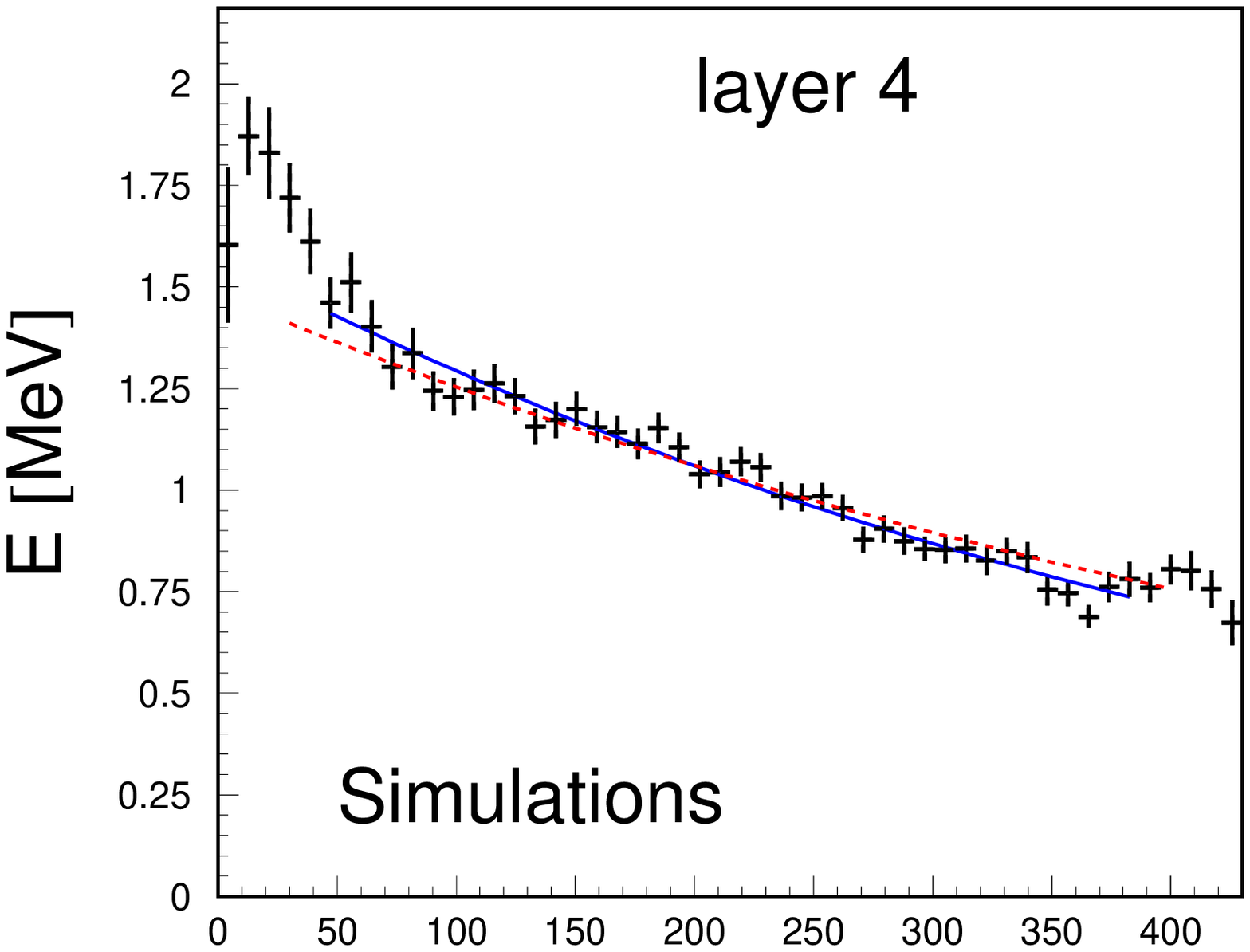}
\includegraphics[width=0.46\textwidth]{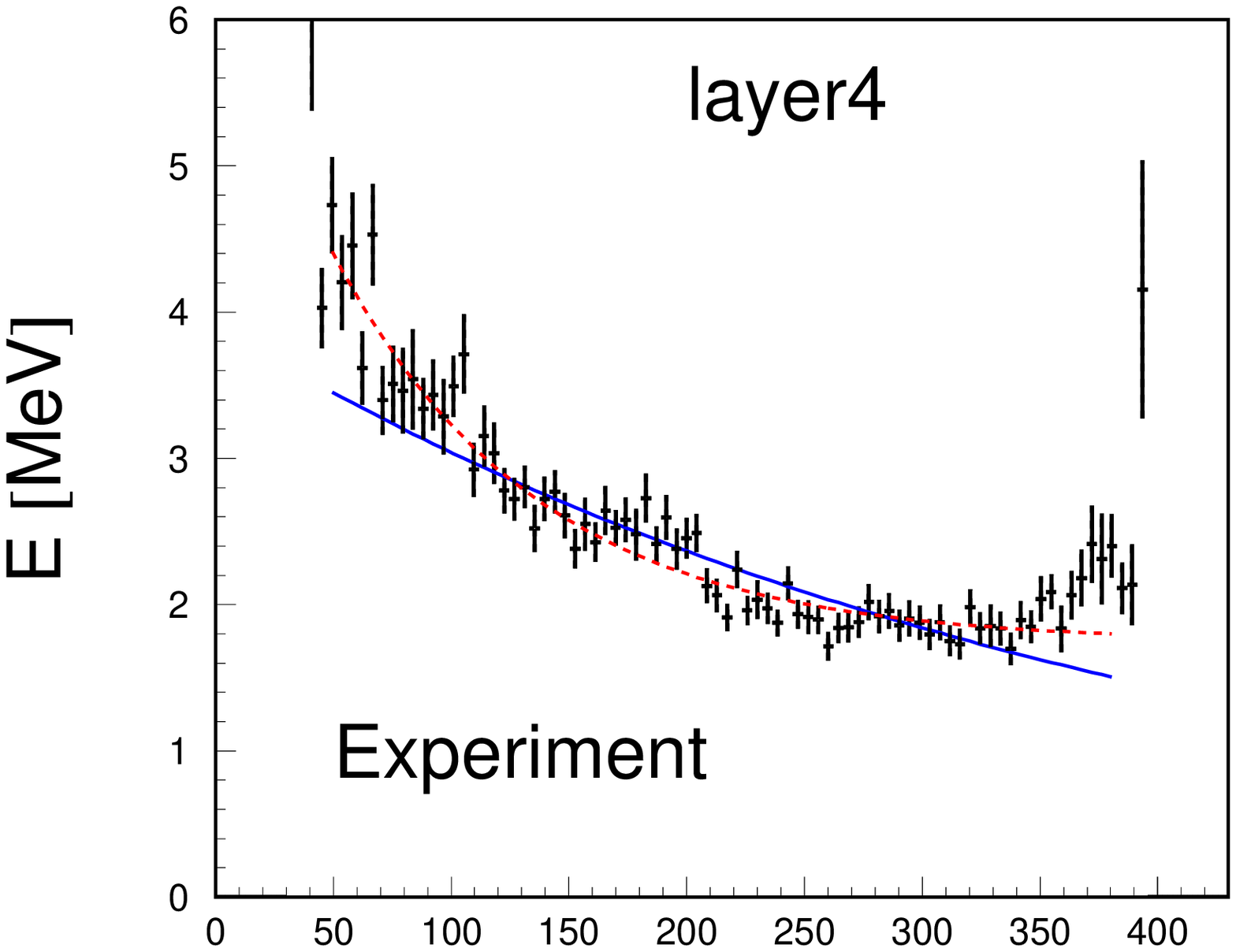} 

\vspace{-2.45cm}
\hspace{0.7cm}
\includegraphics[width=0.46\textwidth]{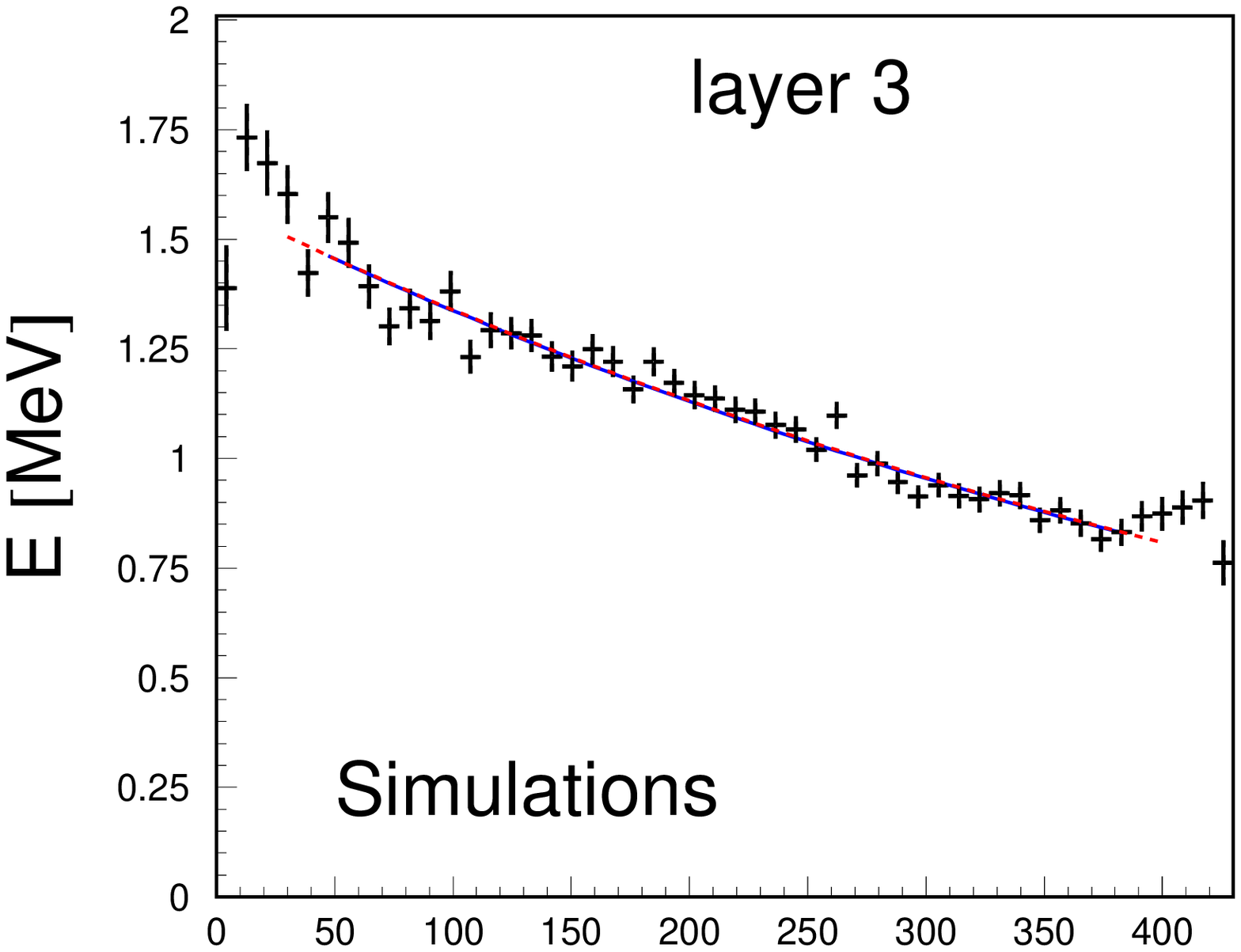}
\includegraphics[width=0.46\textwidth]{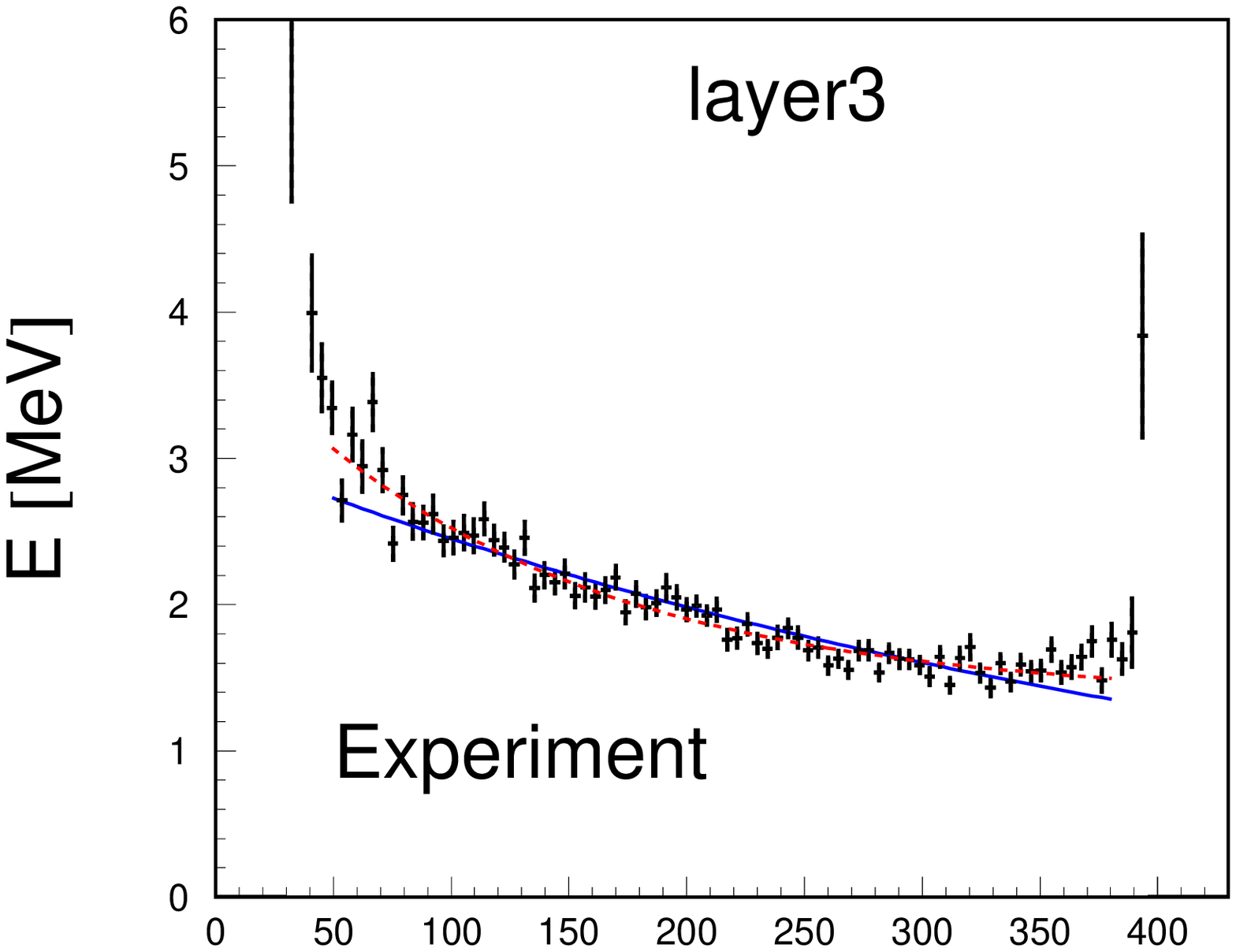} 

\vspace{-2.45cm}
\hspace{0.7cm}
\includegraphics[width=0.46\textwidth]{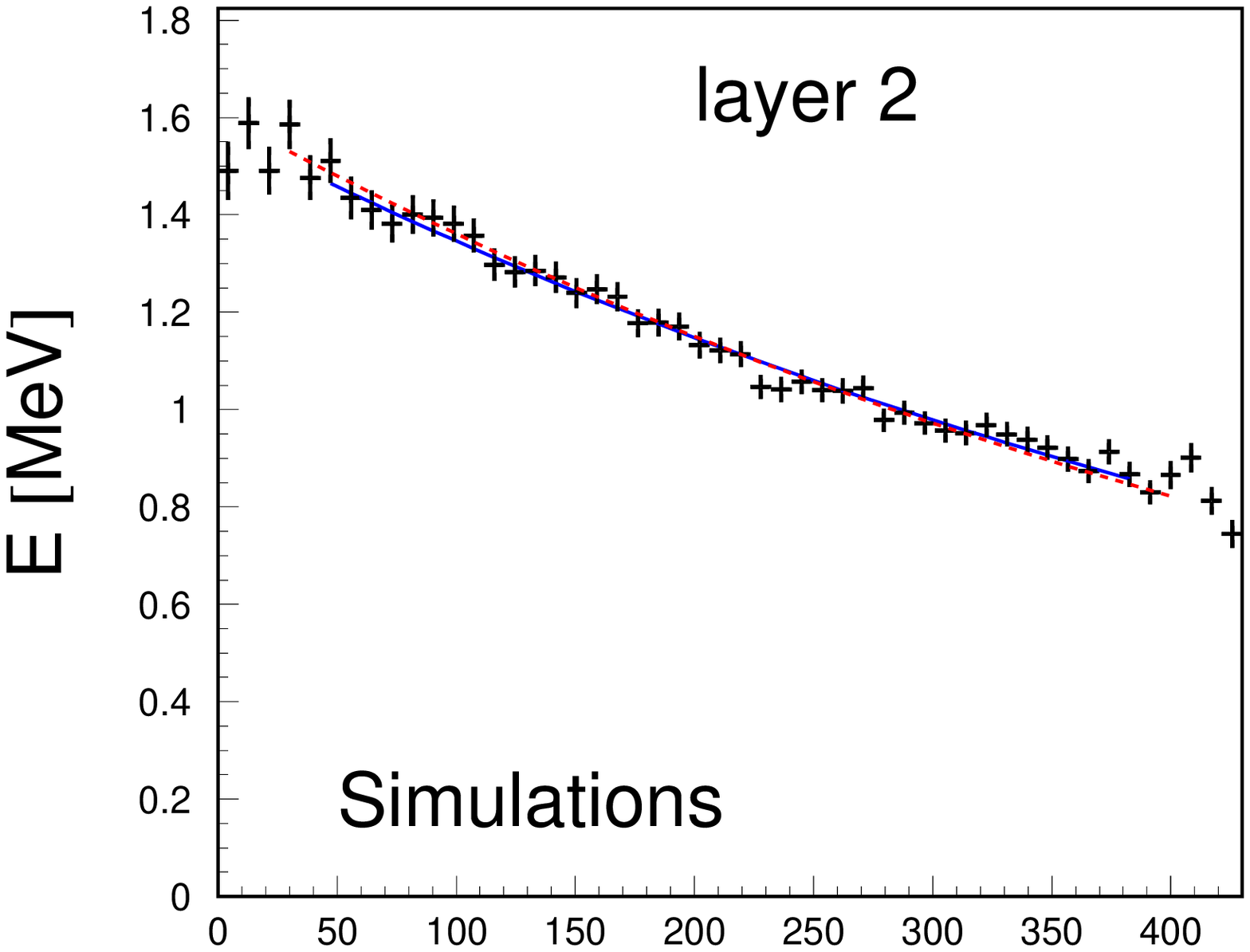}
\includegraphics[width=0.46\textwidth]{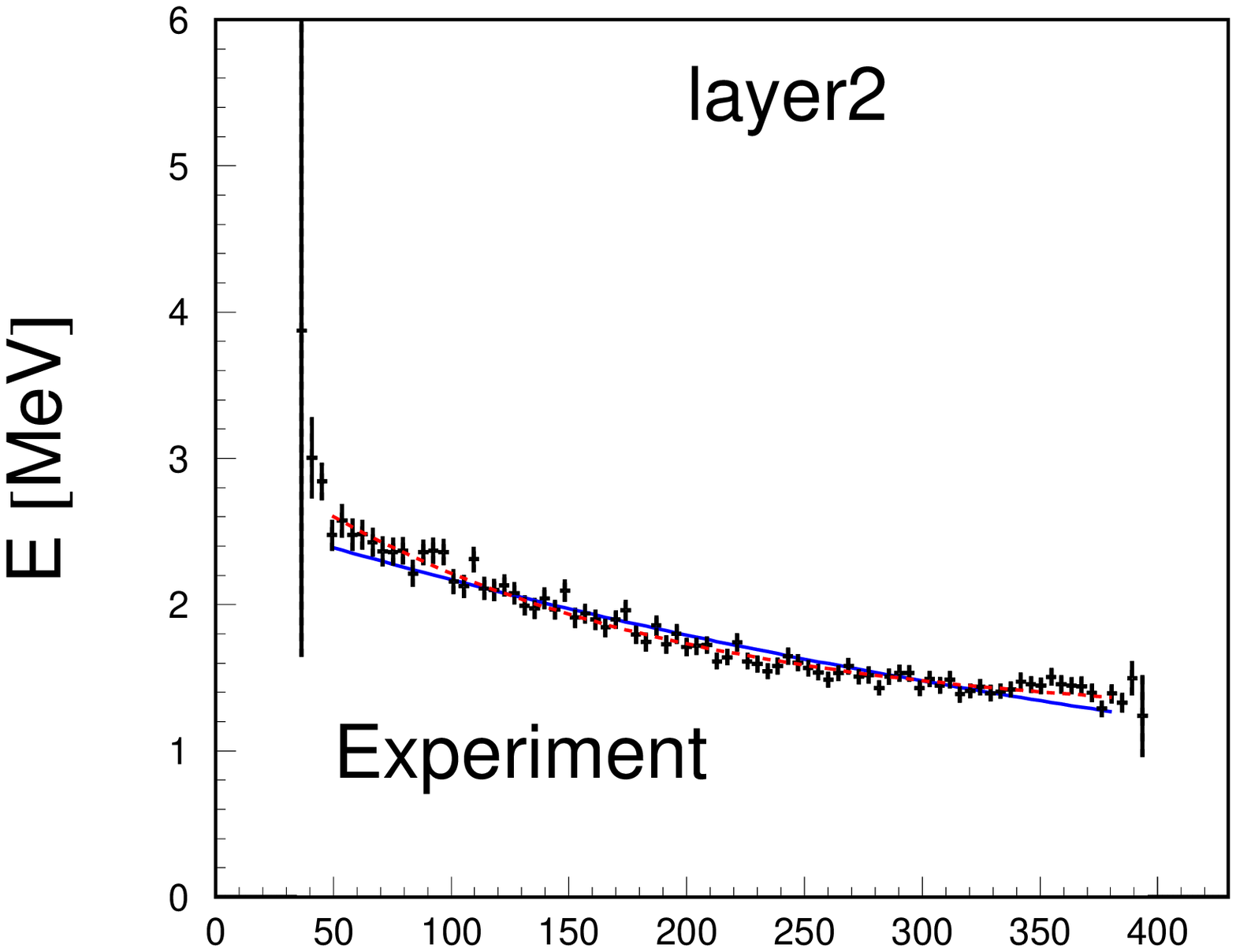} 

\vspace{-2.45cm}
\hspace{0.7cm}
\includegraphics[width=0.46\textwidth]{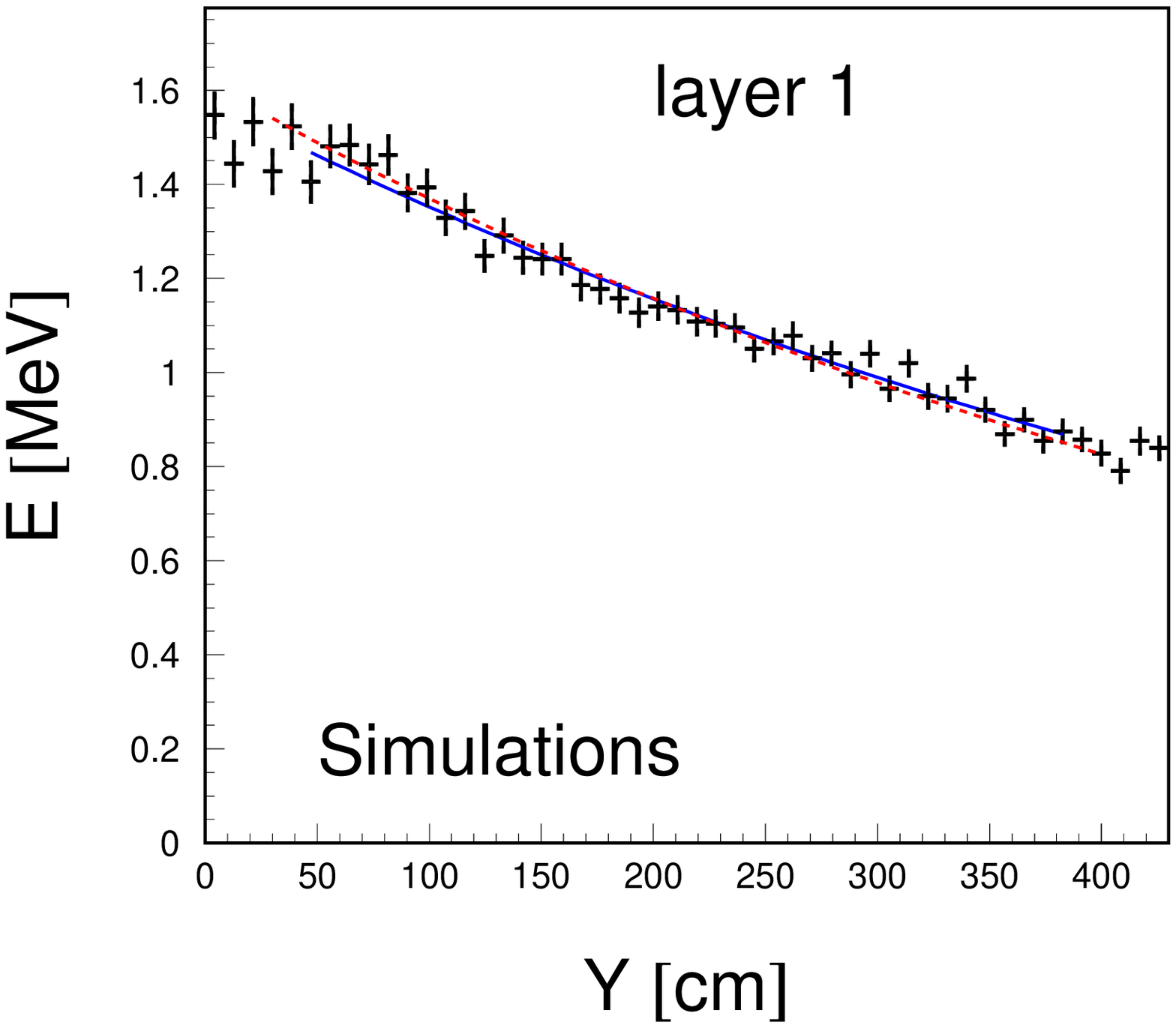}
\includegraphics[width=0.46\textwidth]{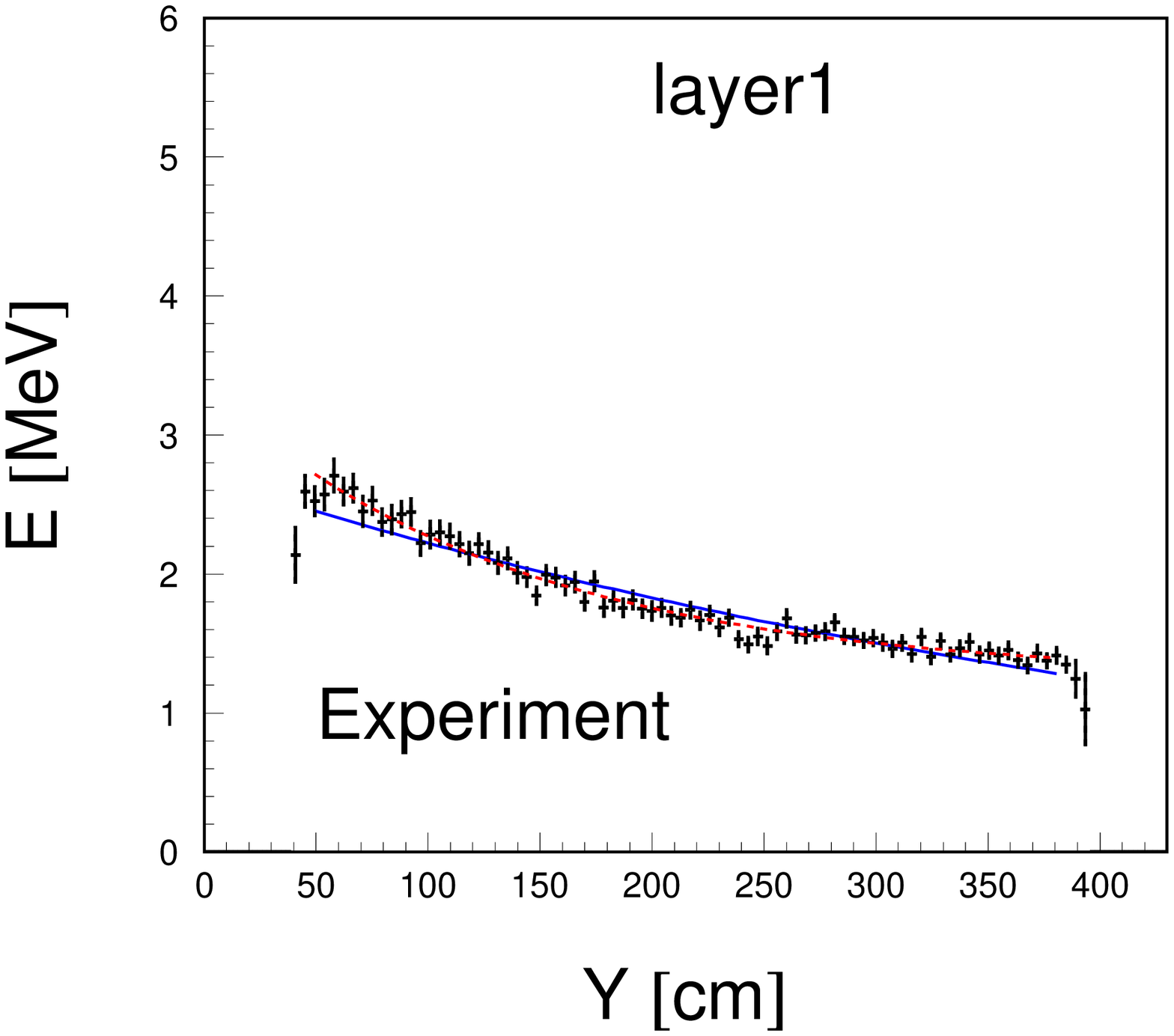} 
\vspace{-1.0cm}
\caption{The energy distributions as seen at one side of the module: Monte Carlo (left column) and  
 experimental data (right column). The curves are described in the text. Courtasy of Eryk Czerwi\'nski and B. Di Micco \cite{erykbiagio}.}  
\label{attenuation_studies}
\end{figure}
\indent In order to achieve a uniform, energy deposition along the module a simulated energy deposit distributions  
 (Fig.~\ref{energy_5layers_studies}) were fit with second order polynominal functions which were then used to normalize the experimental spectra 
and for the check also simulated spectra were normalized in the same way. 
Fig.~\ref{attenuation_studies} presents an energy of signals registered at one side of the module as a function of 
the distance between the edge of the module and the hit position.        
In order to establish a value of the light attenuation length ($\lambda$) we fit to the spectra from Fig.~\ref{attenuation_studies} 
 (right column) an exponential function: 
\begin{eqnarray}
E(y) = \alpha \cdot e^{-\frac{y}{\lambda}}~, 
\end{eqnarray}
with $\alpha$ and $\lambda$ being a free parameters \cite{erykbiagio}. 
The result of the fit is shown as dark blue, continuous line in Fig.~\ref{attenuation_studies} (right column). 
The weighted averages of attenuation length determined at both sides (A and B) amounts to 522$\pm$7 cm and  407$\pm$9 cm, for 
1,2,3 layers and for 4,5 layers, respectively.  
The result is shown as a red, dashed curve in Fig.~\ref{attenuation_studies} (right). \\ 
\indent As a next step the determined attenuation length were used to simulate the light output 
at the edges of the module with the DIGICLU program. The result is shown in Fig.~\ref{attenuation_studies} (left). 
These studies were made to check if results from simulations are in agreement with results which were achieved 
from the experimental data sample. 
 The code of the lines is the same as in the right panel. 
The simulated and experimental spectra are in very good agreement except a region below 50~cm and above 400~cm. 
This is due to the fact that this part of the barrel is situated behind the endcap modules. And in the simulations presented 
in this thesis only a barrel part of the calorimeter is taken into account.  
Furthermore we tested also a fit with a function with two attenuations lengths \cite{FLUKA_simulation_KLOE_EMC}: 
\begin{equation}
E(z) = \alpha_{1} \cdot e^{-\frac{z}{\lambda_{1}}} + \alpha_{2} \cdot e^{-\frac{z}{\lambda_{2}}}~.     
\end{equation}
The result is presented in Fig.~\ref{attenuation_studies} where left column indicates results from simulations 
and right column shows experimental data.
As one can see both presented fits describe the data well.  
\vspace{-0.2cm}
%
%
% In this figure one can see also fits with polynominal functions to the Monte Carlo results (left panel).
% These studies were made to check if a results from simulations are agreement with the results which were
% achieved with the KLOE experimental data sample.    
%
% The black line is a fit with the following formula to the simulation points: \\
% \begin{equation}
% fit_att = par(1) \cdot e^{\frac{-x}{par(2)}} \\
% \end{equation}
%
% and the red line represents fit with the next formula with the values 
% of the attenuation length which were used in DIGICLU: \\
% \begin{equation}
% fit_att = par(1) \cdot (0.35 \cdot e^{\frac{-x}{595.}}) \\ 
% \end{equation}
% for layers from 1 to 3. \\
% For the layers 4 and 5 we used: 
% \begin{equation}
% fit_att = par(1) \cdot (0.35 \cdot e^{\frac{-x}{443.}}) \\
% \end{equation}
%
% where "par" is also the free fit parameter. \\
%
% \indent We achieved a very good agreement between experimental data distribution and Monte Carlo simulations
% after implemented a new value of attenuation factor. 
% after correction with uniform distribution of the energy deposits in cells. 
%

\subsection{Parametrization of the energy threshold function}
\begin{figure}[H]
\hspace{2.5cm}
\parbox[c]{1.0\textwidth}{\includegraphics[width=0.69\textwidth]{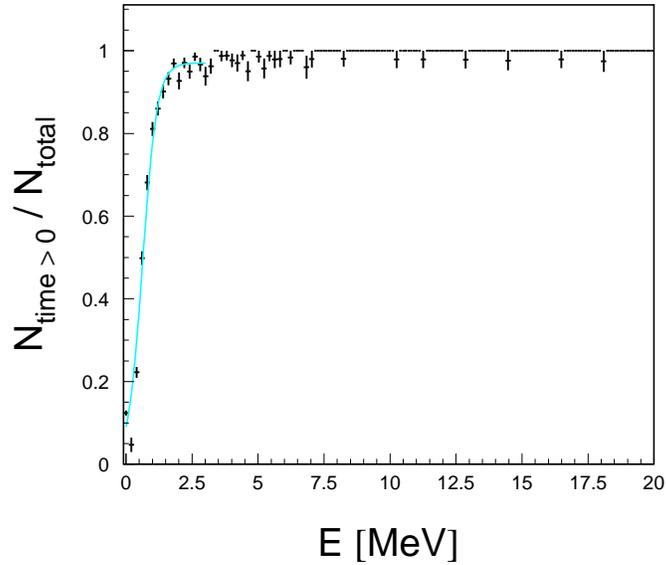}}
\caption{Ratio of number of cells with both time and energy signal to total number of cells with an energy signal. 
Courtasy of E. Czerwi\'nski and B. Di Micco \cite{erykbiagio}.}
\label{threshol_fit}
\end{figure}

\hspace{\parindent}
In the experiment the signals which amplitude is lower than a given threshold are not 
read by the TDC board and 
 information about their times is missing. In order to account for this effect in 
simulations   
the threshold curve was determined using the ratio between the number of cells with time information 
over all cells as a function of the cell energy.  
This ratio is shown in Fig.~\ref{threshol_fit}. 
%
% In order to validate an experimental data sample from the experiment and Monte Carlo simulation tools, we had to
% conduct threshold studies. In experiment is set-up value for threshold parameter, it assures that all signals 
% which have got lower energy than threshold value won't be rejected. 
%
%
% To agreement simulations with data we 
% implemented a threshold formula into source code of the DIGICLU program. 
% tylko do digiclu chyba ??????????? - tak na pewno do digiclu :-)
%
%
%
\newpage
One can see that the distribution is not a step-like function but rather a smooth increase of the ratio in the range from 0 to 2 MeV 
is observed. The shape of this increase can be described by the Fermi-Dirac distribution \cite{stefano_misceti_communication,fermi_dirac_net,
strzalkowski}: 
\begin{equation}
f_{th}(E) = (1 + e^{-\frac{E-\mu}{\sigma}})^{-1}~,
\label{threshold_formula}
\end{equation}
with free parameters $\mu$ and $\sigma$.  
%
% Our studies relied on making fit (using threshold formula Equation~\ref{threshold_formula}) to experimental data and estimate of the parameters $\mu$ and $\sigma$ values.
% \indent We performed the plot of distribution of energy which was deposited in cells in one layer, subsequently 
% we made the same plot but with condition that 
% reconstructed time is greater than zero (we take only events without a signal from electronic background), afterwards  
% we can divide the second plot by the first.
% The result of this operation is shown in (Figure~\ref{threshol_fit}).
%
% We made only fit for range of energy near the threshold, one can see for bigger range this ratio is equal 1. 
% After this procedure we were able to estimate values of the parameters $\mu$ and $\sigma$ \cite{erykbiagio}. \\
%
A fit for the range of energy near the threshold resulted in the following values: 
\begin{eqnarray}
\nonumber
\sigma & = & 0.2644\pm0.0068 \text{~MeV}~, \\
\nonumber
\mu & = & 0.5648\pm0.0087 \text{~MeV}~.  
\end{eqnarray}
\subsection{Calibration of the ionization deposits}
\begin{figure}[H]
\vspace{-1.4cm}
\hspace{0.7cm}
\includegraphics[width=0.46\textwidth]{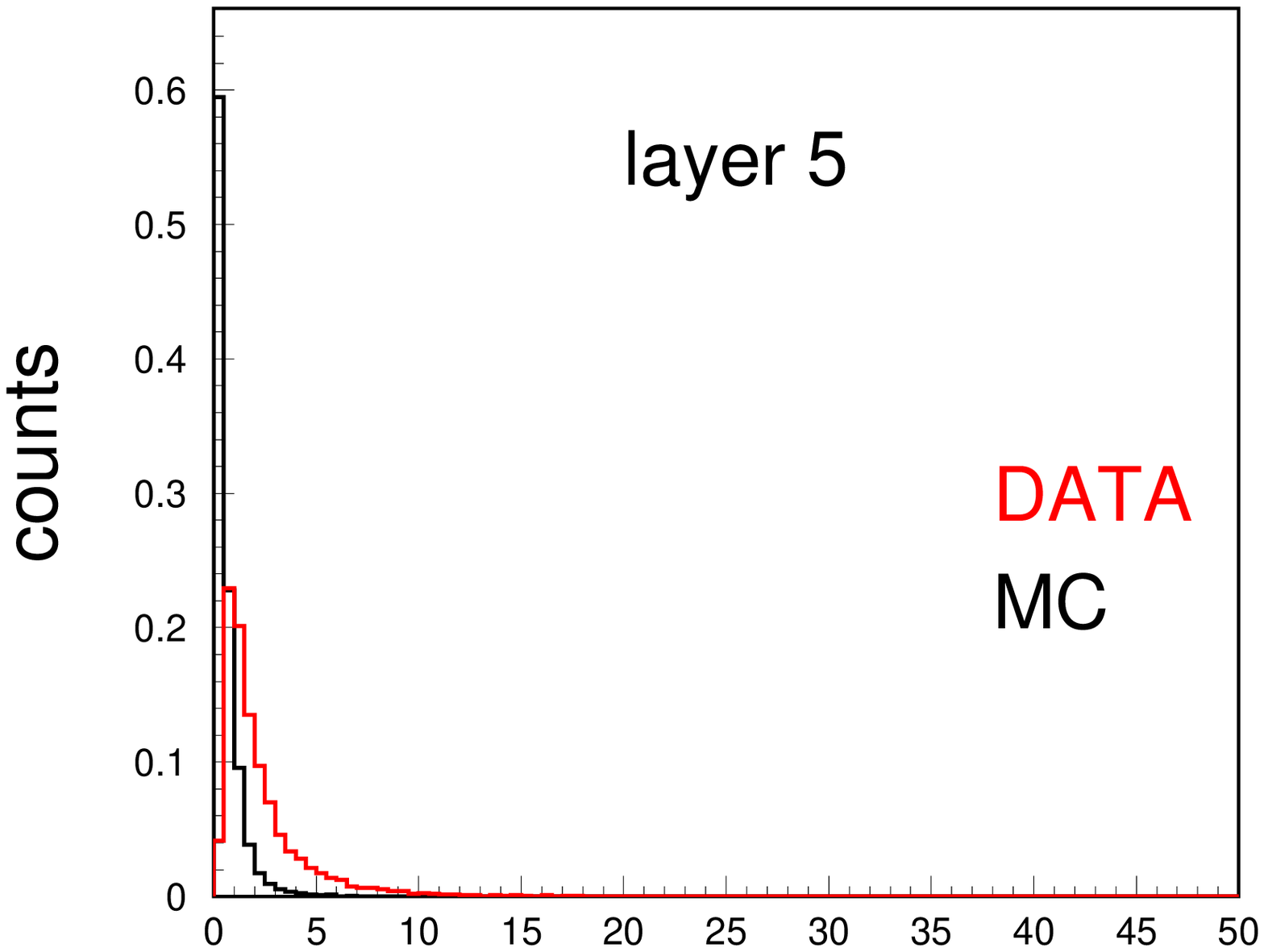}
\includegraphics[width=0.46\textwidth]{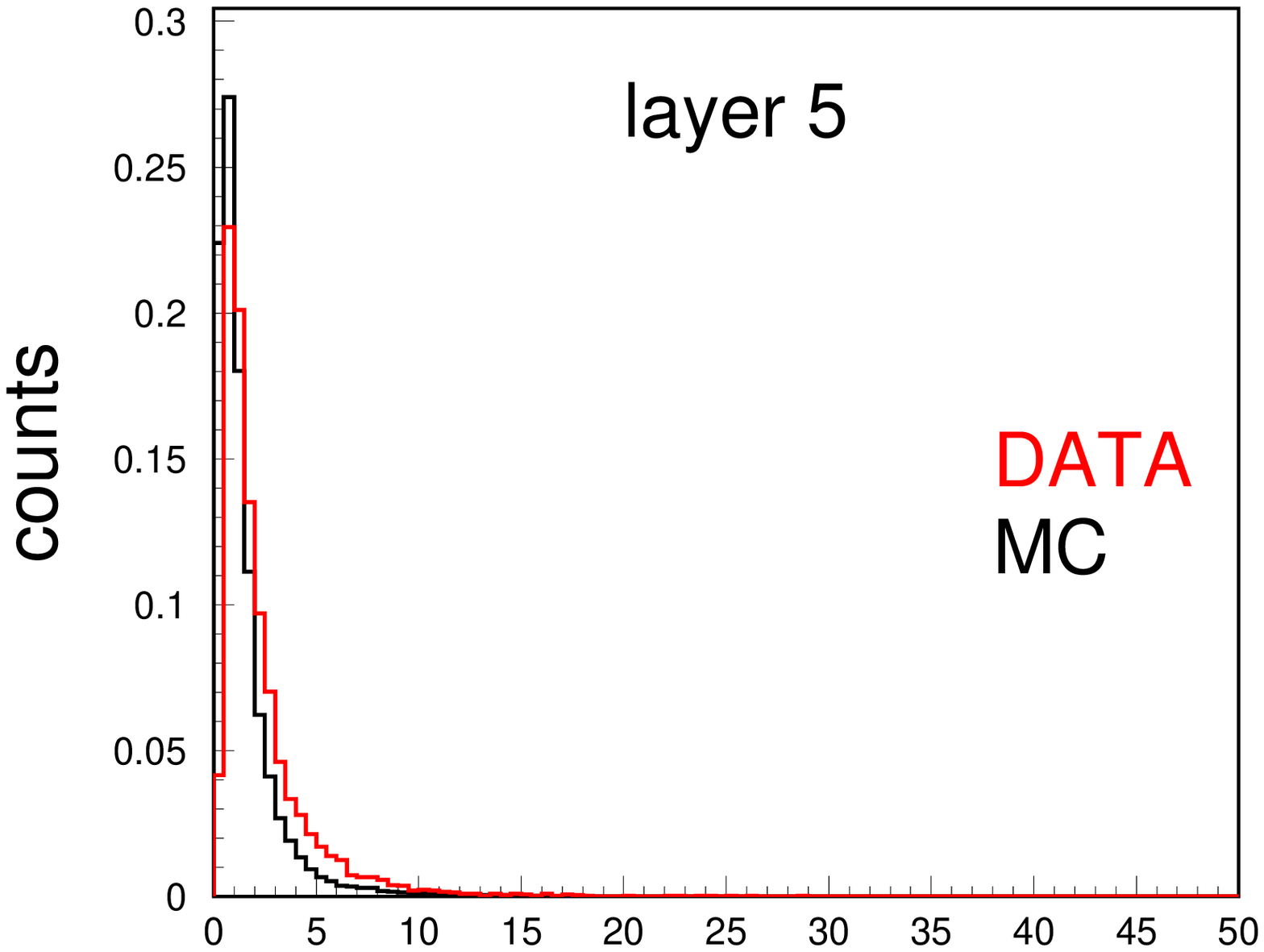} 

\vspace{-2.45cm}
\hspace{0.7cm}
\includegraphics[width=0.46\textwidth]{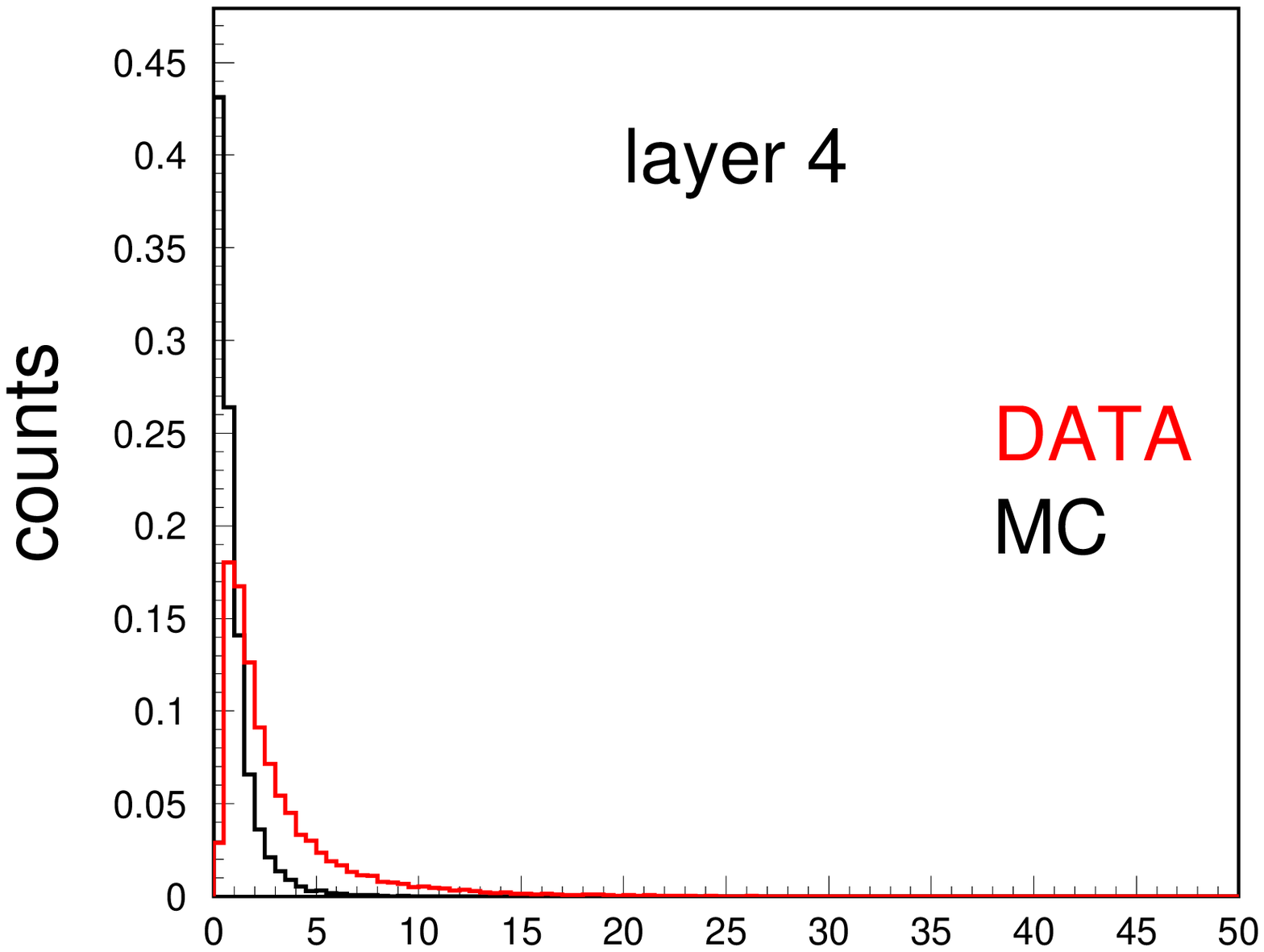}
\includegraphics[width=0.46\textwidth]{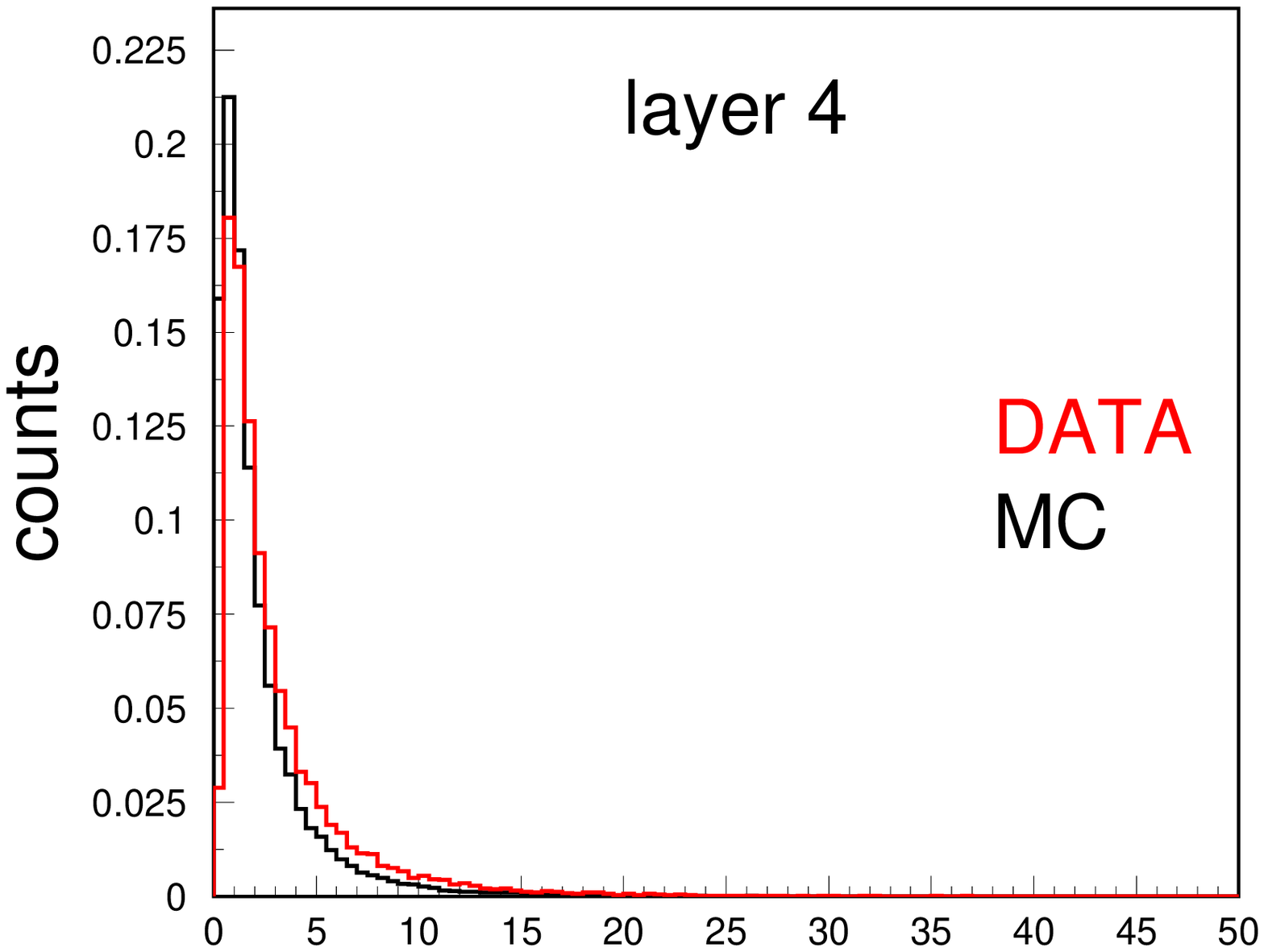}

\vspace{-2.45cm}
\hspace{0.7cm}
\includegraphics[width=0.46\textwidth]{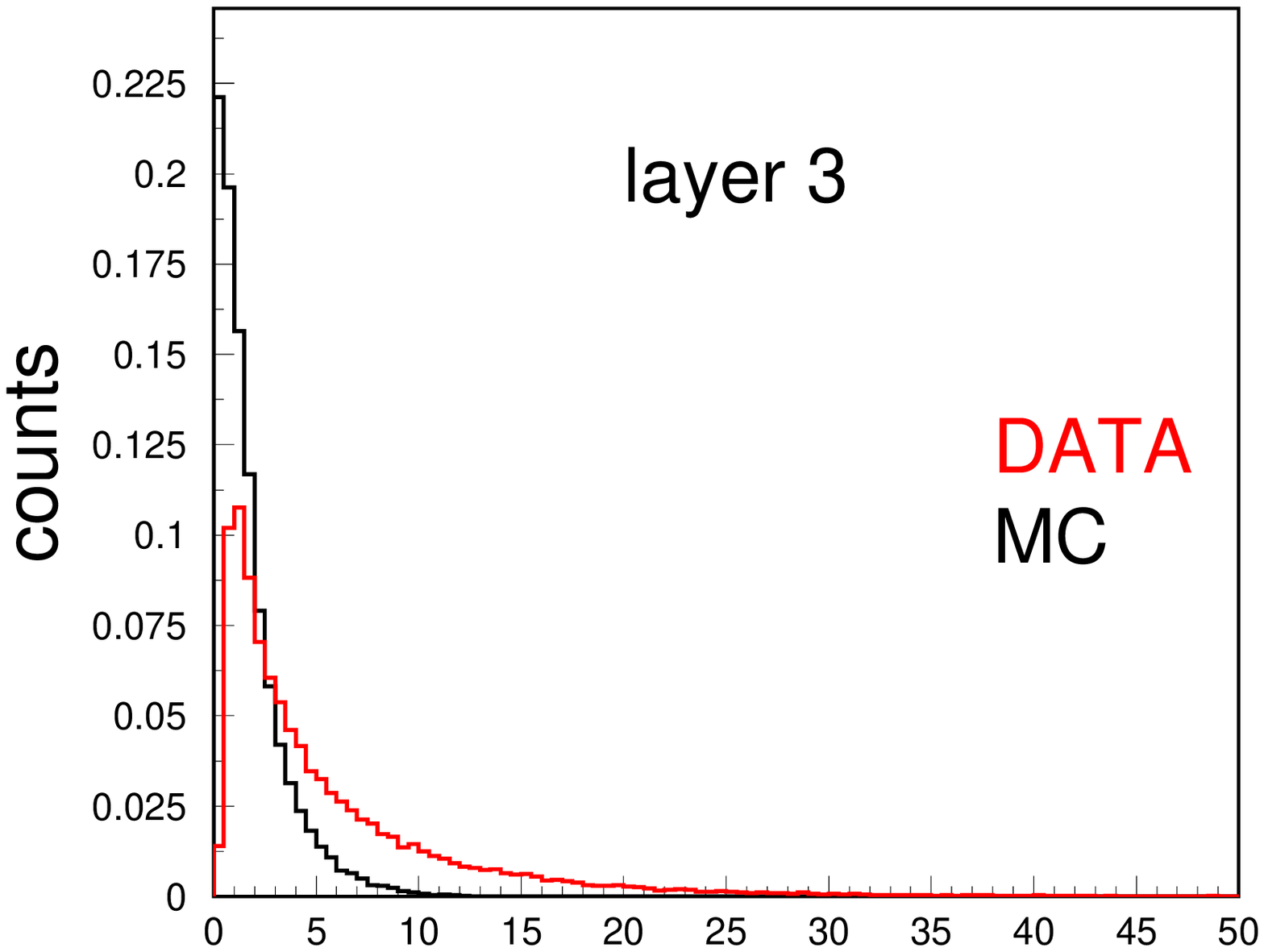}
\includegraphics[width=0.46\textwidth]{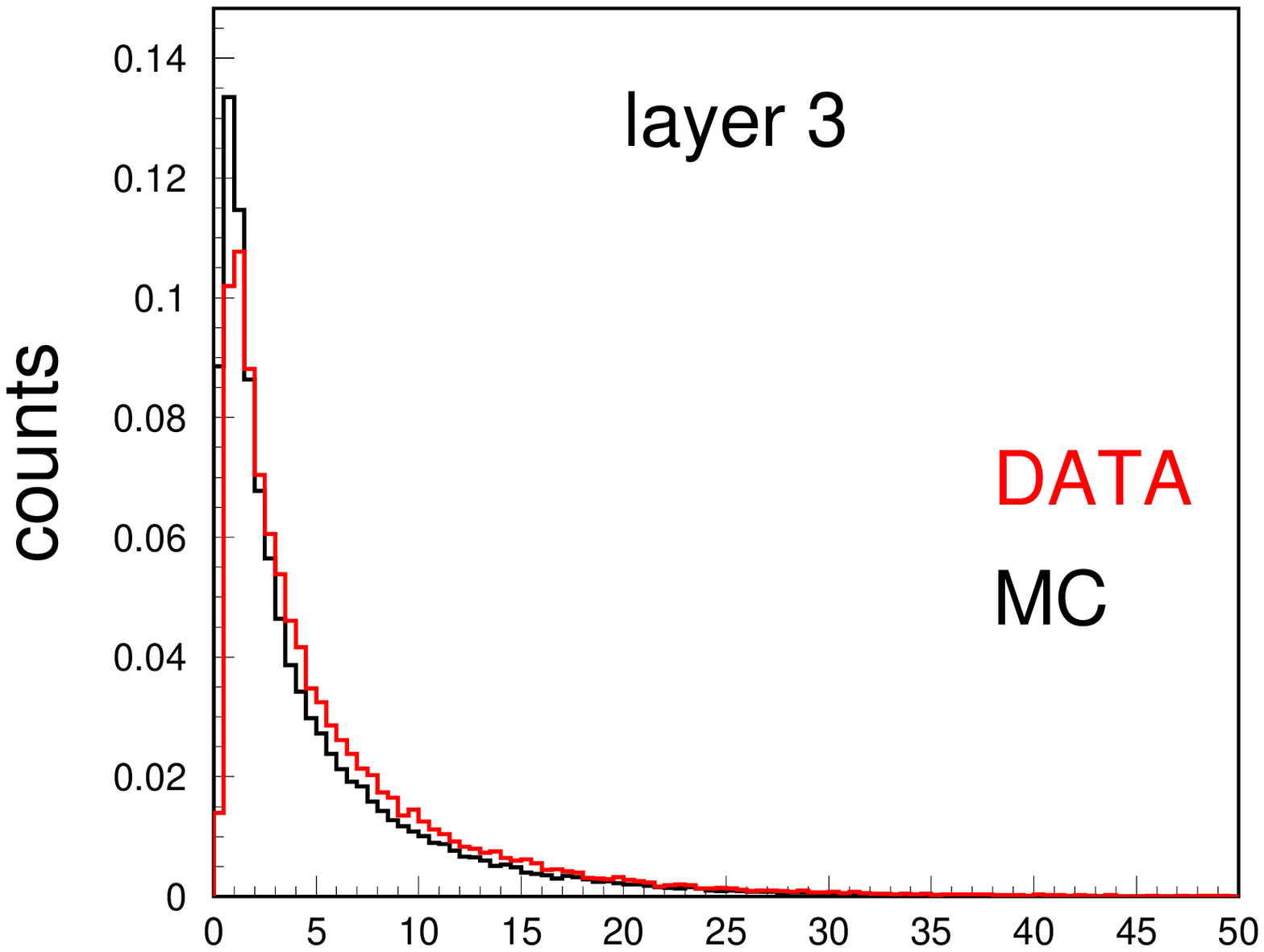}

\vspace{-2.45cm}
\hspace{0.7cm}
\includegraphics[width=0.46\textwidth]{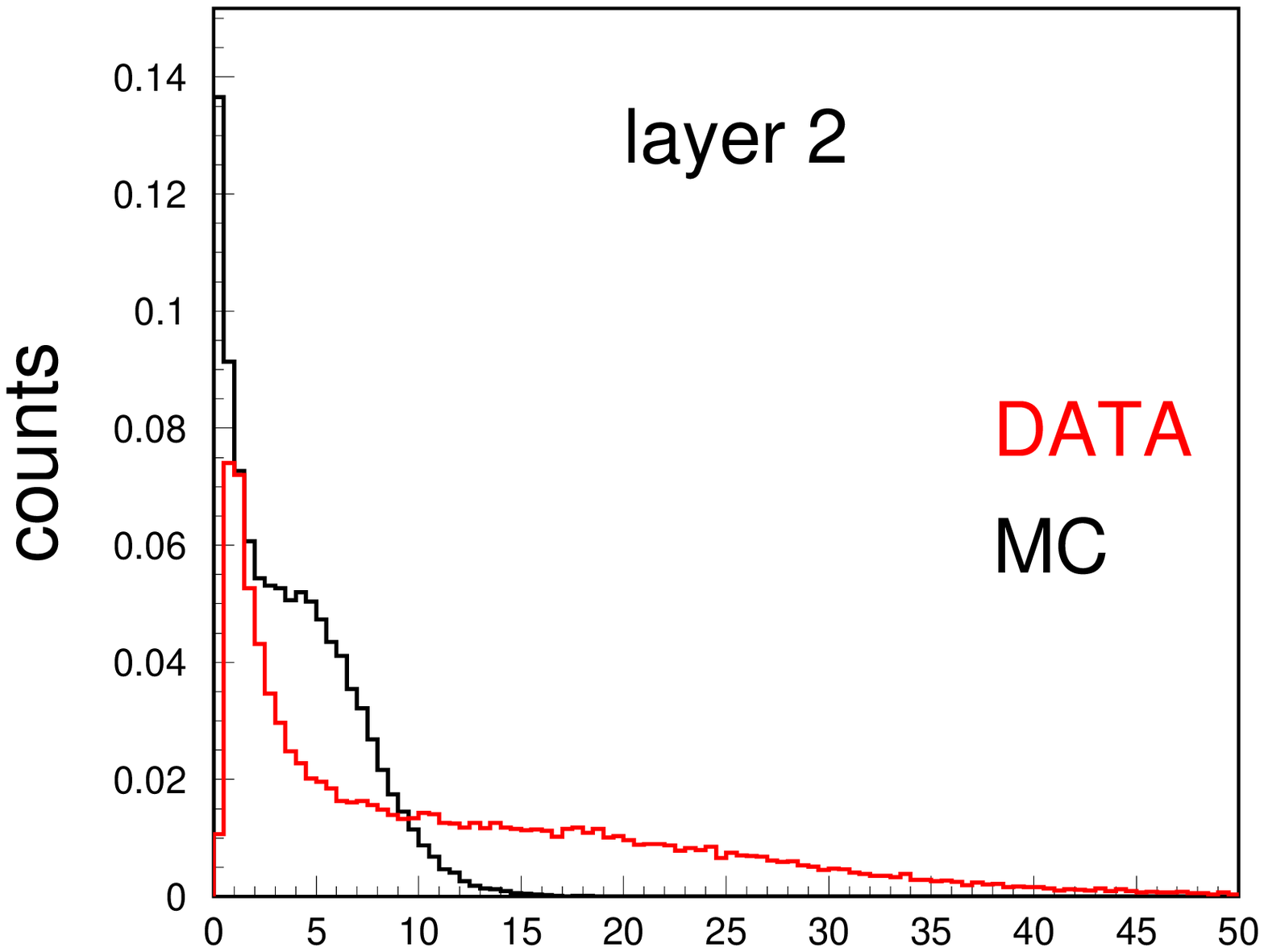}
\includegraphics[width=0.46\textwidth]{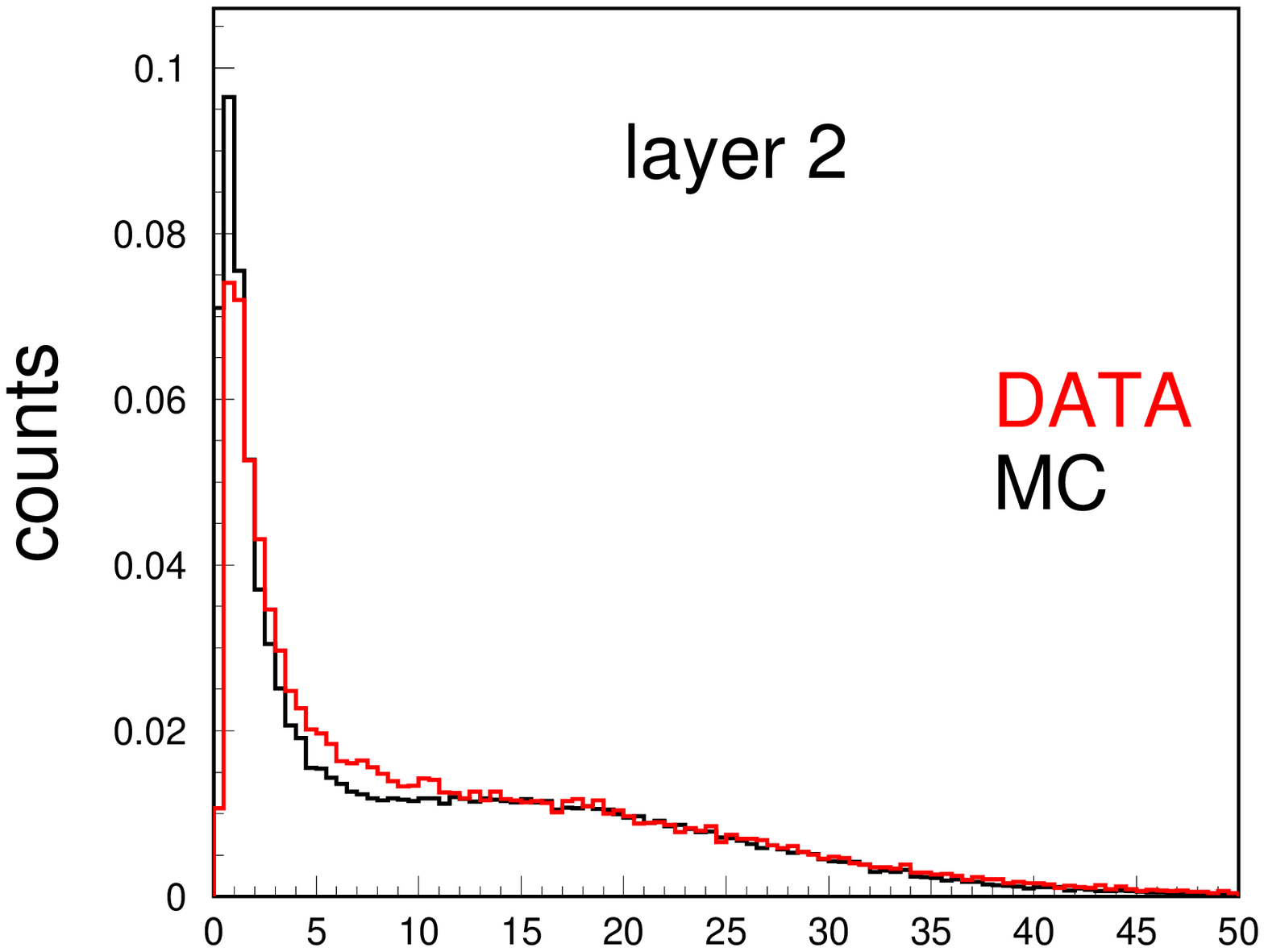}

\vspace{-2.45cm}
\hspace{0.7cm}
\includegraphics[width=0.46\textwidth]{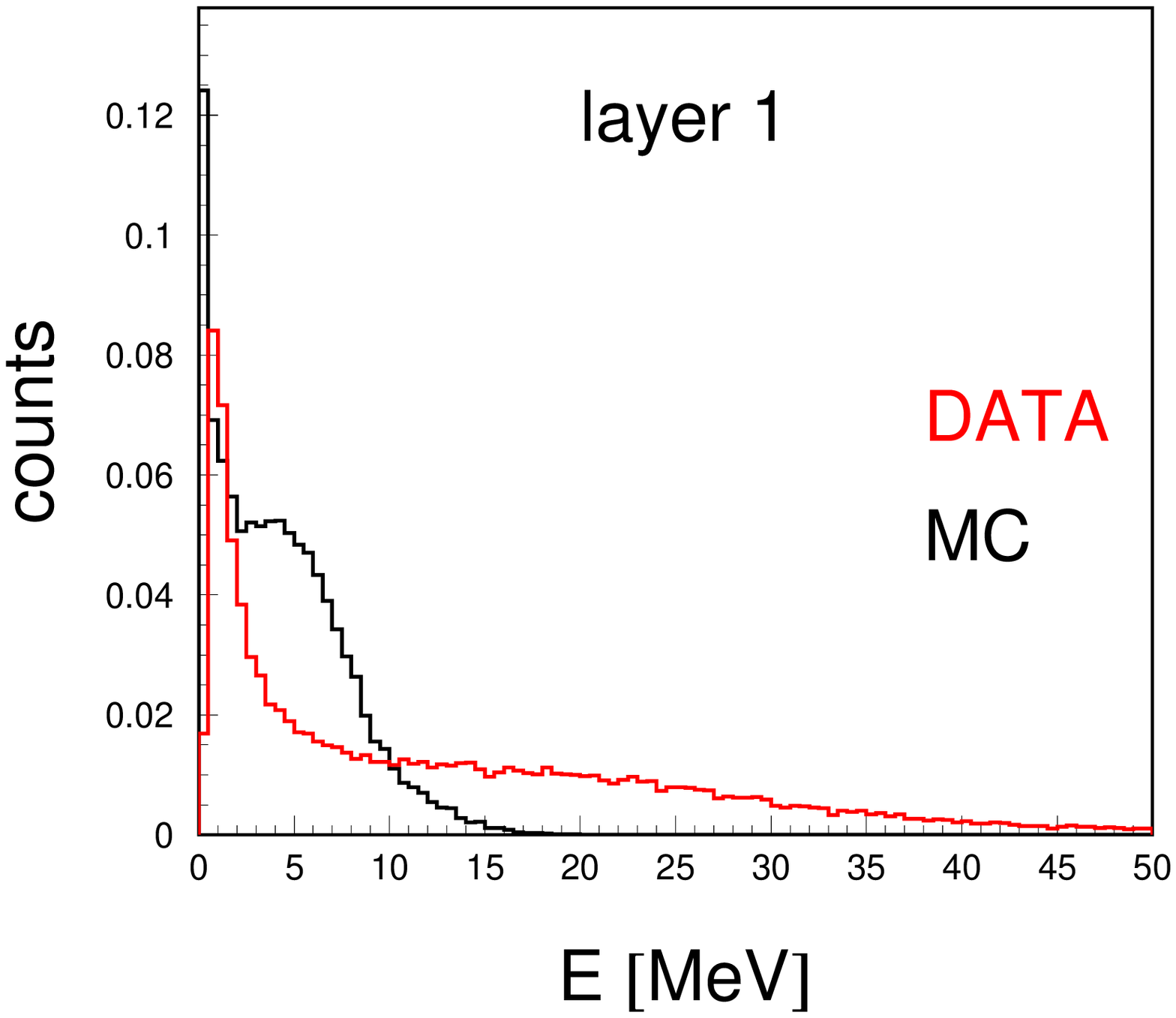}
\includegraphics[width=0.46\textwidth]{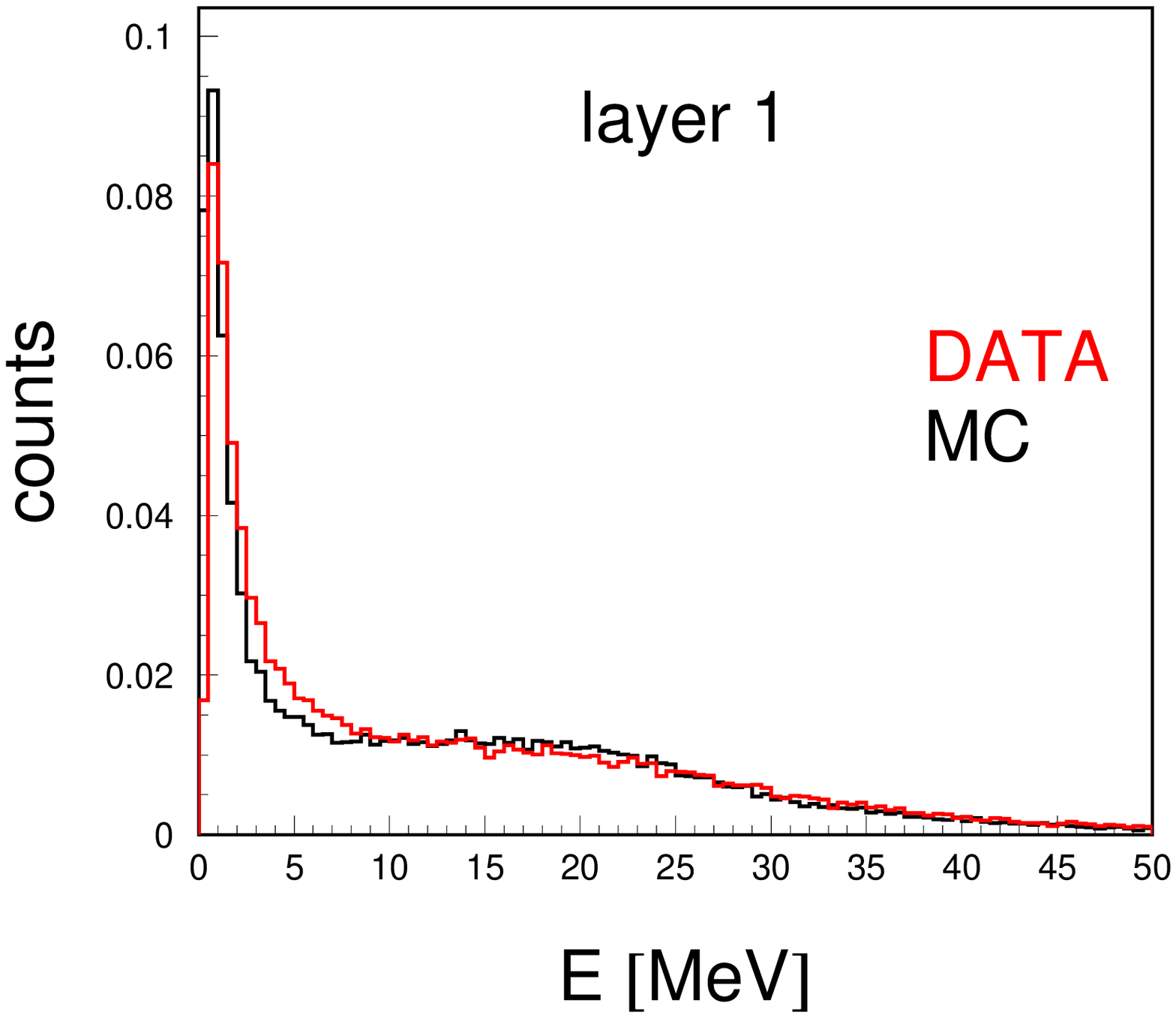}
\vspace{-1.0cm}
\caption{Comparision of energy deposits in 5 layers as determined from simulation (black line) and experiment (red line). 
The presented results before (left column) and after (right column) calibration. Courtasy of E. Czerwi\'nski and B. Di Micco \cite{erykbiagio}.}
\label{jjpluto_vs_data}
\end{figure}

\begin{figure}[H]
\hspace{2.5cm}
\parbox[c]{1.0\textwidth}{\includegraphics[width=0.69\textwidth]{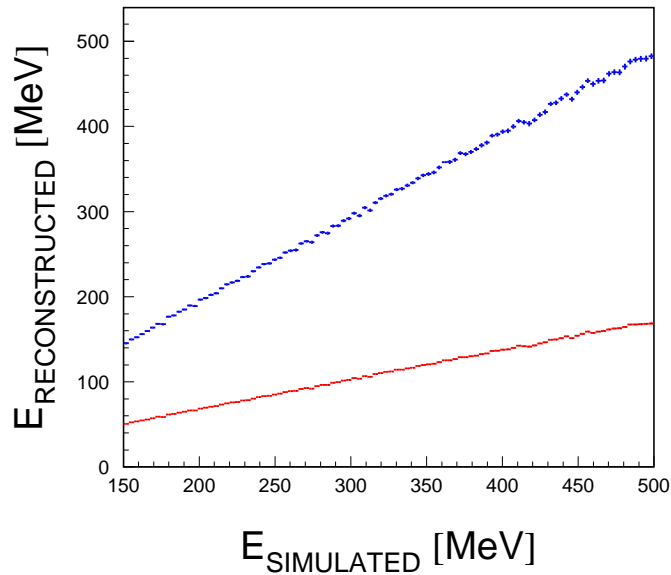}}
\caption{Calibration of the simulated and reconstructed energy. Courtasy of E. Czerwi\'nski and B. Di Micco \cite{erykbiagio}.}
\label{linearity_jjpluto_data}
\end{figure}

\hspace{\parindent}
The next step in the tuning of the simulation program is the adjustment of the absolute scale of the energetic 
response of the calorimeter. For this purpose we compared results of simulations with experimental distribution 
of energy deposits in the calorimeter module. The results for five layers are shown in Fig.~\ref{jjpluto_vs_data} (left column). 
\newpage
One sees that the energy deposits reconstructed for simulated events are much smaller then for the experimental data. 
In order to correct this discrepancy, the simulated values of energy deposits were multiplied by the factor of 3.74. 
The shape of the corrected spectra is in a very good agreement with experimental distributions for each layer (Fig.~\ref{jjpluto_vs_data} right). \\ 
\indent Next, correction was needed because the clustering algorithm not always is able to reconstruct
the entire energy deposited in the clusters. That's why sum of energy deposits which were simulated is greater than sum 
of energy in reconstructed clusters. For this case a linearity correction was needed and it was obtained with 
requiring straight line with slope 1 (blue line in Fig.~\ref{linearity_jjpluto_data}).
The red curve represents the situation before calibration (slope parameter is not equal 1) and blue after. 
Applying the corrections described in this section we achieved  
very good agreement between data and simulations of energy response of the calorimeter when using FLUKA   
and DIGICLU procedures \cite{erykbiagio}.  
%
%
%
%
%\begin{figure}[h]
%\hspace{1.5cm}
%\parbox{0.35\textwidth}{\centerline{\epsfig{file=1reaction-time-position.eps,width=0.30\textwidth}}}
%\caption{
%First reaction-time-position.
%}
%\end{figure}
%
%\begin{figure}[h]
%\hspace{3.5cm}
%\parbox{0.35\textwidth}{\centerline{\epsfig{file=2reaction-resultsonsurface-40000events.eps,width=0.30\textwidth}}}
%\caption{
%Second reaction - energy.
%}
%\end{figure}
%
%\begin{figure}[h]
%\hspace{3.5cm}
%\parbox{0.35\textwidth}{\centerline{\epsfig{file=2reaction-time-position.eps,width=0.30\textwidth}}}
%\caption{
%Second reaction - time-position.
%}
%\end{figure}
%
%
%
\chapter{Estimations of background due to the merging and splitting of photon clusters}
\hspace{\parindent}
%
%\section{Influence of the merging and splitting effects for reactions: -> napisac jakie}
%Przyklad problemu, na podstawie danych z KLOE (motywacja Biagio)
%
%\section{Background of the $\eta \to \pi^{0} \gamma \gamma$ (merging of $\eta \to 3\pi^{0}$)}
%
The main motivation of this work is the estimation of merging and splitting effects for reactions 
 like e.g. $\eta \to 3\pi^{0}$ or $K_{S} \to 2\pi^{0}$ which may constitue background in studies of 
the $\eta \to \pi^{0}\gamma\gamma$ and $K_{S} \to 3\pi^{0}$ reactions, respectively. 
%
% where $\eta$ meson decays to channels with $\pi^{0}$ mesons and $\gamma$ quantas. We perform the simulations due 
% to improve efficiency of the reconstruction products from reactions: $\eta \to \pi^{0}\gamma\gamma$ and 
% $K_{S} \to 3\pi^{0}$ and minimazing 
% background from the channel $\eta \to 3\pi^{0}$ in the first case and from $K_{S} \to 2\pi^{0}$ in the second, 
% which influence to main signal. \\
%
 In Table~\ref{br_spli_merg} we present a Branching Ratios (BR) for the main investigated reactions 
and for the background channels.
\vspace{-0.5cm}
\begin{table}[H]
\begin{center}
\begin{tabular}{|l|c|c|c|c|}  \hline \hline
\emph{Main Decay} & 
                {\emph{BR (\%)}} & {\emph{Background Channel}} & {\emph{BR (\%)}} & {\emph{Effect}} \\ \hline
$\eta \to \pi^{0}\gamma\gamma$ & (4.4 $\pm$ 1.6) $\cdot 10^{-6}$ & $\eta \to 3\pi^{0}$ & (32.51 $\pm$ 0.28) & merging  \\ \hline
$K_{S} \to 3\pi^{0}$ & < 1.2 $\cdot 10^{-9}$ & $K_{S} \to 2\pi^{0}$  & (30.69 $\pm$ 0.05) & splitting  \\ \hline
$K_{L} \to 2\pi^{0}$ & (8.69 $\pm$ 0.04) $\cdot 10^{-2}$ & $K_{L} \to 3\pi^{0}$ & (19.56 $\pm$ 0.14) & merging  \\  \hline \hline
\end{tabular}
\end{center}
\vspace{-0.3cm}
\caption{Branching Ratios for examples of studied reactions and the background decays  
\cite{pdg_niebieska}.}
\label{br_spli_merg}
\end{table}
\vspace{-0.2cm}
In this Table one can see that all the background reactions have much bigger BR (many orders of magnitude) than the investigated channels. 
% Thus, these 
% reactions will appear very often during the $\phi$ meson decays in detector and the problem with reconstruction 
% these events will be relevant.  
Therefore, even small merging and splitting effects may obscure considerably observation of signals 
from $\eta \to \pi^{0}\gamma\gamma$, $K_{S} \to 3\pi^{0}$ or $K_{L} \to 2\pi^{0}$ decays.

\section{Merging of clusters from $\eta \to 3\pi^{0}$ as a background for the $\eta \to \pi^{0} \gamma \gamma$ channel}
\hspace{\parindent}
The example of the merging effect, inducing a wrong cluster counting, comes from analysis of the $\eta \to \pi^{0}\gamma\gamma$ 
events \cite{Biagio_thesis}. 
The Branching Ratio (BR) for $\eta \to 3\pi^{0}$   
  is by several orders of magnitude more frequent than 
$\eta \to \pi^{0}\gamma\gamma$ (see Table~\ref{br_spli_merg}). 
%
% \hspace{0.5cm} 
% \vspace{0.25cm}
\begin{figure}[H]
\hspace{3.5cm}
\parbox[c]{1.0\textwidth}{
\parbox[c]{0.55\textwidth}{\includegraphics[width=0.49\textwidth]{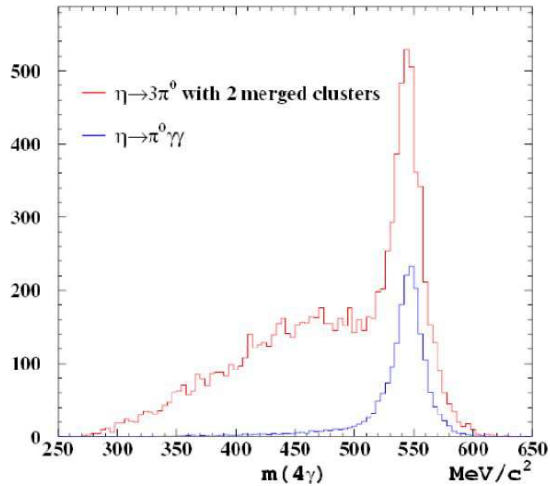}}
}
\caption{Invariant mass of four photons for the $\eta \to \pi^{0}\gamma\gamma$ events and for the 
$\eta \to 3\pi^{0}$ background with two merged clusters \cite{Biagio_thesis}.}
\label{cluster_merging_effect}
\end{figure}
%
% Jeszcze jakis z innej pracy.
%
%%%%%%%%%%%%%%%%%%%%%% OSZACOWANIE WIELKOSCI PROBLEMU %%%%%%%%%%%%%%%%%%%%%%%%%%%%%%%%%%%%%%%%%%

\begin{figure}[H]
\hspace{2.5cm}
\parbox[c]{0.50\textwidth}{\epsfig{file=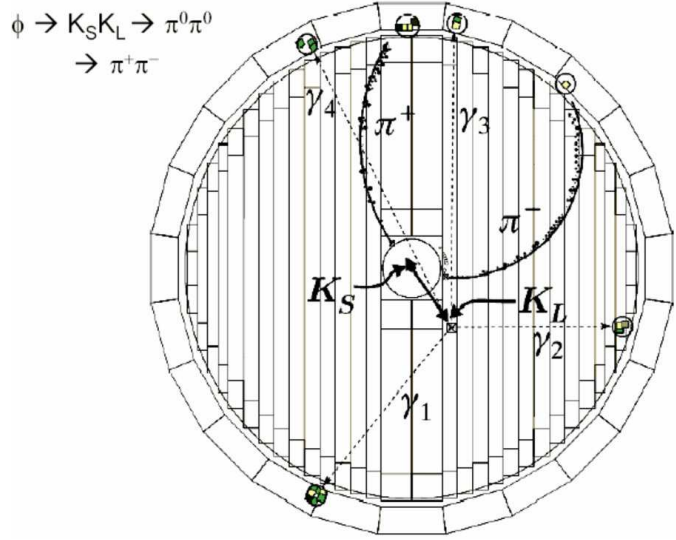, width=0.59\textwidth}}
\caption{A CP-violating $\phi \to K_{S}K_{L}$ event. The $K_{S}$ decays into $\pi^{+}\pi^{-}$ 
very close to the interaction point. The $K_{L}$ decays into $\pi^{0}\pi^{0}$ further away 
resulting in four photons which are detected by the calorimeter \cite{kloe_electromagnetic_nim2}.}
\label{decay_channel}
\end{figure}

%Zrobic symulacje jak czesto uderzaja w 1, 2 i 3 moduly dwie gammy dla calego barrel !!!
%(JJpluto + Flukaexample = alfa06 lub teraz we FLUCE).
%
In Fig.~\ref{cluster_merging_effect} a consequence of the merging of clusters is presented for the case  
when 6$\gamma$ quanta from 3$\pi^{0}$ decay were reconstructed as 4 clusters. 
% 
% 
% This efect is related with the fact that 
% 6$\gamma$ quanta from 3$\pi^{0}$ decay were reconstructed as 4 merging clusters, thus the final topology 
%  of this reaction is interpreted with 4 reconstructed particles in the finale state. 
In this case the topology of $\eta \to 3\pi^{0}$ becomes equal to that of the $\eta \to \pi^{0}\gamma\gamma$ reaction. 
% So in this case 
% decay $\eta \to 3\pi^{0}$ becomes a background channel for the signal. In the figure~\ref{cluster_merging_effect} is shown
% the huge background surviving the kinematical fit, due to $\eta \to 3\pi^{0}$ decays with double merged
% clusters. 
And as a result on 4$\gamma$ invariant mass distribution one can see a big peak from background channel exactly at the mass 
of the $\eta$ meson \cite{Letter_of_intent}. \\ 
\indent The next very interesting example of reactions for which problems with merging of
 clusters is significant and which can be studied with the KLOE 
calorimeter~\cite{kloe_electromagnetic_nim2} is CP violating decay $K_{L} \to 2\pi^{0}$. An example of such 
event is presented in Fig.~\ref{decay_channel}.  
% are kaons decays.
% For example the following reaction $\phi \to K_{S}K_{L}$ is interesting to study due to the fact that 
% CP symmetry is not conservated (see Figure~\ref{decay_channel}). 
%
%Dla ktorej moze wystapic problem merging i splitting i ze ona jest bardzo ciekawa bo 
%lamie CP.
%
%
% \hspace{.5cm} 
%\vspace{-0.2cm}
%
%\vspace{-0.2cm}
The $K_{S}$ meson decays into $\pi^{+}\pi^{-}$ near the collission point. The $K_{L}$ meson
decays into $\pi^{0}\pi^{0}$ and then these pions decaying through electromagnetic interaction
to 4$\gamma$ quanta. Those four photons are detected by the EmC. 
% The background channel for 
% $K_{L} \to 2\pi^{0}$ decay is $K_{L} \to 3\pi^{0}$ reaction. 
Thus $K_{L} \to 3\pi^{0} \to 6\gamma$ with two merged clusters may obscure a signal from the 
 $K_{L} \to 2\pi^{0}$ reaction. \\   
%
%
%
%\chapter{Clustering algorithm tests}
%
%\chapter{Influence of the quantum efficiency}
%
%
\newpage
\indent Implementation of the full barrel geometry into the FLUKA program allowed us to study the effect of merging 
and splitting quantitatively. And hereafter we will present the results for the reconstruction of clusters 
from the $\eta \to 3\pi^{0}$ and $K_{S} \to 2\pi^{0}$  
in view of the merging and splitting which estimation is crucial for the study of the $\eta \to \pi^{0}\gamma\gamma$ and 
$K_{S} \to 3\pi^{0}$ decays. 
\vspace{-0.3cm}
\begin{figure}[H]
\hspace{2.8cm}
\parbox[c]{1.0\textwidth}{
\parbox[c]{0.535\textwidth}{\includegraphics[width=0.54\textwidth]{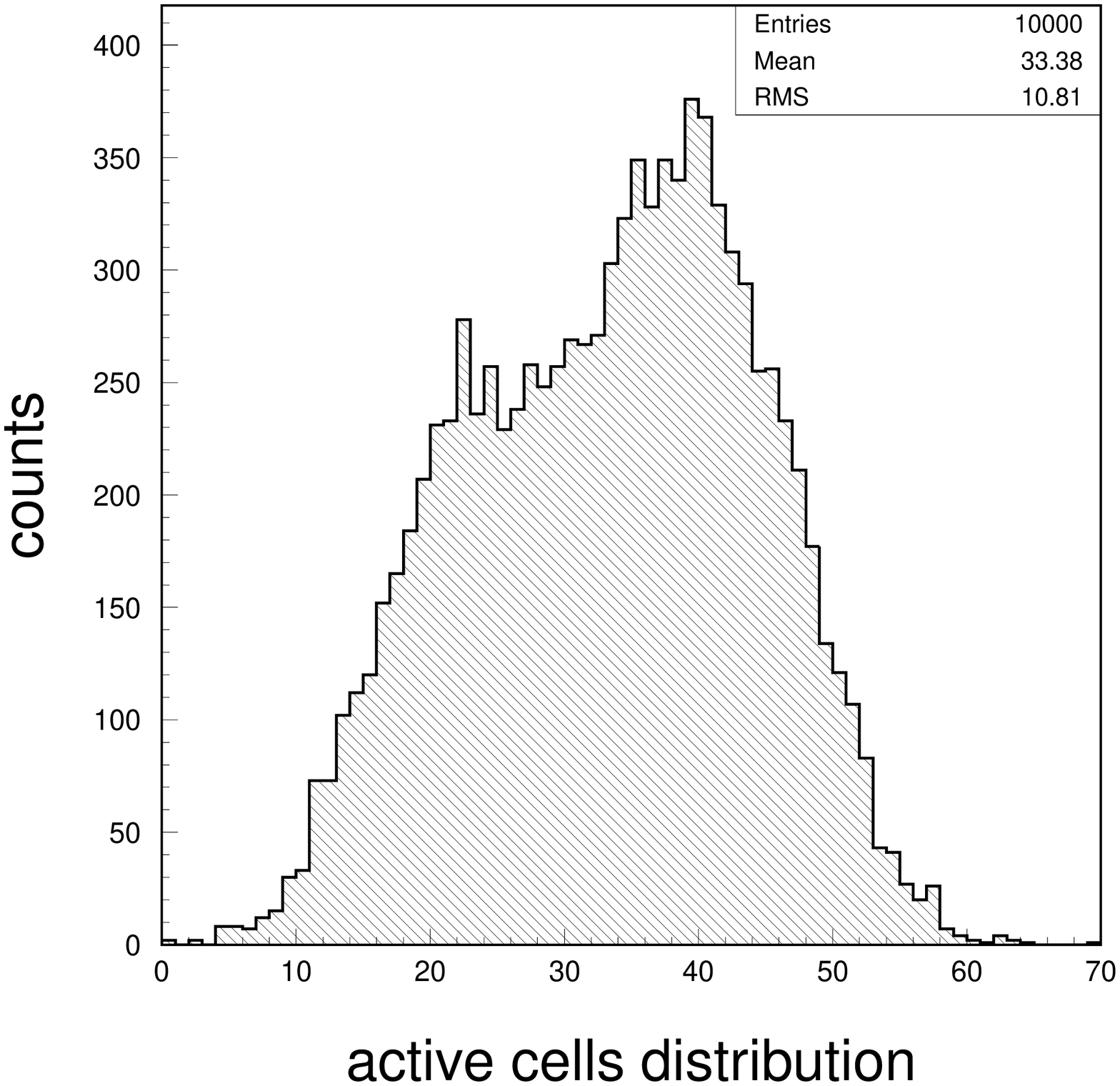}}
}
\caption{Distribution of the number of active cells  
 for reaction: $e^{+}e^{-} \to \phi \to \eta\gamma \to 3\pi^{0}\gamma \to 7\gamma$. Statistics is equal to 10 000 events.}
\label{clusters_24_cells}
\end{figure}
\vspace{-0.2cm}
The Fig.~\ref{clusters_24_cells} shows a distribution of the total number of cells which given signal.   
% 
%
% the middle a distribution of 
Whereas distribution of the total number of preclusters and the total number of 
reconstructed clusters are presented in the next Figure. 
\begin{figure}[H]
\hspace{-0.5cm}
\parbox[c]{1.0\textwidth}{
\parbox[c]{0.5\textwidth}{\includegraphics[width=0.55\textwidth]{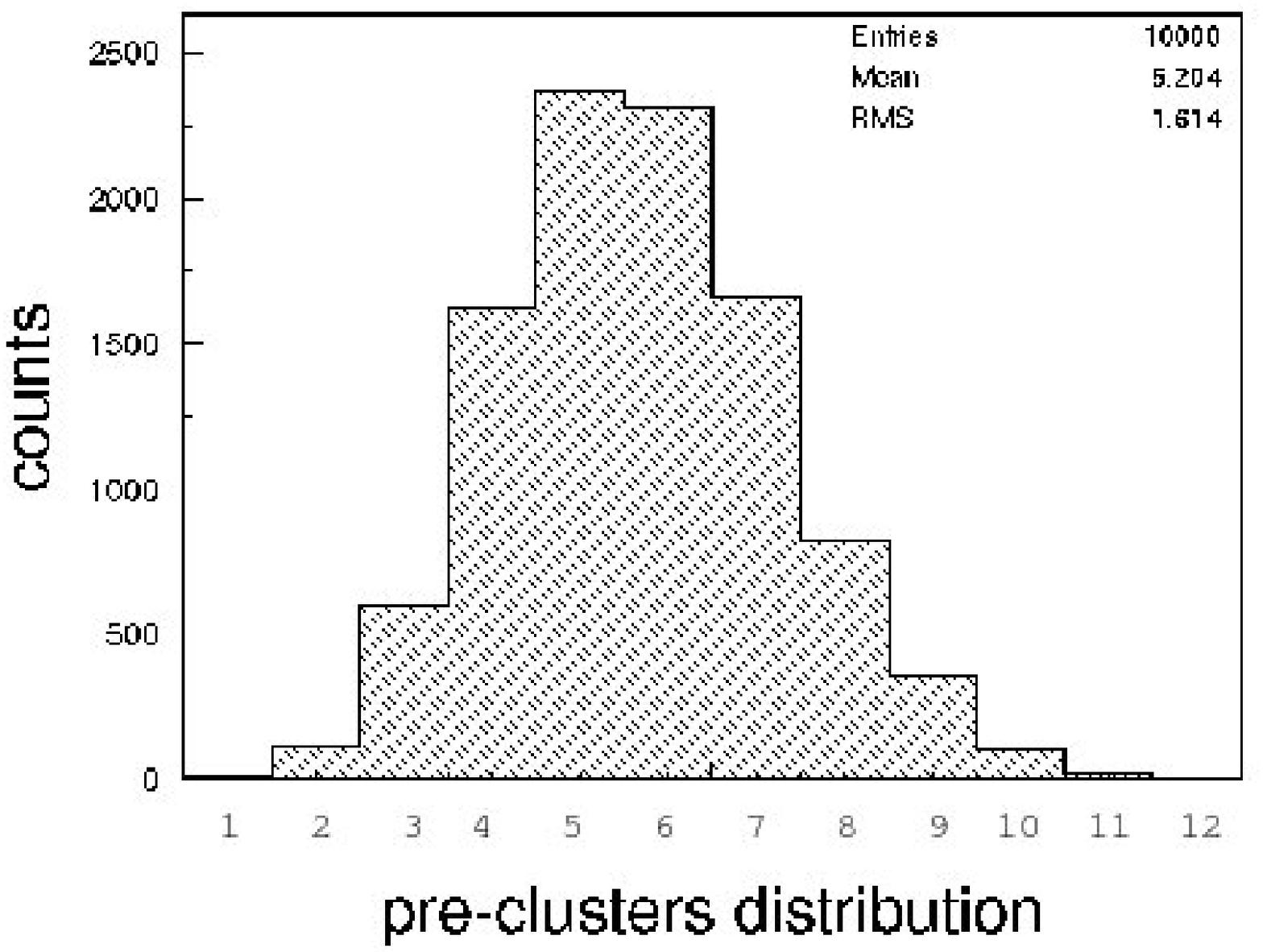}}
\hspace{0.2cm}
\parbox[c]{0.5\textwidth}{\includegraphics[width=0.55\textwidth]{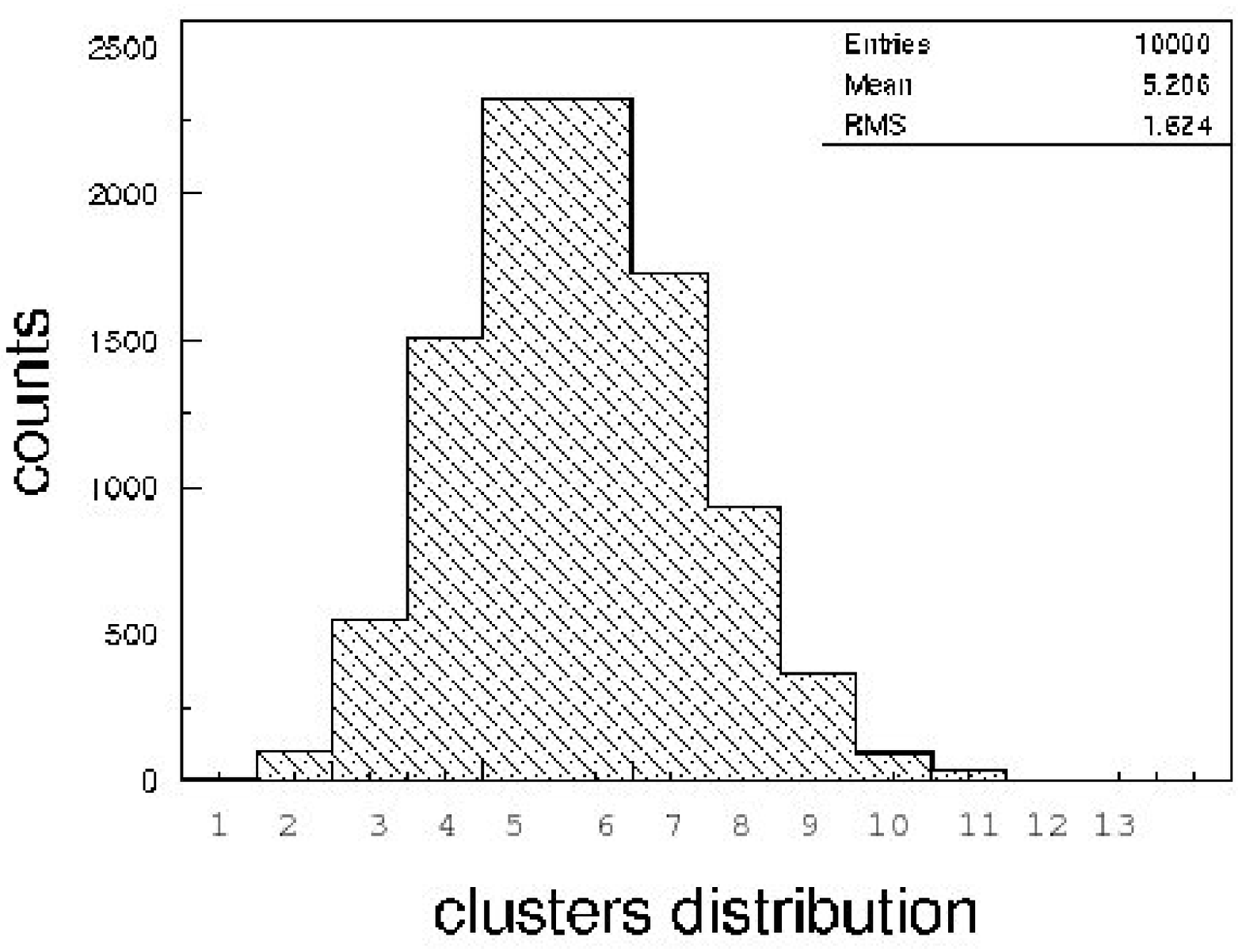}}
}
\caption{Distributions of the reconstructed pre-clusters (left) and total reconstructed clusters (right), 
 for reaction: $e^{+}e^{-} \to \phi \to \eta\gamma \to 3\pi^{0}\gamma \to 7\gamma$. Statistics is equal to 10 000 events.}
\label{clusters_24}
\end{figure}
The reconstruction possibilities of the clustering algorithm 
are presented in Fig.~\ref{clusters_24}. 
One can see that the most   
probably the clustering algorithm reconstructs 5 or 6 clusters even though 7 photons had hit the detector.   \\   
%
% Also interesting is that 
% algorithm didn't reconstructed a clusters between modules, on the edges of modules. We suppose that this is the reason why efficiency of
% $\gamma$ quanta on the whole barrel is equal only 75.28\%. 
%
The frequencies of reconstructed clusters are presented in  
Table~\ref{clusters_reaction_2}.   
\begin{table}[H]
\begin{center}
\begin{tabular}{|l|c|c|}  \hline \hline
\emph{Number of reconstructed} & \emph{Frequency for} &  
                {\emph{Frequency for}} \\ % \hline
\emph{preclusters and clusters} & \emph{preclusters (\%)} &  
                {\emph{clusters (\%)}} \\ \hline
 1  &  0.12    &  0.12   \\
 2  &  1.11    &  1.02   \\
 3  &  5.99    &  5.50   \\ 
 4  &  16.22   &  15.13  \\ 
 5  &  23.70   &  23.23  \\ 
 6  &  23.09   &  23.26  \\  
 7  &  16.60   &  17.27  \\
 8  &  8.25    &  9.35   \\
 9  &  3.59    &  3.71   \\
 10 &  1.06    &  0.97   \\
 11 &  0.22    &  0.36   \\
 12 &  0.27    &  0.05   \\  
 13 &  0.00    &  0.03   \\  \hline \hline
\end{tabular}
\end{center}
\caption{Efficiency of the reconstructed pre-clusters and clusters for  $e^{+}e^{-} \to \phi \to \eta\gamma \to 3\pi^{0}\gamma \to 7\gamma$ channel. 
Statistics equal to 10 000 events.}
\label{clusters_reaction_2}
\end{table}
%\vspace{0.5cm}
%
One can see that the topology of this channel could be interpreted both with 4 $\gamma$ quanta (15.13\%) as well as with 6 photons (23.26\%). 
In the first case this channel could constitute a background for $\eta \to \pi^{0} \gamma \gamma$ reaction (merging effect). 
% and 
% in the second four reconstructed photons from this channel could constitue a background for reaction $K_{S} \to 3\pi^{0}$. 
%
%
\section{Splitting of clusters for $K_{S} \to 2\pi^{0}$ as a background for $K_{S} \to 3\pi^{0}$}
\begin{figure}[H]
\hspace{3.6cm}
\parbox[c]{1.0\textwidth}{
\parbox[c]{0.55\textwidth}{\includegraphics[width=0.48\textwidth]{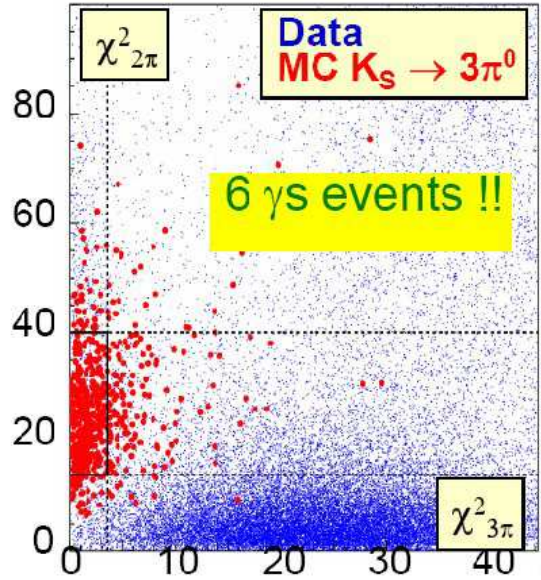}}
}
\caption{Disribution of the $\chi^{2}$ variable determined under $3\pi^{0}$ and $2\pi^{0}$ hypotheses of the MC 
$K_{S} \to 3\pi^{0}$ signal and of the data \cite{Letter_of_intent}.}
\label{cluster_splitting_effect}
\end{figure}

\begin{figure}[H]
\hspace{3.0cm}
\parbox[c]{1.0\textwidth}{
\parbox[c]{0.535\textwidth}{\includegraphics[width=0.56\textwidth]{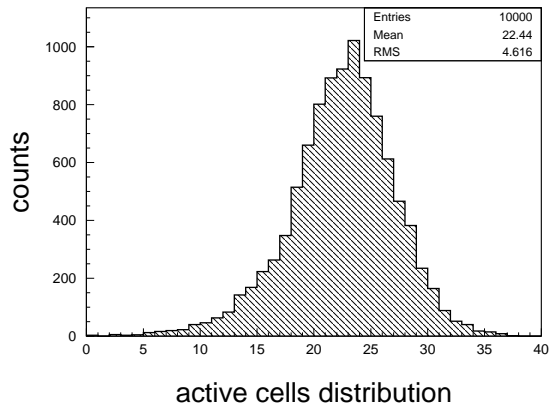}}
}
\vspace{-0.5cm}
\caption{Distribution of the number of active cells 
 for $e^{+}e^{-} \to \phi \to K_{L}K_{S} \to K_{L} 2\pi^{0} \to K_{L} 4\gamma$ reaction. Statistics is equal to 10000 events.}
\label{clusters_24_cells_2}
\end{figure}

\begin{figure}[H]
\parbox[c]{1.0\textwidth}{
\parbox[c]{0.5\textwidth}{\includegraphics[width=0.52\textwidth]{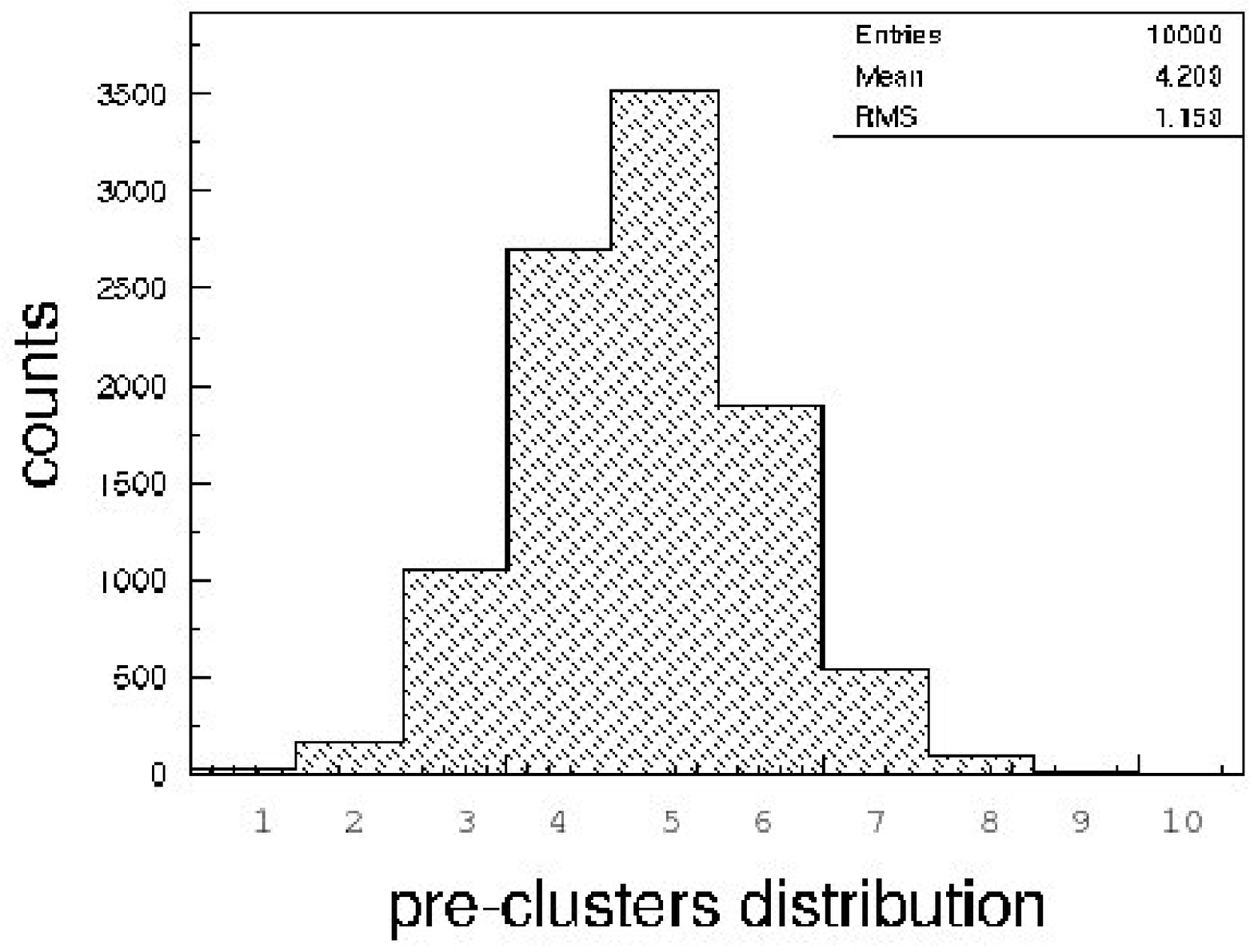}}
%\hspace{0.1cm}
\parbox[c]{0.5\textwidth}{\includegraphics[width=0.52\textwidth]{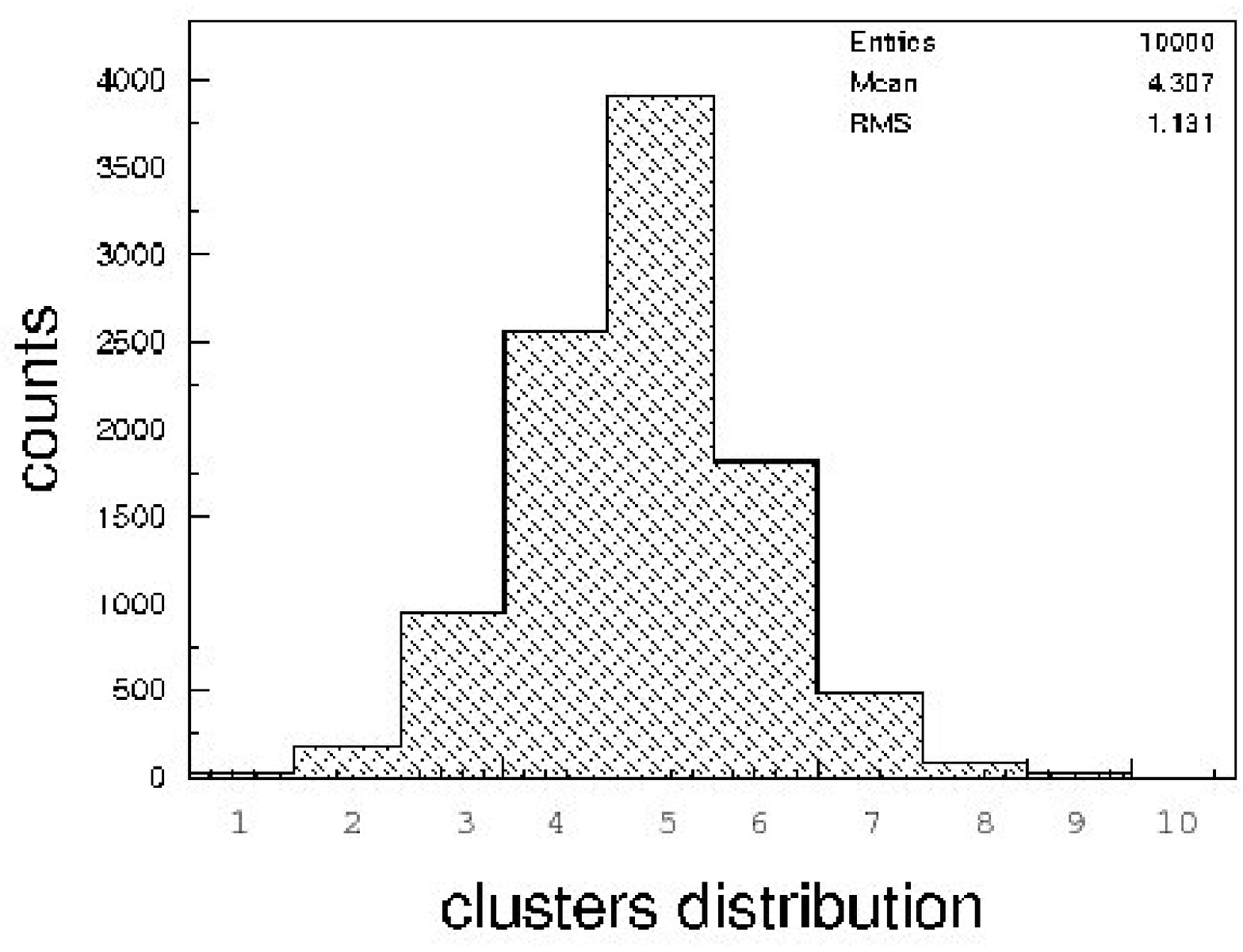}}
}
\caption{Distributions of the number of reconstructed pre-clusters (left panel) and total number of reconstructed clusters (right panel),
 for $e^{+}e^{-} \to \phi \to K_{L}K_{S} \to K_{L} 2\pi^{0} \to K_{L} 4\gamma$ reaction. Statistics is equal to 10000 events.
\label{clusters_24_2}
}
\end{figure}

\hspace{\parindent}
Splitting is related with the features of the EmC detector and with reconstruction possibilities of the KLOE clustering algorithm which not always
correctly reconstructs positions and energies of particles interacting with detector
material and may assign more then one cluster to signals induced by a single particle. \\
\indent Fig.~\ref{cluster_splitting_effect}  
% the influence of the splitting effect is shown.
% The background for $K_{S} \to 3\pi^{0}$ is decay $K_{S} \to 2\pi^{0}$ withdouble-splitting 
% of the photon clusters \cite{Letter_of_intent}.
shows that data with 6$\gamma$ quanta in the final state gives in most cases much lower $\chi^{2}$ 
when fit under the hypothesis of 2$\pi^{0}$ then for the 3$\pi^{0}$ hypothesis whereas for a real 
6$\gamma$ events originating from the $K_{S} \to 3\pi^{0}$ decay one expects $\chi^{2}_{3\pi}$ 
to be small and $\chi^{2}_{2\pi}$ to be large as it is seen in the sample of the MC $K_{S} \to 3\pi^{0}$ 
data (Fig.~\ref{cluster_splitting_effect}).  
\newpage
\vspace{-0.3cm}
Therefore, the $K_{S} \to 2\pi^{0} \to 4\gamma$ decay with two splitted clusters 
gives the same event topology as expected for the $K_{S} \to 3\pi^{0} \to 6\gamma$ reaction.     
%
%
% The populations of the $\chi^{2}$ variables in 3$\pi^{0}$ and 2$\pi^{0}$ hypotesis
% for the MC signal and for the data events are different \cite{Letter_of_intent}. \\
%
\indent We investigated $e^{+}e^{-} \to \phi \to K_{L}K_{S} \to K_{L} 2\pi^{0} \to K_{L} 4\gamma$ reaction with condition that 
all four $\gamma$ quanta hit the barrel calorimeter. We performed approximation that K$_{S}$ meson decays exactly in 
 the collision point. However, we suppose that this 
approximation hasn't significant influence on the results because a distance between collision point and hit position at calorimeter 
surface is at least 2 meters. We present the distribution of the active cells in Fig.~\ref{clusters_24_cells_2}.  
\vspace{-0.8cm}
We also investigated the distribution of pre-clusters and clusters for this channel (Fig.~\ref{clusters_24_2}). 
One can see that in 18.08\% a clustering algorithm reconstructed 6$\gamma$ quanta. Therefore for this case 
the topology of investigated channel with 6 reconstructed photons (two splitted clusters) is the same as for K$_{S} \to 3\pi^{0}$ reaction and 
it constitute a background for this decay. 
%
%
%\hspace{-2.6cm}
%
%
% One can see that for 3$\gamma$ quanta in final state which hit the barrel, the algorithm mostly reconstructed 3.234 clusters. 
% It consists 107.8\% of total number of particles. We suppose that this fact is caused by splitting of clusters. \\
%
\vspace{-0.2cm}
The fractions of numbers of reconstructed clusters are presented in the Tab.~\ref{clusters_reaction_2_2}.
\vspace{-0.4cm}
\begin{table}[H]
\begin{center}
\begin{tabular}{|l|c|c|}  \hline \hline
\emph{Number of reconstructed} & \emph{Frequency for} &  
                {\emph{Frequency for}} \\ % \hline
\emph{preclusters and clusters} & \emph{preclusters (\%)} &  
                {\emph{clusters (\%)}} \\ \hline
 1   &  0.26      &  0.26   \\
 2   &  1.70      &  1.72   \\
 3   &  10.47     &  9.38   \\
 4   &  26.97     &  25.59  \\
 5   &  35.23     &  39.05  \\
 6   &  18.94     &  18.08  \\
 7   &  5.42      &  4.84   \\
 8   &  0.94      &  0.86   \\
 9   &  0.06      &  0.21   \\
 10  & 0.01       &  0.01   \\  \hline \hline
\end{tabular}
\end{center}
\caption{Efficiency of the reconstructed pre-clusters and clusters for $e^{+}e^{-} \to \phi \to K_{L}K_{S} \to K_{L} 2\pi^{0} \to K_{L} 4\gamma$ channel.}
\label{clusters_reaction_2_2}
\end{table}
%\vspace{0.5cm}
%
One can see that correct reconstruction of four clusters constitutes only about 26\% of total number of events.
%
% For this channel we also investigated a distribution of clusters as a function of the azimuthal angle. 
% This distribution is presented in the figure~\ref{clusters_24_3}. 
%
% \begin{figure}[H]
%\hspace{3.5cm}
%\parbox[c]{1.0\textwidth}{
%\parbox[c]{0.5\textwidth}{\includegraphics[width=0.52\textwidth]{1reaction_ATAN2_XCLUYCLU.eps}}
%}
%\caption{Distribution of the azimuthal angle for reconstructed clusters on the barrel, 
%for reaction: $e^{+}e^{-} \to \phi \to \eta\gamma \to 3\gamma$. Statistic = 1000 events.}
%\label{clusters_24_3}
%\end{figure}
%
%
 Both merging and splitting effects could be reduced by the better shower shape reconstruction induced
by a finer read-out granularity. But also the identification of particles based on the shower shape analysis
would greatly take advantage of an increase in the calorimeter granuality \cite{Letter_of_intent}. \\

\newpage
\section{Reconstruction efficiency with the KLOE clustering algorithm}
\hspace{\parindent}
In order to estimate the
reconstruction efficiency of the present KLOE clustering algorithm we  
% we performed the simulations.
% We tried to reproduce a situation when photons hitting a surface of the module of
% the calorimeter. 
 investigated detector response for two $\gamma$ quanta with momentum of 500 MeV/c. 
The response was generated as a function of the distance between the $\gamma$ quanta. 
% which hitting 
% with the different distance on surface between them. 
Subsequently we have reconstructed a total 
number of clusters which were 
recognized by clustering algorithm. 
% The result from our studies is that the positions of hitting particles 
% have got relevant influence
% for the reconstruction possibilities of the present KLOE algorithm.
The two investigated situations are  
 shown in Fig.~\ref{2gammy_powierzchnia_corel} and in Fig.~\ref{2gammy_powierzchnia2_corel}.
In the first case the distance between photons on x axis is 0 cm and the distance on y axis 
was varried between 0 and 200~cm.     
\mbox{
\hspace{-0.2cm}
\parbox{0.50\textwidth}{
\begin{figure}[H]
\includegraphics[width=0.50\textwidth]{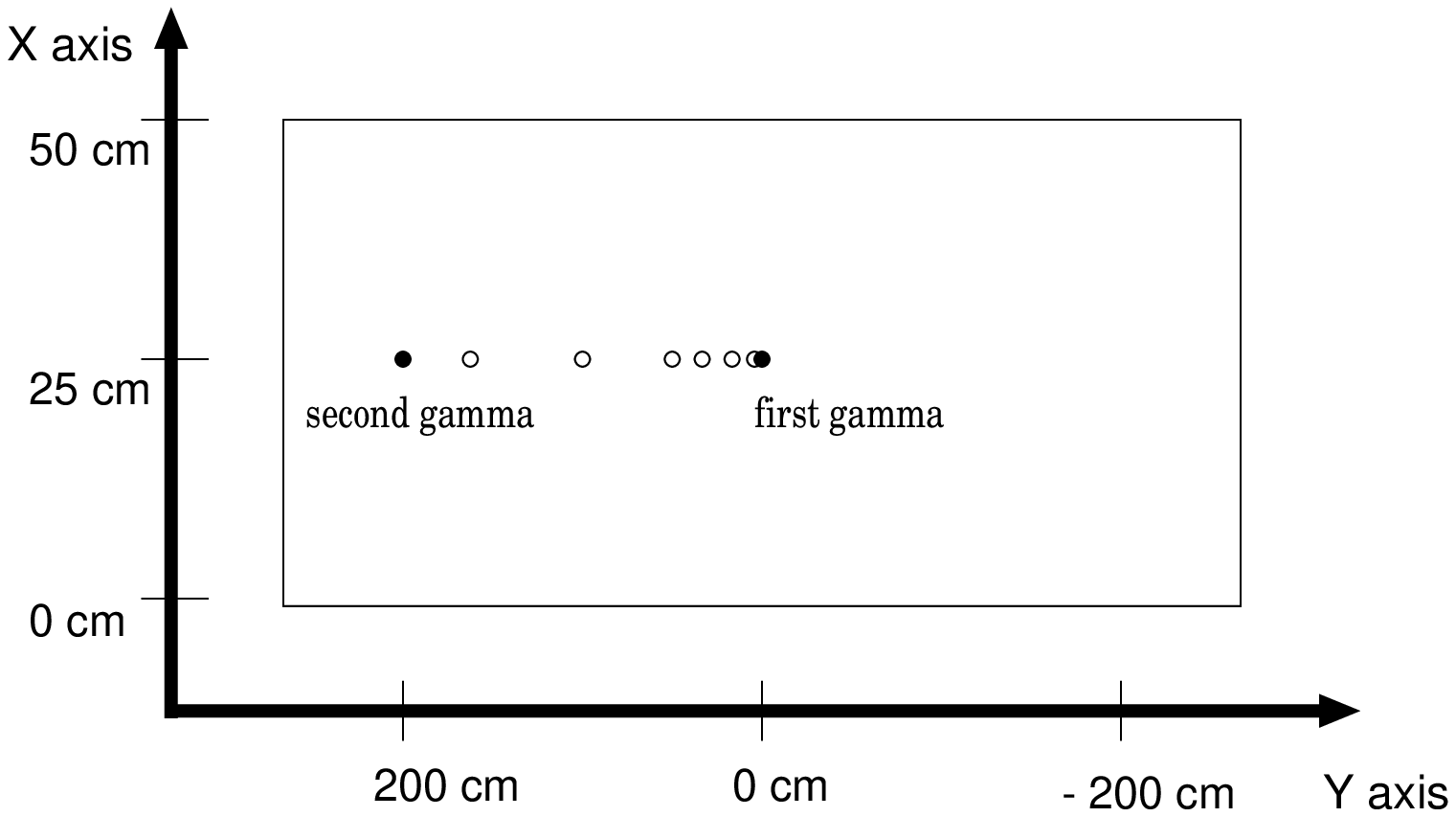}
\caption{Hit positions of $\gamma$ quanta for $\Delta$X~=~0~cm.}
\label{2gammy_powierzchnia_corel}
\end{figure}
}
\hspace{0.0cm}
\parbox{0.50\textwidth}{
\begin{figure}[H]
%\vspace{-0.4cm}
\vspace{0.4cm}
\includegraphics[width=0.50\textwidth]{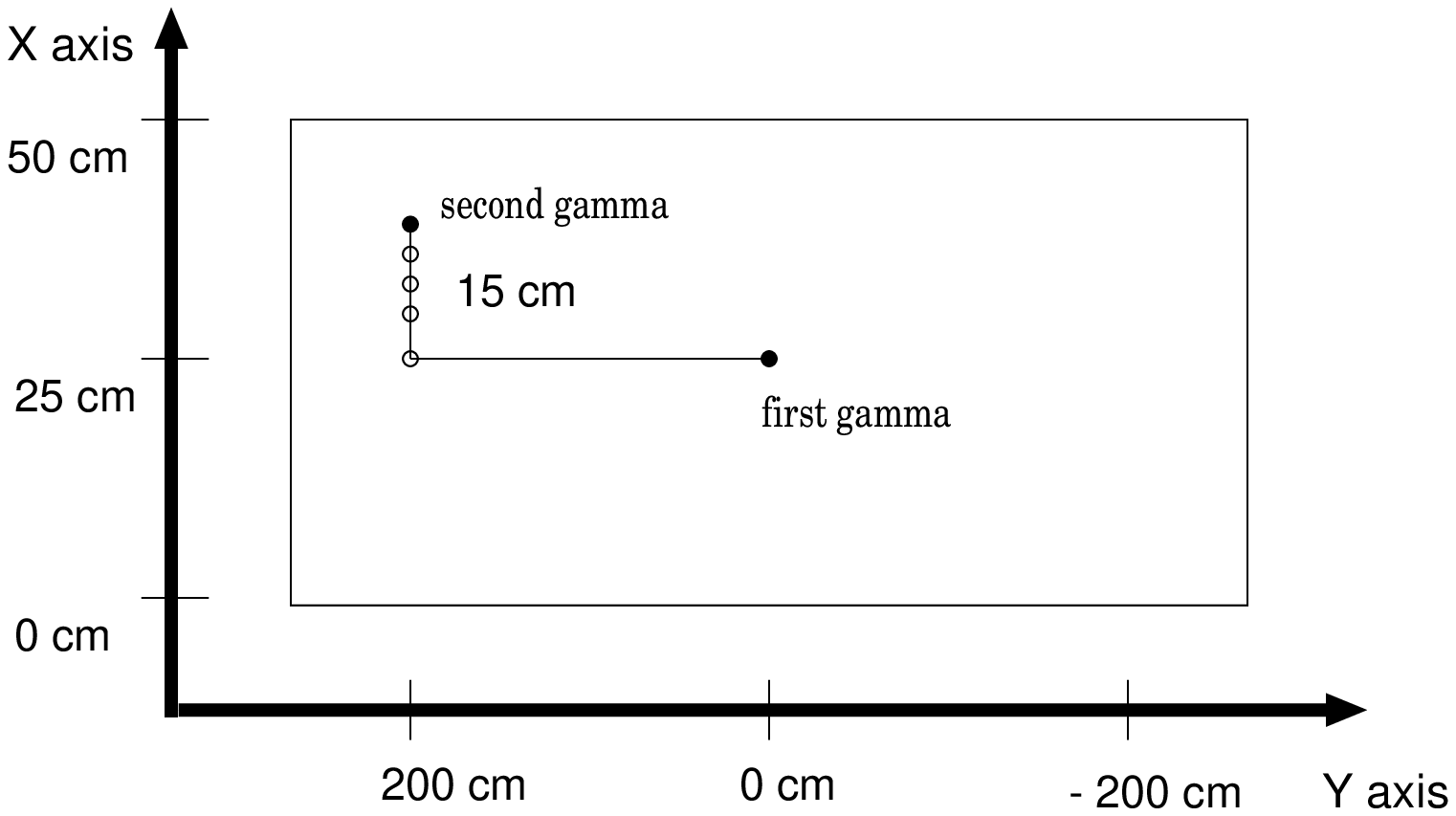}
\caption{Hit positions of $\gamma$ quanta for $\Delta$Y~=~200~cm.}
\label{2gammy_powierzchnia2_corel}
\end{figure}
}
}

If the reconstruction efficiency of the KLOE algorithm was 100\%, 
this algorithm would always reconstruct two clusters which we interpret as two photons  
which hit the module.  
But results of simulations show (see Fig.~\ref{Efficiency_of_the_reconstruction_on_x_distance} and 
Fig.~\ref{Efficiency_of_the_reconstruction_on_y_distance}) that reconstruction 
efficiency is changing as a function of distance between these two particles. 
%On y axis is shown a ratio for 1 reconstructed cluster
%to situation where 2 clusters were reconstructed properly. One can see that for distance between particles less than 
% 400 cm, one bad cluster is 
%reconstructed about five times more frequently than two clusters. This result can be interpreted as influence 
%of the merging effect.  
%For distance less than 50 cm the reconstruction efficiency of this
%algorithm is very low. This situation is causing that if we reconstucting one cluster we can suppose for example 
% that only one $\gamma$ quanta 
%hit in this place with a higher energy.    
%

\mbox{
\parbox{0.44\textwidth}{
\begin{figure}[H]
\hspace{-1.1cm}
\vspace{-0.1cm}
\includegraphics[width=0.49\textwidth]{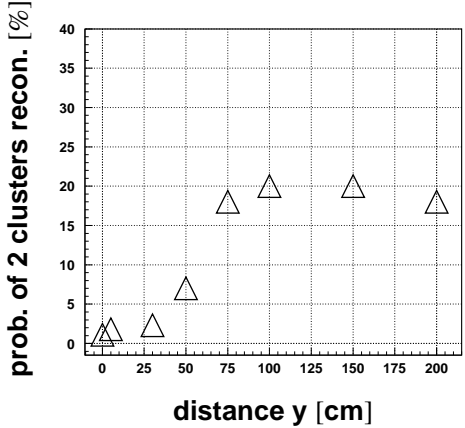}
\caption{The efficiency of the two clusters reconstruction as a function of the distance along y axis provided that x coordinate for both photons 
was the same.}
\label{Efficiency_of_the_reconstruction_on_y_distance}
\end{figure}
}
\hspace{0.5cm}
\parbox{0.44\textwidth}{
\begin{figure}[H]
\vspace{-0.1cm}
\includegraphics[width=0.49\textwidth]{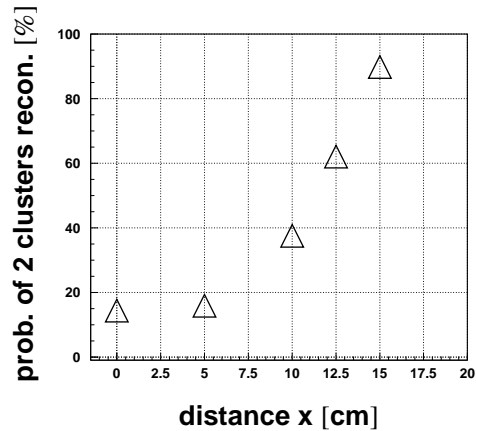}
\caption{The efficiency of the two clusters reconstruction as a function of the distance between photons along the x axis.}
\label{Efficiency_of_the_reconstruction_on_x_distance}
\end{figure}
}
}
%

% The second simulated situation is shown in the Fig.~\ref{2gammy_powierzchnia2_corel}. 
As expected the reconstruction efficiency is increasing
with increasing distance between two $\gamma$ quanta.  
 This result shows that the biggest problem with reconstruction
is in the case when particles hit with the small difference in the x coordinate. 
% This efficiency of reconstruction photons is interpreted as the influence of merging of clusters. 
% 
%
% lose themselves on the x axis. The distributions of the distance 
% for $\gamma$ quanta on surface was
% simulated either with vertex generator.
%
%
%%%%%%%%%%%%%%%%%%%%%%%%%%%%%%%%%%%%%%%%%%%%%%%%%%%%%%%%%%%%%%%%%%%%%%%%%%
%%%%%%%%%%%%%% REConstruction efficiency vs quantum efficiency %%%%%%%%%%%
%%%%%%%%%%%%%%%%%%%%%%%%%%%%%%%%%%%%%%%%%%%%%%%%%%%%%%%%%%%%%%%%%%%%%%%%%%
%
\section{Reconstruction efficiency as a function of the photocathode quantum efficiency}
\hspace{\parindent}
%
%
% Studies of the quantum efficiency. 
\vspace{-0.3cm}

We have performed investigations of the influence of the quantum efficiency of the photocatode  
on the merging and splitting effects. 
% In order to achieve this we simulated 
% the reconstruction efficiency of clusterisation
% as a function of the increasing quantum efficiency in photomultipliers. 
We carried out our simulations using a vertex generator for reactions: 
 $e^{+}e^{-} \to \phi \to \eta \gamma \to 3 \gamma$
 and 
 $e^{+}e^{-} \to \phi \to \pi^{0} \gamma \to 3 \gamma$. 
Then we simulated the calorimeter response using FLUKA program, 
reconstructed PM response by means of the DIGICLU procedure and finally reconstructed clusters 
by means of the present KLOE clusterisation algorithm. 
The studies presented in this section were performed for one module only. 
The efficiency is defined by the following formula: 
%
% The definition of the efficiency factor presents a following formula:
%
\begin{equation}
\epsilon = \frac{N(\geq 1 cluster)}{N(\geq 1 hit)}~, 
\label{efficiency_formula}
\end{equation}
where: N is the total number of events with a condition in parenthesis. 
This condition assures that we take events for which the module was hit by at least one $\gamma$. 
\begin{figure}[H]
\hspace{3.0cm}
\parbox[c]{1.0\textwidth}{
\vspace{-1.6cm}
\parbox[c]{0.5\textwidth}{\includegraphics[width=0.64\textwidth]{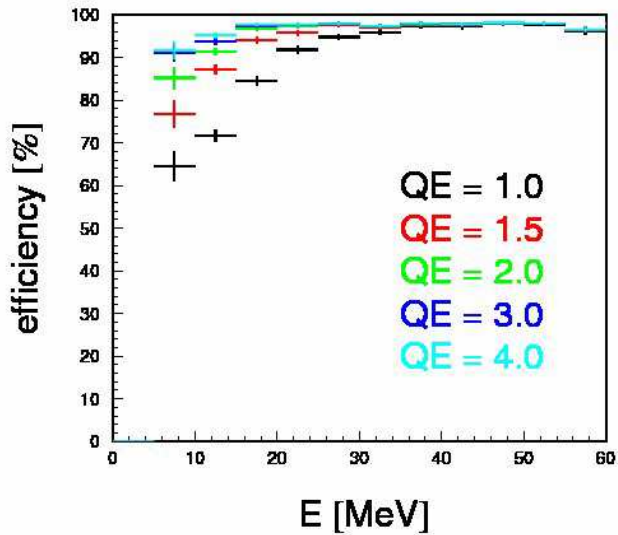}}
}
\vspace{-0.3cm}
\caption{Reconstruction efficiency as a function of the energy and quantum efficiency. Courtasy of E. Czerwi\'nski \cite{eryk}.}
\label{quantum_efficiency}
\end{figure}

\vspace{-0.2cm}
% The efficiency is a ratio for all events where was reconstructed at least 1 cluster to all events
% where at least 1 $\gamma$ quanta hit the module \cite{erykbiagio}.
The results of simulations for the single calorimeter module are shown in Fig.~\ref{quantum_efficiency}.
We have studied \cite{erykbiagio} the reconstruction efficiency as a function of the energy in the cluster 
for five values of the quantum efficiency as depicted inside the Figure. QE = 1 represents results for 
a present quantum efficiency of photomultipliers used at KLOE, amounting to 23\% \cite{eryk,didomenico_krakow}.   
Remaining values of QE indicates a factors by which the efficiency was increased in our investigations. 
%
%
% The black points present a 1.0 * PQE (with present a photomultipliers quantum efficiency),
% red present a 1.5 * PQE and green 2.0 * PQE \ref{quantum_efficiency}. 
%
From these studies 
% of the efficiency of 
% the reconstruction clusters with the present KLOE algorithm one 
 one can see that as expected with the increasing 
value of the quantum efficiency
for photomultipliers the efficiency of reconstruction is increasing too. 
%a quantum efficiency two times give the best
%reconstruction for the $\gamma$ quanta.    

\section{Merging and splitting probabilities as a function of the quantum efficiency}
\begin{figure}[H]
\hspace{-0.8cm}
\parbox[c]{1.00\textwidth}{
\parbox[c]{0.59\textwidth}{\includegraphics[width=0.59\textwidth]{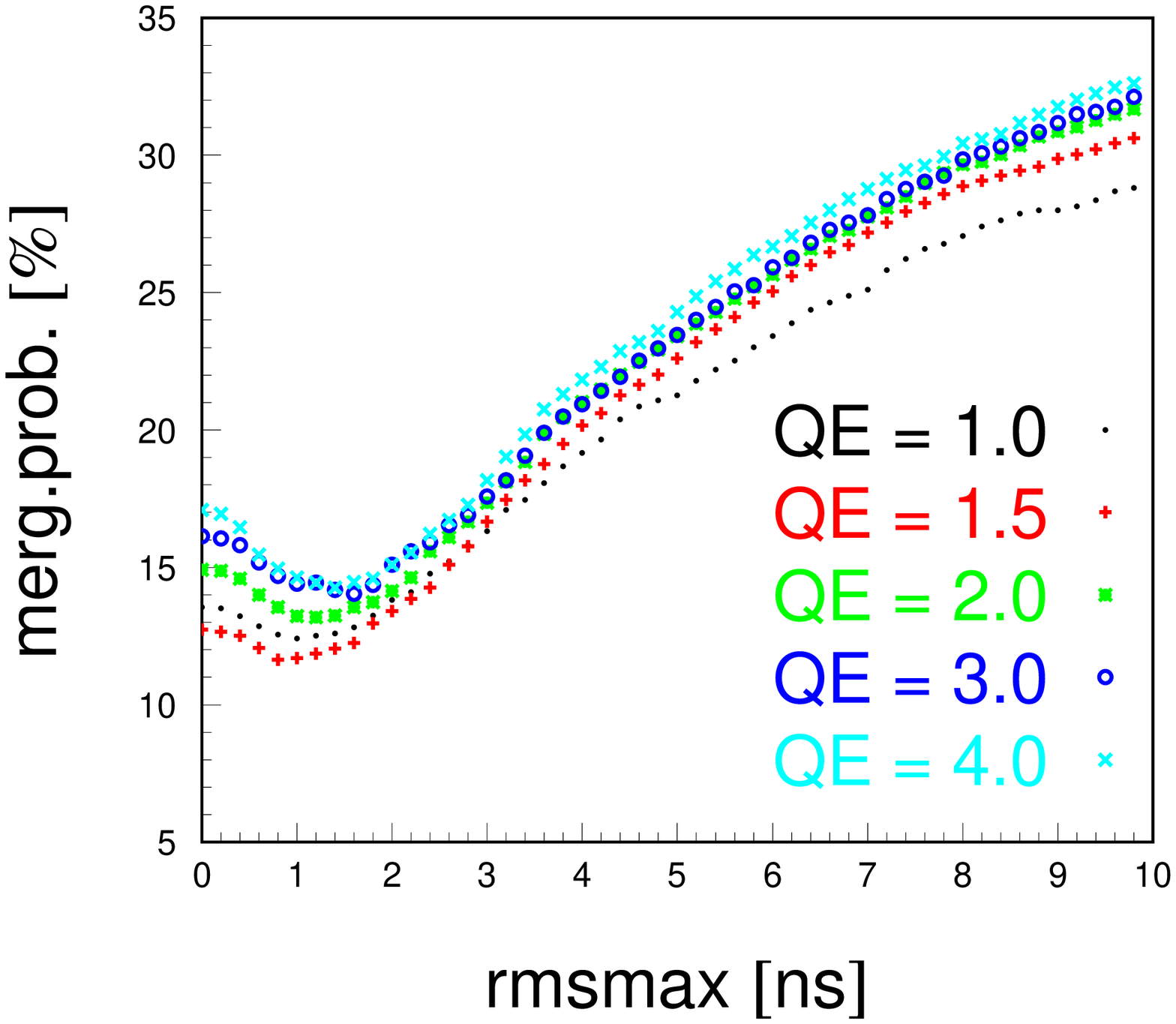}}
\hspace{-0.9cm}
\parbox[c]{0.59\textwidth}{
\vspace{-0.7cm}
\includegraphics[width=0.59\textwidth]{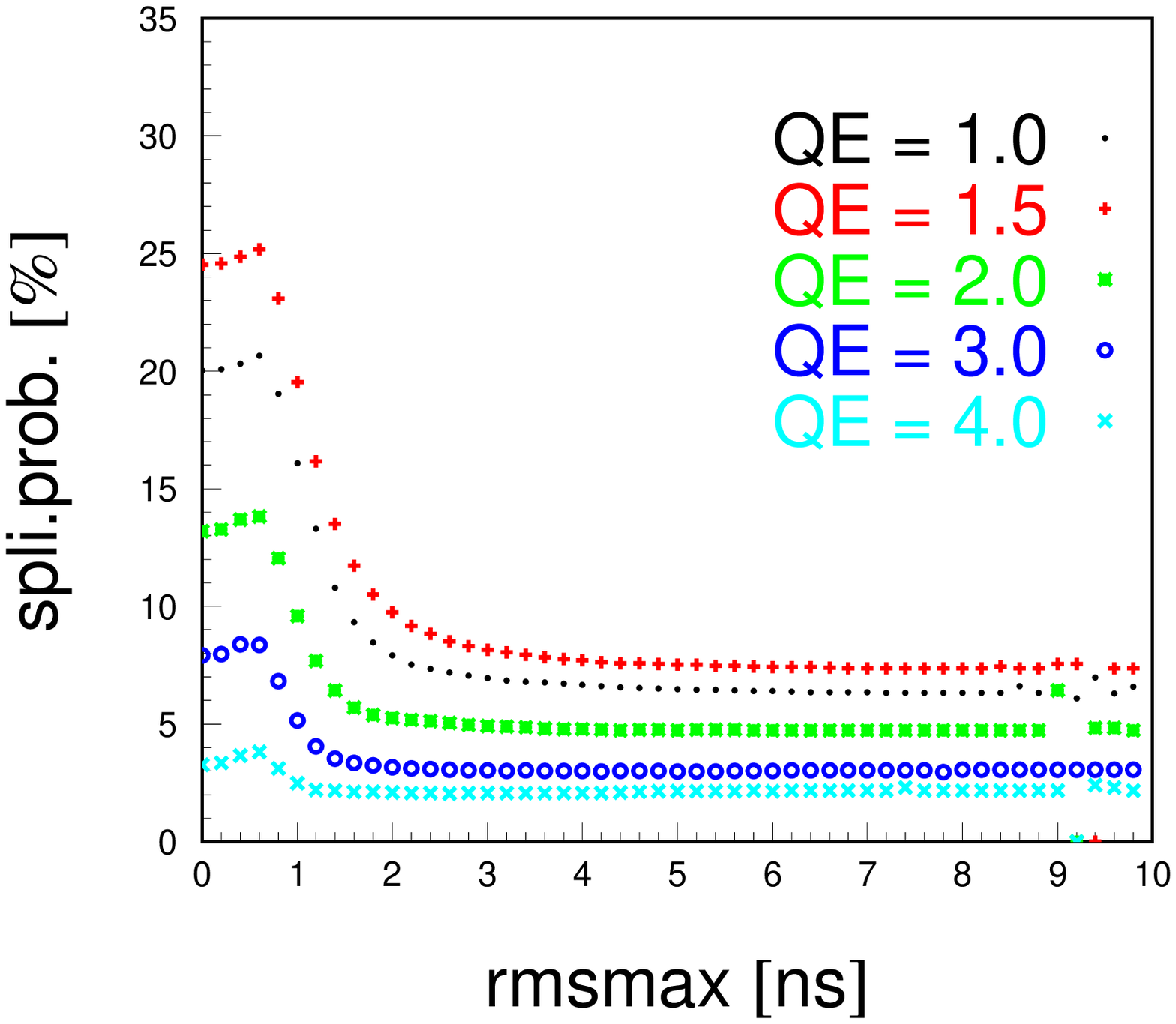}}
}
\caption{Merging and splitting studies with single EmC module.  
Rmsmax is a value of the width (RMS) of the time distribution
of the cells above which the clustering algorithm tries to break a cluster \cite{erykbiagio}. 
At present KLOE algorithm this value is equal to 2.5 ns. Courtasy of E. Czerwi\'nski \cite{eryk}.}        
\label{merg_split_1module}
\end{figure}

\hspace{\parindent}
Subsequently, we determinated the influence of value of the quantum efficiency on the magnitude 
of the merging and splitting
effects. The results of these simulations are shown in Fig.~\ref{merg_split_1module}. 

The merging probability was defined according to the following formula:
\begin{equation}
\text{merg. probability} = \frac{\text{N}(\text{1 cluster and} \geq \text{2 hits})}{\text{N}(\geq\text{1 cluster and} \geq\text{2 hits})} 
\label{merg_formula}
\end{equation}
and the splitting probability was defined as:
\begin{equation}
\text{split. probability} = \frac{\text{N}(\geq\text{2 clusters and 1 hits})}{\text{N}(\geq\text{1 cluster and 1 hits})} 
\label{split_formula}
\end{equation}
Fig.~\ref{merg_split_1module} shows that the splitting probability increase with increasing of the quantum efficiency of  
photomultipliers \cite{erykbiagio} but one can see that merging effect is much less sensitive to the changing of the 
quantum efficiency. 
%
%
%
%\chapter{Influence of the upgrated clustering algorithm}
%
% \chapter{Calorimeter efficiency for $\gamma$ quanta}
% \hspace{\parindent}
%
%We simulated the energetic response of the barrel calorimeter and reconstructed of clusters due to 
%investigate merging and splitting of clusters.
%
% \section{Energy deposition efficiency of the cluster reconstruction in the barrel calorimeter}
% \hspace{\parindent}
%
%
%
%
%

\chapter{Conclusions}
\hspace{\parindent}
We have built a full simulation of the Barrel Calorimeter of the KLOE detector with FLUKA Monte Carlo package.
 This implementation of the geometry was used for simulations of the physical energy response of the calorimeter. The Monte Carlo
 cooperates with vertex generator which reproduces kinematics of the physical reactions. \\
%
% reproduces $\phi$ meson decays with hight multiplicity of $\gamma$ quanta in 
% the final state of the reaction. \\
%
\indent We estimated influence of merging and splitting of clusters for reconstruction possibilities of the particles  
on the whole barrel calorimeter with a present clustering algorithm. In particular we estimated these effects as a function of the 
 quantum efficiency of photomultipliers \cite{erykbiagio}. 
As a results we found that increase of the quantum efficiency does not improve significantly the cluster reconstruction in view 
of merging and splitting effects. \\ 
%
% However, studies on increasing of Quantumm Efficiency for the KLOE photomultipliers \cite{erykbiagio} shows that merging and splitting of clusters 
% aren't significant reduced. \\ 
%
%
\indent The prepared program with geometry of the 24 modules of the barrel calorimeter enables fast simulations of a choosen reaction channel 
and facilitates studies of the detector response to different reactions separately. \\
\indent This geometry may be used to test response of photomultipliers with higher granularity of cells, 
especially for testing performances of the multianode photomultiplier. \\
\indent Implemented geometry of the barrel calorimeter in FLUKA reproduces the merging and the splitting effects,   
 and permits also to study the efficiency of clusterisation algorithm 
 at the edges of the modules in the barrel calorimeter.  
 These all usefull options come out from the fact that the smallest parts of detector 
(the fibers structure) were realistically implemented in FLUKA Monte Carlo. \\ 
%
% \indent Using FLUKA with detailed description of barrel calorimeter geometry we are able to achieve an accurate 
% reproduction of the physical response for particle spectra from 1 GeV to 2 GeV. \\ 
%
\indent We investigated a shapes of peaks for time distribution of reconstruction of clusters, the shapes are changing as a function 
 of distance between photons. This information could be very useful for the upgrade of the clustering 
algorithm \cite{moskal}. \\  
\indent We estimated also the influence of merging and splitting of clusters for 
$e^{+}e^{-} \to \phi \to K_{L}K_{S} \to K_{L} 2\pi^{0} \to K_{L} 4\gamma$ and 
 $e^{+}e^{-} \to \phi \to \eta\gamma \to 3\pi^{0}\gamma \to 7\gamma$ reactions. The studies were made determining the response of the 
whole barrel calorimeter with DIGICLU program. \\ 
\indent Finally, it is worth mentioning that our study encouraged the authors of the FLUKA to upgrade the program 
and in the near future a new version permitting to 
replicate the cells more than once will be available. \\

\appendix

\chapter{Kinematic fit procedure}
\hspace{\parindent}
The kinematic fit procedure is used to correct measured quantities $x_{i}$ ($x_{1} ... x_{i} ... x_{n}$)  
in case of redundancy in the measured set of variables. In most cases $x_{i}$ is normal distributed around $x_{i}^{true}$
with standard deviation $\sigma_{i}$.
\begin{eqnarray}
\sigma(x_{i}) = \frac{1}{N-1} \sum_{i}^{N} (x_{i} - \bar{x})^{2} \\
%\nonumber
%\text{where:} \\
%\nonumber
%\text{ $\bar{x}$ - is a average of the all x values}
\end{eqnarray}
where $\bar{x}$ is a average of the all x$_{i}$ values.
Variables $x_{i}^{true}$ could be constrained by some physical law (in our investigations by energy and momentum conservation)
in this case they must satisfy k equations, which can in general be written in the form:
\begin{equation}
F_{j}(x_{1}^{true},...,x_{n}^{true}) = 0 ~~~~ \text{j = 1,...,k}
\label{prawo_zachowania}
\end{equation}
The main purpose of using fit procedure is to find a new approximation ($\mu_{i}$) of $x_{i}^{true}$ value
by minimizing the quantity:
\begin{equation}
\chi^{2} = \sum_{i} \frac{(x_{i} - \mu_{i})^{2}}{\sigma_{i}^{2}}
\end{equation}
and imposing the (\ref{prawo_zachowania}) constraints on $\mu_{i}$.
This purpose can be achieved using langrangian multipliers method, that consists of minimizing the quantity:
\begin{equation}
\chi^{2} = \sum_{i} \frac{(x_{i} - \mu_{i})^{2}}{\sigma_{i}^{2}} + \sum_{j}\lambda_{j}F_{j}(\mu_{1},...,\mu_{n})
\end{equation}
related with the variables $\mu_{i}$ and $\lambda_{j}$ \cite{Biagio_thesis}.
To minimize a $\chi^{2}$ value an iterative procedure is needed to be used. \\
\indent The detailed description of the $\chi^{2}$ distribution can be found e.g. in \cite{nowak_statystyka}.

\chapter{Monte Carlo Methods}
\hspace{\parindent}
Monte Carlo simulations are methods of solving numerical assignments with the aid of adequate constructed statistical computations.   
They are frequently used in experimental and theoretical physic \cite{zielinski_monte_carlo}. 
% These methods are concerned with performing 
% computer calculations and simulations for solving physical problems. 
 Although computer memory and processor 
performance have increased dramatically over the last two decades, many physical problems are too complicated to be solved 
without approximations of the physics processes, quite apart from the approximations inherent in any numerical method. Therefore, 
most calculations done in computational physics involve some degree of approximation \cite{computational_physics}. \\ 
\indent Solving a physical problem often requires to solve an ordinary or partial differential equation. This is the case in for example  
classical mechanics, electrodynamics, quantum mechanics and fluid dynamics. On the other hand in statistical physics we must calculate 
sums or integrals over large numbers of degrees of freedom. \\
\indent Whatever type of problem we try to solve, it is only in very few cases that analytical solutions are possible. 
In most cases we therefore resort to numerical calculations to obtain useful results \cite{computational_physics}. \\     
\indent These methods are used in physic for the last sixty years \cite{zielinski_monte_carlo}. 
\chapter{Energy distributions for $\gamma$ quanta}
\begin{figure}[H]
\parbox[c]{1.0\textwidth}{
\hspace{0.8cm}
\parbox[c]{0.45\textwidth}{\includegraphics[width=0.49\textwidth]{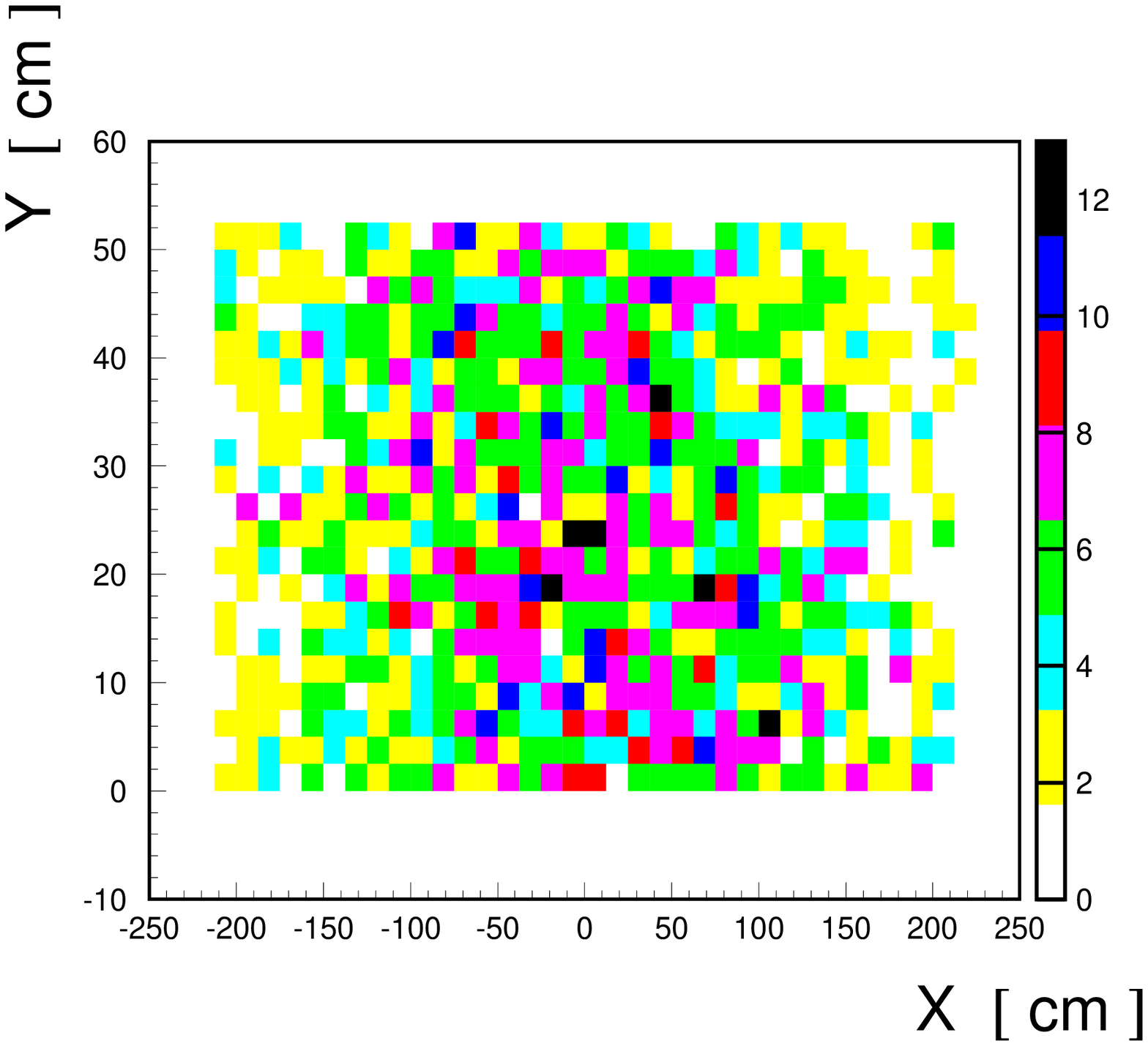}}
\parbox[c]{0.45\textwidth}{\includegraphics[width=0.49\textwidth]{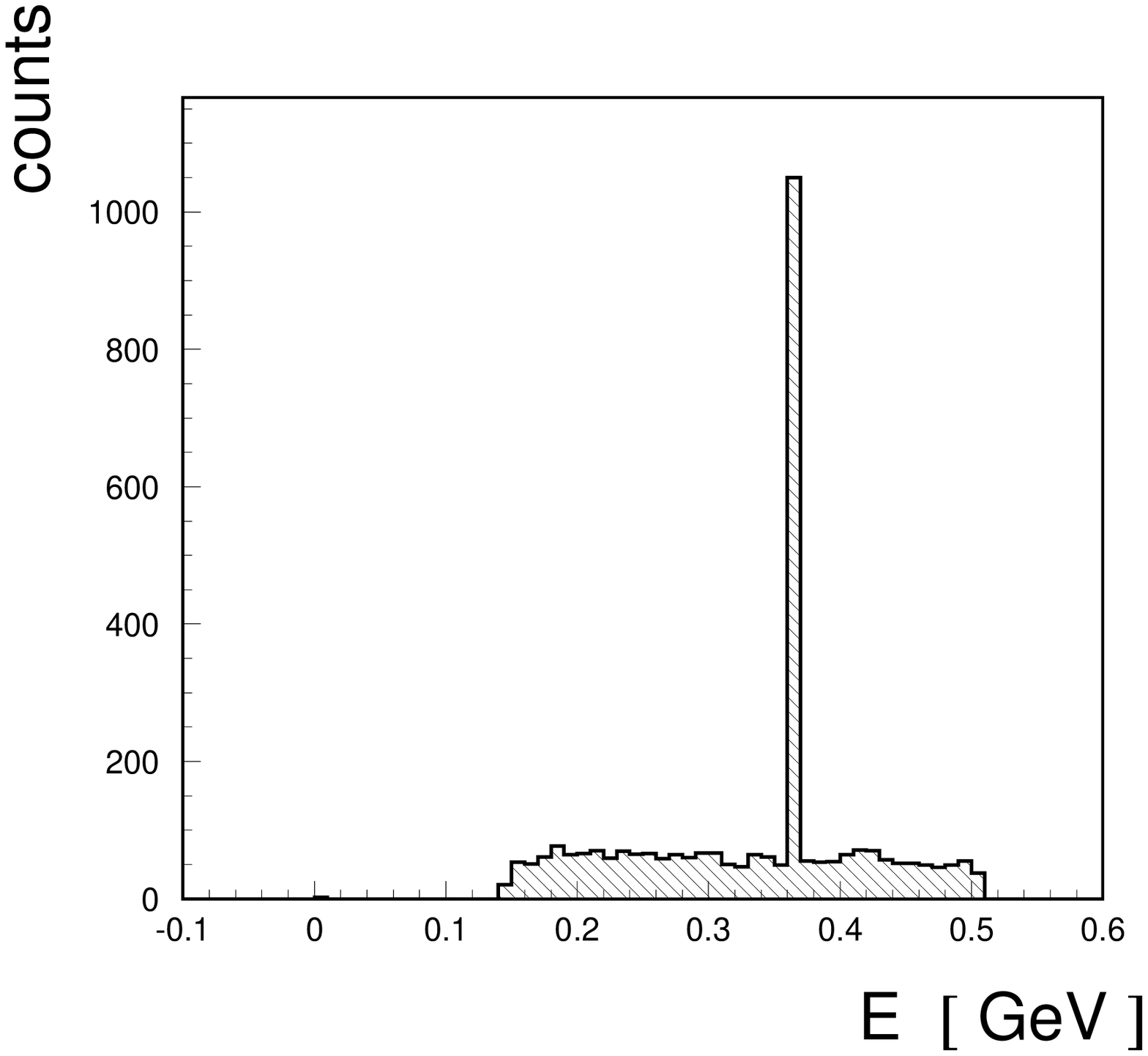}}
}
\parbox[c]{1.0\textwidth}{
\vspace{-1.4cm}
\hspace{0.8cm}
\parbox[c]{0.45\textwidth}{\includegraphics[width=0.49\textwidth]{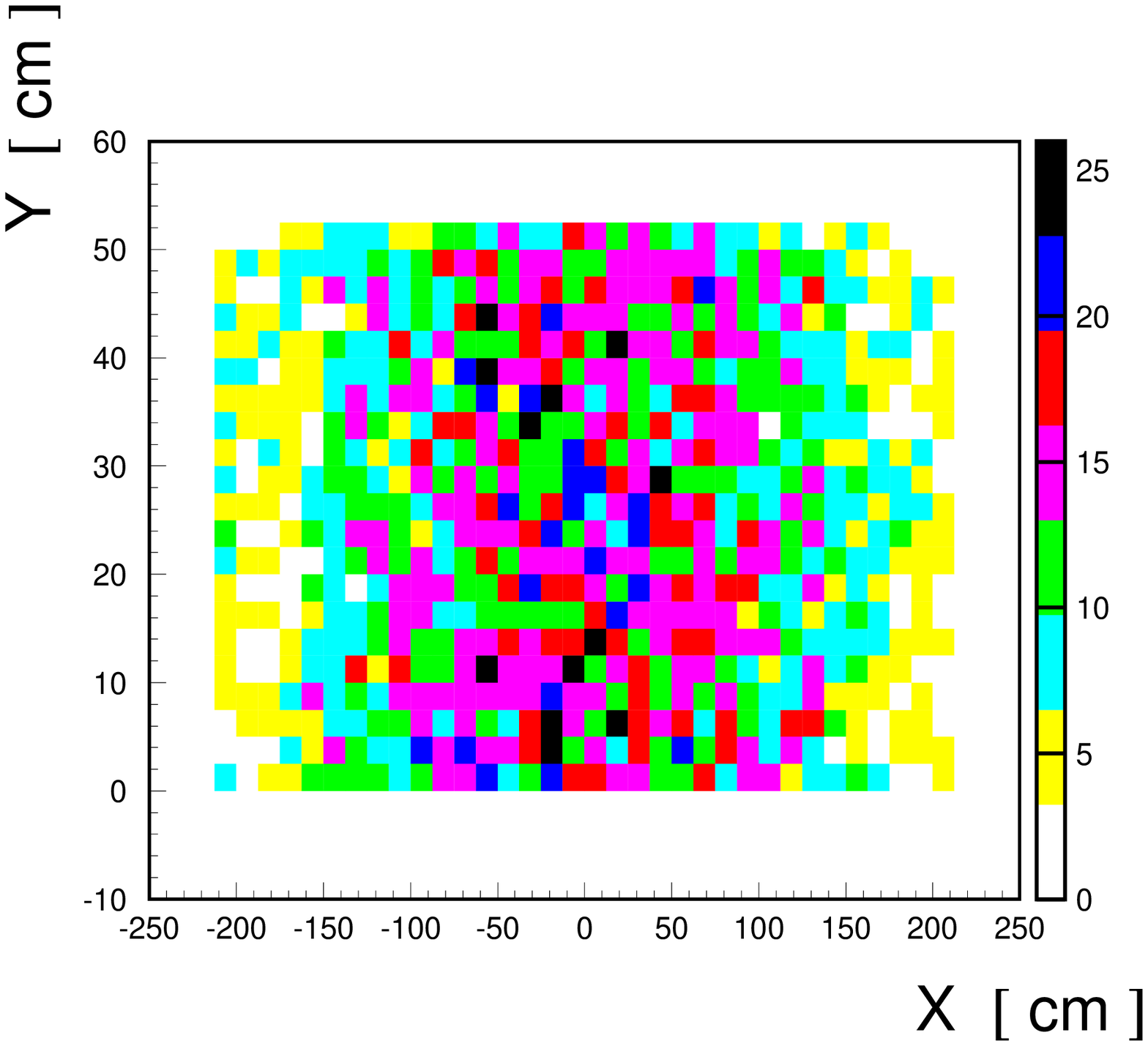}}
\parbox[c]{0.45\textwidth}{\includegraphics[width=0.49\textwidth]{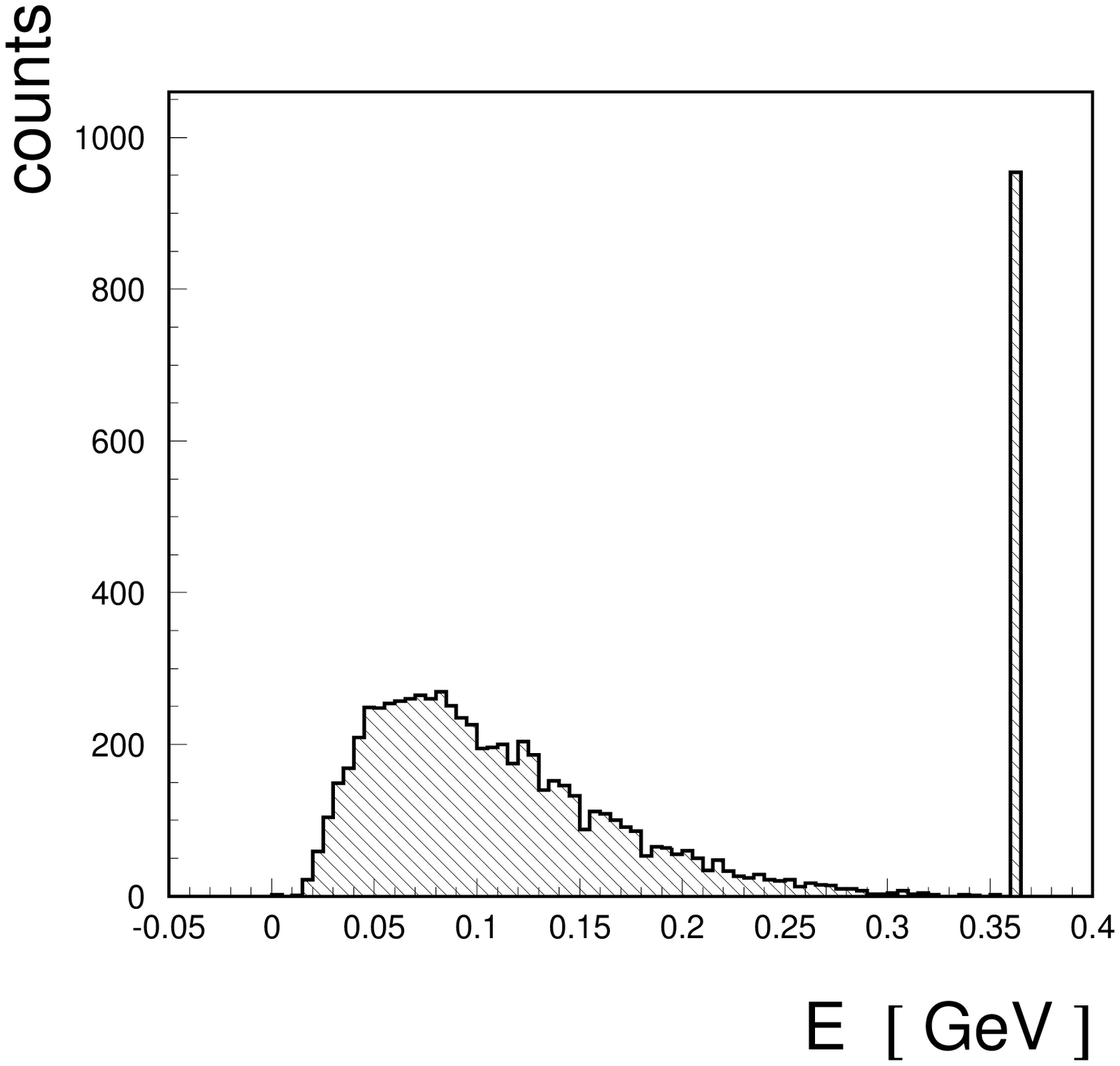}}
}
\caption{Simulated distributions of the hit position at the surface of one calorimeter module (left panel) and energy spectra 
 (right panel) of $\gamma$ quanta from the reactions $e^{+}e^{-} \to \phi \to \eta \gamma \to 3 \gamma$ (upper panel) and 
$e^{+}e^{-} \to \phi \to \eta \gamma \to 3 \pi^0 \gamma \to 7 \gamma$ (lower panel).}
\label{gamma_on_surface_energy}
\end{figure}

\begin{figure}[H]
\parbox[c]{1.0\textwidth}{
\mbox{
%\hspace{0.3cm}
\parbox{0.33\textwidth}{
\begin{figure}[H]
\includegraphics[width=0.36\textwidth]{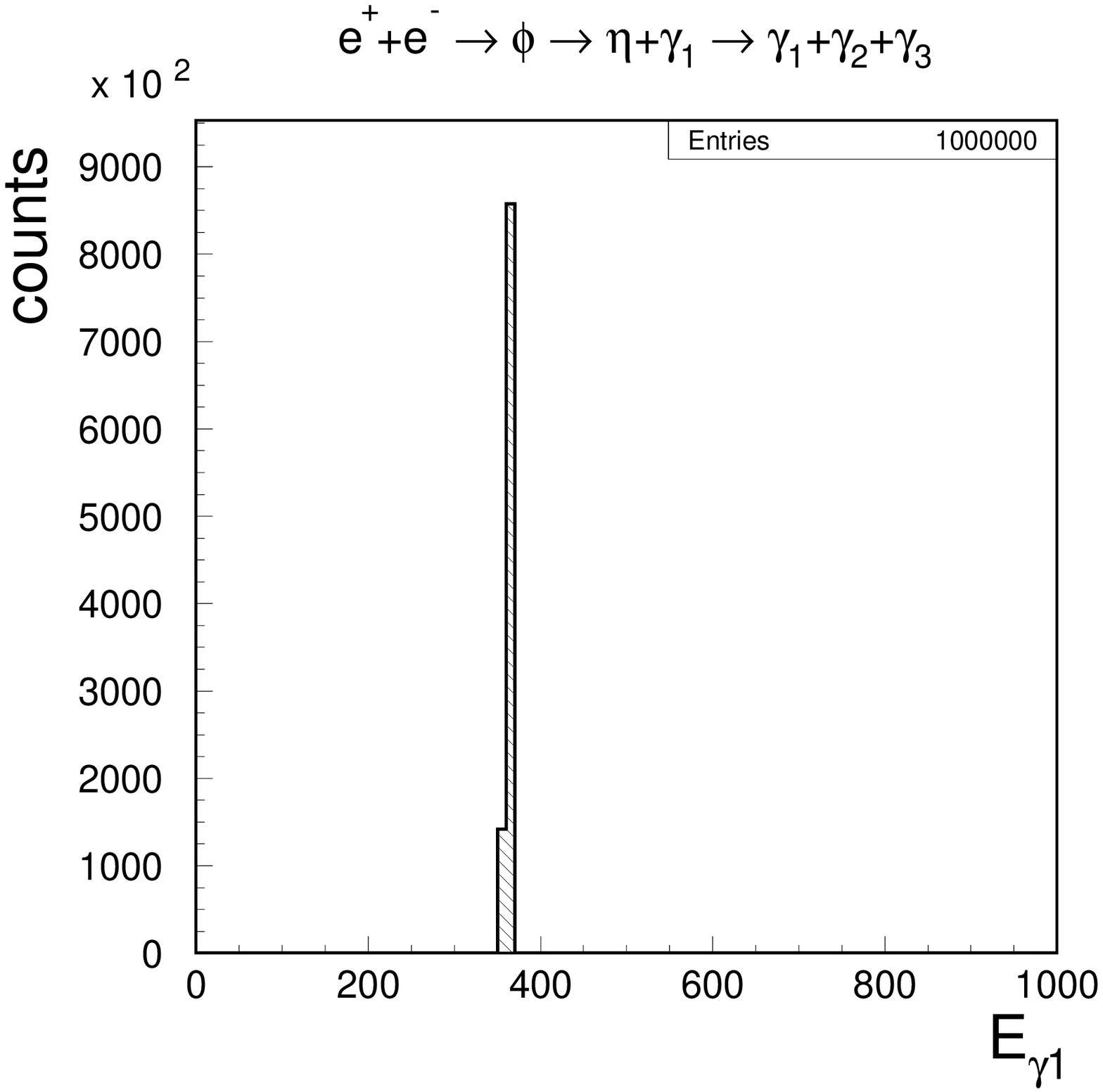}
\end{figure}
}
\hspace{-0.2cm}
\parbox{0.33\textwidth}{
\begin{figure}[H]
\includegraphics[width=0.36\textwidth]{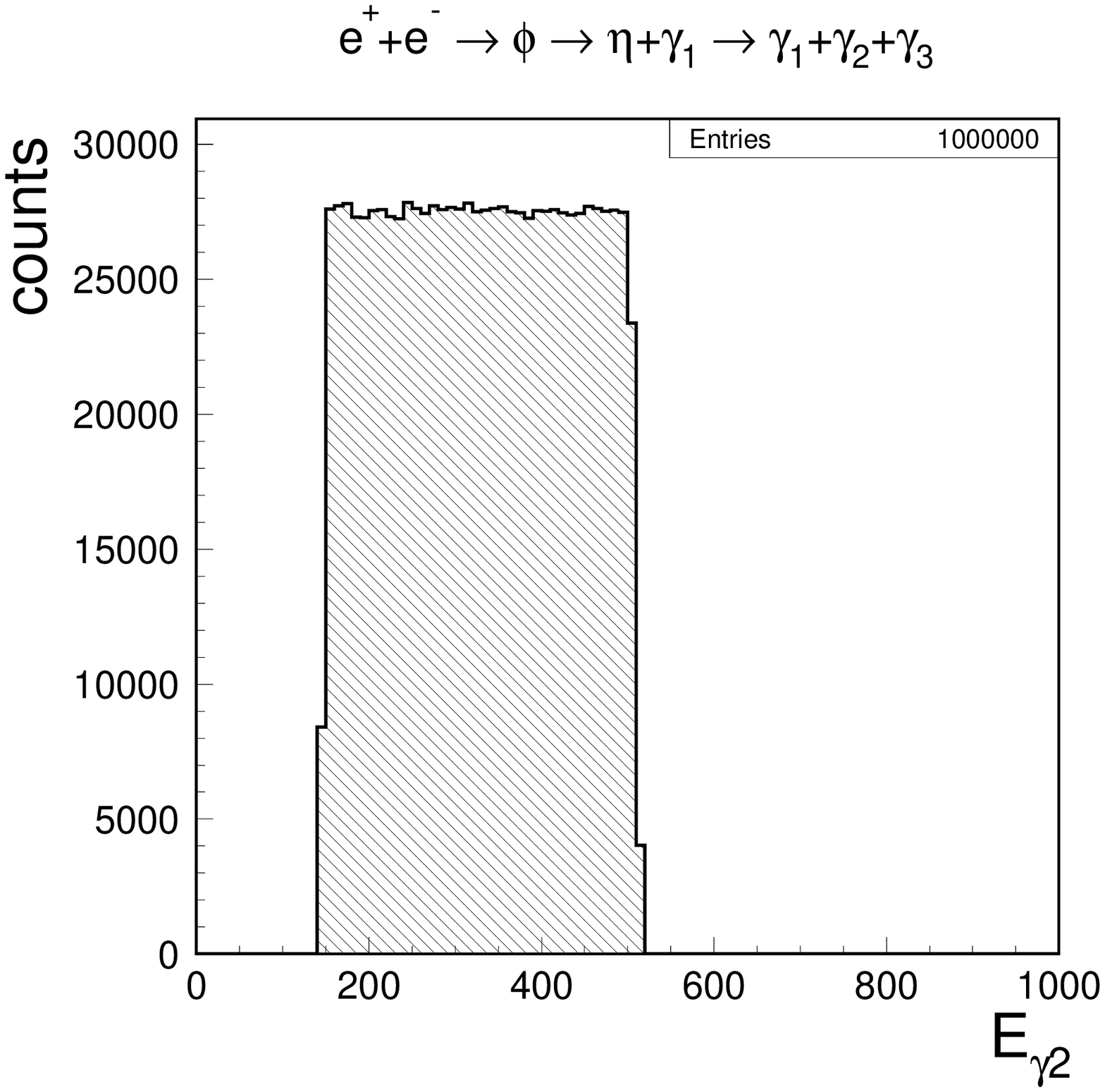}
\end{figure}
}
\hspace{-0.2cm}
\parbox{0.33\textwidth}{
\begin{figure}[H]
\includegraphics[width=0.36\textwidth]{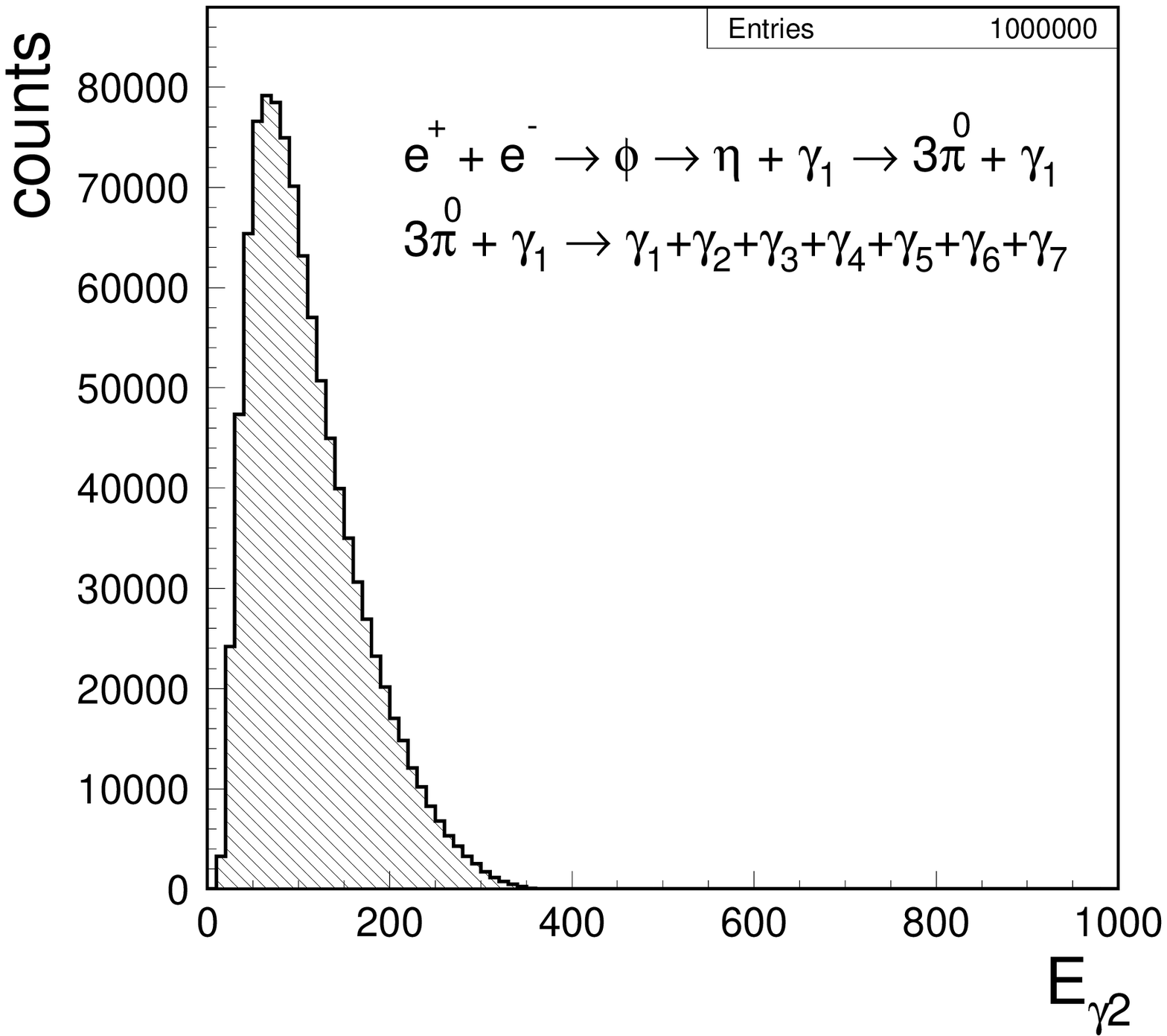}
\end{figure}
}
}
}
\vspace{-0.3cm}
\caption{The energy distribution for photons from reactions: $e^{+}e^{-} \to \phi \to \eta \gamma \to 3\gamma$ and 
$e^{+}e^{-} \to \phi \to \eta \gamma \to 3 \pi^0 \gamma \to 7 \gamma$.} 
\label{reaction_parameters}
\end{figure}

\hspace{\parindent}
We have investigated\footnote{We used the BXDRAW subroutine in this investigations \cite{FLUKA_homepage}.} 
the distribution of the energy and the position of the $\gamma$ quanta on
the bottom surface of a single calorimeter module. We performed simulations for reactions: 
$e^{+}e^{-} \to \phi \to \eta \gamma \to 3 \gamma$ 
% (upper panel in the Fig.~\ref{gamma_on_surface_energy} and for the 
and $e^{+}e^{-} \to \phi \to \eta \gamma \to 3 \pi^0 \gamma \to 7 \gamma$~.  
 The result is presented in Fig.~\ref{gamma_on_surface_energy} in upper and lower panel, respectively.
In the left column in Fig.~\ref{gamma_on_surface_energy}, one can see hit distributions for particles 
 entering the module. 
%
% We also want to present that the distribution of $\gamma$ quanta on surface of the module has relevant influence for the 
% energetic response of this detector. One can see 
%
In the right panels in this figure the  
sharp signals orginate from a monenergetic $\gamma$ quanta from $\phi$ meson decay with energy equal to 363 MeV. 
In the case of the $e^{+}e^{-} \to \phi \to \eta \gamma \to 3 \pi^0 \gamma \to 7 \gamma$ 
reaction we observe more events at the lower energy range. These signals orginate from 
$\gamma$ quanta from $\pi^{0}$ meson decays.
%
%
% This are results with statistic of 40000 events. 
% One can see also that for channel with 7$\gamma$ quanta in the final state more photons hit the module. \\
% \indent In order to compare a results of the energy distribution on the enetering of the module, we present the energy distribution for 
% these two reactions simulated with vertex generator.
% 
As a further example in Fig.~\ref{reaction_parameters} we present energy spectra separately for the radiative photon and for 
the photon from the $\eta$ and pion decays.
\vspace{-0.3cm}

In the left panel the monoenergetic signal for photon from $\phi$ meson decay is shown. 
% The energy distribution for this $\gamma$ quanta for these two reactions is the same, because it is the same particle from 
% the $\phi$ meson decay. 
 The middle panel presents energy distribution for photons from the $\eta$ meson decay and  
 the third panel presents the energy distribution for $\gamma$ quanta from $\pi^{0}$ mesons decays.  
%
%One can see the monoenergetic signal for photon from $\phi$ meson decay directly and signals from photons 
%at the lower energy range from $\pi^{0}$ mesons decays. \\   
%
% \indent The conclusion from our investigations is that energy distribution in layers is caused by some relevant factors like
% solid angle, angle between surface and particle track, and at the final with the energy of $\gamma$ quanta on the surface calorimeter.  
%
\chapter{Time distribution for single and multi-gamma hits}
\begin{figure}[H]
\hspace{0.4cm}
\parbox[c]{1.00\textwidth}{
\parbox[c]{0.35\textwidth}{\includegraphics[width=0.49\textwidth]{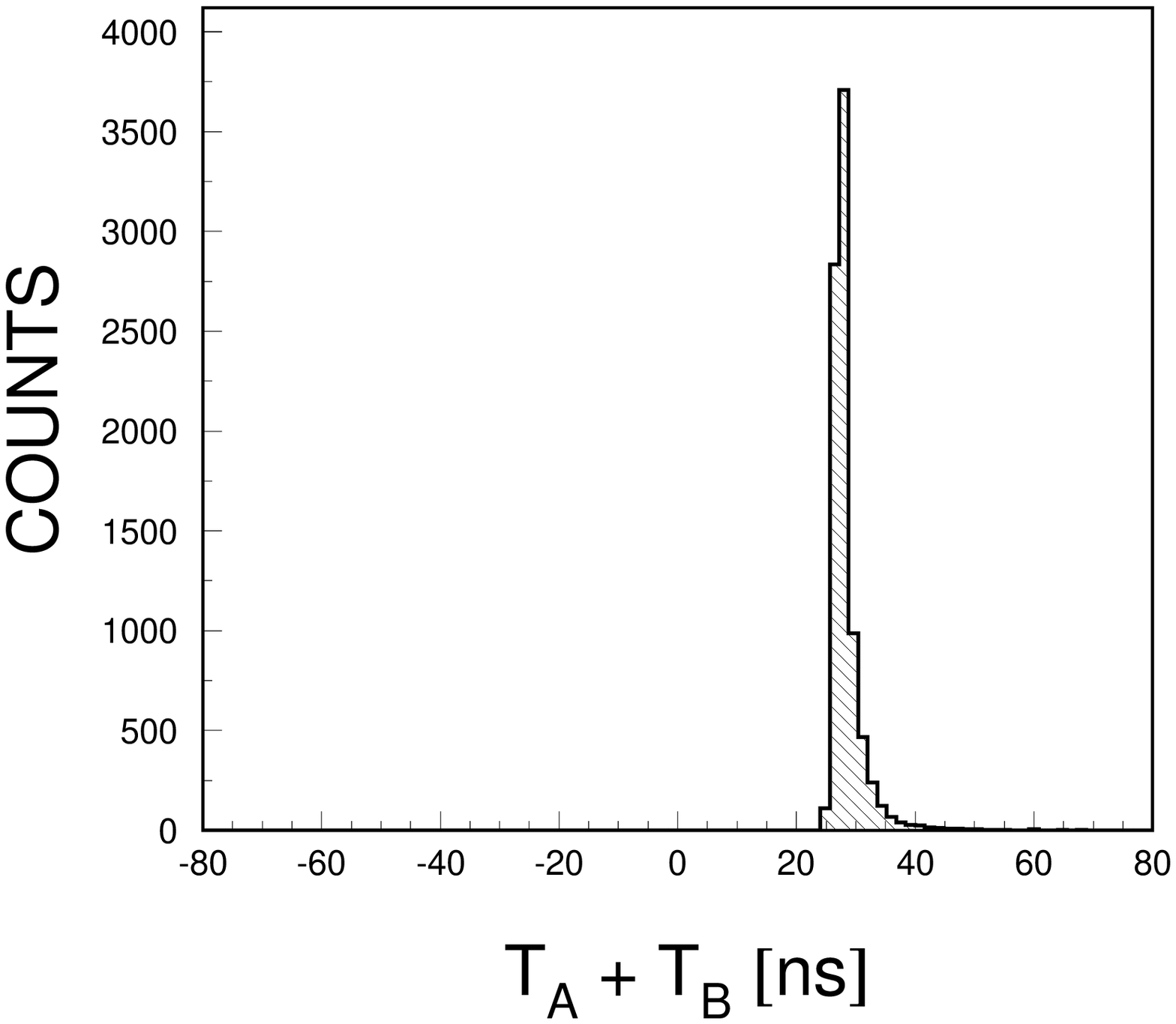}}
\hspace{1.1cm}
\parbox[c]{0.35\textwidth}{\includegraphics[width=0.49\textwidth]{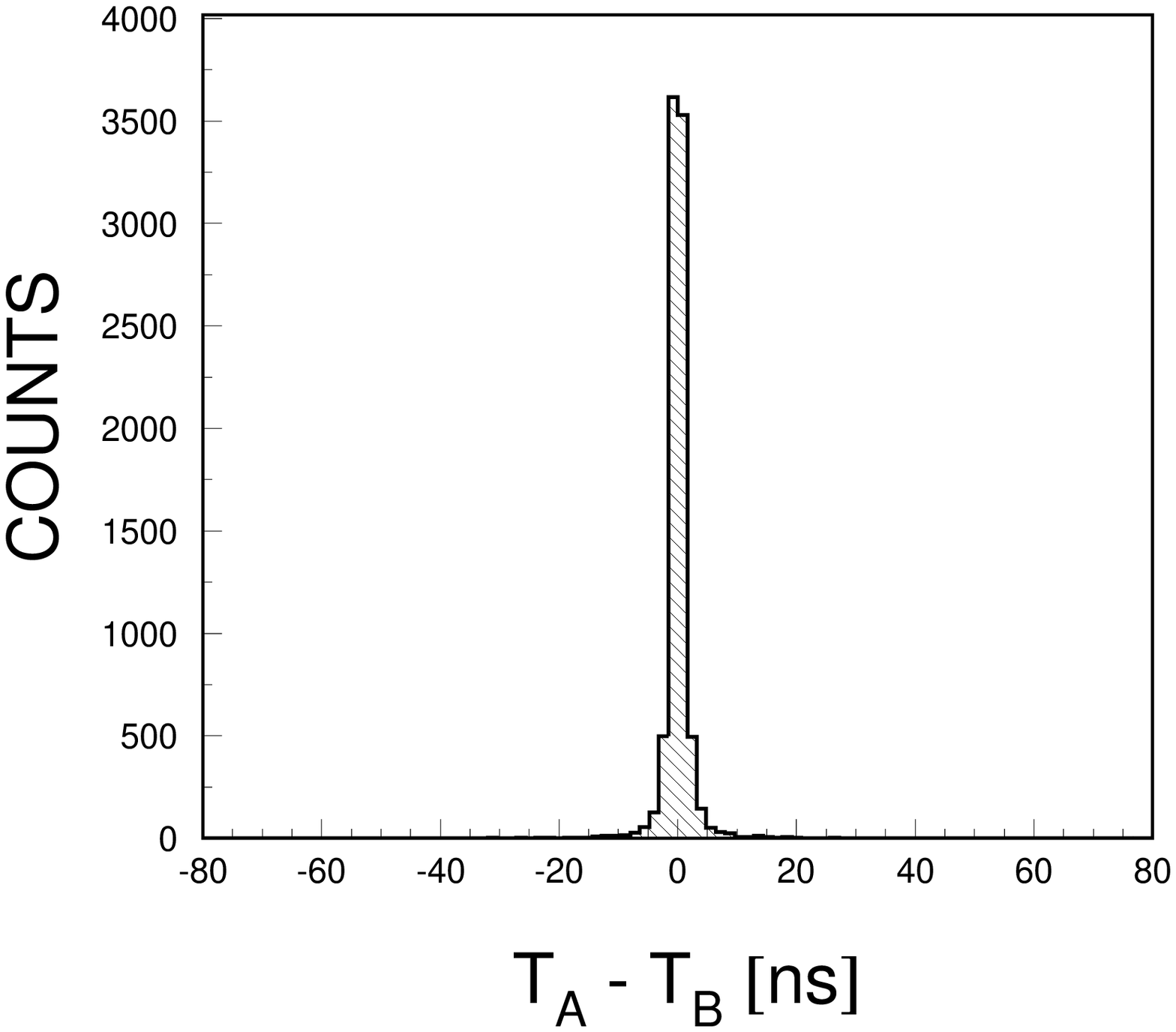}}
}
\caption{Distribution of sum and difference of time for one $\gamma$ quanta. $T_{A}$ and $T_{B}$ denote the time where a signal
from scintillating fibers comes into a photomultipliers on side A and side B, respectively.}
\label{Time_distribution_studies_for_1_gamma}
\end{figure}

\begin{figure}[H]
\hspace{0.4cm}
\parbox[c]{1.00\textwidth}{
\parbox[c]{0.35\textwidth}{\includegraphics[width=0.49\textwidth]{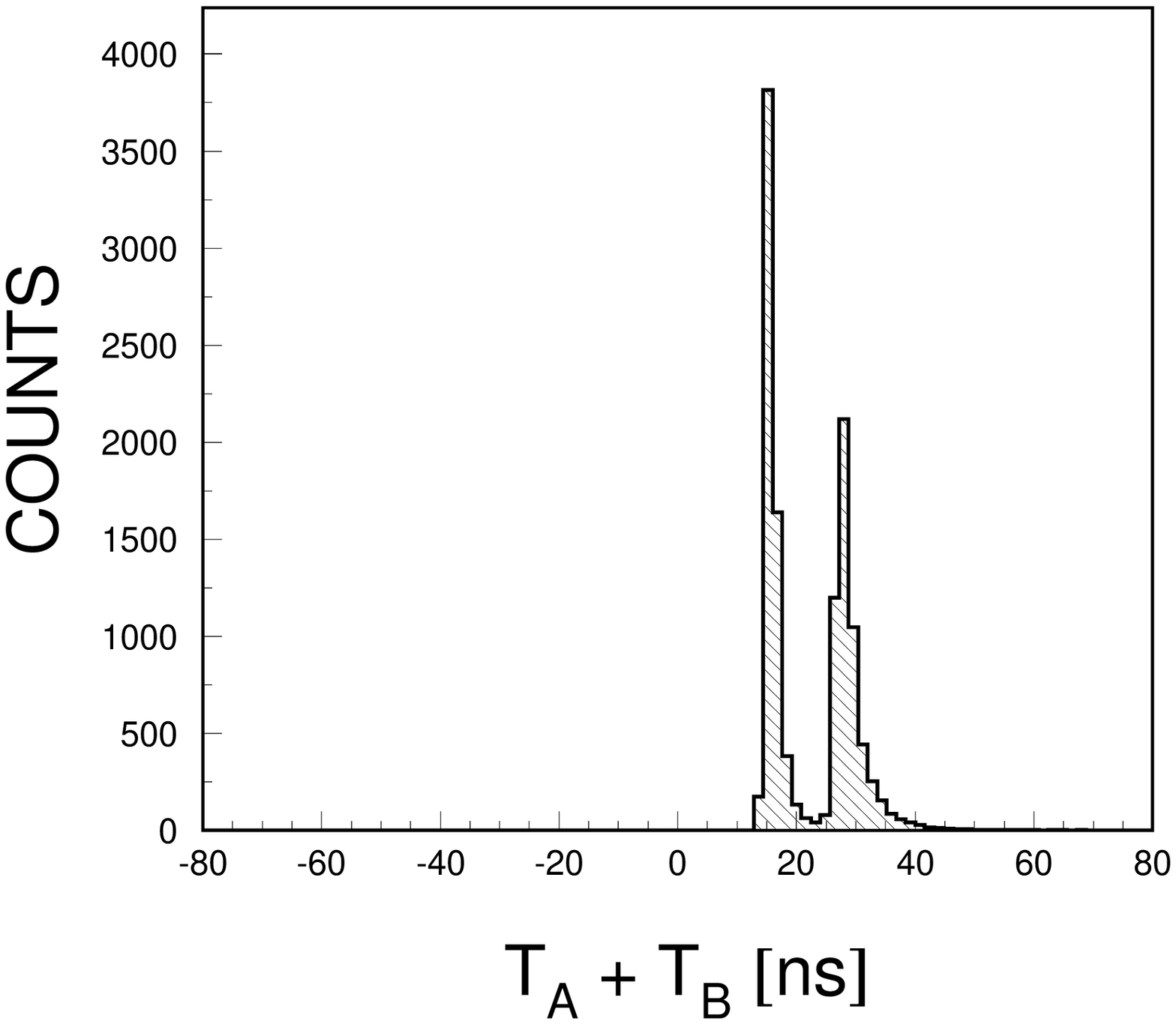}}
\hspace{1.1cm}
\parbox[c]{0.35\textwidth}{\includegraphics[width=0.49\textwidth]{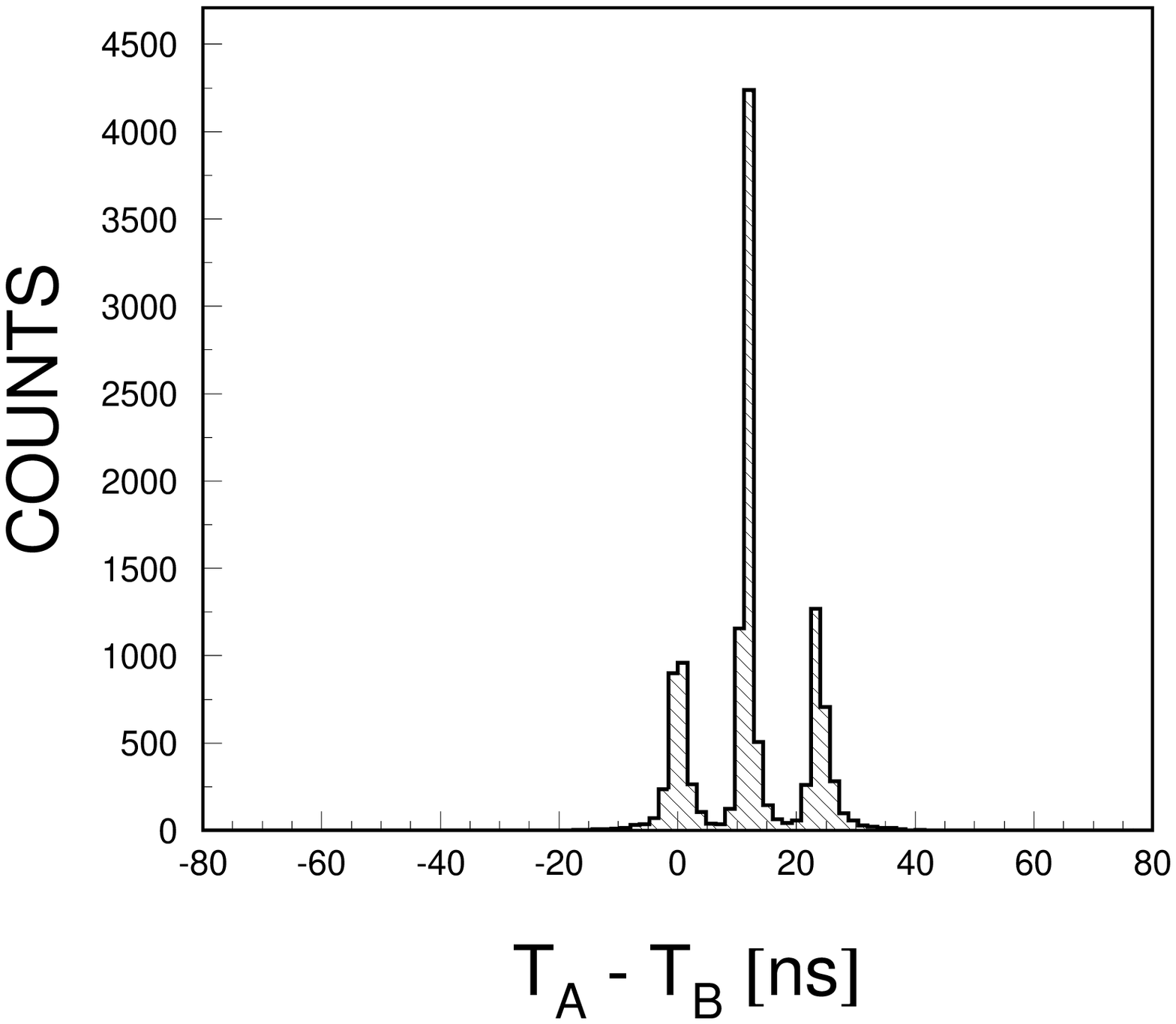}}
}
\caption{Distribution of sum and difference of time for two $\gamma$  at the same x position but different y.}
\label{Time_distribution_studies_for_2_gamma_on_y_axis}
\end{figure}

\begin{figure}[H]
\hspace{0.4cm}
\parbox[c]{1.00\textwidth}{
\parbox[c]{0.35\textwidth}{\includegraphics[width=0.49\textwidth]{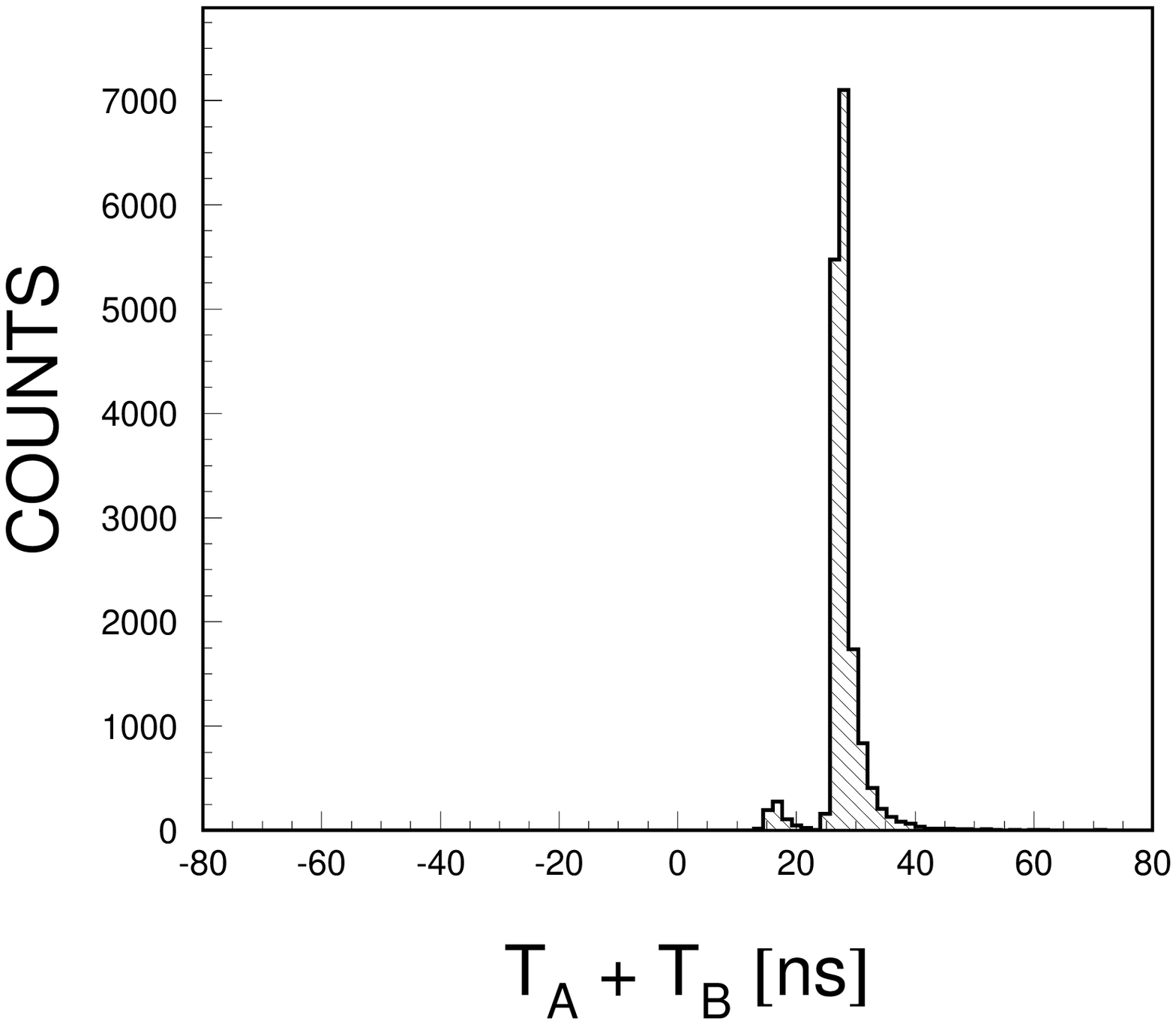}}
\hspace{1.1cm}
\parbox[c]{0.35\textwidth}{\includegraphics[width=0.49\textwidth]{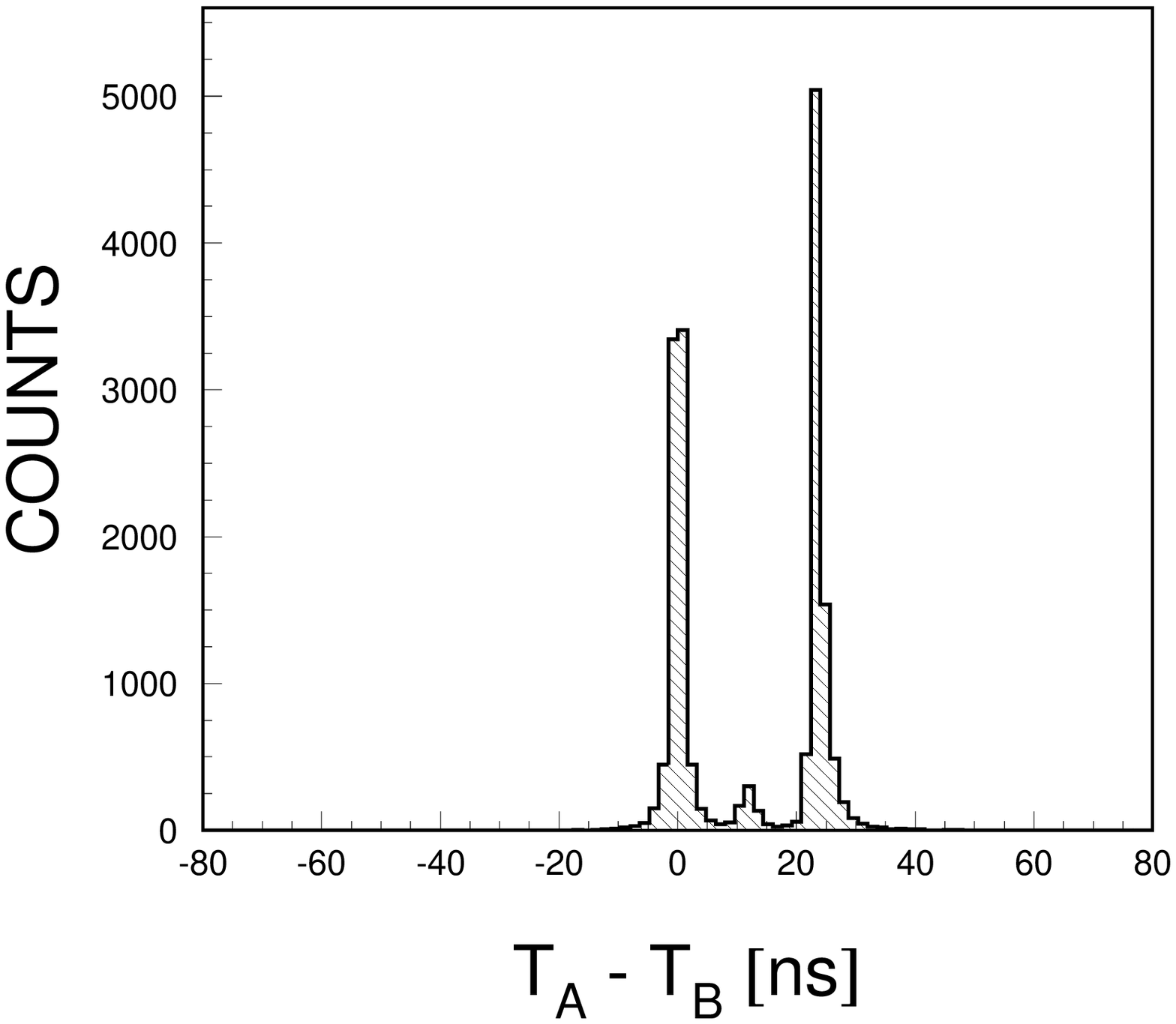}}
}
\caption{Distribution of sum and difference of time for two $\gamma$ hitting the module with a distance of 15~cm along the x axis.}
\label{Time_distribution_studies_for_1_gamma_on_x_axis}
\end{figure}

\hspace{\parindent}
In this appendix 
 we investigated distributions of the sum and difference of times of light signals from the sides of the module.  
 Our studies show that distributions 
 are changing as a function of the distance between $\gamma$ quanta hitting the module. This information could be used to
improve efficiency of the KLOE clustering algorithm in the near future \cite{moskal}.  
In Fig.~\ref{Time_distribution_studies_for_1_gamma} result for the situation where one 
$\gamma$ quanta 
hit in the middle of the calorimeter module is presented. One can see only one peak for both difference and 
sum of time of signals from sides A and B. 
% on the one calorimeter module was reconstructed. 
%
%
\vspace{-0.7cm}
For the two $\gamma$ quanta which hit the 
surface\footnote{Two studing cases of hit positions of photons entering 
 the calorimeter surface are presented in Fig.~\ref{2gammy_powierzchnia_corel} and Fig.~\ref{2gammy_powierzchnia2_corel}. Positions of $\gamma$ 
quanta are   
 marked as black circles.} 
  at the same x but different y    
the result is different 
(see Fig.~\ref{Time_distribution_studies_for_2_gamma_on_y_axis}). One can see that distribution of difference of times for 
signals from side A and B consists of three separated peaks and the distribution of the sum of times from sides A and B shows 
two maxima. 
\newpage
\vspace{-0.7cm}
%
% This information can be use to improve reconstruction efficiency of the clustering algorithm due to 
% the fact that these two time distributions of the sum and difference of the times are different. 
Next we investigated the situation where distance on x axis between two $\gamma$ quanta is equal to 15~cm. 
In this case we observe only one maximum in the distribution of the sum and two maxima in the distribution of the difference of 
times (Fig.~\ref{Time_distribution_studies_for_1_gamma_on_x_axis}). 
%
% (see Fig.~\ref{Efficiency_of_the_reconstruction_on_x_distance}), we rejected two peaks
% each from one photon (Fig.~\ref{Time_distribution_studies_for_1_gamma_on_x_axis}). 
% It is to easy to see for the distribution of the time difference.
%
\vspace{-0.7 cm}
%
% One can see in this case the shape of distributions is dfferent too.
% Subsequently, 
In the following we present an example of results of cell distributions simulated for several distances 
% the simulations of the photomultipiers response for several cases with different distance on
on x axis between places where photons hit the module and for several angle between a surface and a photon direction. 
% on the calorimeter module 
% and particle track before hitting.
The results of the sum of energy deposited in cells for side A and B we present using a displayer 
program\footnote{This program written in a Fortran code~\cite{erykbiagio} 
is based on the CERN library tool~\cite{paw_reference_cern}.}. 
 Fig.~\ref{1_gamma_displayer} presents a photomultipliers energetic response for one $\gamma$ quanta which hit the 
middle of the module at an angle of 90 degrees.  

\mbox{
\parbox{0.43\textwidth}{
\begin{figure}[H]
\includegraphics[width=0.43\textwidth]{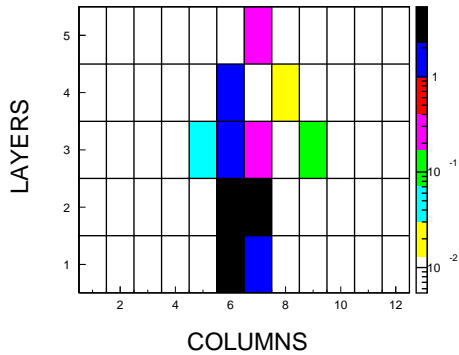}
\vspace{-0.2 cm}
\caption{Distribution of cells for one $\gamma$ quantum hitting with 90 degrees in the middle of the module.}
\label{1_gamma_displayer}
\end{figure}
}
\hspace{0.2cm}
\parbox{0.43\textwidth}{
\begin{figure}[H]
\includegraphics[width=0.43\textwidth]{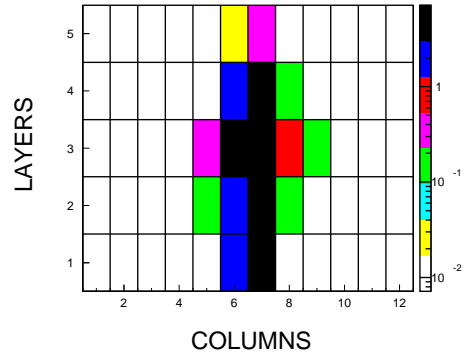}
\vspace{-0.2 cm}
\caption{Distribution of cells for two $\gamma$ quanta hitting with 90 degrees with a distance on y axis equal to 200~cm and with the same 
coordinate on x axis.}
\label{2_gamma_displayer}
\end{figure}
}
}

Comparing this result with the case (see Fig.~\ref{2_gamma_displayer}) where two $\gamma$ quanta hit the module at a distance on y axis equal to 
200~cm but with the same  
 x coordinate,  
 we can see that it is not possible to distinguish these two situations due to the fact that the second case can be 
interpreted for example as one particle which
hit the module but with higher energy.
However, we can easily identify two particle events when a distance on the x axis 
between hit positions is larger than 15~cm (see also Fig.~\ref{Efficiency_of_the_reconstruction_on_x_distance}). 
In Fig.~\ref{2_gamma_displayer_distance_onx_25cm} a distribution for the case when the  
 distance between two photons was equal to 25~cm is presented. 
% and it was
% enough to precise distinguish a two particles tracks in the reconstruction process.

\mbox{
\parbox{0.43\textwidth}{
\begin{figure}[H]
\includegraphics[width=0.43\textwidth]{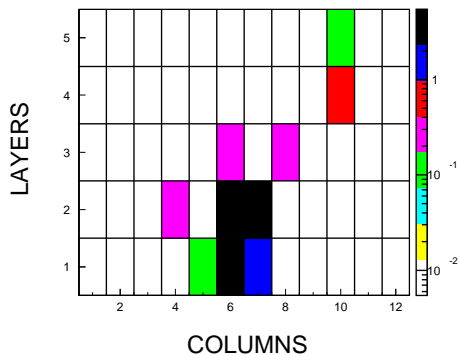}
\caption{The distribution of cells for one $\gamma$ quantum hitting the module with the angle equal to 30 degrees with respect to the surface.}
\label{1_gamma_displayer_30degrees}
\end{figure}
}
\hspace{0.2cm}
\parbox{0.43\textwidth}{
\begin{figure}[H]
\includegraphics[width=0.43\textwidth]{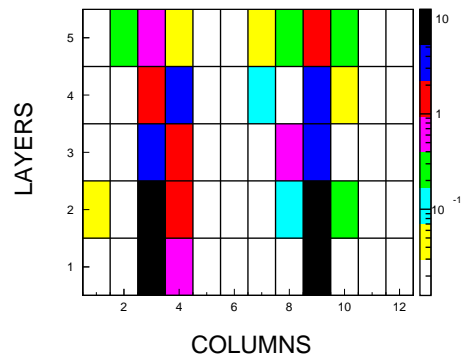}
\caption{The distribution of cells for two $\gamma$ quanta hitting the module with 25~cm distance along x axis.}
\label{2_gamma_displayer_distance_onx_25cm}
\end{figure}
}
}

% We present also an example of the splitting effect of clusters. 
We simulated also a   
 situation where photon hit the module with the angle equal to   
 30 degrees (see Fig.~\ref{1_gamma_displayer_30degrees}) with respect to 
the surface.  
One can see two cells on the right side on this plot which aren't connected to the main cluster.
It is worth to mention that such distributions may be interpreted as two hits.

\chapter{Estimation of probability for multi-gamma hits at a single calorimeter module}
\begin{figure}[H]
\parbox[c]{1.0\textwidth}{
\hspace{1.0cm}
\parbox[c]{0.445\textwidth}{\includegraphics[width=0.445\textwidth]{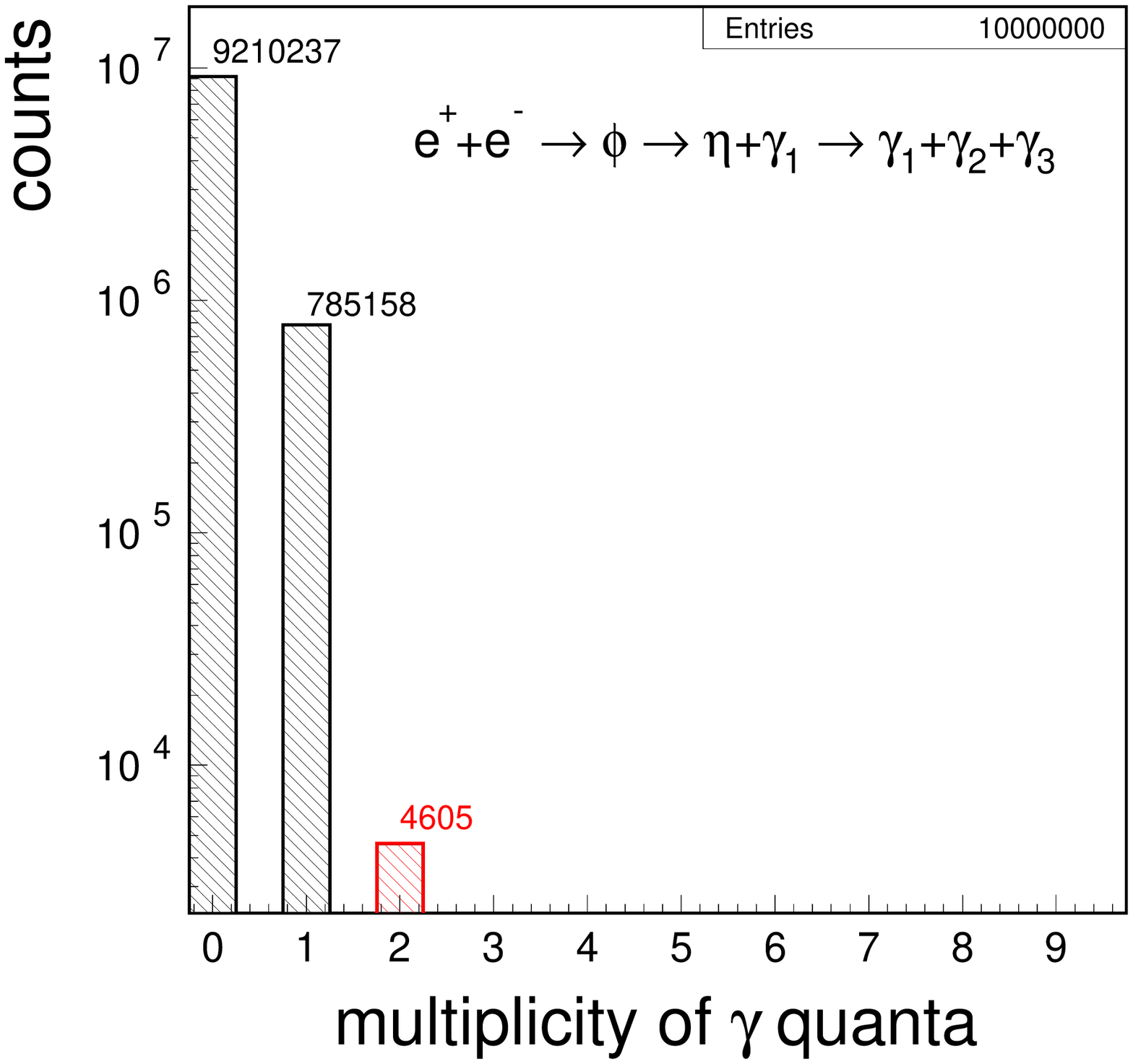}}
\parbox[c]{0.445\textwidth}{
\vspace{-0.45cm}
\includegraphics[width=0.445\textwidth]{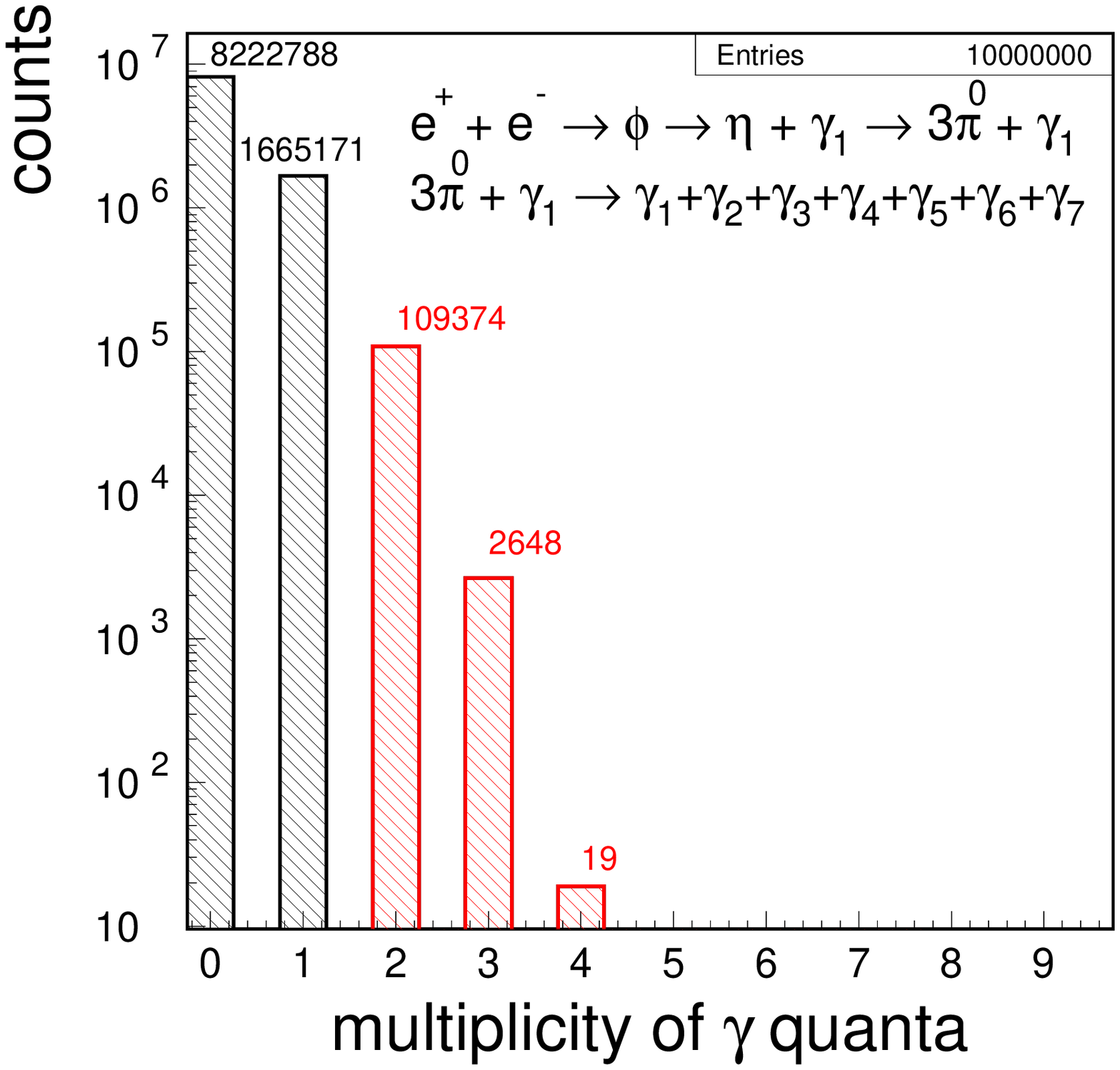}}
}
\parbox[c]{1.0\textwidth}{
\hspace{1.0cm}
\parbox[c]{0.445\textwidth}{\includegraphics[width=0.445\textwidth]{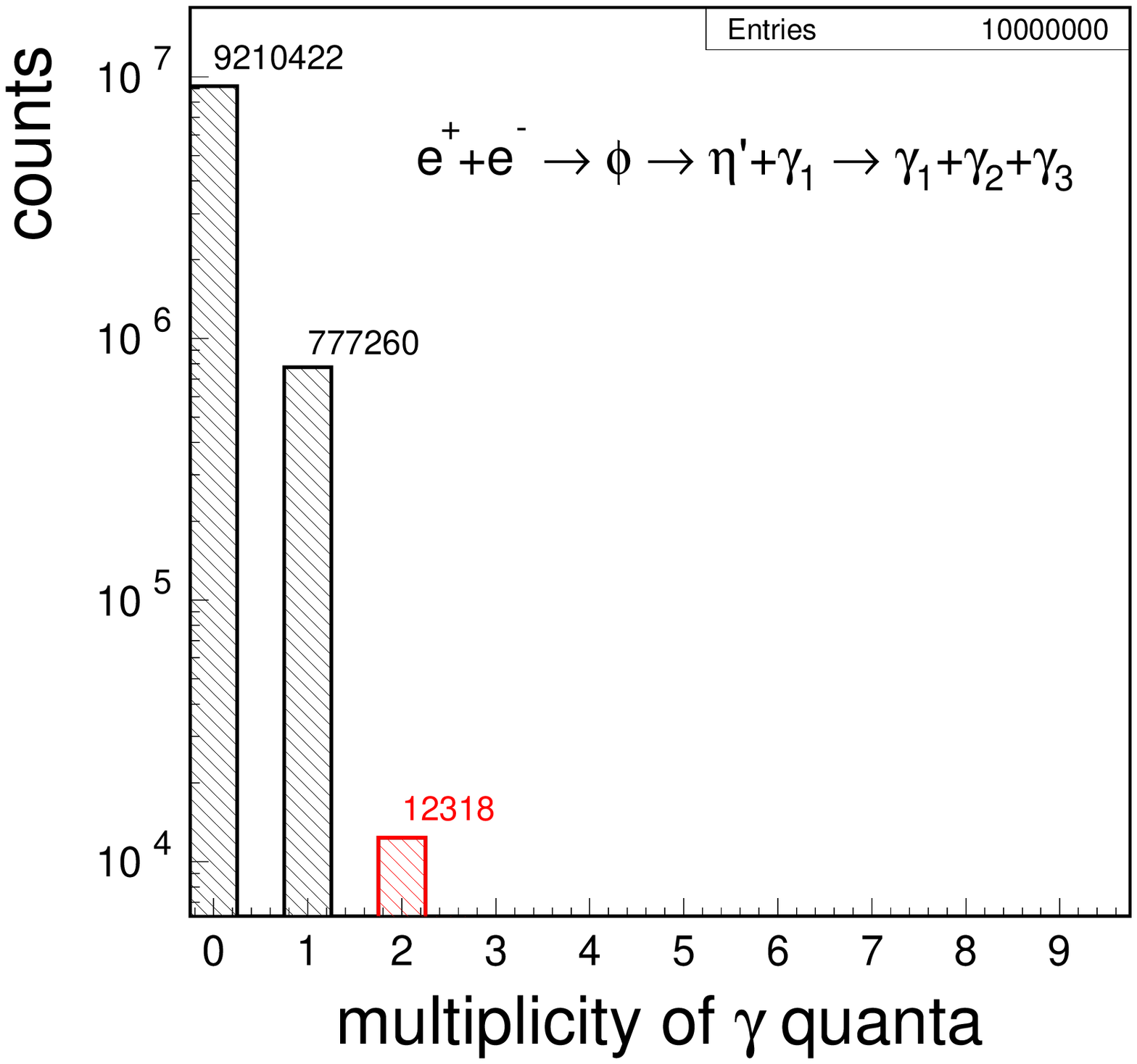}}
\parbox[c]{0.445\textwidth}{\includegraphics[width=0.445\textwidth]{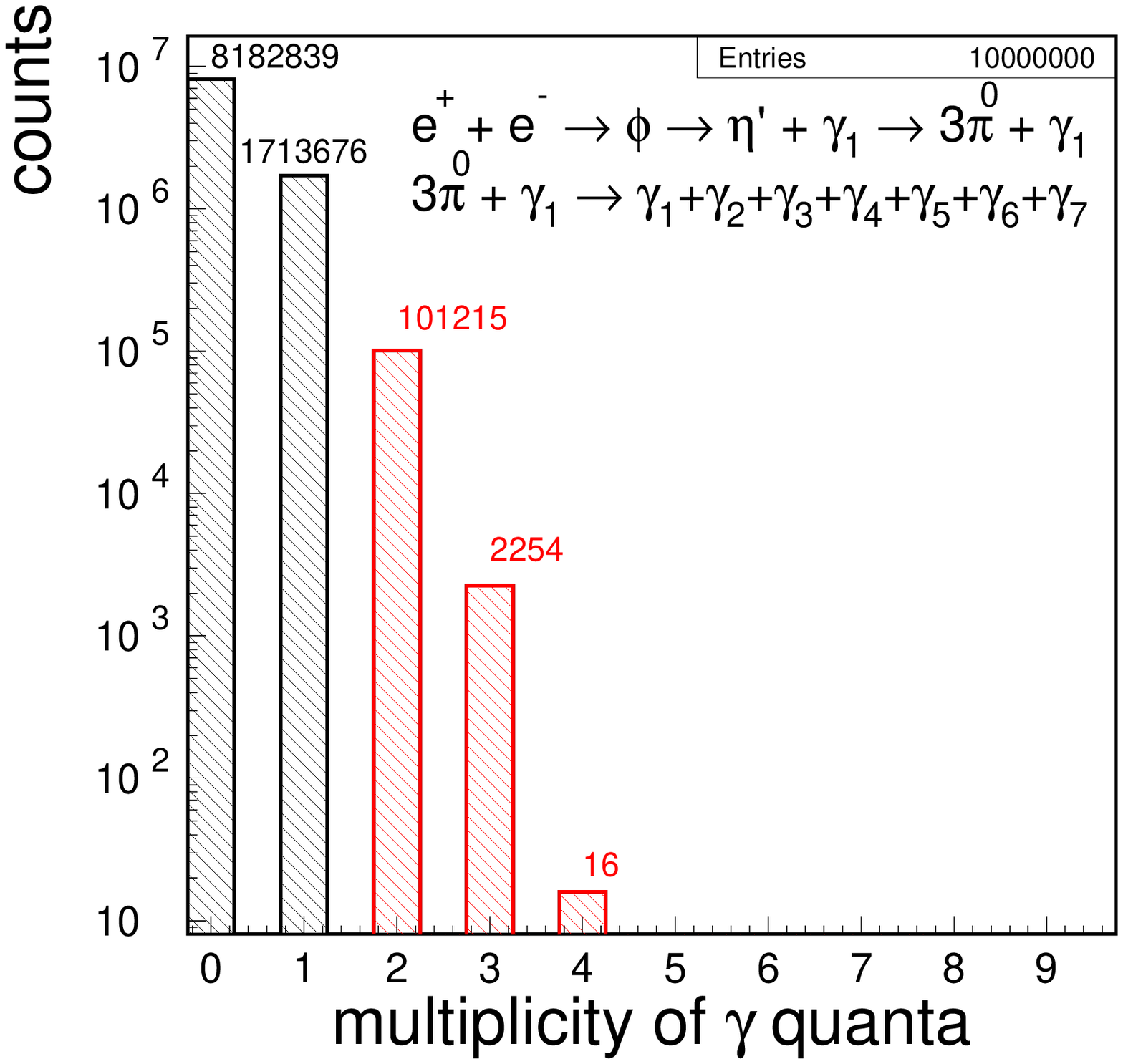}}
}
\caption{The multiplicity of the $\gamma$ quanta on a single calorimeter module. For reactions described in the figure.}
\label{multiplicity_of_the_gamma_quanta}
\end{figure}

\begin{figure}[H]
\parbox[c]{1.0\textwidth}{
\parbox[c]{0.325\textwidth}{
\hspace{-0.2cm}
\includegraphics[width=0.36\textwidth]{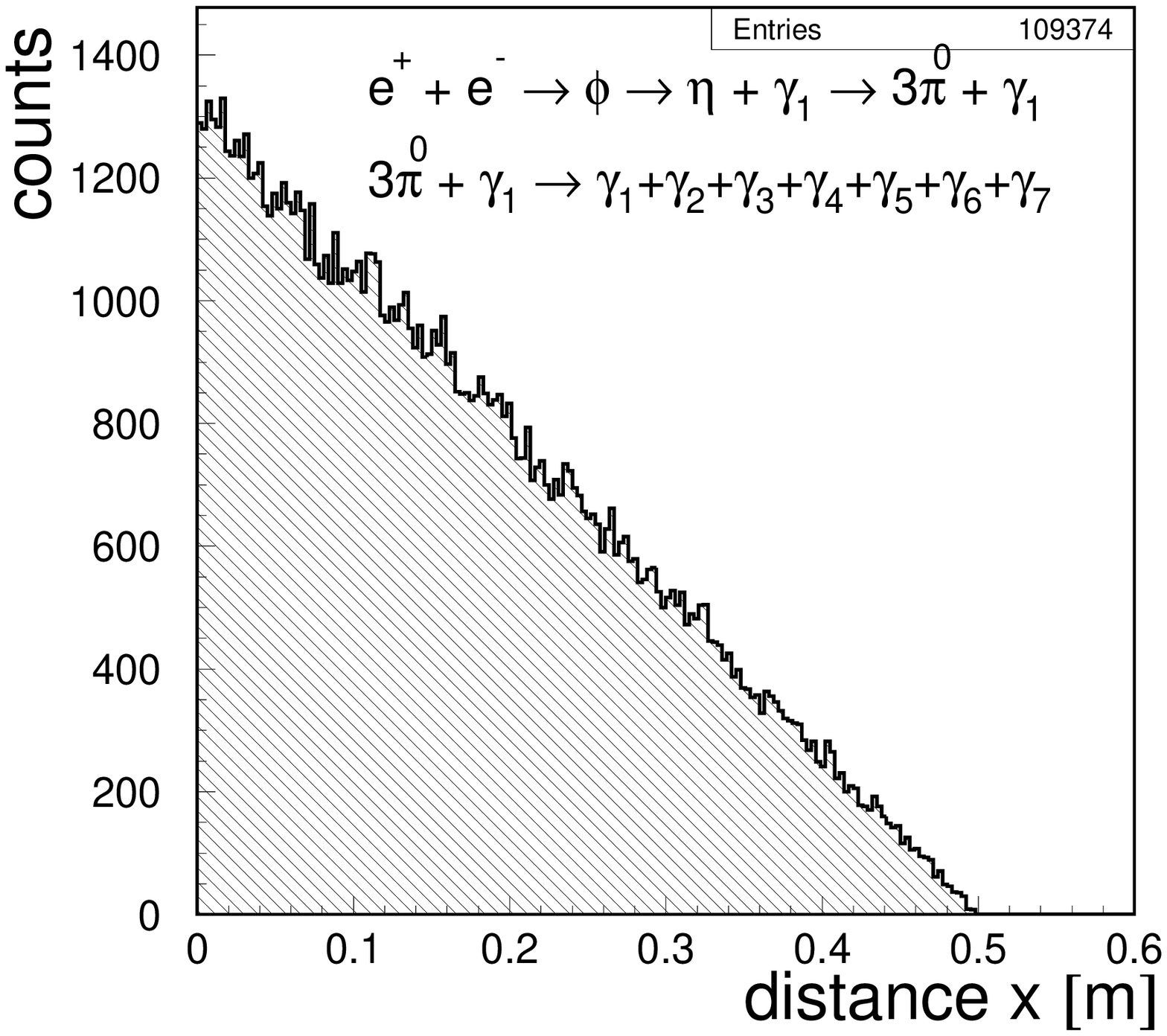}}
\parbox[c]{0.325\textwidth}{
\hspace{-0.3cm}
\includegraphics[width=0.36\textwidth]{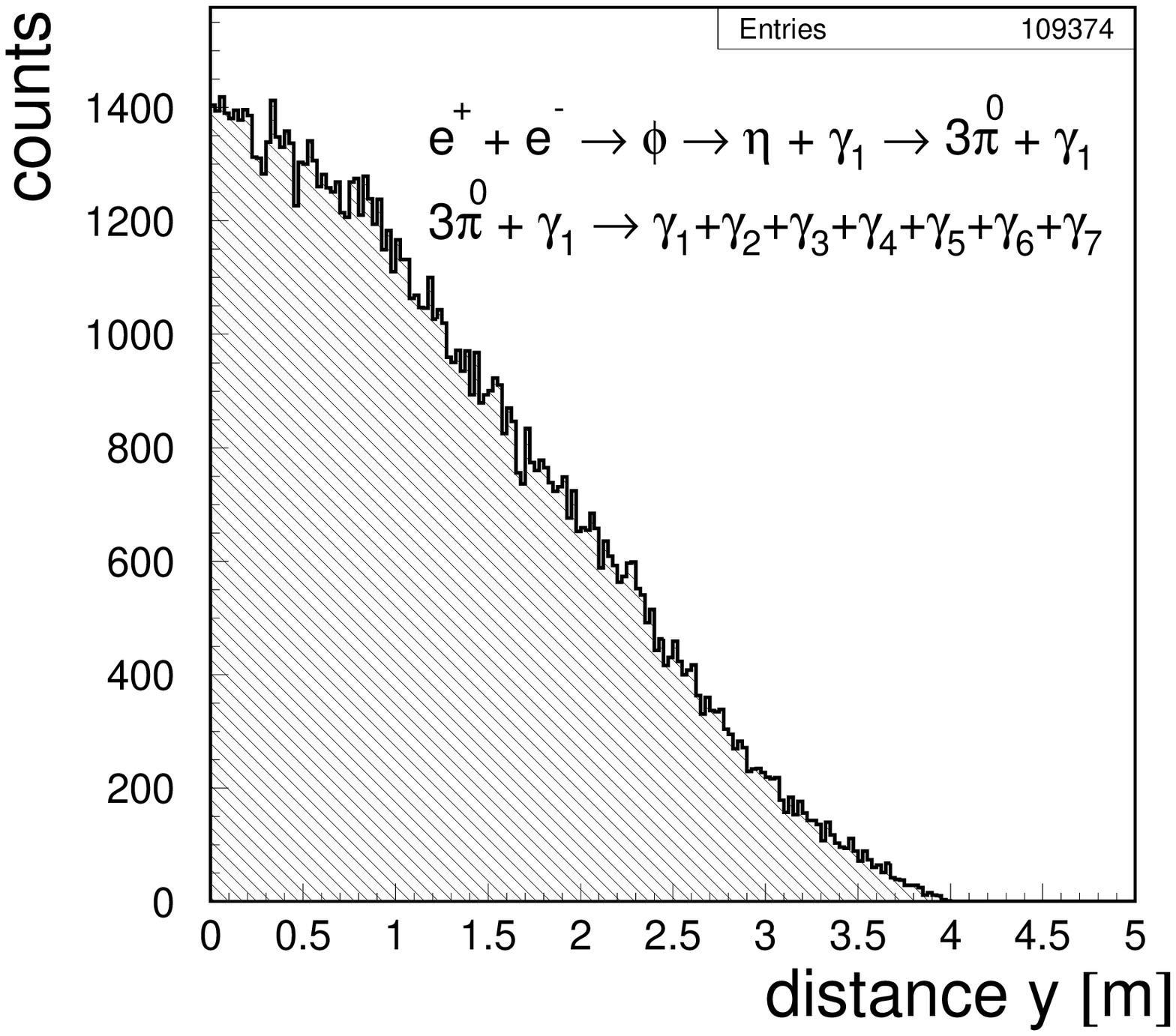}}
\parbox[c]{0.325\textwidth}{
\hspace{-0.3cm}
\includegraphics[width=0.36\textwidth]{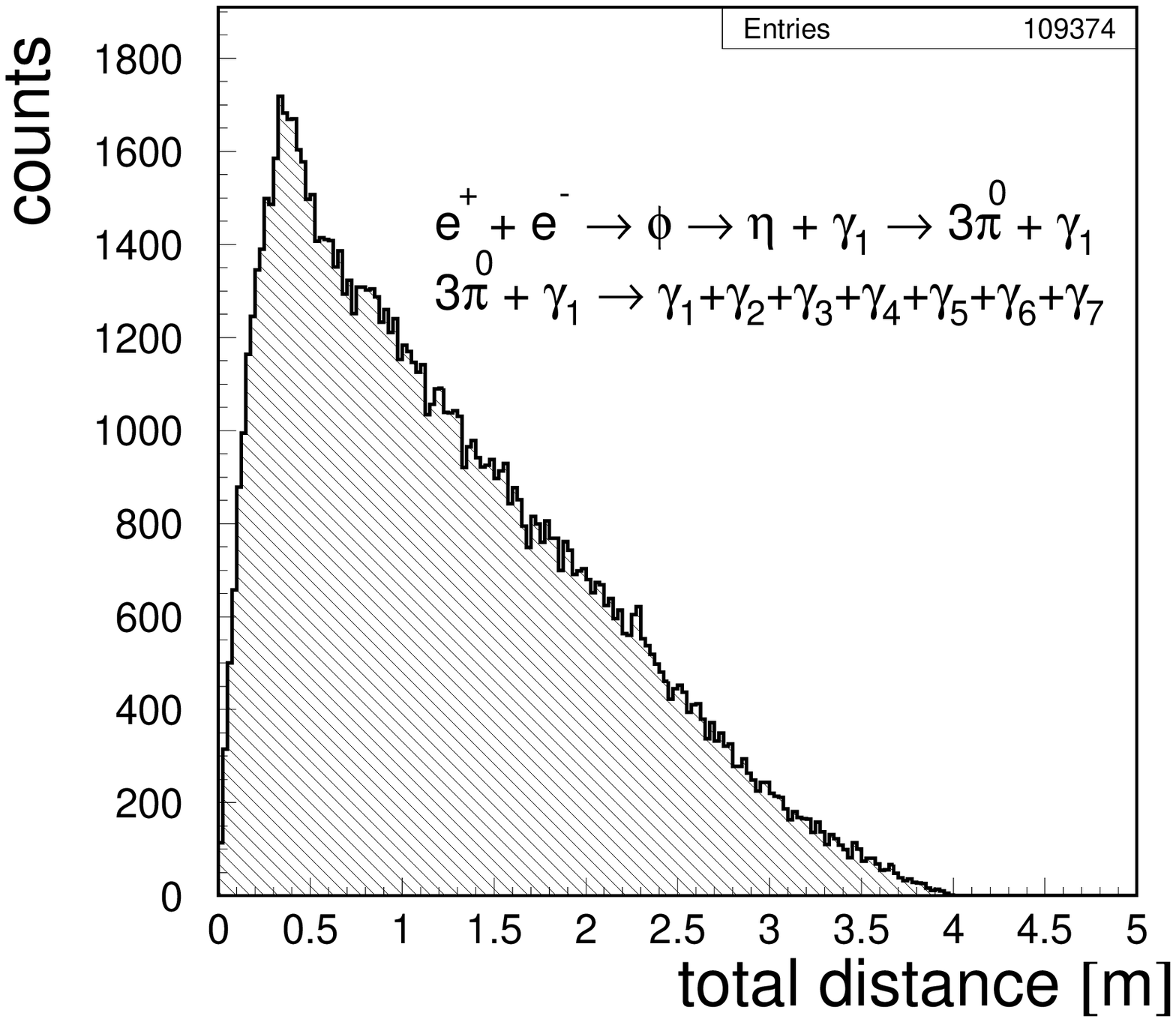}}
}
\caption{Distance between two gamma quanta from $\eta \to 3\pi^{0}$ in one calorimeter module.}
\label{2reaction_distance}
\end{figure}

\begin{figure}[H]
\parbox[c]{1.0\textwidth}{
\parbox[c]{0.325\textwidth}{
\hspace{-0.2cm}
\includegraphics[width=0.36\textwidth]{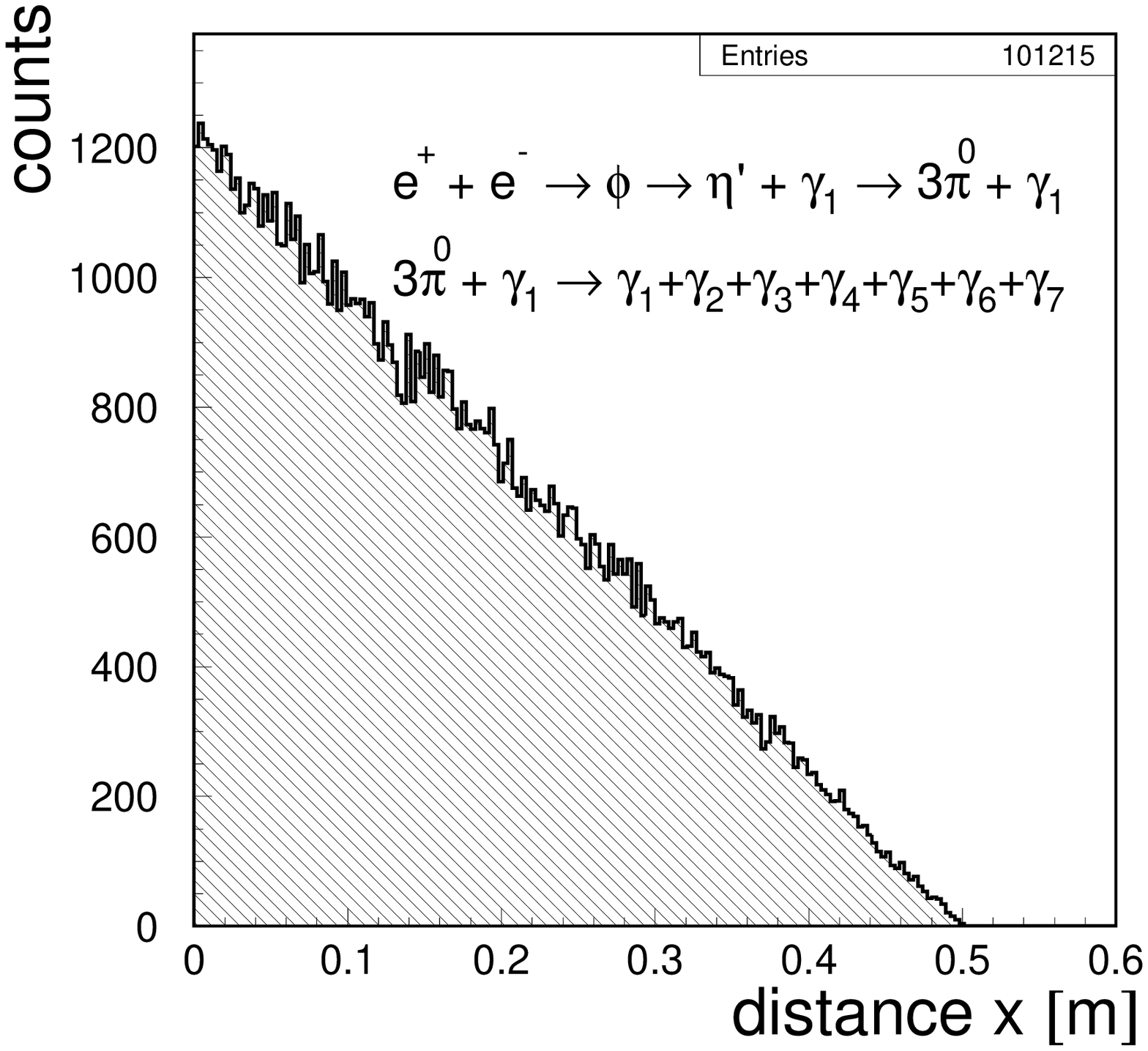}}
\parbox[c]{0.325\textwidth}{
\hspace{-0.3cm}
\includegraphics[width=0.36\textwidth]{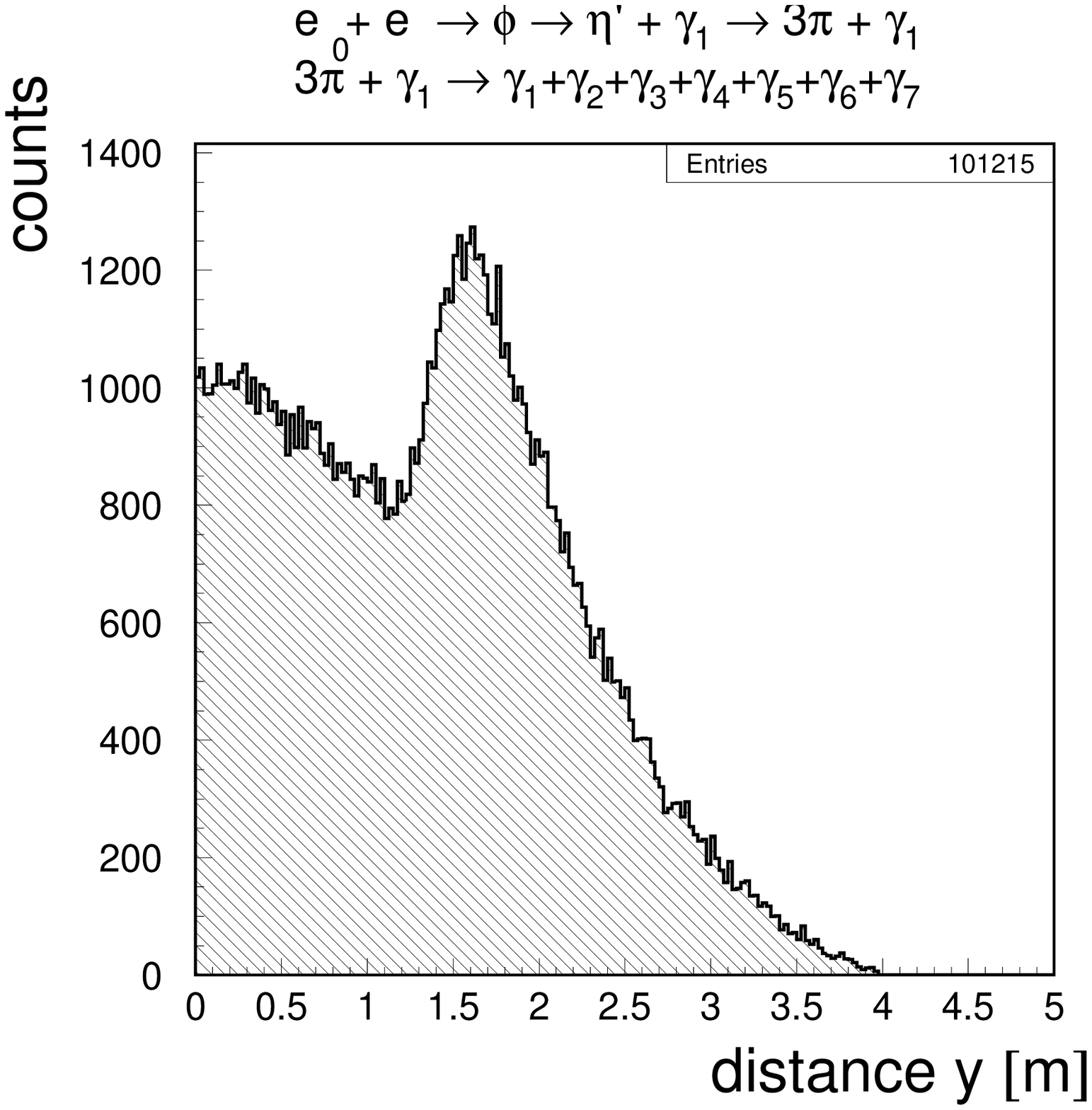}}
\parbox[c]{0.325\textwidth}{
\hspace{-0.3cm}
\includegraphics[width=0.36\textwidth]{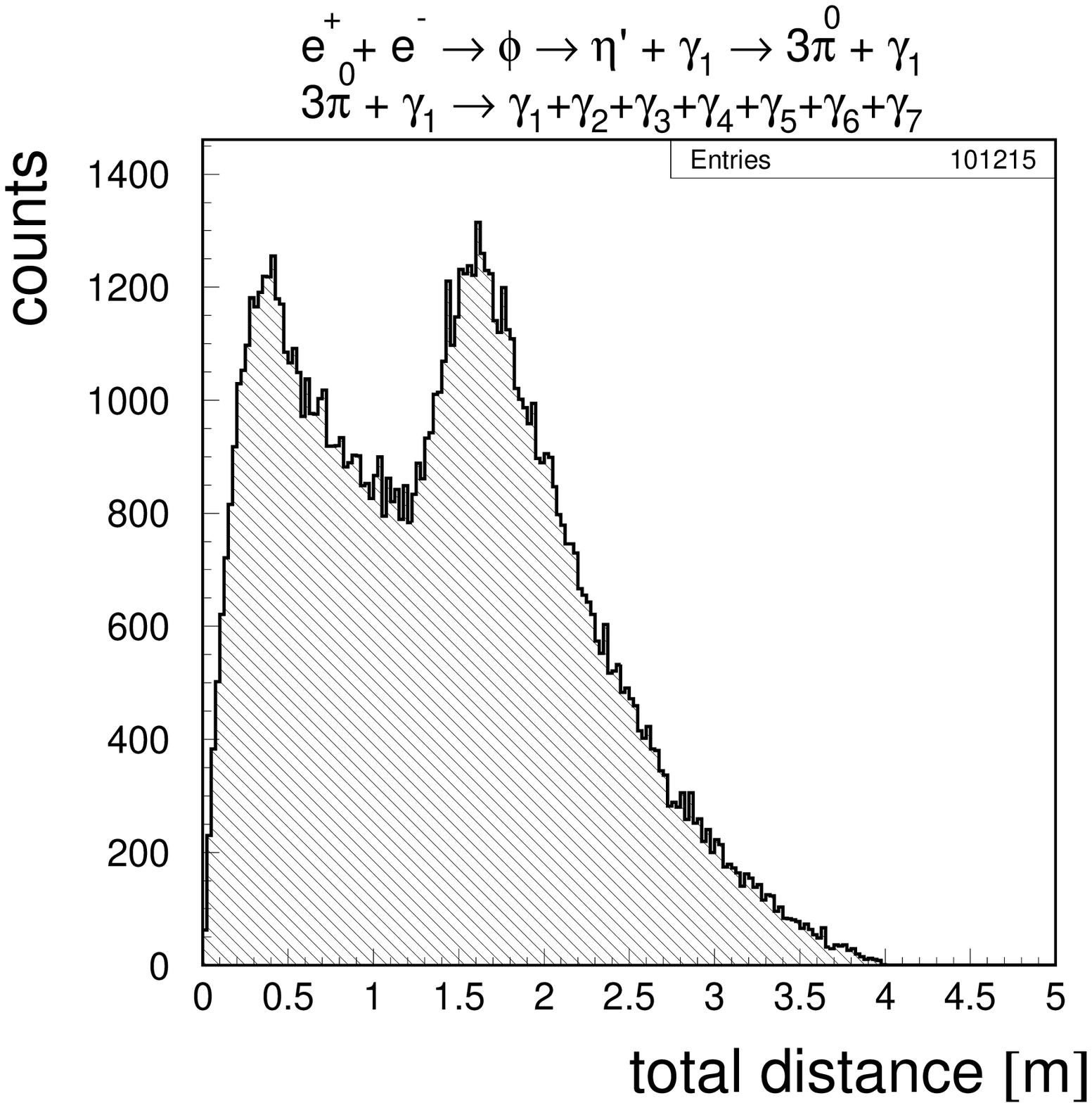}}
}
\caption{Distance between two $\gamma$ quanta from reaction $\eta^{\prime} \to 3\pi^{0}$ on one calorimeter module.}
\label{4reaction_distance}
\end{figure}

\hspace{\parindent}
%
% I would like to present some results from simulations FLUKA Monte Carlo with vertex generator.
%
Using FLUKA Monte Carlo with vertex generator we were able to estimate a probability on the  
multi-gamma hits at calorimeter module orginating from the $\eta$ meson decays.
%
%The first four figures presente the angular and energy distributions
%of the gamma quanta from the first two above listed reactions.
%The figures are self-explaining.
%
%
We estimated a multiplicity of the $\gamma$ quanta on one module of the KLOE 
calorimeter. By multiplicity we define the number of $\gamma$ quanta hitting
one module (Fig.~\ref{multiplicity_of_the_gamma_quanta}).  
\noindent If it is equal to zero, it signifies that our tested
module did not seen any $\gamma$, if it is equal to one than the module
registered one gamma quantum. Due to the axial symmetry of KLOE
this result is valid for all detection modules.
Thus, we see that for the $\phi \to \eta\gamma \to 3\gamma$ reaction
 only for  0.6\%  of the total number of events the 2$\gamma$ quanta hit one module and in the rest 99.4\%
only one $\gamma$ is registered by one module.
However, for the reaction $\phi \to \eta\gamma \to 7\gamma$ where $\eta$ meson decays into 3$\pi^{0}$
the situation changes drastically and already in 6.3\% cases there is two
or more $\gamma$ quanta hitting the module. \\
\indent In the situation where more than one photon hit the calorimeter module it is possible that merging effect
 will appear for some distances between hit positions of particles on the surface. The estimations of the order
 of multiplicity effect was shown in Fig.~\ref{Efficiency_of_the_reconstruction_on_x_distance} and
Fig.~\ref{Efficiency_of_the_reconstruction_on_y_distance}.
%
%
%
% Thus, using vertex generator we were able to estimate a situation when two $\gamma$ quanta 
% hit one calorimeter module (Fig.~\ref{multiplicity_of_the_gamma_quanta}). 
% This was independent way to discriminate a influence
% of the multiplicity effect. \\
%
 In the following we will describe the distribution of the distance
   between the $\gamma$ quanta in the case when two $\gamma$ hits the module. 
%   This was achieved
%   using the vertex generator simulation tool.
   This result is relevant in view of the reconstruction possibilities of the clustering algorithm
 (see Fig.~\ref{Efficiency_of_the_reconstruction_on_x_distance} and  
 Fig.~\ref{Efficiency_of_the_reconstruction_on_y_distance}).
   The result for the $\eta \to 3\pi^{0}$ is shown in Fig.~\ref{2reaction_distance}.
   The corresponding reactions are
   written inside the figures.
%
%   The distributions for reactions 1, 2 and 3 looks very similar (as is shown in the Fig.~\ref{2reaction_distance}),
%    with the maximum at about 1.7 meters.
%
%\begin{figure}[h!]
%\hspace{-4cm} przesuwa parboxa w lewo !!!
%\parbox[c]{1.15\textwidth}{
%\parbox[c]{0.55\textwidth}{\includegraphics[width=0.59\textwidth]{1reactiondistancex.eps}}
%\parbox[c]{0.55\textwidth}{\includegraphics[width=0.59\textwidth]{1reactiondistancey.eps}}
%}
%\parbox[c]{1.15\textwidth}{
%\parbox[c]{0.55\textwidth}{\includegraphics[width=0.59\textwidth]{1reactiondistancetotal.eps}}
%}
%\caption{ Distance betwen two gammas quanta on 1 calorimeter module }
%\label{odnosnik}
%\end{figure}
\vspace{-0.9cm}
%
%\begin{figure}[h!]
%\parbox[c]{1.15\textwidth}{
%\parbox[c]{0.55\textwidth}{\includegraphics[width=0.59\textwidth]{3reactiondistancex.eps}}
%\parbox[c]{0.55\textwidth}{\includegraphics[width=0.59\textwidth]{3reactiondistancey.eps}}
%}
%\parbox[c]{1.15\textwidth}{
%\parbox[c]{0.55\textwidth}{\includegraphics[width=0.59\textwidth]{3reactiondistancetotal.eps}}
%}
%\caption{ Distance betwen two gammas quanta on 1 calorimeter module }
%\label{odnosnik}
%\end{figure}
%
%
%\vspace{-0.3cm}
 And the result for the $\eta^{\prime} \to 3\pi^{0}$ is shown in Fig.~\ref{4reaction_distance}. 
 The distribution determined for the $\eta^{\prime}$ is not very intuitive.  
%   However, for the reaction where $\eta^{\prime} \to 3\pi^{0}$ the result 
%   is very not intuistic (see Fig.~\ref{4reaction_distance}) and 
   The maximum at 1.7~m is due to the fact 
   that the pions from the $\eta{\prime}$ decay are very fast and
    both $\gamma$ quanta from one pion can
   hit one module. The pion decays immediately in the target 
   and the distance between the two $\gamma$ quanta from its decay at the
   calorimeter surface is indeed around 1.7 m.
\vspace{-0.7cm}
\newpage
%\vspace{-0.7cm}
\begin{figure}[H]
\hspace{-0.0cm}
\parbox[c]{1.0\textwidth}{
\parbox[c]{0.325\textwidth}{
\hspace{-0.2cm}
\includegraphics[width=0.36\textwidth]{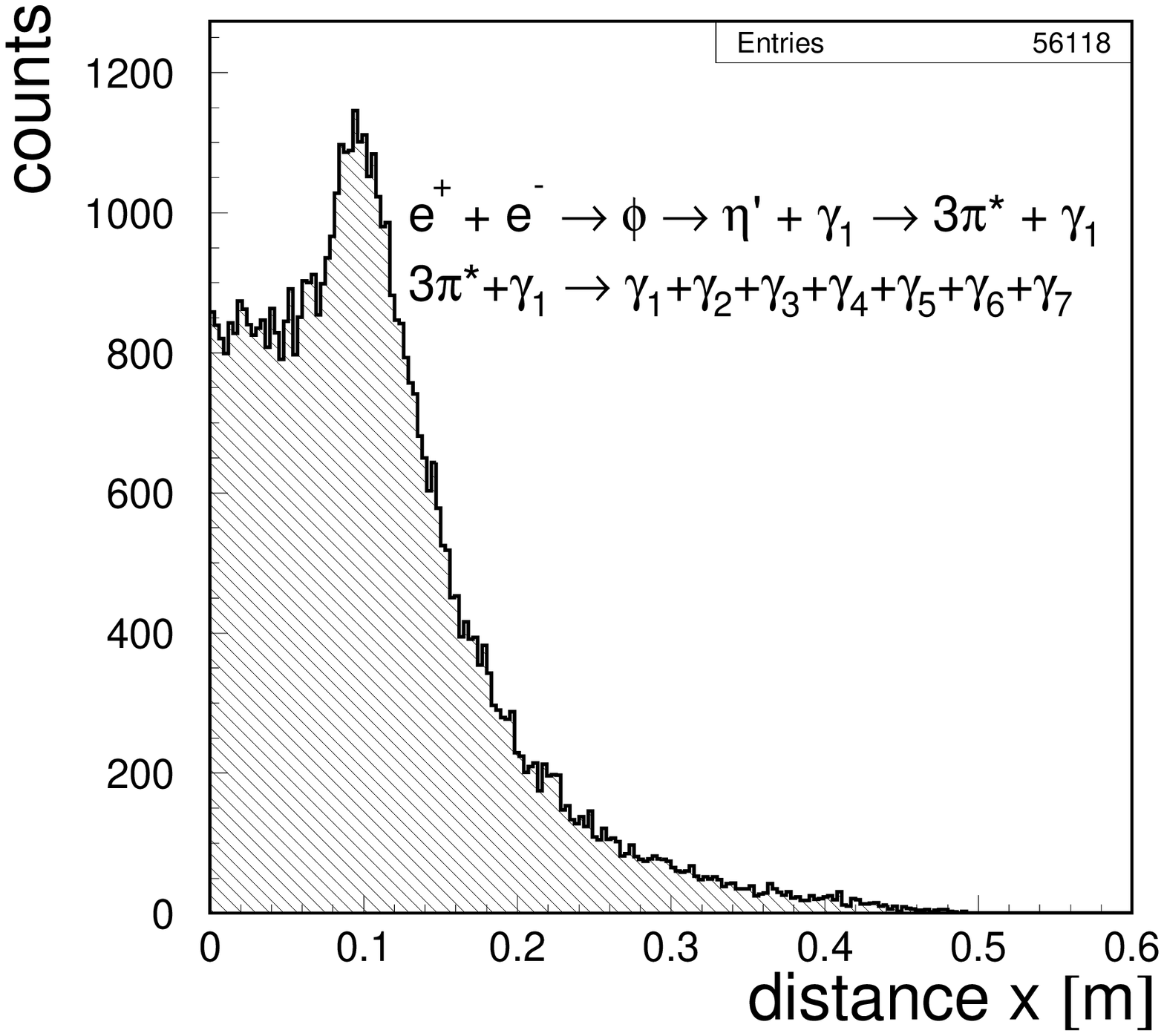}}
\parbox[c]{0.325\textwidth}{
\hspace{-0.3cm}
\includegraphics[width=0.36\textwidth]{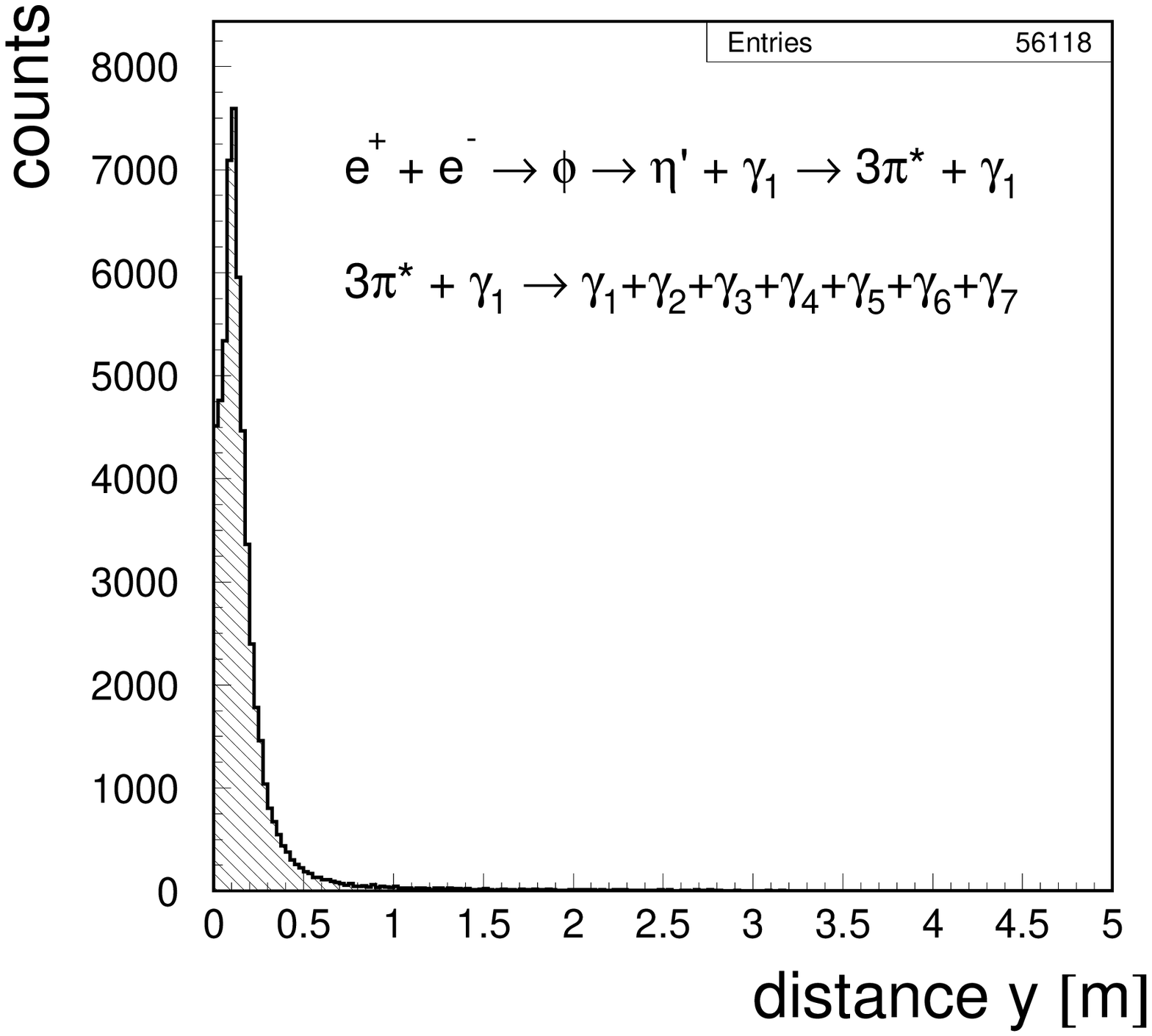}}
\parbox[c]{0.325\textwidth}{
\hspace{-0.3cm}
\includegraphics[width=0.36\textwidth]{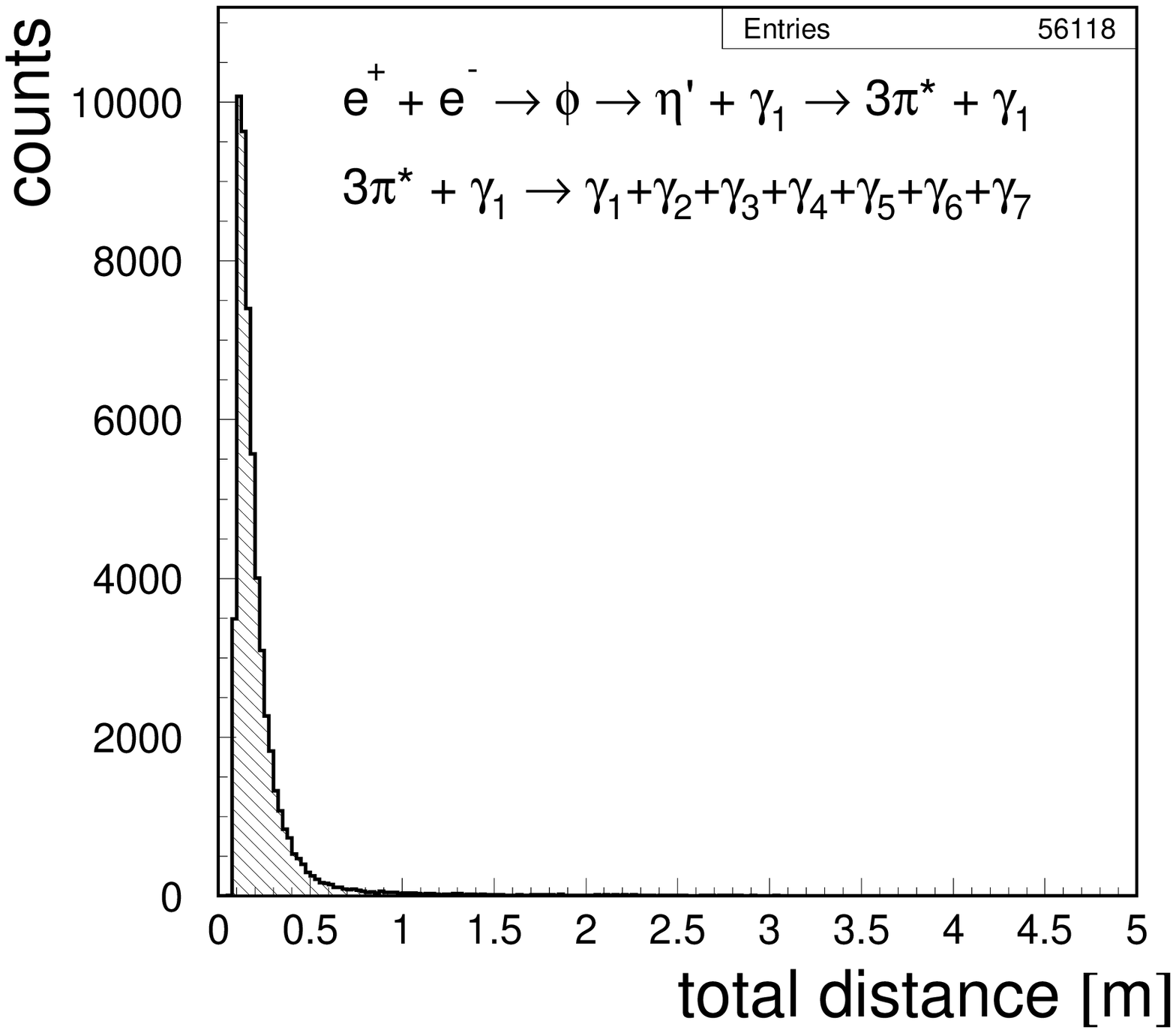}}
}
\caption{Distance between two $\gamma$ quanta on one calorimeter module for reactions as depicted in the figures assuming 
that the "pion" mass is equal to 10 MeV.}
\label{4reaction_distance_10mevpion}
\end{figure}
   We did also one more check by decreasing in the simulations the mass of the pion
   to 10 MeV. In such a case these "pions" are much faster and the laboratory angle
   between gamma quanta is decreasing significantly,
   and the distance between them in
   the calorimeter is as expected to be much smaller. The result of this simulation is shown in 
 Fig.~\ref{4reaction_distance_10mevpion}.

% Thus, we estimated a probability of the frequency of case when 2$\gamma$ hit one module and distribution of distance 
% between these 2$\gamma$. These studies were done due to determine the multiplicity effect probability. 
%
%

\chapter{Example of the event reconstruction for $e^{+}e^{-} \to \phi \to \eta\gamma \to 3\pi^{0}\gamma \to 7\gamma$ reaction}
\begin{figure}[H]
\hspace{1.5cm}
\parbox[c]{1.0\textwidth}{
\parbox[c]{0.75\textwidth}{\includegraphics[width=0.79\textwidth]{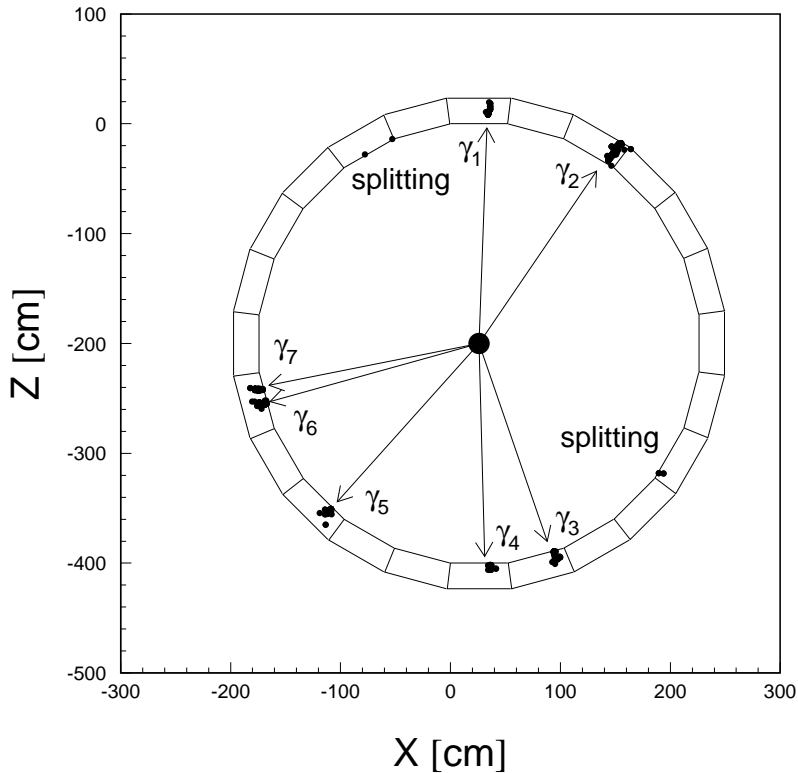}}
}
\caption{Splitting of energy deposits in the barrel calorimeter.}
\label{24_modules_1event}
\end{figure}

\hspace{\parindent}
%
% Before clusterisation process we are able to observe a influence of the merging and splitting of signals  
% using our 24 modules geometry in simulations with a vertex generator. 
%
The capabilites of the prepared program to reproduce a merging and splitting effects
 in light output from scintilating fibers are ilustrated in Fig.~\ref{24_modules_1event}.  
One can see seven significant energy deposits belonging to 7 photons from the  
$e^{+}e^{-} \to \phi \to \eta\gamma \to 3\pi^{0}\gamma \to 7\gamma$ reaction, and also some small deposits 
of energy which don't orginate directly from photons are well seen. 
%
% , these deposits will be related with splitting effect of clusters because
% a the first cells and clusters are build by clustering algorithmn in based on energy depositions in scintillating fibers.
%
The splitted energy deposits may cause reconstruction of the false clusters. 
On the other hand 
% we observe the influence of the merging effect too because 
two very close tracks of photons  
(near z = -250~cm) may be reconstructed as one merged cluster since the distance 
between particles is only about 4~cm, leading with a large probability to false reconstruction (see Fig.~\ref{24_modules_1event}).   
%
% 
% as was shown (see Fig.~\ref{Efficiency_of_the_reconstruction_on_x_distance}) when we describing the 
% reconstruction possibilities of the present clustering algorithm, this situation is very
% hard for correct reconstruction two separated particles. \\    
%
 After simulating the energy response of the scintillating fibers, on 24 modules with FLUKA, we extended appropriately the DIGICLU program 
in order to perform reconstruction of signals from photomultipliers on the whole barrel calorimeter. \\
% \indent Also we modified our vertex generator with condition we take into account only events where 
%  7$\gamma$ quanta hit the barrel calorimeter per one event.    
%
\begin{figure}[H]
\hspace{2.3cm}
%\parbox[c]{1.0\textwidth}{
%\parbox[c]{0.5\textwidth}{\includegraphics[angle=0,width=0.44\textwidth]{clusters4_2.eps}}
%\hspace{-1.2cm}
\parbox[c]{0.7\textwidth}{\includegraphics[width=0.70\textwidth]{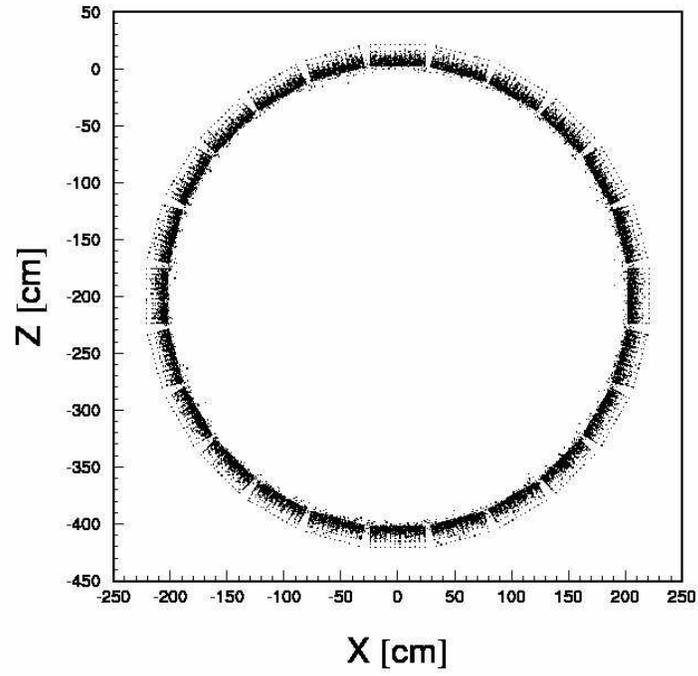}}

\caption{Reconstruction of clusters for reaction: $e^{+}e^{-} \to \phi \to \eta \gamma \to 3\pi^{0} \gamma \to 7\gamma$. 
Statistics is equal to 1000 events.
}
\label{clusters_24_2_3}
\end{figure}
%
% In the Fig.~\ref{clusters_24_2_3} is presented a distribution of reconstructed clusters on the barrel calorimeter. The left panel includes 
% results for one event, for the same event for which was presented a light output from scintillating fibers in the Fig.~\ref{24_modules_1event}. 
% The right panel presents a distribution of reconstructed clusters on the barrel calorimeter (statistics is equal to 1000 events).
%
% For this example event, the clustering algorithm reconstructed only five clusters, when we simulated seven 
% photons in the final state which hit the barrel.  
%
Fig.~\ref{clusters_24_2_3} presents distribution of reconstructed clusters on the barrel part of the calorimeter 
for the 1000 $e^{+}e^{-} \to \phi \to \eta\gamma \to 3\pi^{0}\gamma \to 7\gamma$ events simulated with vertex generator and FLUKA program. One 
can see that the clusters aren't reconstructed on the edges of modules.  

\chapter{Energy deposition as a function of azimuthal angle}
\begin{figure}[H]
\vspace{-0.5cm}
\hspace{3.5cm}
\parbox[c]{0.5\textwidth}{
\includegraphics[width=0.59\textwidth]{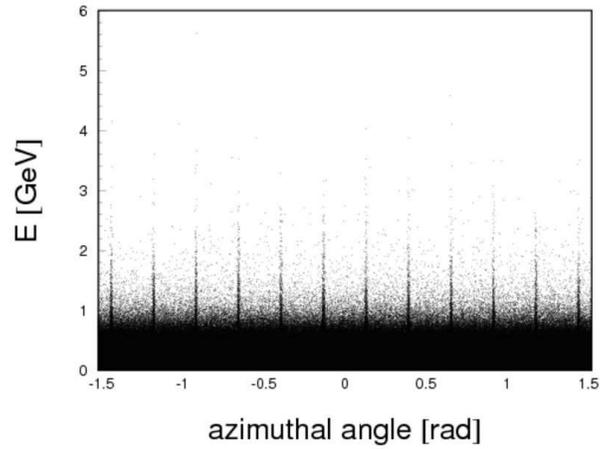}
}
\caption{Energy deposits on the upper half of the barrel calorimeter as a function of cylindrical angle.}
\label{energy_vs_phi}
\end{figure}

\hspace{\parindent}
We studied energy deposits in scintillating fibers as a function of the azimuthal angle of $\gamma$ quanta producted in the 
interaction region. 
The results for the upper part of barrel calorimeter is shown in Fig.~\ref{energy_vs_phi}. 
%
%The order of magnitude for deposited energy is presented in GeV (dac na rysunek).
%
% In Fig.~\ref{energy_vs_phi} presents a distribution of energy deposited in scintilating 
% fibers as a function of the cylindrical angle. These are the results of the simulations with FLUKA
%  Monte Carlo and the KLOE barrel calorimeter.  
%
%
%
%
%
%One can see that most energy is not deposited by photons in the middle parts of the calorimeter modules
% but rather near their edges (upper panel). 
%On the lower a energy distribution as a function of x axis for upper half of barrel is presented. The higher 
%energy depositions were appearing on the edges of modules. 
% This effect is related with angular distribution which on the one hand determined 
% length of particle's track in the material of module (Fig.~\ref{drogi_w_module_od_kata}) but also influence to amount 
% of the calorimeter material which "see" a particle \ref{krawedz_kat_phi0}. 
%
%
\begin{figure}[H]
\hspace{3.5cm}
\parbox[c]{0.5\textwidth}{\includegraphics[width=0.59\textwidth]{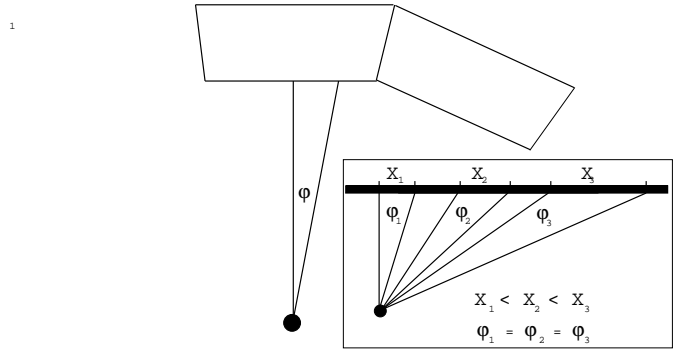}}
\caption{Schematic view of the $\phi$ angle distribution.}
\label{krawedz_kat_phi0}
\end{figure}
One can see that at the edges of the module a maximum of the energy deposits appears. 
To some extent it may be explained by the fact that  
on the edges the value of the solid angle  
 for the same $\phi$ angle interval  
 is larger than in the middle (see Fig.~\ref{krawedz_kat_phi0}).  

% and 
% the same distance on surface is being "seen" (Fig.~\ref{krawedz_kat_phi0}) with the bigger andle value, thus more 
% photons hit this whole region.   
%
%
%
\begin{figure}[H]
\hspace{3.5cm}
\parbox[c]{1.0\textwidth}{
\parbox[c]{0.5\textwidth}{\includegraphics[width=0.59\textwidth]{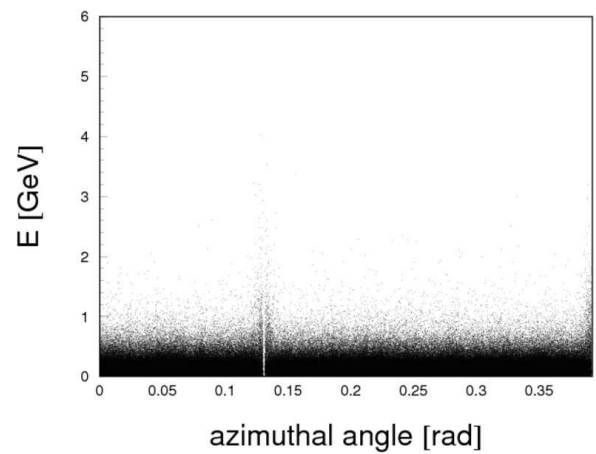}}
\hspace{0.5cm}
%\parbox[c]{0.5\textwidth}{\includegraphics[width=0.59\textwidth]{E_vs_xril_10000ev.eps}}
}
\caption{Energy deposits on the barrel calorimeter as function of the cylindrical angle in the range from 0 to 22.5 degree.}
\label{energy_vs_phi_male}
\end{figure}
In the simulations we did not include fibers which were near the edge since in a real detector the trapezoid 
shape 
%
% 0ne can see also effect of fact that there is not so many fibers near the edge (we taking into account this effect in our simulations 
% see Fig.~\ref{24_details_of_implementation}),    
% accordingly to the trapezoid shape 
of the module was acomplished by cutting some parts from rectangular piece \cite{kloe_electromagnetic_nim2} 
and hence the   
% at begining when the modules were produced in technology center \cite{kloe_electromagnetic_nim2} and some 
scintillating fibers at the sides  
were broken. In Fig.~\ref{energy_vs_phi_male} one can see that for value of an angle around 0.15 rad particles 
didn't deposited energy. This is because this 
position is exactly between the edge of the first and second trapezoid modules. 
% On the right panel is shown this area more precisely 
% but as a function of distance. One can see that on the egde (52.5 cm) more energy was deposited. It doesn't seen a empty region 
% between the modules in this distribution because a 
% modules aren't parallel but they touch at the small angle, and x coordinate is parallel to the first.    
%
%
%
\chapter{Definition of the coordinate system}
\begin{figure}[H]
\hspace{0.5cm}
\parbox{1.0\textwidth}{\centerline{\epsfig{file=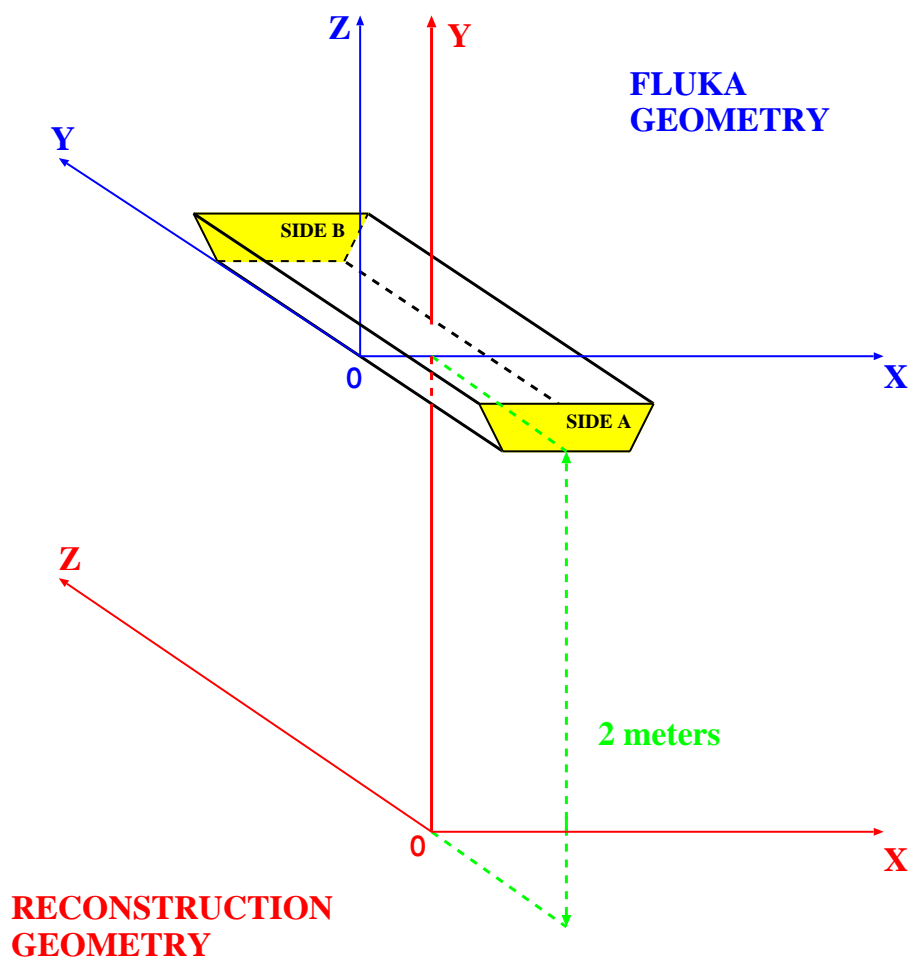,width=0.80\textwidth}}}
\caption{
 The coordinate systems used in simulations.
}
\label{coordinate_system}
\end{figure}

\hspace{\parindent}
  The coordinate systems which were used in simulations
   are shown in Fig.~\ref{coordinate_system}. The blue axis denote the frame which was used 
   for simulations with FLUKA and with the VERTEX GENERATOR. The red coordinate system which is the same as coordinate frame 
of the KLOE detector has been used 
   in simulations of clusterization effects and simulations of the photomultipliers response with
DIGICLU\footnote{Digiclu frame is called the reconstruction geometry system.}.
%
%
%For some channels with hight multiplicity of the $\gamma$ quanta in the final
%state (see \ref{2reaction_distance}). 
%
Throughout this thesis we plot all figures using the coordinate frame of the FLUKA.

\newpage
~\\
\newpage
\thispagestyle{empty}
~\\
~\\
~\\
~\\
~\\
~\\
~\\
~\\
~\\
~\\
~\\
~\\
~\\
~\\
~\\
~\\
~\\
~\\
~\\
~\\
~\\
~\\
~\\
\hspace{-1.5cm}
\mbox{
\parbox[l]{0.5\textwidth}{
The world is given to me only once, not one existing and one perceived. 
Subject and object are only one. The barrier between them cannot be said to 
have broken down as a result of recent experience in the physical sciences, 
for this barrier does not exist. 
}
\hspace{1.0cm}
\parbox[r]{0.5\textwidth}{
\'Swiat dany mi jest tylko raz, nie ten istniej\c{a}cy i nie ten do\'swiadczany.
 Podmiot i obiekt s\c{a} jednym. Bariera pomi\c{e}dzy nimi nie mo{\.z}e by\'c 
z{\l}amana jako wynik ostatnich do\'swiadcze\'n w naukach fizycznych, bariera ta nie 
istnieje. 
}
}
~\\
~\\
~\\
\hspace{5.5cm}
\parbox[l]{1.0\textwidth}{
~~~~~~~~~~~~~~~~~~~~~~~~~~~~~~~~~~~~~~~~~~~~~~~~~~~~~~~~~~~~~~~~~~~~~~~~~~~ Erwin Schr{\"o}dinger (1887 - 1961) 
}
~\\
~\\
~\\
~\\
~\\
\hspace{-1.5cm}
\mbox{
\parbox[l]{0.5\textwidth}{
Education is what remains after one has forgotten everything he learned in school. 
}
\hspace{1.0cm}
\parbox[r]{0.5\textwidth}{
Edukacja jest tym co zostaje jak zapomni si\c{e} wszystko czego nauczy{\l}o si\c{e} w szkole.   
}
}
~\\
~\\
~\\
\hspace{-2.5cm}
\parbox[l]{1.0\textwidth}{
~~~~~~~~~~~~~~~~~~~~~~~~~~~~~~~~~~~~~~~~~~~~~~~~~~~~~~~~~~~~~~~~~~~~~~~~~~~~ Albert Einstein (1879 - 1955)
}
%
%%%%%%%%%%%%%%%%%%%%%%% MONTE CARLO TOOLS %%%%%%%%%%%%%%%%%%%%%%%%%%%%%
% \chapter{Monte Carlo - tools}
%
% \section{FLUKA}
%\begin{figure}[h!]
%\parbox[c]{0.2\textwidth}{\includegraphics[width=1.05\textwidth]{1reaction-resultsonsurface-40000events.eps}}
%\caption{ First reaction - energy.}
%\label{odnosnik}
%\end{figure}
%
%\begin{figure}[h]
%\hspace{1.5cm}
%\parbox{0.35\textwidth}{\centerline{\epsfig{file=1reaction-resultsonsurface-40000events.eps,width=0.30\textwidth}}}
%\caption{
%First reaction energy. 
%}
%\end{figure}
%
%\begin{figure}[h]
%          \centerline{
%             \epsfig{file=1reaction-resultsonsurface-40000events.eps, scale=0.5}
%             }
%          \caption{Picture without equations}
%          \label{fig-eg2}
%       \end{figure}
%
%
% ------------------------------------------
%                Acknowledgment 
% -----------------------------------------

\newpage
~ \\
\thispagestyle{empty}

%\fancyhead[RO]{\textbf{\sffamily{{{\thepage}}~}}}
%\fancyhead[LO]{\bf\footnotesize{{\nouppercase{}}}}
%\fancyhead[RE]{\textbf{\sffamily{{{\thepage}}~}}}
%\fancyhead[LE]{\bf\footnotesize{{\nouppercase{}}}}

\newpage
~ \\
\thispagestyle{empty}
\begin{center}
\LARGE{
\textbf{Acknowledgment}
}
\end{center}

\hspace{\parindent}

\indent At first I wish to thank Prof. dr hab. Pawe{\l} Moskal for organizing for me an additional place in  
travel to COSY Summer School in the year 2006. Since this event my real adventure with physic has begun. I also would like to thank you Pawe{\l} 
for great patience in checking this diploma thesis, for giving me a chance to study a wonderfuld world of particle physics, for learning me 
to ask myself why, and in which purpose I do my present job. And for many suggestions in the moments when I didn't know what I should 
do in the next step. And of course for coffees and discussions about many interesting topics. \\
\indent I would like to thank dr Biagio Di Micco for many hours of working together with FLUKA geometry, with reconstruction of particles 
on the 24 modules of the barrel calorimeter. I am greatly indebted to you for sitting in Frascati till late evening and trying to explain me 
where I should look for mistakes, and of course for great patience. \\
\indent I would like to thank dr Anna Ferrari for explaining me Birk effect and details related with machinery of FLUKA Monte Carlo. 
Also I am very grateful to you for checking this thesis and for reassuring when I was very tired. \\
\indent I am very grateful to Dr Antonio Passeri for financial support of my stay in Frascati, for very nice atmosphere during the work and 
for help with numbering of channels in cosmic ray displayer. \\
\indent I wish to thank Dr Catherina Bloise for letting me work in Laboratori Nazionali di Frascati and for financial support of my stays in Frascati. 
Also I am very grateful for nice atmosphere in the work. \\
\indent I would like to thank Dr Federico Nguyen for patience when I was talking with Biagio all the time and disturbed him in his work, and 
for talkig during the lunch about football players and situation in the group during eliminations. \\            
\indent I wish to thank Master of Science Simona Bocchetta for nice atmosphere during the work. \\  
\indent I would like to thank Prof. dr hab. Bogus{\l}aw Kamys for financial support of my stays in J{\"u}lich and in Frascati and for 
teaching me how to understand nucleus complex reactions. \\
\index I am very gratefull to Prof. dr hab. Lucjan Jarczyk for pertinent suggestions and remarks on thursdays meeting in Nuclear Physics Division of 
the Jagiellonian University. \\ 
\indent I ma very grateful to Master of Science Marcin Zieli\'nski for two years of common work in 222 room in Nuclear Physics Division. 
For these two years I had got always great support from your side in all situations in my life. I would like to thank for discussions, coffees 
and wonderful atmosphere during the work in this room. \\ 
\indent I wish to thank Master of Science Wojciech Krzemie\'n for many discussions and help in my professional and private life. Also I am very 
grateful for working with you. \\ 
\indent I would like to thank Dr Marek Jacewicz for discussions and coffees. \\
\indent I thank Master of Science Joanna Przerwa and Master of Science Pawe{\l} Klaja for coffees, cakes and discussions about 
military weapons used during the Second World War. \\
\indent I thank Master of Science Eryk Czerwi\'nski for many explanations related with his work and patience in precise checking of this diploma thesis. \\
\indent I would like to thank all the people from COSY-11 group and working in room 222 for wonderful atmosphere during the work.  
  Especially, 
I have in my mind: Master of Science Dagmara Rozp\c{e}dzik, Master of Science Barbara Rejdych,  
Dr Rafa{\l} Czy{\.z}ykiewicz, Ewelina Czaicka, Jakub Bo{\.z}ek and Micha{\l} Silarski. \\
\indent On my personal side I would like to thank my family: my parents Lucyna and J\'ozef Zdebik and my brother Rafa{\l} for love, patience and support 
during many last years. \\
\indent I thank very much Master of Science Izabela Peru{\.z}y\'nska for being close to me in hard moments, in my life. \\
\indent And at the end I would like to thank all people without whom this thesis would never be accomplished, and with whom I could work with pleasure  
in Krak\'ow, Frascati and J{\"u}lich. \\
%

%
%\newpage
%~ \\
%\thispagestyle{empty}

\newpage
~ \\
\thispagestyle{empty}
\begin{center}
\LARGE{
\textbf{Podzi\c{e}kowania}
}
\end{center}

\hspace{\parindent}
\thispagestyle{empty}

\indent Na pocz\c{a}tku chcia{\l}em podzi\c{e}kowa\'c Prof. dr hab. Paw{\l}owi Moskalowi za zorganizowanie mi dodatkowego miejsca 
na Szkole Letniej COSY Summer School w 2006 roku. Poczynaj\c{a}c od tego wydarzenia rozpocz\c{e}{\l}a si\c{e} moja prawdziwa przygoda z fizyk\c{a}. 
Jak równie{\.z} chcia{\l}em Ci podzi\c{e}kowa\'c Pawle za wielk\c{a} cierpliwo\'s\'c, kt\'ora Ci towarzyszy{\l}a podczas sprawdzania tej pracy, 
za danie szansy 
badania wspania{\l}ego \'swiata cz\c{a}stek elementarnych, za uczenie mnie stawiania sobie celu w swojej bie{\.z}\c{a}cej pracy. Jak r\'ownie{\.z} za wiele 
sugestii w momentach gdy nie wiedzia{\l}em, w~kt\'orym kierunku dalej p\'oj\'s\'c, oraz oczywi\'scie za wsp\'olne kawy i interesuj\c{a}ce dyskusje. \\
\indent Chcia{\l}bym podzi\c{e}kowa\'c doktorowi Biagio Di Micco za wiele godzin wsp\'olnej pracy z geometri\c{a} programu FLUKA, jak 
równie{\.z} z rekonstrukcj\c{a}
cz\c{a}stek na ca{\l}ym "barrel calorimeter". Jestem bardzo wdzi\c{e}czny za siedzenie p\'o\'znymi wieczorami w Instytucie we Frascati 
i t{\l}umaczenia gdzie powinienem szuka\'c b{\l}\c{e}d\'ow, oraz za wielk\c{a} cierpliwo\'s\'c. \\   
\indent Chcia{\l}bym podzi\c{e}kowa\'c Pani doktor Annie Ferrari za wyt{\l}umaczenie mi efektu Birka i detali funkcjonowania program Fluka Monte Carlo. 
Jak r\'ownie{\.z} jestem bardzo wdzi\c{e}czny za sprawdzenie tej pracy i za s{\l}owa otuchy gdy by{\l}em bardzo zm\c{e}czony. \\
\indent Jestem bardzo wdzi\c{e}czny doktorowi Antonio Passeri za wsparcie finansowe moich pobyt\'ow we Frascati, za mi{\l}\c{a} atmosfer\c{e} 
podczas pracy oraz 
za pomoc z numeracj\c{a} kana{\l}\'ow w wy\'swietlaczu promieni kosmicznych. \\   
\indent Chcia{\l}bym podzi\c{e}kowa\'c Pani doktor Catherine'a Bloise za umo{\.z}liwienie mi pracy w Laboratorium Narodowym we Frascati 
 oraz za wsparcie finansowe 
moich pobytów we Frascati. Jak równie{\.z} jestem bardzo wdzi\c{e}czny za mi{\l}\c{a} atmosfer\c{e} w pracy. \\ 
\indent Chcia{\l}bym podzi\c{e}kowa\'c doktorowi Federico Nguyen za cierpliwo\'s\'c w momentach gdy rozmawia{\l}em z Biagio i 
przeszkadza{\l}em mu w jego pracy, 
 jak r\'ownie{\.z} za rozmowy o zawodnikach pi{\l}karskich oraz o sytuacji w grupie podczas eliminacji. \\   
\indent Chcia{\l}bym podzi\c{e}kowa\'c Pani magister Simonie Bocchetta za mi{\l}\c{a} atmosfer\c{e} podczas pracy. \\  
\indent Chcia{\l}bym podzi\c{e}kowa\'c Prof. dr hab. Bogus{\l}awowi Kamysowi za wsparcie finansowe moich pobyt\'ow w J{\"u}lich i Frascati oraz za 
t{\l}umaczenia 
reakcji j\c{a}drowych przez j\c{a}dro z{\l}o{\.z}one. \\
\indent Jestem bardzo wdzi\c{e}czny  Prof. dr. hab. Lucjanowi Jarczykowi za trafne uwagi i sugestie podczas czwartkowych spotka\'n w Zak{\l}adzie Fizyki J\c{a}drowej 
Uniwersytetu Jagiello\'nskiego. \\
\indent Jestem bardzo wdzi\c{e}czny magistrowi Marcinowi Zieli\'nskiemu za dwa lata wsp\'olnej pracy w~pokoju 222 w~Zak{\l}adzie Fizyki J\c{a}drowej. 
Przez te dwa lata mia{\l}em zawsze wielkie wsparcie z~Twojej strony we wszystkich sytuacjach w moim {\.z}yciu. Chcia{\l}em podzi\c{e}kowa\'c za 
dyskusje, kawy 
i wspania{\l}\c{a} atmosfer\c{e} podczas pracy w tym pokoju. \\  
\indent Chcia{\l}bym podzi\c{e}kowa\'c magistrowi Wojciechowi Krzemieniowi za wiele dyskusji i pomoc w~zawodowym i prywatnym {\.z}yciu. 
R\'ownie{\.z} jestem bardzo 
wdzi\c{e}czny za prac\c{e} z Tob\c{a}. \\
\indent Chcia{\l}em podzi\c{e}kowa\'c doktorowi Markowi Jacewiczowi za dyskusje i kawy. \\ 
\indent Chcia{\l}em podzi\c{e}kowa\'c Pani magister Joannie Przerwie i magistrowi Paw{\l}owi Klaji za kawy, ciastka i dyskusje o 
broni wojskowej u{\.z}ywanej podczas Drugiej Wojny \'Swiatowej. \\
\indent Chcia{\l}bym podzi\c{e}kowa\'c magistrowi Erykowi 
Czerwi\'nskiemu za wiele wyja\'snie\'n odno\'snie jego pracy i cierpliwo\'s\'c w dok{\l}adnym sprawdzaniu tej pracy 
magisterskiej. \\
\indent Chcia{\l}bym podzi\c{e}kowa\'c wszystkim cz{\l}onkom grupy COSY-11 i pracuj\c{a}cym w pokoju 222, za wspania{\l}\c{a} atmosfer\c{e} podczas pracy. 
Szczeg\'olnie, mam na my\'sli: magister Dagmar\c{e} Rozp\c{e}dzik, magister Barbar\c{e} Rejdych, doktora Rafa{\l}a Czy{\.z}ykiewicza, 
 Ewelin\c{e} Czaick\c{a}, Jakuba Bo{\.z}ka i Micha{\l}a Silarskiego. \\   
\indent Z mojej prywatnej strony chc\c{e} podzi\c{e}kowa\'c mojej rodzinie: moim rodzicom Lucynie i J\'ozefowi Zdebikom i bratu Rafa{\l}owi, za 
mi{\l}o\'s\'c, 
cierpliwo\'s\'c i wsparcie przez wiele ostatnich lat. \\
\indent Bardzo dzi\c{e}kuj\c{e} magister Izabeli Peru{\.z}y\'nskiej za bycie blisko mnie w trudnych momentach, w~moim {\.z}yciu. \\
\indent I na ko\'ncu chcia{\l}em podzi\c{e}kowa\'c wszystkim ludziom, bez kt\'orych ta praca nigdy by nie powsta{\l}a, i z kt\'orymi praca 
by{\l}a przyjemno\'sci\c{a} w Krakowie, Frascati i J{\"u}lich. \\

\newpage
~ \\
\thispagestyle{empty}
\end{document}